\documentclass[12pt]{article}
\usepackage[letterpaper, margin=1.5in]{geometry}
\usepackage{setspace}
\usepackage[latin1]{inputenc}
\usepackage{amsthm,amsmath,amsfonts} 
\usepackage{amssymb}
\usepackage{mathrsfs}

\usepackage{eurosym}
\usepackage{verbatim} 
\usepackage[english]{babel}
\usepackage[T1]{fontenc}
\usepackage{bbm,bm}

\usepackage{lmodern}
\usepackage{blindtext}

\usepackage{tabularx}
\usepackage[table,xcdraw]{xcolor}
\usepackage{float}

\usepackage{graphicx, floatflt} 
\usepackage[labelformat=simple]{subcaption}

\DeclareCaptionLabelFormat{subcaptionlabel}{\normalfont(\textbf{#2}\normalfont)}
\usepackage{color}
\usepackage[export]{adjustbox}

\usepackage{enumitem}

\usepackage{multirow}
\usepackage{booktabs}
\usepackage{adjustbox}
\usepackage{blindtext}
\usepackage{esint}
\usepackage[title]{appendix}
\theoremstyle{definition}

\graphicspath{{Figures/}}

\usepackage{authblk}

\title{\bf ESG-Valued Portfolio Optimization and Dynamic Asset Pricing}
\author[1,*]{Davide Lauria}
\author[1]{W. Brent Lindquist}
\author[2]{Stefan Mittnik}
\author[1]{Svetlozar T. Rachev}

\affil[1]{\small Department of Mathematics \& Statistics, Texas Tech University, Lubbock TX 79409-1042, U.S.A.
davide.lauria@ttu.edu; zari.rachev@ttu.edu}
\affil[2]{\small Department of Statistics, Ludwig Maximilians Universit\"{a}t M\"{u}nchen,
80799 M\"{u}nchen, DE. and Scalable Capital, 80538 M\"{u}nchen, DE. stefan@scalable.capital}

\affil[*]{Corresponding author, brent.lindquist@ttu.edu}

\begin{document}
\maketitle

\noindent\textbf {Abstract}
ESG ratings provide a quantitative measure for socially responsible investment.
We present a unified framework for incorporating numeric ESG ratings into dynamic pricing theory.
Specifically, we introduce an ESG-valued return that is a linearly constrained transformation of financial return and ESG score.
This leads to a more complex portfolio optimization problem in a space governed by reward, risk and ESG score.
The framework preserves the traditional risk aversion parameter and introduces an ESG affinity parameter.
We apply this framework to develop ESG-valued: portfolio optimization; capital market line; risk measures;
option pricing; and the computation of shadow riskless rates.

\section{Introduction}\label{sec:intro}

Socially responsible investing (SRI)  is a broad approach advocating investment in activities and companies
that produce positive impact on the environment and society.
The definition of SRI is quite general; its practices span a heterogeneous system of policies and scopes 
\cite{Daugaard_2020, Widyawati_2020}.
The heterogeneity within the SRI movement and forces driving differing SRI outlooks have been studied on four
levels: terminological, definitional, strategic and practical \cite{Hochstadter_2015, Sandberg_2008}.
Within the vast realm of SRI, a more focused approach in investment is emerging based upon \textit{ESG ratings},
wherein each company is separately evaluated in three categories: environmental, social and governance. 
Evaluation is performed on the basis of factors within each category.
The total ESG score of a company is then determined by a combination of three category scores,
providing a ``sustainability ordering'' between firms.

ESG investing is experiencing fast growth as institutional investors, such as mutual and pension funds, now provide SRI-related financial products.
(See for instance the recent report from the Organization for Economic Co-operation and Development on ESG investing \cite{OECD_ESG_2020},
and references therein.)
Different approaches have been taken by asset managers to integrate ESG factors into investment decisions \cite{Zadeh_2018, Duuren_2016}.
These approaches generally fall into three strategy groups.
The first strategy, an \textit{exclusionary} approach, ``filters out certain companies based upon products or certain corporate behavior''
\cite{Berry_2013}.
The second strategy, the \textit{inclusionary} approach, ``involves adjusting the weights of an investment in a firm according to whether
its behavior is more or less socially responsible'' (ibid).
There is a degree of subjectivity in both approaches.
In the third, \textit{empirical}-based strategy,  asset managers seek for more exposure in companies with higher ESG scores,
thus avoiding the subjectivity of the exclusionary and inclusionary approaches through the use of an empirical measure that
does not exclude participation in less virtuous entities.
According to the work of Berry at al. \cite{Berry_2013},
most socially responsible financial products focus on the exclusionary approach.
In contrast, they found that investors prefer to reward firms who display overall positive social policies,
rather than exclude the less responsible through negative screening.
A similar conclusion was reached by Amel-Zadeh and Serafeim \cite{Zadeh_2018}.
These results could explain the slow reception of SRI, and drive new approaches to implement the third class of strategies.

Research has highlighted two main reasons behind the push for ESG products:
ethical beliefs and improved financial performance \cite{Hartzmark_2019}.
Investigation of the connection between  stakeholder preferences (ethical beliefs) and responsible investing includes
the work of B\'{e}nabou and Tirole \cite{Benabou_2010}, Kr\"{u}ger \cite{Kruger_2015}, Lins et al. \cite{Lins_2017},
Liang and Renneboog \cite{Liang_2017}, and Starks at al. \cite{Starks_2017}.
Kr\"{u}ger investigated how stock markets react to positive and negative events connected with
corporate social responsibility, finding a strong reaction to negative events.
Starks at al. found that long-term investors are more attracted toward higher rated ESG companies,
``tilting their portfolios towards firms with high-ESG profiles''.

The literature has been more focused on the performance of ESG investing compared to conventional return-centric
investing.
An extensive literature review of over 2000 articles \cite{Friede_2015} concludes that 90\% of the studies
``find a nonnegative ESG [to] corporate financial performance relation''
and that this relationship appears to be stable over time.
Brooks and Oikonomou \cite{Brooks_2018} obtain similar conclusions after reviewing a large set
of studies focusing on the connection between ESG disclosure and performance and their effects on firm value.
They found ``a positive and statistically significant but economically modest link'' between
corporate social performance and firm financial perfomance.
They also found a strong negative relationship between financial risk, whether systematic or idiosyncratic,
and corporate social performance. 
Bello \cite{Bello_2005} performed a comparison between socially responsible and traditional mutual funds
to attempt to quantify the impact of ethical screening on portfolio diversification and overall performance.
He reported that the two classes of mutual funds do not differ statistically in the characteristics of assets they hold,
in diversification, or in overall performance.
In constrast, a similar study \cite{Abate_2021} on European mutual funds reports a superior ``efficiency'' of funds that
focus on high ESG rated securities.
Giese et al. \cite{Giese_2019_A} studied ESG causality on (as opposed to ``correlation with'') financial
performance by considering transmission channels of systematic and idiosyncratic risk within a standard discounted
cash flow model, through which ESG ratings data (specifically MSCI ESG ratings data) impact valuation and
performance.
An interesting outcome of their study relates to the intensity and longevity of ESG signals (ESG rating changes)
compared with factors such as momentum or volatility.

A component of the literature has focused on the effect of specific components of the three ESG categories.
In the area of governance,
Pan et al.  \cite{Pan_2022} found a negative relationship between returns and  high pay ratios.
They claim that, in 2018, inequality-averse investors decreased the relative amount invested in
stocks with a high pay ratio.
Krueger at al. \cite{Krueger_2020} surveyed institutional investors on whether, and how,
they integrate climate risk into their financial decisions.
The majority of those interviewed consider climate change to have relevant implication for financial performance
and have consequently taken action in their practices to accommodate such risk.
G\"{o}rgen et al. \cite{Gorgen_2020} have investigated the impact of carbon risk on equity prices of ``brown''
and ``green'' firms.
Their research uncovered two opposing effects: while brown firms display higher average returns,
companies that become less green show a drop in returns.

Other relationships between ESG ratings and the financial system have been studied.
Examples include the effect of ESG scores on credit ratings \cite{Bannier_2022, Fernandez_2015}
and on the fixed income market \cite{Inderst_2018, Jang_2020, Kanamura_2021}.
The analysis of ESG factors associated with a fixed income security,
especially one issued by a government,
involves dealing with complex methodological and technical implications \cite{Inderst_2018}.
The literature identifies two main approaches to translating ESG ratings into the fixed income market. 
The first is to issue products specifically designed to fund projects having positive environmental
or social impact.
Examples of securities issued with the specific purpose of financing projects related to
environmental and climate change objectives are green and blue bonds\footnote{
	The first green bond was issued by the European Investment Bank in 2007.
	\textit{https://www.eib.org/en/investor-relations/cab/index.htm}} \cite{Dorfleitner_2021,Ehlers_2017}.
Ehlers et al. \cite{Ehlers_2017} have documented lower credit spreads\footnote{
	The spread of yield at issuance over the yield curve of an appropriate treasury security
	of the same maturity at issuance date.}
on borrowing green bonds compared to that of a traditional counterpart issued by the same institutions.
Polbennikov at al. \cite{Polbennikov_2016} found that ESG corporate bond portfolios have experienced
a modest positive incremental return.
The second approach has a broader view.
It consists of assigning the ESG rating of the company to their fixed income products, imposing a tighter connection
between the global business activity of the issuer and the bond rating \cite{Alessandrini_2021, PRI_2013}.
A related issue is the ESG valuation of general fixed income securities backed by governments or municipalities \cite{Badia_2019}.

A major concern of ESG ratings is the variability of scores among different rating agencies,
which raises the issues of reliability and potential bias selection; see for instance 
\cite{Berg_2019, Billio_2021, Chatterji_2016, Christensen_2021, Gibson_2021}, and references therein.
Berg et al. \cite{Berg_2019} analyzed ESG scores from  six large providers.
They decomposed the spread among scores into three main sources:
\begin{itemize}[noitemsep,topsep=0pt]
\item[i)] scope (56\%), defined by the range of company attributes that the score intends to measure;
\item[ii)] measurement (38\%), the methodology used to compute a particular attribute; and
\item[iii)] weights (6\%), how the attribute values are combined to form a final score;
\end{itemize}
and were able to attribute the percentage weight (quoted above) to each source.
The main source of variability found among ESG scoring systems was the selection of attributes (the scope).
The authors warn of possible bias effects involved in the assignment of ESG scores as well as in research attempting
to study the link between ESG scores and financial performance.
For example, a ``rater effect'' was discovered under which
``a rater's overall view of a firm influences the measurement of specific categories''.
Bilio et al. \cite{Billio_2021} found substantially low agreement in the assignment of ESG scores by four rating agencies.
They investigated the impact of this variability on financial performance by comparing the Jensen-alpha
for two equi-weighted portfolios:
one formed by those companies that were rated by all four ESG rating agencies and
the other composed of companies that were rated by none of the agencies.
They found no evidence for a difference in financial performance between these two portfolios.

The construction of optimal portfolios is one corner-stone of modern finance theory.
The traditional approach is based on two measures: financial reward and financial risk.  
The reward measure is usually represented by the expected value of future portfolio returns.
Several measures have been proposed to quantify the risk related to the unpredictable component
of the variability of portfolio returns;
the most common of these are: the standard deviation \cite{Markowitz_1952},
the absolute deviation \cite{Konno_1991},
the value at risk (VaR) \cite{Lwin_2017},
and the conditional value at risk (CVaR) \cite{Rockafellar_2000}.
A common optimization approach consists of minimizing the reward-risk trade-off on the basis of a
statistical model for future asset returns, see for instance \cite{Kolm_2014} and \cite{Mansini_2007}.
The idea of incorporating ESG information or more general SRI aspects in this optimization problem
was first pursued in the framework of negative screening.
An example is the portfolio strategy approach proposed by Geczy and Guerard \cite{Geczy_2021},
which combined negative screening with a factor model and mean-variance optimization.
 
A more complex set of approaches have added ESG information directly to the portfolio selection
and/or optimization problem.
Bilbao at al.  \cite{Bilbao_2013} introduced a two step optimization model.
In the first step, an efficient frontier is computed using expected value of wealth as the
reward measure and either mean-variance or CVaR as the risk measure.
An investor social behavior satisfaction index is then computed on the efficient frontier.
In the second step, the portfolio showing the best financial and social behavior is selected.
The approach taken by Hirschberger et al. \cite{Hirschberger_2013} and Utz et al. \cite{Utz_2015}
involves portfolio selection in which sustainability is modeled, after risk and return, as a third selection
criterion.
Gasser at al. \cite{Gasser_2017} have incorporated an ESG score as a linear term in the
classical mean-variance optimization.
Chen at al. \cite{Chen_2021} proposed a three step approach:
the first step applies data envelopment analysis to the scores in each of the three ESG categories
to obtain a more informative total  score;
the second step creates a restricted investment universe using financial attributes and the ESG scores obtained
from the first step;
the third step applies standard mean-variance optimization to the restricted portfolio. 
Pedersen et al. \cite{Pedersen_2021} integrated the relative ESG score of each asset into the
objective function and studied the differential impact of environmental and governance scores
on portfolio performance.
Schmidt \cite{Schmidt_2020} added the portfolio ESG value to the objective function.
Asset weights in a long-only portfolio were then optimized in terms of return, mean variance and ESG value.
 In Cesarone et al.  \cite{Cesarone_2022}, the ESG dimension is added as a hard constraint
 under mean-variance optimization with long-only positions solved numerically under data-driven scenarios.
Their analysis is conducted on four  investment universes:
DJIA, Euro Stocxx 50, FTSE100, NASDAQ100 and S\&P500.
They find that optimal portfolios with highest ESG constraints gave better results only within the
DJIA and S\&P500 universes.

The discussion above underscores the fact that SRI investing and ESG ratings are ``here to stay'', and that
societal pressures will require that the optimization of financial returns be considered within a larger SRI framework.
We note that ESG scores are an ``SRI measure'' that, to first-order approximation,
is independent of the traditional return and risk-measures used in portfolio optimization.\footnote{
	Literature reviews such as those done by Friede et al. \cite{Friede_2015} and
	Brooks and Oikonomou \cite{Brooks_2018} have revealed less-than-conclusive results regarding
	the strength of correlation between ESG scores and financial performance.}
It is our goal, in this and further work, to produce a consistent, ESG-based, asset pricing framework upon which pricing,
optimization, and risk management practices can be based.
Our objective is to add the ESG score as a third dimension,
while preserving the main machinery developed in dynamic asset pricing theory
(see for instance \cite{Duffie_2001}).
As a result, we will be able to find optimal reward-risk-ESG asset allocations,
but also be able to hedge the ESG-valued risk by pricing any contingent claims traded in financial markets.
Our work has commonality with approaches discussed above in that we propose a linearly constrained
combination of asset returns and ESG scores to produce \textit{ESG-valued} returns.
Our work then builds consistently upon this ESG-valued framework,
extending its development into principle components of modern finance theory.

In Section \ref{sec:ESG_rtn} we present this linearly constrained transformation;
in  Section \ref{sec:ESG_PO} we develop its application to portfolio optimization,
and  analyze empirically the ESG-valued efficient frontier using a portfolio of stocks
from the Dow Jones Industrial Average.
In Section \ref{sec:ESG_PM} we consider ESG-valued measures appropriate for evaluating the
performance of ESG-valued optimizations.
In Section \ref{sec:ESG_TP}, we extend the results by developing an ESG-valued risk-free rate to
enable the identification of tangent portfolios to the ESG-valued efficient frontier,
thus formulating an ESG-valued capital asset pricing model (CAPM).
The extension to an ESG-valued model for valuing European contingent claims is covered in Section \ref{sec:ESG_OP}.
A numerical study is also presented.
In Section \ref{sec:ESG_SR} we investigate the effect of the ESG-adjustment on the
computation of a riskless rate in markets having no riskless asset; the so-called shadow
riskless rate as defined by Rachev et al. \cite{Rachev_2017}.
Final discussion is presented in Section \ref{sec:Disc}.

\section{ESG-Valued Returns} \label{sec:ESG_rtn}

We define the ESG-valued return as an affine combination of the ESG score for any asset and its return.
Let  $\text{ESG}_{i,t}$  be the ESG score for asset $i$ on day $t$,
and  $\varsigma_{i,t}$ denote the value of $\text{ESG}_{i,t}$ appropriately normalized to the range $[-1,1]$.\footnote{
	This normalization is discussed in Section \ref{sec:ESG_scale}.}
Since ESG scores are usually provided as positive values,
the normalization to the range $[-1,1]$ produces a positive (negative) shift in $\varsigma_{i,t}$ for those
assets with ESG scores above (below) some determined ``average'' value of the original scale.
For any $\lambda \in [0,1]$, we then define the ESG-valued return
\begin{equation}
\zeta_{ i, t }( \lambda ) = \lambda \frac{ \varsigma_{i,t} }{ c } + \left( 1-\lambda \right) r_{i,t}. \label{eq:esg_transform}
\end{equation}
The constant $c \in R$ is needed to insure that the ranges of $ \varsigma_{i,t} / c$ and  $r_{i,t}$ are comparable.
The value of $c$ will depend on the return time period;
for the daily returns considered here, we used the value $c = 255$
to approximate the number of trading days in a year.
Note that $\lambda$ is chosen by the investor;
for our purposes, we consider it to be independent of $i$ and $t$.
The ESG-valued return defined in (\ref{eq:esg_transform}) assigns more weight to the ESG score
of each asset as $\lambda \uparrow 1$.
In the limit $\lambda = 1$, the investor abandons any concern regarding the financial risk-reward trade-off
and concentrates solely on the ESG risk-reward impact on the portfolio.
Conversely, when $\lambda = 0$, the investor abandons any interest in ESG.

To avoid confusion, the phrases ``ESG-valued return'' and ``adjusted return''
will always refer to $\zeta_{ i, t }( \lambda )$.
The word ``return'' used alone will refer to the financial return $r_{i,t}$.

\subsection{Scaling of ESG Values} \label{sec:ESG_scale}

To use the ESG-valued return \eqref{eq:esg_transform}, the investor has to determine a scoring system that
reflects desired ESG criteria.
As discussed in the Introduction,
preference might, for example, be given to considering environmental implications over social aspects;
the choice may be driven by ethical values or by consideration of profit maximization.\footnote{
	Some literature has focused on the higher returns provided by portfolios based on certain ESG criteria.
	Pedersen at al. \cite{Pedersen_2021} found that governance factors are more linked to financial performance
	than those related to environmental impact.}
Choice of criteria will direct choice of scoring methodology and perhaps scoring agency.

Given a user-determined decision on the scoring system,
we first consider the issue related to the time scale $\Delta t$ embedded in \eqref{eq:esg_transform}.
Currently, ESG scores are provided on an annual basis,\footnote{
	Although some providers distribute monthly scores, these data are usually expensive
	and are not provided on a regular basis.}
whereas the time scale (hence, the return data) required in \eqref{eq:esg_transform} may vary from  monthly
to intra-day.
Proceding with our choice of $\Delta t = 1$ day, we discuss the issue of assigning values to $\text{ESG}_{i,t}$
on a daily basis.
A straightforward approach is to utilize the last published ESG score for each date
$\tau \in \left\{ t-T+1,\dots,t\right\}$.
While this takes advantage of the most recent ESG information,
the ESG score for asset $i$ will remain constant over the period of a year.
Under this approach, $\varsigma_{i,t}$ in \eqref{eq:esg_transform} is the normalized value of
$\text{ESG}_{i,t_{\text{y}}}$ for asset $i$ at the last yearly issue date, $t_{\text{y}}$, of new ESG values.
A clear drawback to this approach is the introduction of a (potentially large) jump discontinuity in the value of
$\varsigma_{i,t}$ as the ESG year changes,
with a consequent discontinuity in the value of the ESG-valued return.
With ESG scores only available at yearly time points, a fill-in procedure to produce a smoothly varying
daily ESG score would require both interpolation and extrapolation methods.
An alternate approach would be to model the ESG score as a random variable.
This approach, which should be more appropriate as the ESG score is affected by uncertainty,
is challenging to implement given the current, relatively low frequency of ESG data,
especially when compared to that of financial returns.

Taking the above considerations into account,
we invoke the philosophy of using the most current data available
and adopt the following procedure in this paper.
In most financial computations (e.g. portfolio optimization, option pricing),
a ``look-back'' window of $T$ days is required in order to compute statistics summarizing the
random behavior of the assets under consideration.
\textit{We utililze the most current ESG value available for each stock during the look-back period,
regardless of the size of $T$.}
For example, to obtain optimized portfolio weights for 06/06/2019 using a window of 510 days
(5/25/2017 through 6/5/2019),
the most recent ESG scores available from Refinitiv were from 12/31/2018.
These ESG scores were used in \eqref{eq:esg_transform} to compute adjusted returns for each day
in the look-back period.

\begin{figure} [ht]
	\centering
	\includegraphics[width = 0.33\textwidth]{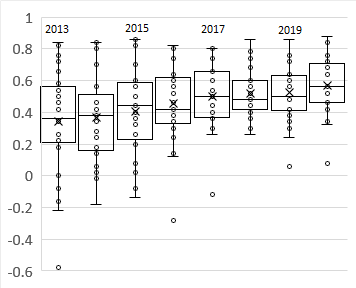}
	\caption{Box-whisker summaries of normalized Refinitiv data for the DJIA data listed in Table~\ref{tab:Ref_DJ_ESG}.}
	\label{fig:refin_data}
\end{figure}

We consider next the issue of normalizing agency-provided ESG scores to the interval $[-1,1]$.
While a linear normalization appears immediate and obvious, this is based upon inherent assumptions of linearity within
the agency's methodology.
To illustrate this issue quantitatively we consider the ESG scores from Refinitiv for 29 of the stocks in the
Dow Jones Industrial Average (DJIA).
These are shown in Table~\ref{tab:Ref_DJ_ESG} for the years 2013 through 2020.
Refinitiv ESG scores are based on a $[0,100]$ scale.\footnote{
	Environmental, Social and Governance Scores from Refinitiv; Feb. 2021.
	https://www.refinitiv.com/content/dam/marketing/en\_us/documents/methodology/refinitiv-esg-scores-methodology.pdf}
We applied the obvious linear transformation to normalize these to $[-1,1]$.
Box-whisker summaries of the resultant transformed scores are shown, by year, in Fig.~\ref{fig:refin_data}.
Very few of the stocks have scaled ESG values that are negative, implying very few (and none for 2018 through 2020) stocks have
a Refinitiv score below 50.
The ESG scores of the lower rated companies improved fairly rapidly over the period 2013 through 2017.
Adjusting the ESG normalization to reflect a yearly change in the median ESG value (which might,
for example, be interpreted as the appropriate percentile for separating ``good''
from ``bad'' ESG ratings) would introduce hidden variability into \eqref{fig:refin_data}.
Addressing this issue in further depth remains outside of the intended scope of this paper;
therefore we proceed with the simple linear transformation to normalize the provided Refinitiv scores from $[0,100]$ to $[-1,1]$.

\section{ESG-Valued Portfolio Optimization}\label{sec:ESG_PO}
We apply the ESG-valued return model to portfolio optimization.
We consider a quite general framework; at the end of day\footnote{
	We develop our formalism for daily returns.
	Reformulation for other time intervals is straightforward,
	assuming appropriate time-varying values for the ESG scores can be found.}
$t$ an investor has to decide how to reallocate
the value $W_{t}$ of a portfolio among a universe ${\cal I}$ of $I$ assets by selecting the fraction (the weight) $\theta_{i}$
to be invested in asset $i= 1, \dots, I$. 
The investor observes the series of $T$ previous asset returns
$\{ r_{i,\tau};\  \tau=t-T+1, \dots, t;\  i=1,\dots,I \}$ and,
based upon some discrete-state, parametric or non-parametric, statistical model,
obtains a series of $S$ one-period-ahead returns $\{ \hat{r}^{s}_{i,t+1};\  s = 1, \dots, S;\  i =1,\dots,I\}$.
Under these assumptions, given  the vector of  portfolio weights
$\theta=\left[ \theta_{1}, \dots, \theta_{I} \right]'$,
the simulated return of the portfolio between $t$ and $t+1$ is defined by the set of $S$ scenarios
\begin{equation}
	\hat{R}_{t+1} =
	\left \{ \hat{R}^s_{t+1} = \sum_{i=1}^{I}\theta_{i}\hat{r}^{s}_{i,t+1};\  s=1,\dots,S \right \}. 		
	\label{eq:RR_const_1}
\end{equation}
The traditional approach is to assign weights by optimizing the trade-off between the financial risk and
financial reward of the portfolio.
The reward is usually measured by the expected value of the future return $\mathbb{E}[ \hat{R}_{t+1}]$,
whereas there are many choices for the risk measure
$\mathbb{V}[ \hat{R}_{t+1}]$.\footnote{
	Artzner at al. \cite{Artzner_1999} proposed an axiomatic approach to define the minimum set of
	propertie that $\mathbb{V}[ \cdot ]$ should satisfy to be a coherent measure of risk.
	Frittelli at al. \cite{Frittelli_2002} have specified requirements to define a convex risk measure.
}
Once the risk measure $\mathbb{V}[ \cdot ]$ has been selected,
the optimal portfolio can be obtained from the following minimization problem,
\begin{equation}
\min_{\theta} \left\{-\alpha \mathbb{E}[ \hat{R}_{t+1}] + (1-\alpha) \mathbb{V}[ \hat{R}_{t+1} ] \right\},
\label{eq:RR_objf}
\end{equation}
subject to the constraints
\begin{align}
&\sum_{i=1}^{I} \theta_{i} = 1,  \label{eq:RR_const_2} \\
&\sum_{i=1}^{I} |\theta_{i} - \bar{\theta}_{i}| \le \gamma,  \label{eq:RR_const_3} \\
&\theta_{i} \ge 0;\  i =1,\dots,I, \label{eq:RR_const_4}
\end{align}
for some parameter $\alpha \in [0,1]$ that represents the risk-aversion attitude of the investor.
Constraint (\ref{eq:RR_const_3}) forces the portfolio turnover to be less than the pre-specified parameter $\gamma$;
here $\bar{\theta}_{i}, i=1,\dots,I$ are the portfolio weights at time $t$.
Constraint (\ref{eq:RR_const_4}) avoids short-selling and can be relaxed when the investor wishes to enter into
such positions.

\subsection{ESG Efficient Frontier} \label{sec:ESG_EF}

The ensemble of ESG-valued return scenarios is
\begin{equation}
	\hat{Z}_{t+1} = \left \{ \hat{Z}^s_{t+1} = \sum_{i=1}^{I}\theta_{i}\hat{\zeta}^{s}_{i,t+1};\  
	s=1,\dots,S \right \},
	\label{eq:ESG_scenarios}
\end{equation}
where
\begin{equation}
	\left\{ \hat{\zeta}^s_{ i, t +1} ( \lambda )\right\}_{s=1}^{S}
	  = \lambda \frac{ \varsigma_{i,t}}{c} + \left( 1-\lambda \right)\left\{ \hat{r}^s_{i,t+1} \right\}_{s=1}^{S}.
	\label{eq:esg_const_transform}
\end{equation}
We consider the portfolio optimization problem (\ref{eq:RR_objf})-(\ref{eq:RR_const_4})
when $\hat{R}_{t+1}$ is replaced by $\hat{Z}_{t+1}$.
We illustrate ESG-valued optimization by application to a specific portfolio consisting of 29 of the 30 stocks\footnote{
     Due to its shorter span of existence, stock from DOW Inc is not included in the portfolio.}
comprising the DJIA.
We use the ESG scores in Table~\ref{tab:Ref_DJ_ESG} provided by Refinitiv for each asset over the period 2013-2020.
We employ the scaling methodology discussed in Section~\ref{sec:ESG_scale} for daily ESG scores.

To generate the ensemble of values $\left\{ \hat{r}^s_{i,t+1}; s = 1, \dots, S \right\}$ in \eqref{eq:esg_const_transform},
we applied an ARMA($p_i, q_i$)-GARCH(1,1) model to the set of historical returns $\{ r_{i,\tau}; \tau=t-T+1, \dots, t; i = 1, \dots, I\}$.
We employed the R package \textit{rugarch} \cite{Ghalanos_2012} to perform the ARMA-GARCH fits.
The function \textit{autoarfima} was used to fit the ARMA parameters $p_i$ and $q_i$
over the ranges $0 \le p_i \le 2$,  $0 \le q_i \le 2$ using the Bayesian information criterion.
Separate fits were performed for each stock.
For each asset, a GARCH(1,1) fit was then performed on the residuals of the best-fit ARMA model using the function \textit{ugarchfit}
and the assumption that the innovations are normal inverse Gaussian (NIG) \cite{BN77,Barndorff_1997}.
In the case in which \textit{ugarchfit} failed to converge,
the autocorrelation $\mathbb{E} [ r_{i,\tau_1}^2,  r_{i,\tau_2}^2 ], \tau_1 \ne \tau_2$
was computed to check for persistence.
If no persistence was found,
the GARCH(1,1) component was dropped in preference of a pure ARMA model.\footnote{
	Persistence was found to be both stock and time dependent;
	even for the same stock, the presence of persistence could change over time.}
The $I \times T$ set of innovations resulting from the ensemble of ARMA($p_i, q_i$)-GARCH(1,1) fits was then
fit to a multidimensional NIG model using the R package \textit{ghyp} \cite{ghyp_2013}.
From the parameterized, multidimensional NIG model, $I \times S$ innovations were randomly generated.
The generated innovations for asset $i$ were then fed into the ARMA($p_i, q_i$)-GARCH(1,1) fit
to generate $S$ sample returns for day $t+1$ for asset $i$.
The choice of the NIG distribution family for the innovations was motivated by its flexibility in modeling potentially
different heavy-tailed behavior for each assset,
while preserving asset return processes as semimartingales, a property that is necessary to perform option pricing.
Since NIG distributions are closed under affine transformations, the ESG-valued returns (\ref{eq:esg_const_transform})
remain within the same distribution class as the unadjusted returns, again preserving the semimartingale property.

Two choices were used for the risk metric  $\mathbb{V}[ \hat{R}_{t+1}]$ in \eqref{eq:RR_objf},
mean-variance (MV) and mean-CVaR${}_\beta$ (mCVaR${}_\beta$) with $\beta = 0.95$ and $0.99$.
Details on solving the minimization problem \eqref{eq:RR_objf}-\eqref{eq:RR_const_4} using each of
these measures are provided in Appendix \ref{sec:App_B}.
Let $\theta^* = [ \theta^*_1 , \dots, \theta^*_I ]$ denote the vector of optimal weights for any
of these solutions.\footnote{
	For brevity we omit labeling the $\lambda$ and $\alpha$ dependence of $\theta^*$.}
We can then define the optimal portfolio return and ESG-valued return,
\begin{align}
	\hat{R}^*_{ t +1} (\lambda) &=  \sum_{i=1}^I \theta_i^* r_{i,t+1}, \label{eq:PortRet} \\
	\hat{Z}^*_{ t +1} (\lambda) &=  \sum_{i=1}^I \theta_i^* \zeta_{i,t+1} ( \lambda ),
	\label{eq:esg_PortRet}
\end{align}
where $r_{i,t+1}$ are the realized asset returns on the close of trading day $t+1$,
from which the ESG-valued returns $\zeta_{i,t+1}(\lambda)$ are computed.
The portfolio price is computed in the usual manner,
\begin{equation}
	P_t(\lambda) = P_0 e^{ R_{\text{cum}}(t;\lambda) };\quad  
	R_{\text{cum}}(t;\lambda) = \sum_{\tau=1}^t \hat{R}^*_\tau(\lambda).
	\label{eq:port_value}
\end{equation}
The ESG-valued numeraire for the portfolio is similarly defined as
\begin{equation}
	P^{(Z)}_t (\lambda) = P^{(Z)}_0 e^{ Z_{\text{cum}}(t;\lambda) };\quad  
	Z_{\text{cum}}(t;\lambda) = \sum_{\tau=1}^t \hat{Z}^*_\tau(\lambda).
	\label{eq:esg_value}
\end{equation}
We assign the initial adjusted value of this numeraire to be $P^{(Z)}_0 = 1$.
For simplicity we refer to $P^{(Z)}_t (\lambda)$ as the ESG-valued price.
For $\lambda = 0$, the portfolio price \eqref{eq:port_value} time series is just a $P_0 / P^{(Z)}_0$ scaled version
of the ESG-valued price \eqref{eq:esg_value}.
Any significant differences only occur for $\lambda > 0$.
The normalized ESG score for the optimized portfolio is
\begin{equation}
	\varsigma^*_{t+1}=\sum_{i=1}^I \theta_i^* \varsigma_{i,t+1}. \label{eq:port_varsig}
\end{equation}
As we have invoked a linear normalization $\varsigma_{i,t} = \mathcal{L}(\text{ESG}_i)$, this is equivalent to
\begin{equation}
	\text{ESG}^*_{t+1}=\sum_{i=1}^I \theta_i^* \text{ESG}_{i,t+1}.\label{eq:port_ESG}
\end{equation}

\begin{figure}[ht]
  \centering
  	\subcaptionbox{}{\includegraphics[width=0.34\textwidth]{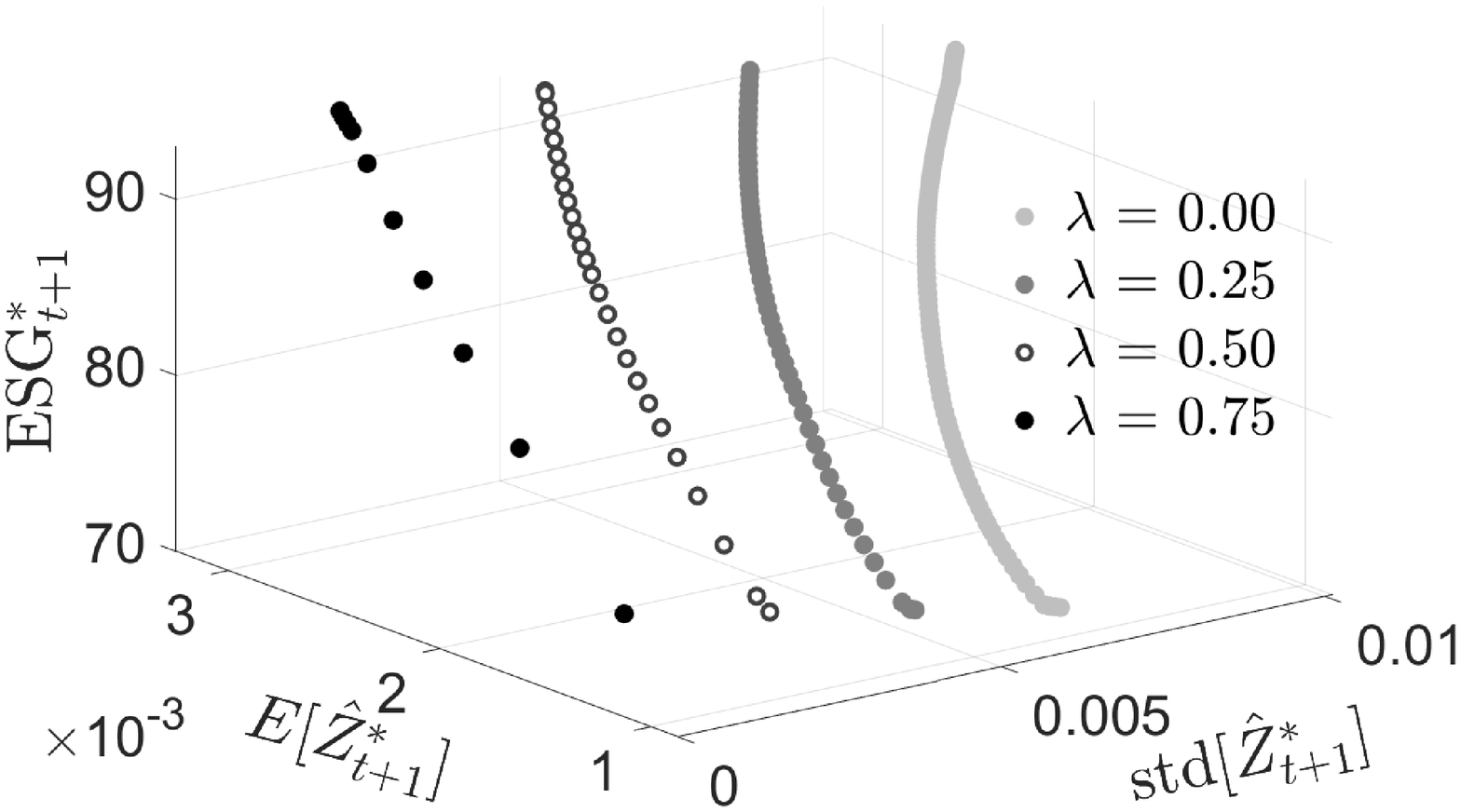}}\hspace{0em}%
	\subcaptionbox{}{\includegraphics[width=0.32\textwidth]{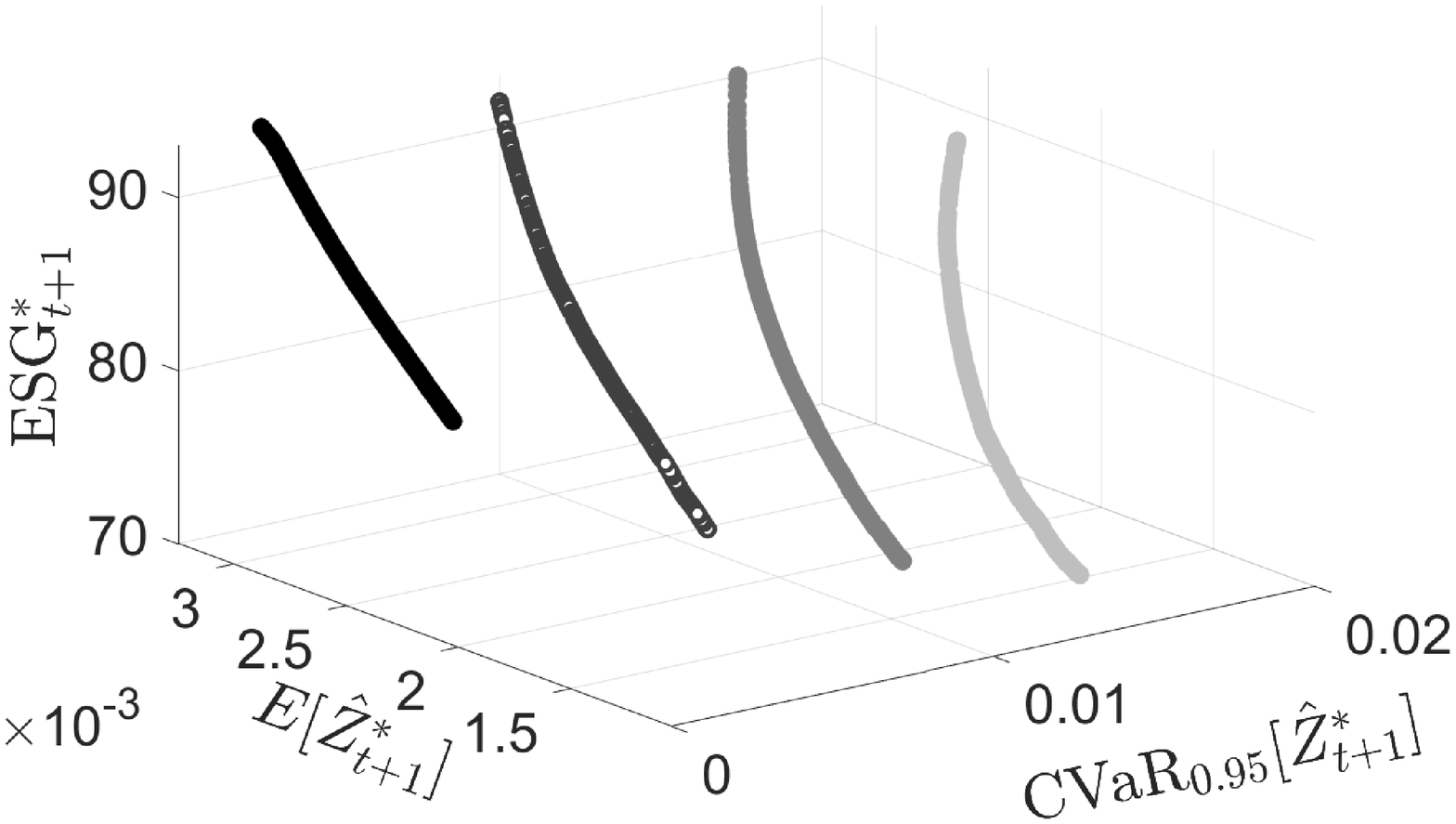}}\hspace{.2em}%
	\subcaptionbox{}{\includegraphics[width=0.32\textwidth]{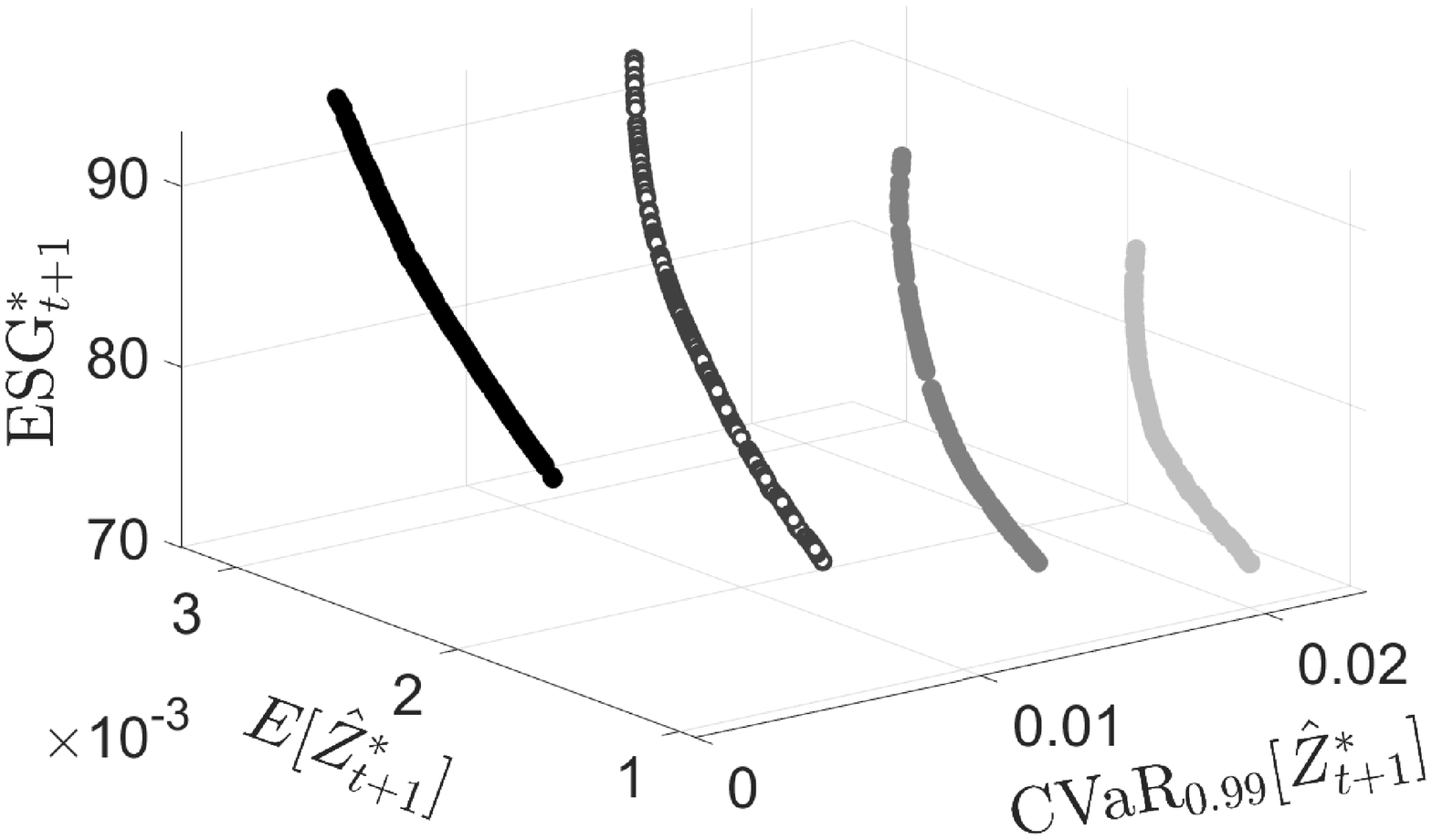}}\hspace{0em}%
	\vspace{.2em}
	\subcaptionbox{}{\includegraphics[width=0.33\textwidth]{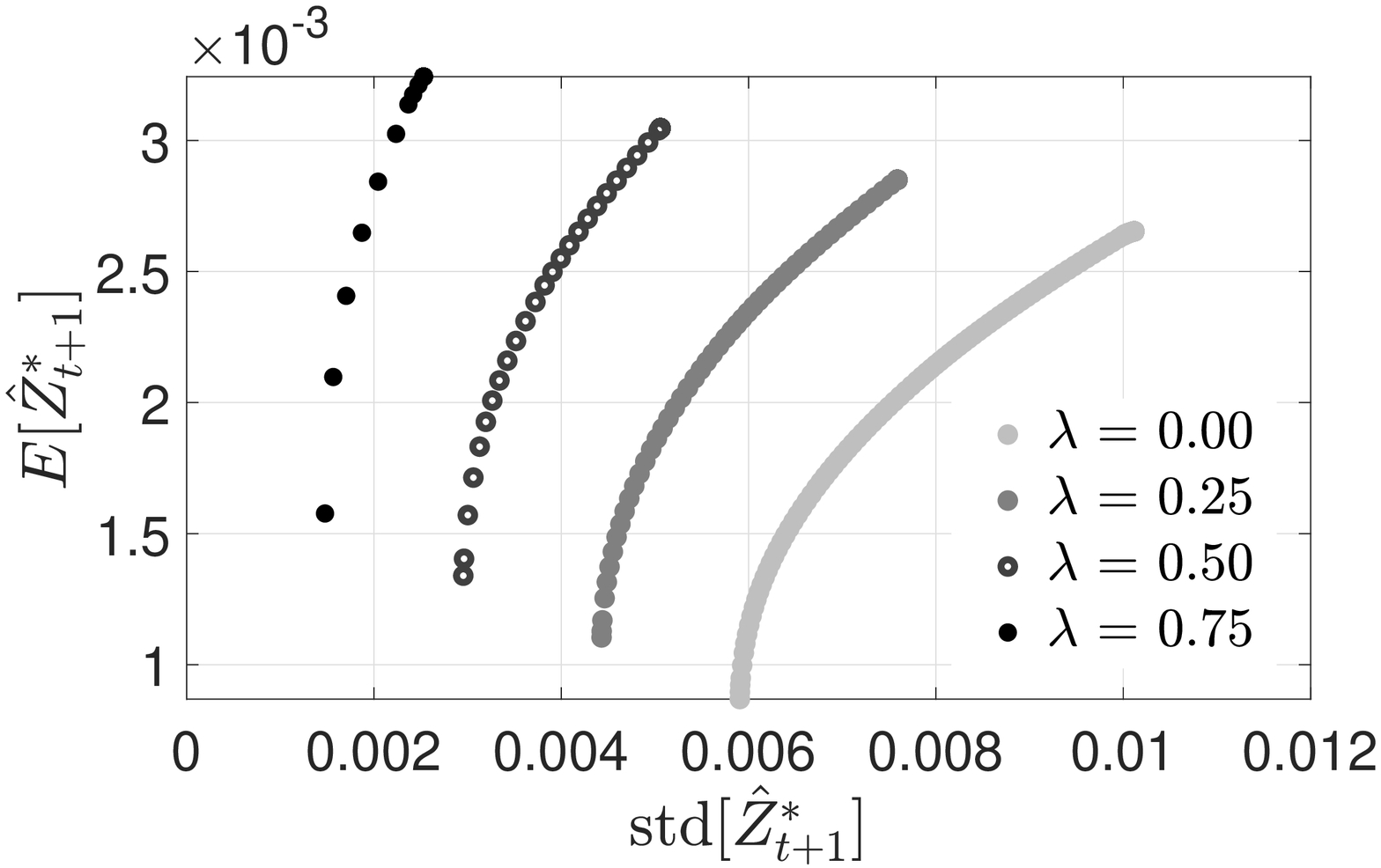}}\hspace{0em}%
	\subcaptionbox{}{\includegraphics[width=0.33\textwidth]{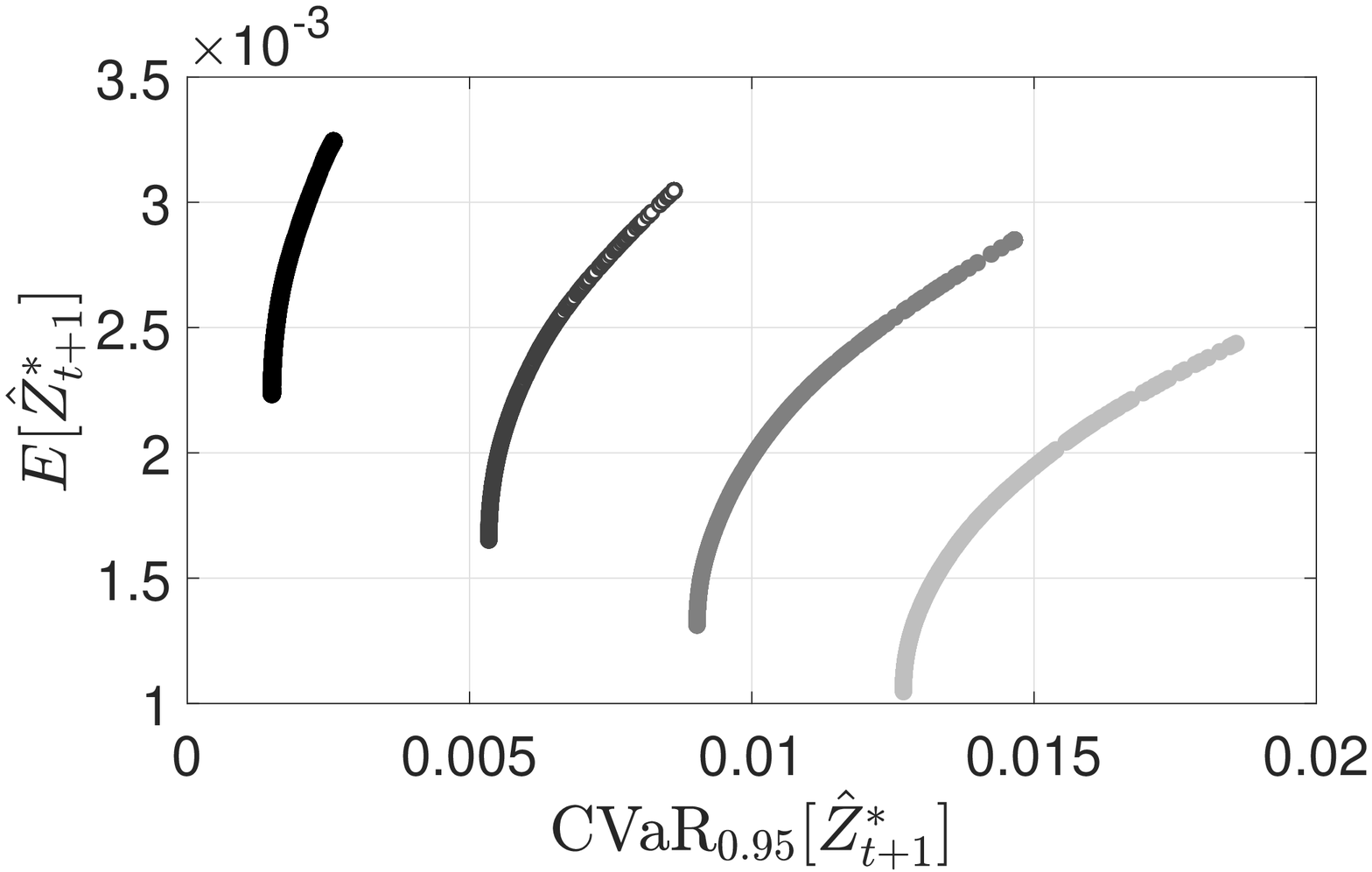}}\hspace{0em}%
	\subcaptionbox{}{\includegraphics[width=0.33\textwidth]{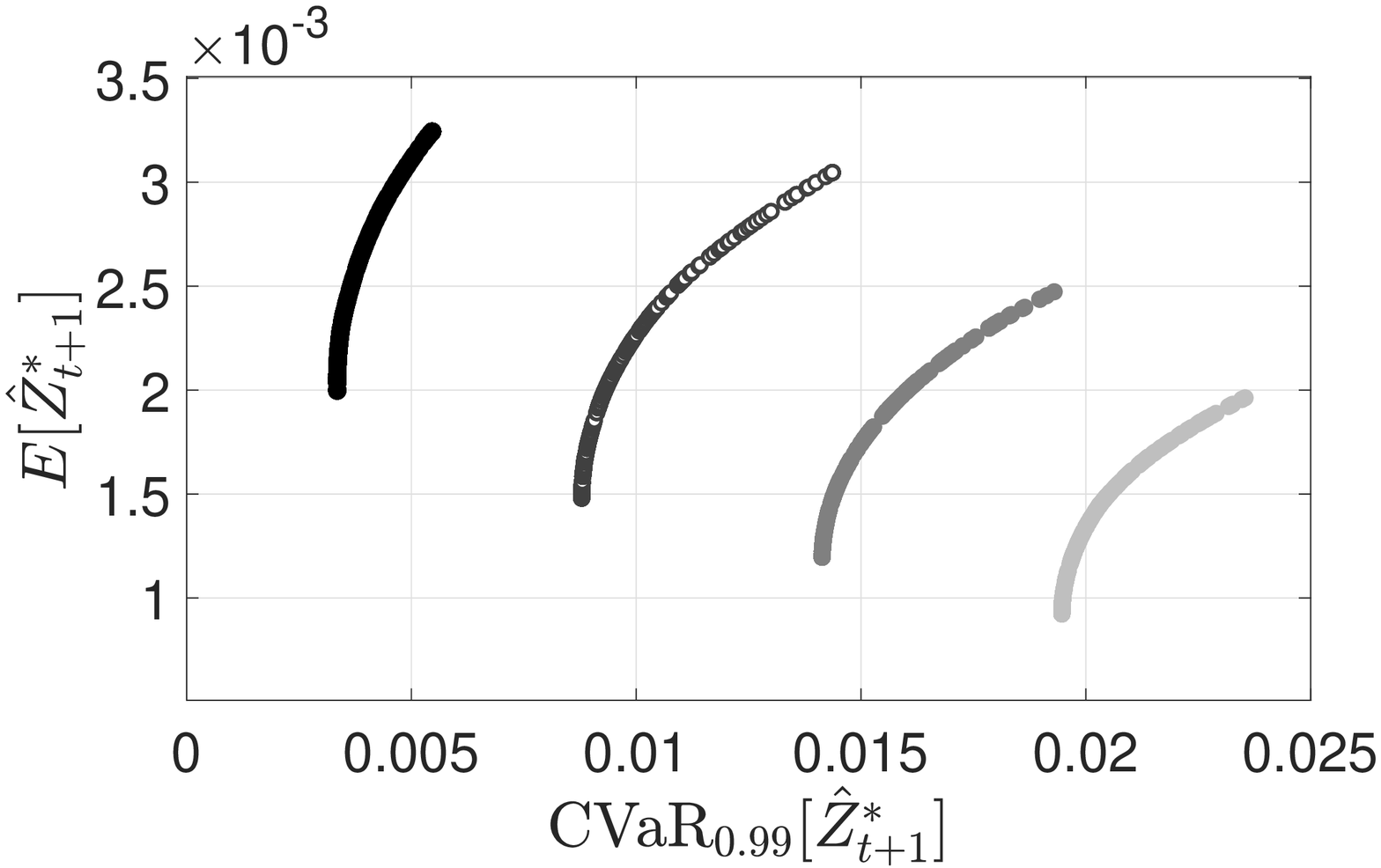}}\hspace{0em}%
	\caption{ESG-valued return efficient frontiers (top) plotted in
		$(\mathbb{V}[ \hat{Z}^*_{t+1}], \mathbb{E}[ \hat{Z}^*_{t+1}],\text{ESG}^*_{t+1})$ space and
		(bottom) projected onto the $(\mathbb{V}[ \hat{Z}^*_{t+1}], \mathbb{E}[ \hat{Z}^*_{t+1}])$ plane.
		Left to right: a,d) MV, b,e) mCVaR${}_{0.95}$ and c,f) mCVaR${}_{0.99}$ optimizations.
		Points on each curve represent equally spaced values of $\alpha \in$ \{0, 0.01, 0.02, \dots, 0.99\}.
	} 
	\label{fig:EF_orig_Fan_n_30122019}	
\end{figure}

We illustrate the optimization procedure by solving for optimal weights for the date 12/30/2019 for our illustrative portfolio.
We use a historical look-back period of $T = 510$ days and generate $S = 10,000$ simulated returns.
Three optimizations were obtained, using MV, mCVaR${}_{0.95}$ and mCVaR${}_{0.99}$ risk measures.
Optimizations were performed for four choices of $\lambda = \left\{ 0,0.25,0.50,0.75\right\}$.
For each $\lambda$, the efficient frontier was computed using the sequence of values
$\alpha \in$ \{0, 0.01, 0.02, \dots, 0.99\}.\footnote{
	For this portfolio, there is essentially no change in the optimized solutions over the range
	$\alpha \in [0.9, 0.99]$.}
The efficient frontier can be examined either in the three-dimensional space
$( \mathbb{V}[ \hat{Z}^*], \mathbb{E}[ \hat{Z}^*], \text{ESG}^*)$
or in $( \mathbb{V}[ \hat{R}^*], \mathbb{E}[ \hat{R}^*], \text{ESG}^*)$.
The efficient frontiers are plotted in
$( \mathbb{V}[ \hat{Z}^*_{t+1}], \mathbb{E}[ \hat{Z}^*_{t+1}], \text{ESG}^*_{t+1})$ space in
Fig.~\ref{fig:EF_orig_Fan_n_30122019}.
At the higher $\lambda$ values, both $\text{ESG}^*$ and  $\mathbb{E}[ \hat{Z}^* ]$ increase rapidly
for low-to-moderate values of the parameter $\alpha$.
This is especially noticeable for the MV portfolios when $\lambda = 0.5$ and $0.75$
(Fig.~\ref{fig:EF_orig_Fan_n_30122019}(a)).
Increasing the value of $\lambda$ decreases the value of the ESG-valued risk measure
$\mathbb{V}[ \hat{Z}^* ]$ and increases the value of the $\mathbb{E}[ \hat{Z}^* ]$.
The increase of $\mathbb{E}[ \hat{Z}^* ]$  with $\lambda$ follows from \eqref{eq:esg_const_transform};
increasing $\lambda$ puts more weight on optimizing the ESG score of the portfolio.
The observed increase of $\text{ESG}^*$ with $\lambda$ follows from the same reasoning.
As we hold ESG scores constant throughout a year's time,
increasing $\lambda$ puts more weight on $\varsigma_{i,t}$ resulting in the observed
decrease of $\mathbb{V}[ \hat{Z}^* ]$.  

\begin{figure}[ht]
    \centering
    	\subcaptionbox{}{\includegraphics[width=0.34\textwidth]{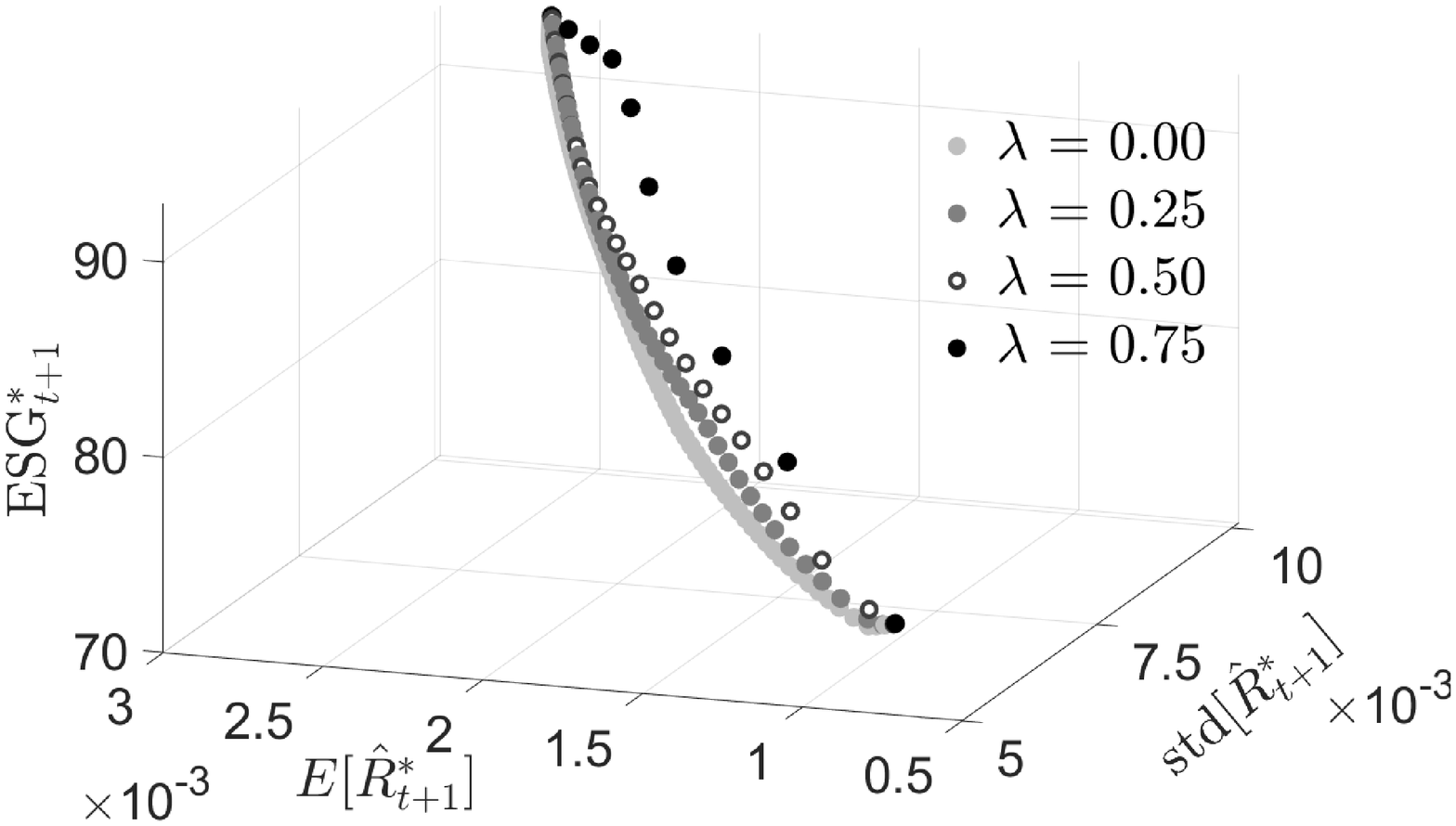}}\hspace{0em}%
	\subcaptionbox{}{\includegraphics[width=0.32\textwidth]{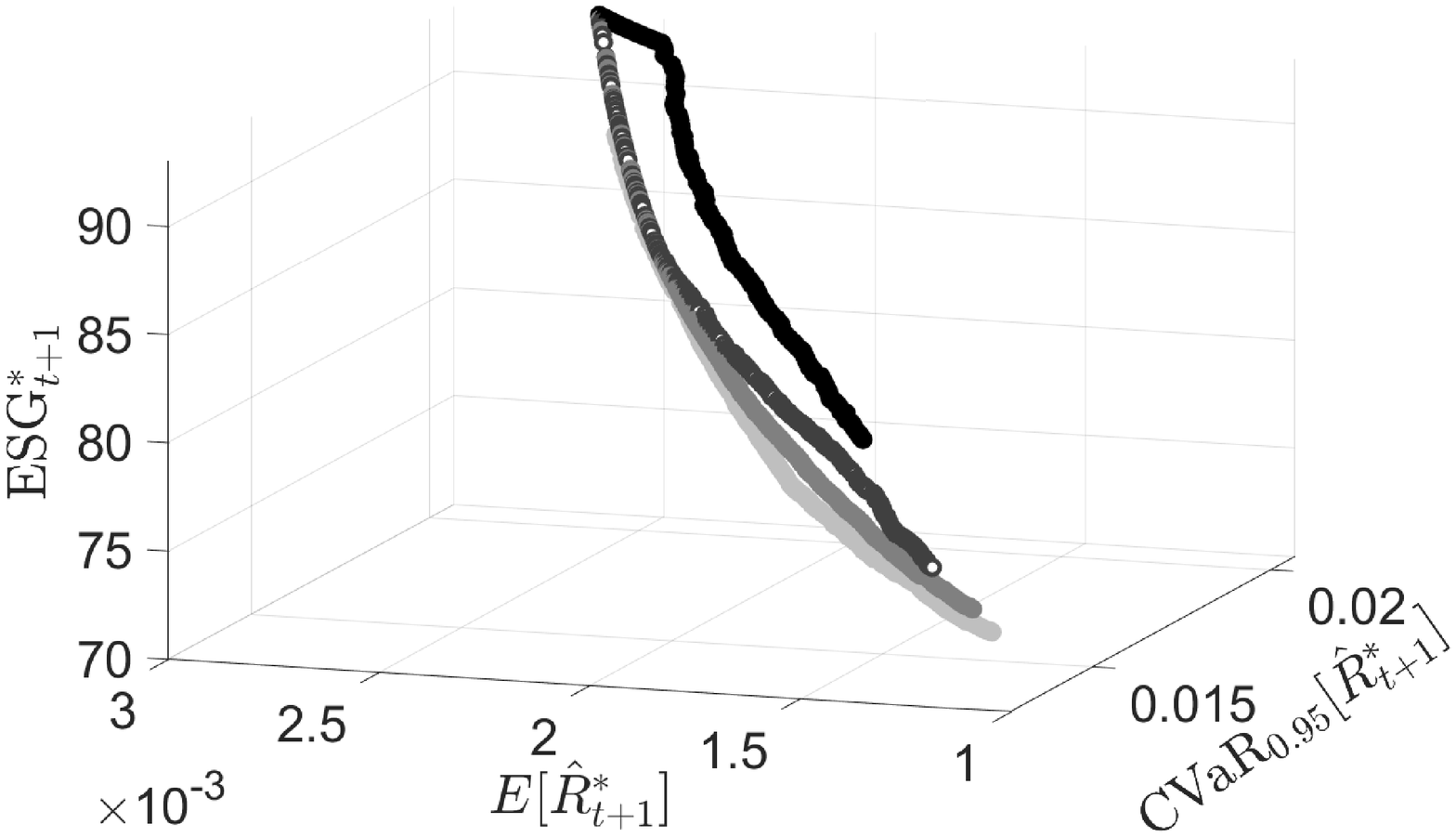}}\hspace{.2em}%
	\subcaptionbox{}{\includegraphics[width=0.32\textwidth]{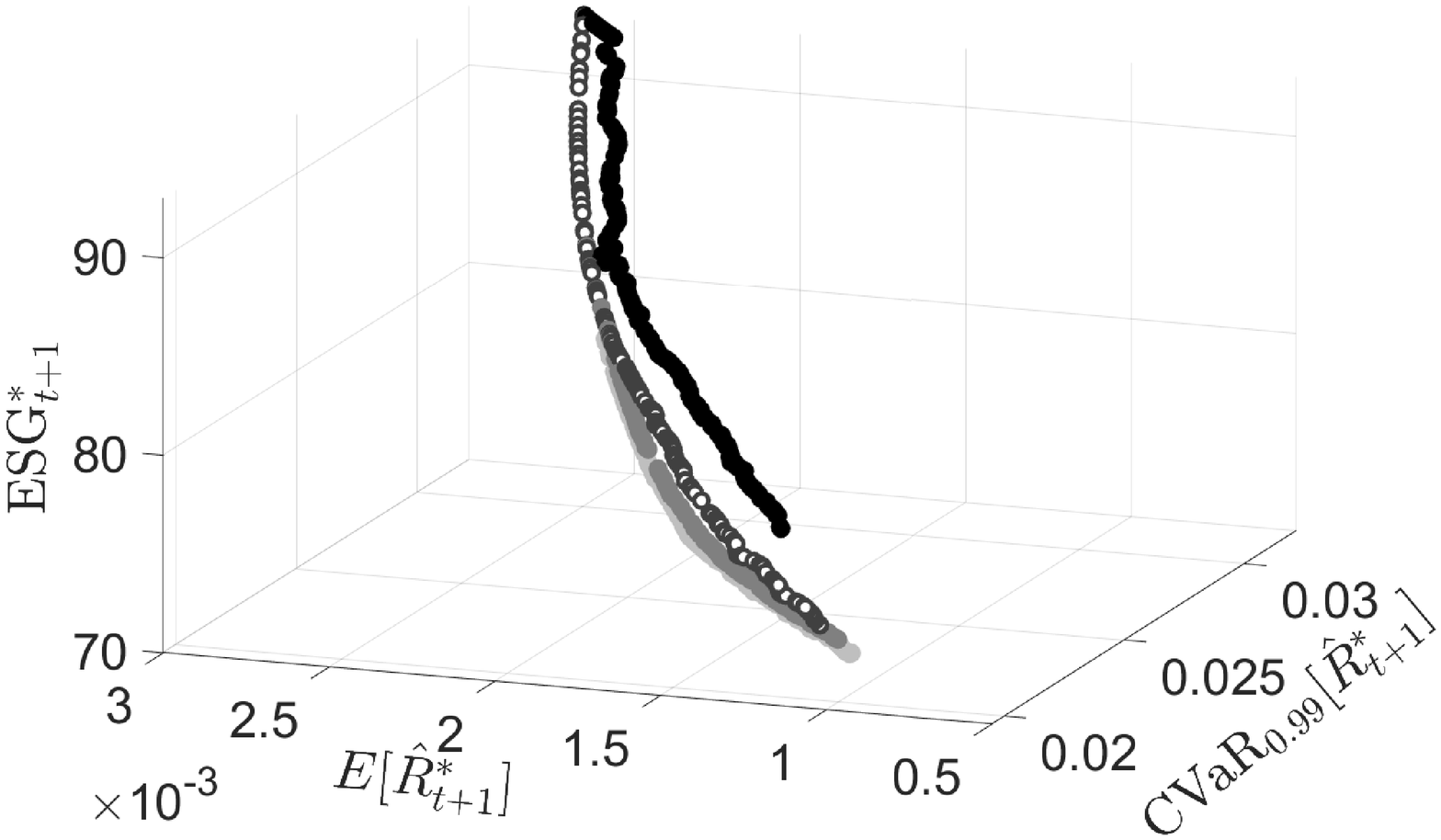}}\hspace{0em}%
	\vspace{.2em}
	\subcaptionbox{}{\includegraphics[width=0.33\textwidth]{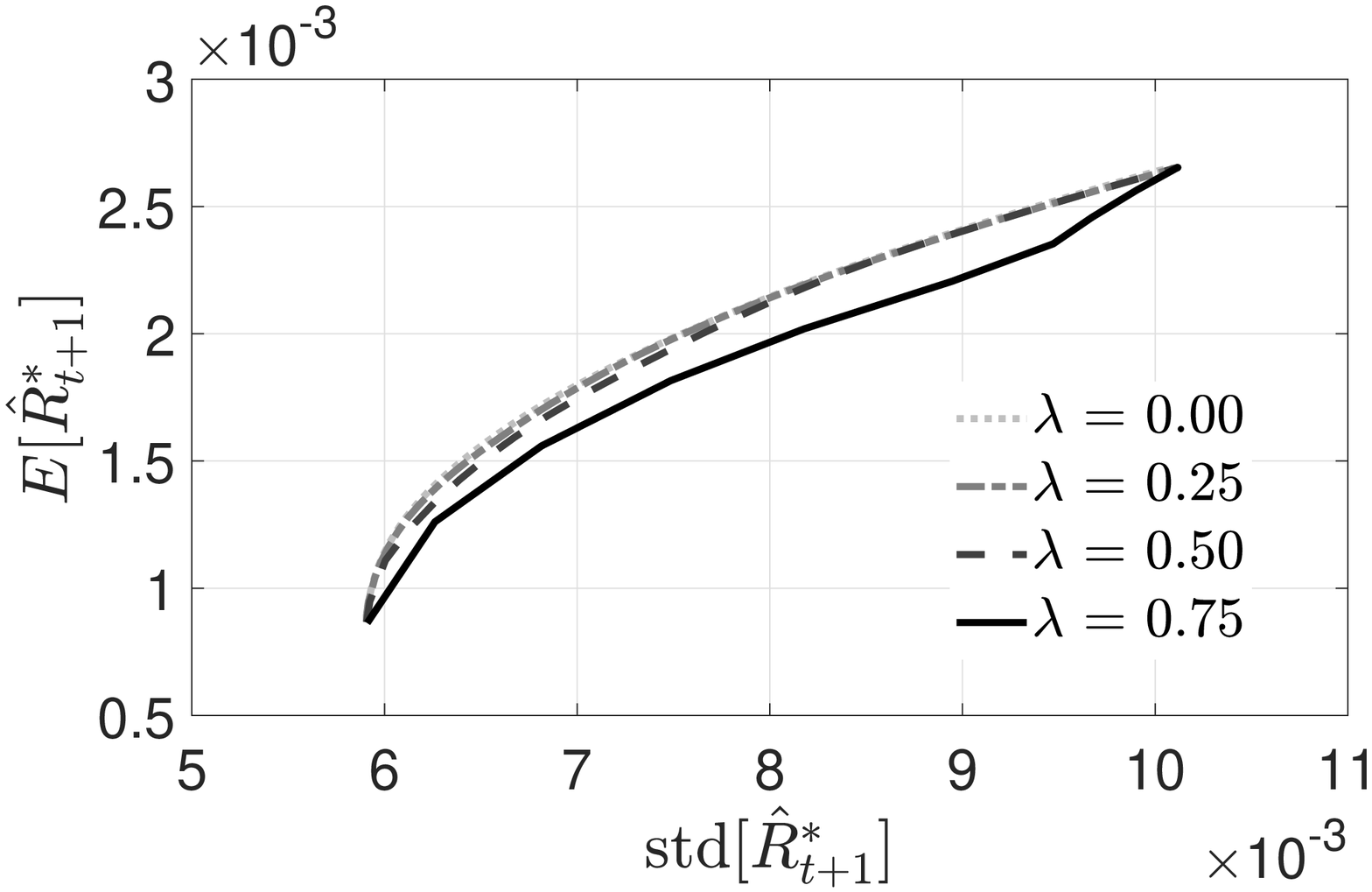}}\hspace{0em}%
	\subcaptionbox{}{\includegraphics[width=0.33\textwidth]{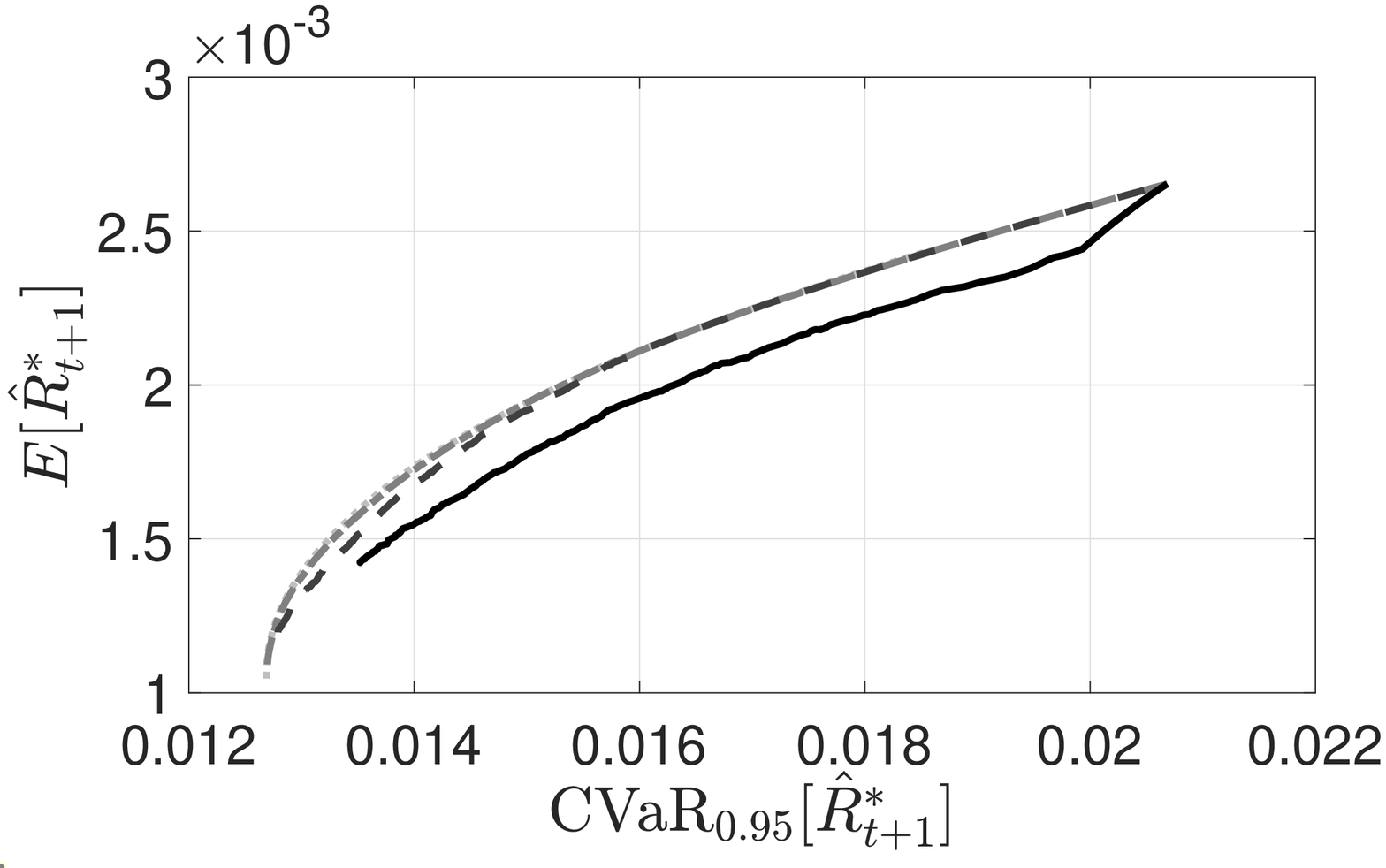}}\hspace{0em}%
	\subcaptionbox{}{\includegraphics[width=0.33\textwidth]{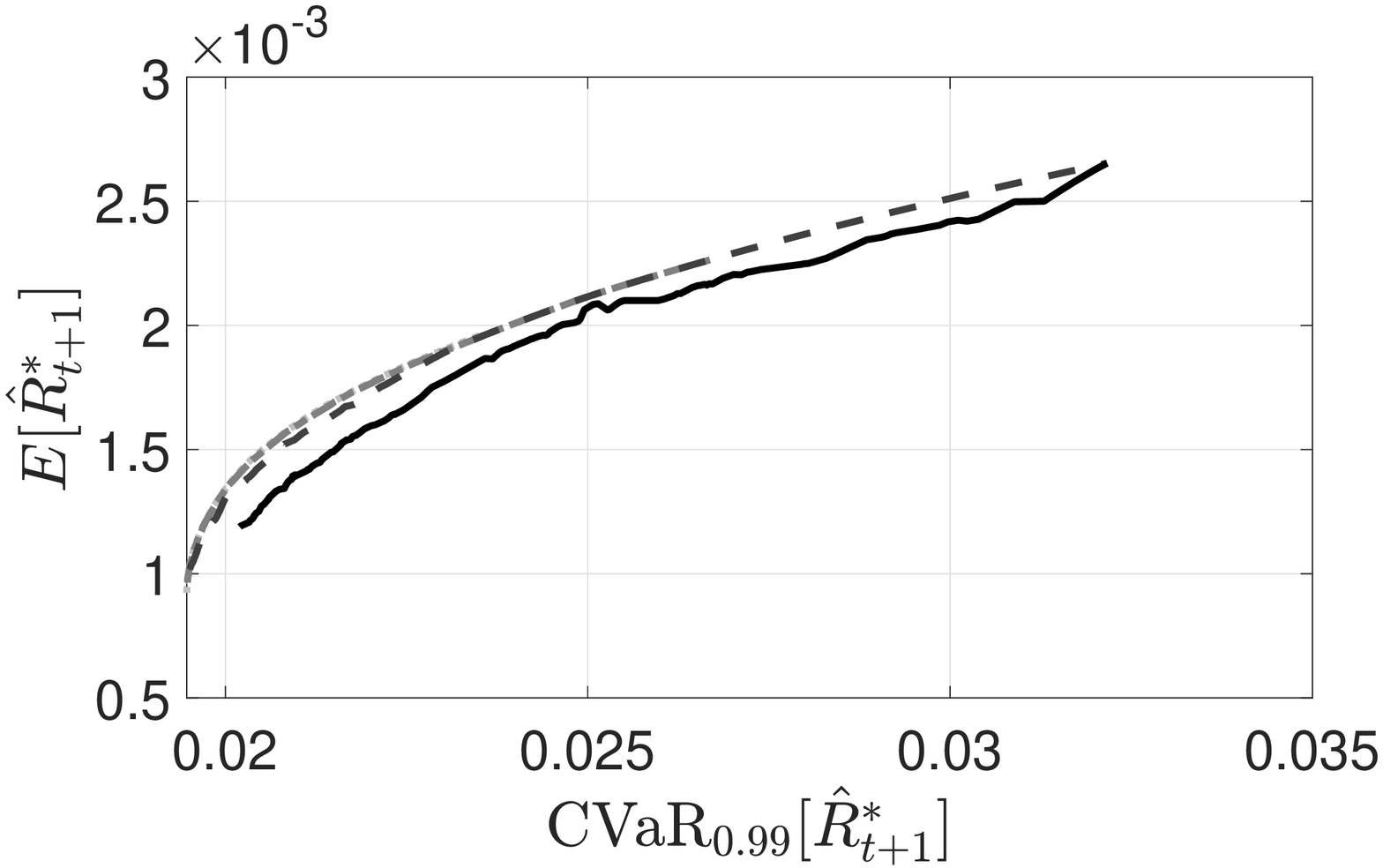}}\hspace{0em}%
	\caption{ESG-valued return efficient frontiers (top) plotted in
		$(\mathbb{V}[ \hat{R}^*_{t+1}], \mathbb{E}[ \hat{R}^*_{t+1}],\text{ESG}^*_{t+1})$ space and
		(bottom) projected onto the $(\mathbb{V}[ \hat{R}^*_{t+1}], \mathbb{E}[ \hat{R}^*_{t+1}])$ plane.
		Left to right: a,d) MV, b,e) mCVaR${}_{0.95}$ and c,f) mCVaR${}_{0.99}$ optimizations.
		Points on each curve in (a)-(c) represent equally spaced values of $\alpha \in$ \{0, 0.01, 0.02, \dots, 0.99\}.
	} 
	\label{fig:EF_Fan_n_30122019}
\end{figure}

Fig.~\ref{fig:EF_Fan_n_30122019} investigates the results in
$( \mathbb{V}[ \hat{R}^*_{t+1}], \mathbb{E}[ \hat{R}^*_{t+1}], \text{ESG}^*_{t+1})$ space,
using the traditional $\mathbb{V}[ \hat{R}]$ and $\mathbb{E}[ \hat{R}]$ axes.
The $\lambda = 0$ curve represents the traditional efficient frontier under which ESG scores are not considered.
For this portfolio and date, on the  $\lambda=0$ frontier $\text{ESG}^*_{t+1}$ increases with increasing values
of $\alpha$;
the choice of more risky portfolios giving rise to a higher ESG score.
This is not a general result as it depends on the distribution of returns estimated at the particular date.\footnote{
	In the supplementary material to this paper we show the $\lambda=0$ efficient frontier computed
	for a different date in which the portfolio ESG score initially increases with $\alpha$,
	but then decreases as $\alpha$ and risk continue to increase.}
For the higher $\lambda$ values, $\mathbb{E}[ \hat{R}^* ]$ also increases rapidly
for low-to-moderate values of the parameter $\alpha$;
again most noticeably for the MV portfolios when $\lambda = 0.5$ and $0.75$
(Fig.~\ref{fig:EF_Fan_n_30122019}(a)).
There is a nonlinear decrease in both $\mathbb{E}[ \hat{R}^*_{t+1}]$ and $\mathbb{V}[ \hat{R}^*_{t+1}]$
as $\lambda$ increases (for any fixed value of $\alpha$),
which is seen most clearly in the projection on the 
$( \mathbb{V}[ \hat{R}^*_{t+1}], \mathbb{E}[ \hat{R}^*_{t+1}])$ plane.
The most significant decrease in these values occurs when $\lambda$ increases from 0.5 to 0.75.
The behavior of this non-linear decrease changes with $\alpha$.
When viewed in the full
$( \mathbb{V}[ \hat{R}^*_{t+1}], \mathbb{E}[ \hat{R}^*_{t+1}], \text{ESG}^*_{t+1})$ space,
it is clear that this shift (at fixed $\alpha$) in the efficient frontier shift is toward larger
$\text{ESG}^*_{t+1}$ values.
Noticeable on the $\lambda=0.75$ efficient frontiers is a ``kink'' in its convex shape.
For mCVaR${}_{0.99}$ optimization, the $\lambda=0.75$ efficient frontier becomes less defined
due to sparsity of data at the 99'th percentile.

\begin{figure}[ht]
	\centering
	\subcaptionbox{MV}{\includegraphics[width=0.34\textwidth]
				{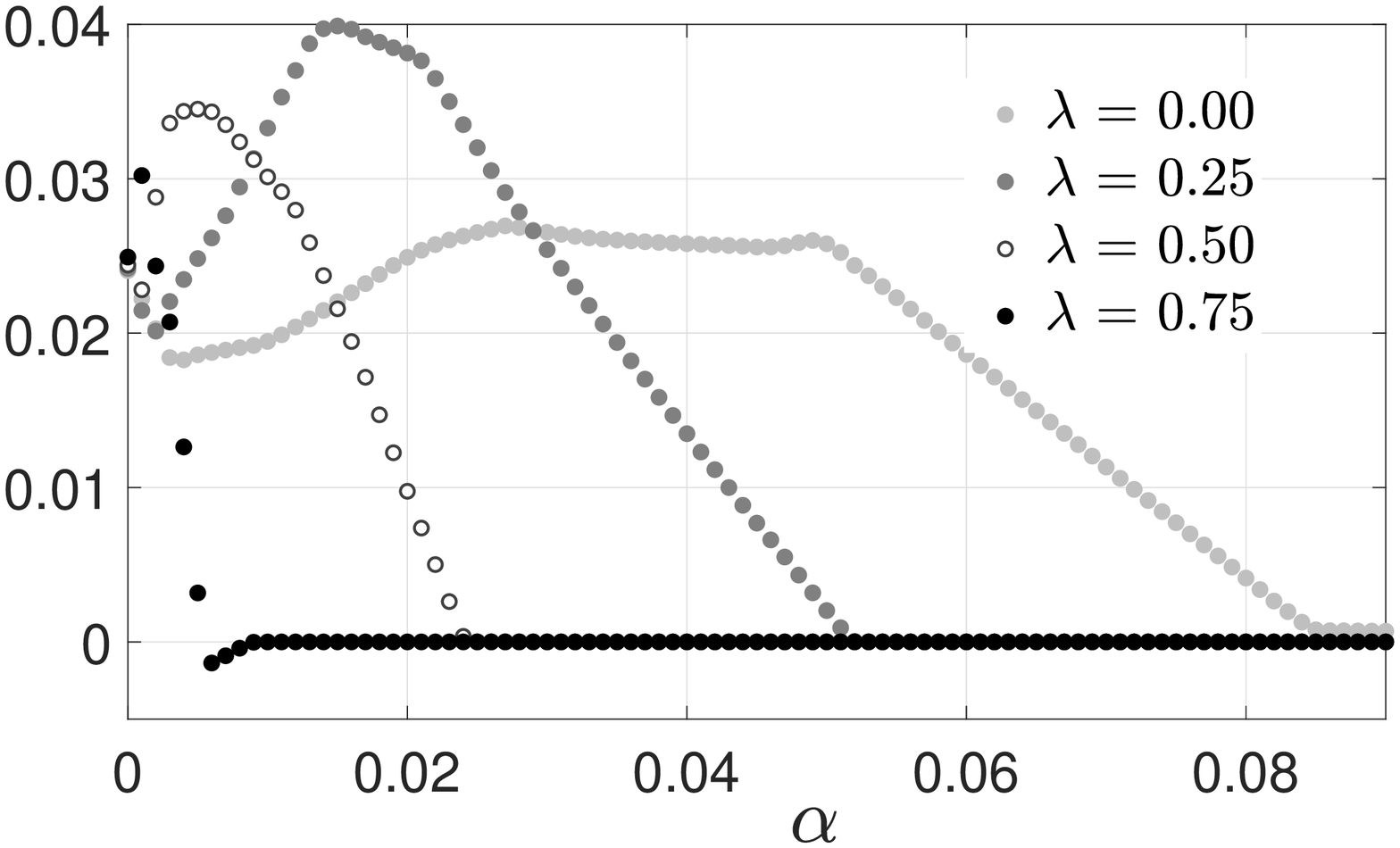}}\hspace{0em}%
	\subcaptionbox{mCVaR${}_{0.95}$}{\includegraphics[width=0.32\textwidth]
				{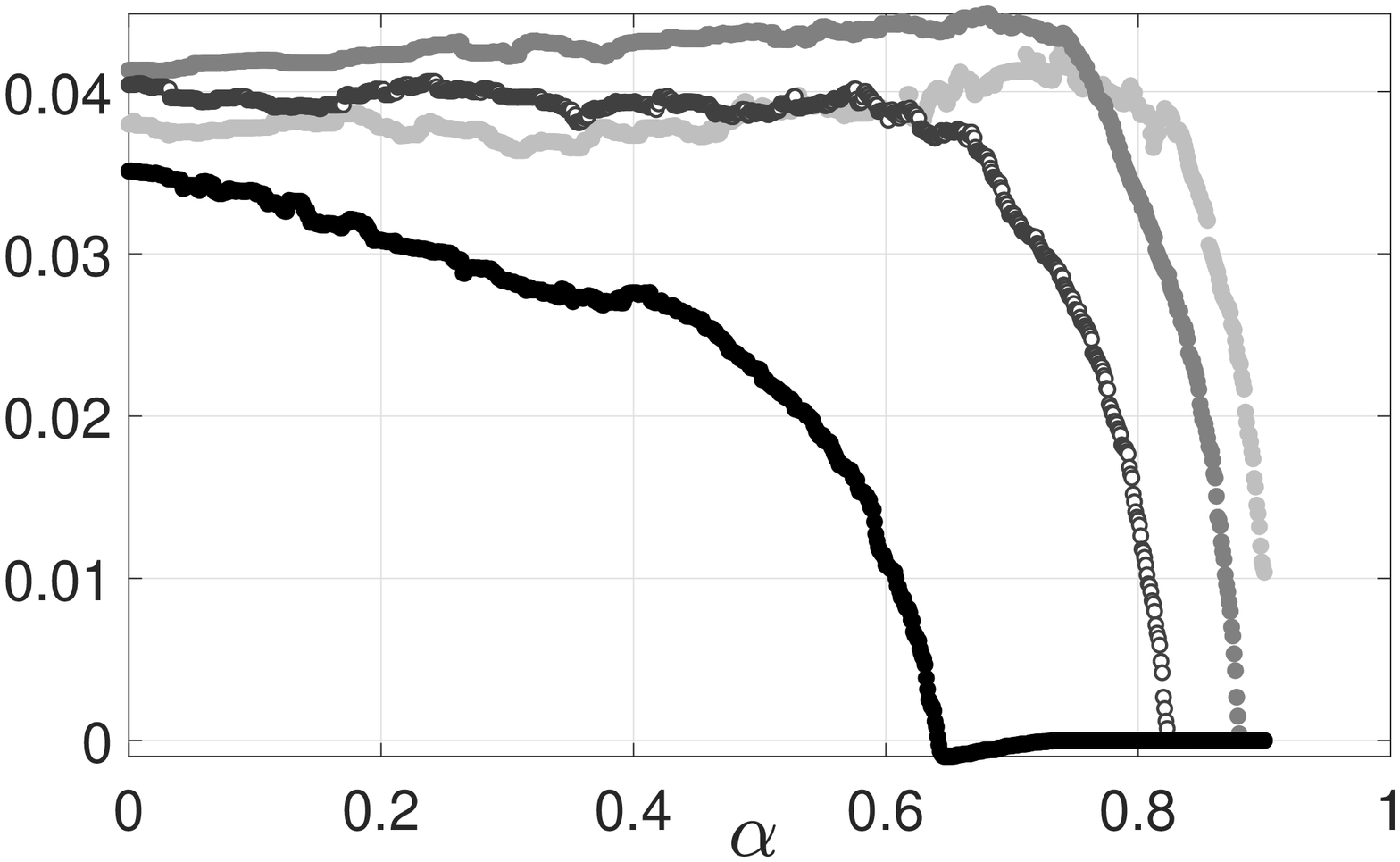}}\hspace{.2em}%
	\subcaptionbox{mCVaR${}_{0.99}$}{\includegraphics[width=0.32\textwidth]
				{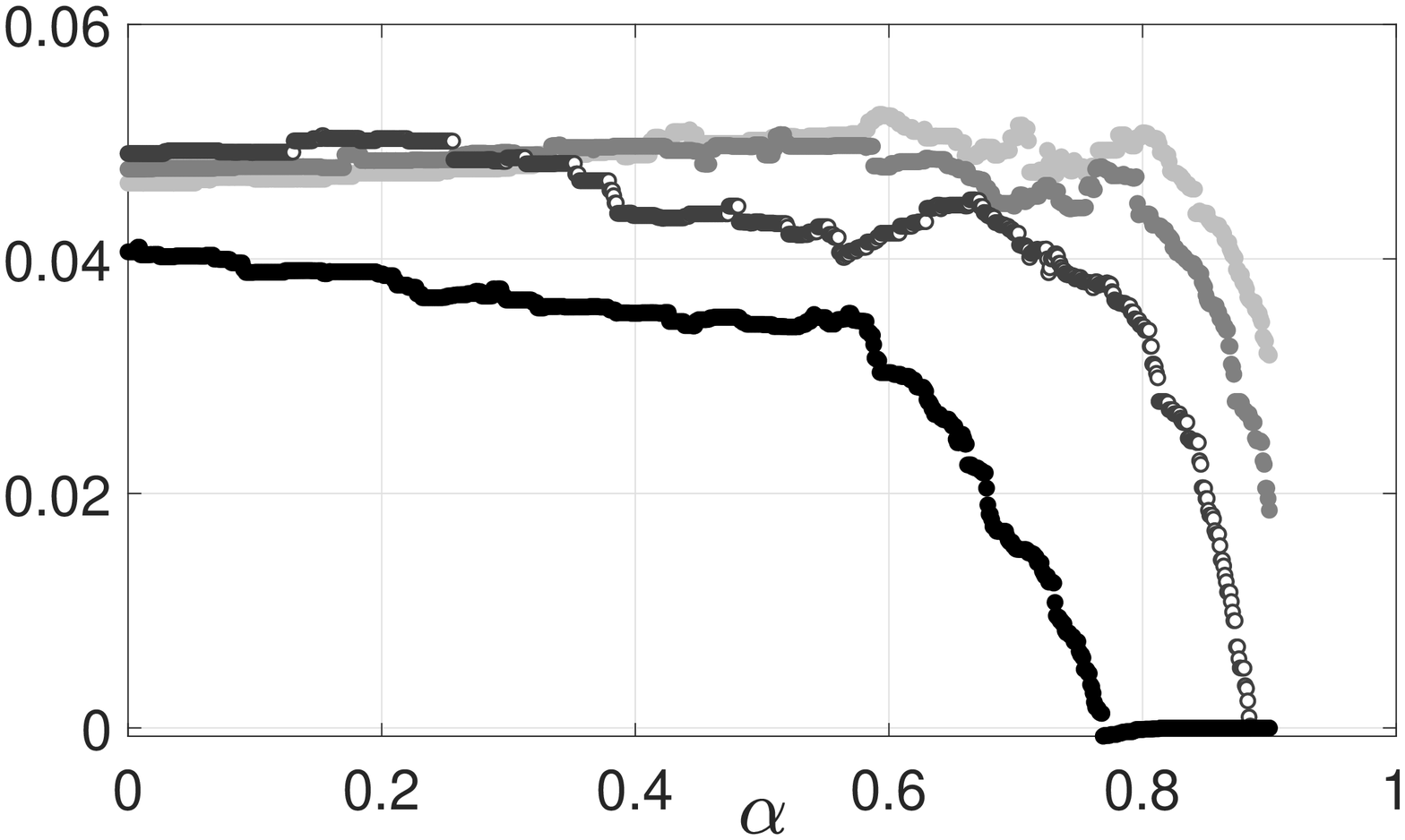}}\hspace{0em}%
	\caption{Relative difference between the ESG score of the 12/30/2019 MV and MCVaR optimized portfolios
	 computed using ESG scores released on 12/31/2019 (out-of-sample) compared with the computations
	 obtained using ESG scores released on 12/31/2018 (in-sample).
	Note the truncated range of the $\alpha$ axis for the MV optimized results.}
	\label{fig:Delta_ESG_Fan_n_30122019}	
\end{figure}

Since new ESG scores arrived on 12/31/2019,
the choice of the optimization date of 12/30/2019 enabled a test of the stability of the optimization results
by comparing the use of in-sample (released on 12/31/2018) versus out-of-sample (released on 12/31/2019)
ESG scores.
Fig.~\ref{fig:Delta_ESG_Fan_n_30122019} plots the relative difference\footnote{
	The relative difference in the portfolio ESG score is defined in the usual manner as
	(ESG(out-of-sample) $-$ ESG(in-sample))/ESG(in-sample).}
between the ESG score of the optimal portfolio computed using individual asset ESG scores released on 12/31/2019
compared with the computations in Figs.~\ref{fig:EF_orig_Fan_n_30122019} and
\ref{fig:EF_Fan_n_30122019} computed using the ESG data released on 12/31/2018.
In general, over all three risk measures, the relative difference does not exceed 5\% percent,
indicating a quite stable ESG portfolio value (at least for this date and these securities).
The relative difference becomes zero as $\alpha$ increases,
approaching zero for smaller values of $\alpha$ as $\lambda$ increases.
For the MV optimized portfolios, the relative ESG difference rapidly disappears over the range $0 \le \alpha \le 0.1$,
The sensitivity of the MV optimal solution to ESG score dramatically decreases for values $\alpha >  0.1$
as the optimized portfolios become extremely concentrated in very few securities.
The relative difference persists longest under mCVaR${0.99}$ optimization, staying roughly constant until $\alpha > 0.6$.

\begin{figure}[h!]
	\centering
	\includegraphics[width = 0.8\textwidth]{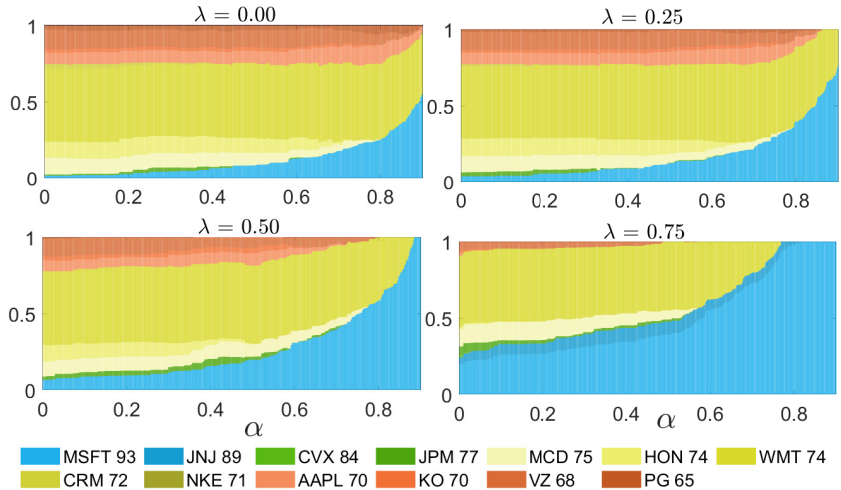}
	\caption{Variation of the weight composition of the mCVaR${}_{0.99}$ optimal portfolios along each efficient
		frontier (as a function $\alpha$).
		Each panel corresponds to a different choice of $\lambda$.
		Assets with lower ESG scores are depicted with warmer colors.
		The legend identifies each stock and the stock's ESG score as assigned by Refinitiv on 12/31/2018.
	}
	\label{fig:Weights_n_MCVaR99_fan_30122019}
\end{figure}

In Fig.~\ref{fig:Weights_n_MCVaR99_fan_30122019} we depict the optimal portfolio weight composition along
the efficient frontiers (i.e. as a function of $\alpha$) of the mCVaR${}_{0.99}$ optimizations for each value
of $\lambda$ considered.
For a given choice of $\lambda$, the diversity of the portfolio decreases as $\alpha$ increases;
at high $\alpha$ the optimizer focuses on those assets with highest expected return.
For a fixed value of $\alpha$,
an increase in $\lambda$ leads to a portfolio composition that decreases the weight of lower ESG-scoring stocks.
The effect is most noticeable for $\lambda = 0.75$.
Similar results (not shown) were obtained for the MV and mCVaR${}_{0.95}$ optimizations.

\section{ESG-Valued Performance Measures} \label{sec:ESG_PM}

A common approach to evaluate the performance of an investment strategy is to use one or more reward-risk ratios (RRRs)
to quantify the trade-off between expected reward and the risk associated with the strategy.
There is an extensive list of such ratios, though not all of them posses the properties of a coherent risk measure;
see Cheridito and Kromer \cite{Cheridito_2013} for a complete overview of RRRs and an analysis of their properties.
For instance, the extensively used Sharpe ratio does not satisfy the monotonicity property.
The development of ESG-valued RRRs involves deep questions concerning the desired
properties that the measures should possess;
for example, the coherent risk measure properties developed by Cherdito and Kromer may comprise a necessary,
but not sufficient, set of properties for ESG-valued RRRs.
The scope of this question is outside of this paper.
We illustrate here a ``straightforward'' strategy for developing ESG-valued RRRs.

\begin{figure}[h!]
	\centering		
	{\includegraphics[width=0.5\textwidth]{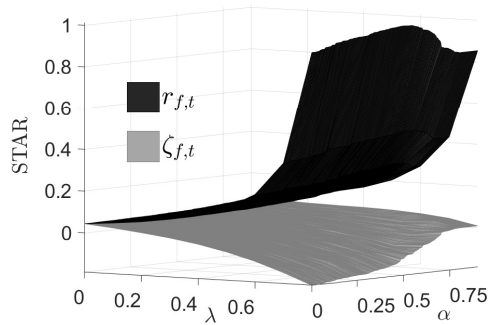}}
	\caption{The ESG-valued STAR ratio computed as a function of the selected $\alpha$ and $\lambda$ values
		for the mCVaR${}_{0.99}$ optimization using equations \eqref{eq:STAR_rf} and \eqref{eq:STAR_zeta}.}
	\label{fig:SR_Fan_n_original_30122019}	
\end{figure}

For any choice of an RRR, one strategy for developing an ESG-valued counterpart involves replacing the portfolio return
$\hat{R}$ in the definition of the RRR with the ESG-valued return $\hat{Z}$.
For illustration, we consider  the stable tail adjusted return (STAR) ratio \cite{Martin_2003},
\begin{equation}
	\text{STAR}_\beta (\hat{R}) = \frac{\mathbb{E}[(\hat{R} - r_{f,t})^+]}{\text{ETL}_\beta[(\hat{R} - r_{f,t})^+]},
\end{equation}
where $r_{f,t}$ is a risk-free rate appropriate for day $t$.
The ESG-valued STAR ratio is then
\begin{equation}
	\text{STAR}_\beta (\hat{Z}) = \frac{\mathbb{E}[(\hat{Z} - r_{f,t})^+]}{\text{ETL}_\beta[(\hat{Z} - r_{f,t})^+]}.
	\label{eq:STAR_rf}
\end{equation}
The ESG-valued STAR ratio satisfies all four properties of a coherent risk measure.
As the STAR ratio will vary along the efficient frontier (i.e. with $\alpha$) and from frontier to frontier
(i.e. with $\lambda$),
in Fig.~\ref{fig:SR_Fan_n_original_30122019}  we summarize the behavior of the STAR ratio as the surface
STAR($\lambda,\alpha$) for the mCVaR${}_{0.99}$ optimization.
(To compute the STAR ratio, we used the 10-year U.S. treasury yield rate to determine daily values for $r_{f,t}$.
Computation of the expectation and ETL in the STAR ratio is performed over the ensemble of ESG-valued
returns generated for day $t+1$ based upon the information available at day $t$.)
As only four values for $\lambda$ were computed while the $\alpha$ grid is much finer,
the smoothly interpolated surface appears coarser in the $\lambda$ direction.
The surface shows a generally convex relationship between  the ESG-valued STAR ratio and $\lambda$
and $\alpha$,
although variation with $\alpha$ is smaller.
The ``kink'' that appears in the $\lambda = 0.75$ efficient frontier produces the non-convexity observed in the surface.

Another strategy would be to replace $r_{f,t}$ with an ESG-valued risk-free rate,
such as that developed in equation \eqref{eq:esg_rf} for defining tangent portfolios.
Fig.~\ref{fig:SR_Fan_n_original_30122019} also shows the surface STAR($\lambda,\alpha$) for the mCVaR${}_{0.99}$
optimization using the formulation
\begin{equation}
	\text{STAR}_\beta (\hat{Z}) =
		\frac{\mathbb{E}[(\hat{Z} - \zeta_{f, t }( \lambda ))^+]}
		      {\text{ETL}_\beta[(\hat{Z} - \zeta_{f, t}( \lambda ))^+]}.
	\label{eq:STAR_zeta}
\end{equation}
with $\zeta_{f, t}( \lambda )$ defined in \eqref{eq:esg_rf}.
Because the risk-free rate $r_{f,t}$ has no $\lambda$ dependence,
as $\lambda$ increases the behavior of  the STAR ratio is effectively that of
$\mathbb{E}[\hat{Z}] / \text{ETL}_\beta[\hat{Z}]$ which increases with $\lambda$.
In contrast, the risk-free rate $\zeta_{f,t}$ has strong $\lambda$ dependence, especially so since this
risk-free rate is assigned an ESG score of 100, larger that that of any assset in the optimized portfolio
and, consequently, larger than that of the portfolio.
As a result, for large $\lambda$ the $\zeta_{f,t}$ term becomes the dominating factor in the numerator
and denominator of \eqref{eq:STAR_zeta}.
Apparently this effect exists even for small, positive values of $\lambda$,
resulting in a monotonic decrease in the STAR ratio as $\lambda$ increases.
In contrast to the $r_{f,t}$ surface, which shows no strong $\alpha$ dependence,
as $\lambda$ increases, the $\alpha$ dependence of the $\zeta_{f,t}$ surface becomes more pronouced
and the STAR ratio decreases with a decrease in $\alpha$ (more risk-averse optimization producing smaller STAR ratios).
The STAR ratio becomes negative under these conditions.

\subsection{Performance Over Time} \label{sec:ESG_Perf}

We now investigate the ESG-optimized investment results over time using the same DJIA-derived portfolio of stocks
as in Section~\ref{sec:ESG_EF}.
We consider a daily investment horizon spanning the period 01/03/2017 through 12/30/2020.
Using the procedure outlined in Section~\ref{sec:ESG_EF} involving 1-day ahead forecasting based upon
two years of previous returns,
for each day $t$ we computed an efficient frontier using the sequence of values $\alpha \in [0,0.001,0.002,...,0.9]$
for each value of $\lambda \in [0,0.25,0.5,0.75]$,
obtaining a set of optimal asset weights for each $\alpha, \lambda, t$ combination.
We employed only the MV and mCVaR${}_{0.99}$ optimizations to examine central and tail behavior.
Transaction costs were included in the evaluation by assuming a cost of 2 basis points for each buy
and each sell order executed in rebalancing the portfolio.
We imposed a daily turnover constraint of 0.4\% ($\gamma=0.004$ in (\ref{eq:RR_const_3}))
to approximate reasonable business practice.
The results are compared to a benchmark consisting of the equally weighted,
buy-and-hold portfolio (EWBH) of the same 29 DJIA stocks.\footnote{
	With reference to equations \eqref{eq:PortRet} and \eqref{eq:esg_PortRet},
	the EWBH portfolio is defined as follows.
	$I = 29$; the stock for Dow Inc. remains excluded.
	No scenario sets are generated; no optimization is performed.
	At $t = 0$, $\theta_i^* = 1/29$.
	For $t > 0$, the number of shares of each stock remains fixed at $n_{i,t} = n_{i,0}$;
	thus $\theta_{i,t+1}^* = n_{i,0} P_i (t) / \sum_{j=1}^I n_{j,0} P_j (t)$,
	where $P_i (t)$ is the price of the asset at the close of trading day $t$.
}

The efficient frontier portfolios were evaluated using standard performance measures:
total return (Tot Rtn), annualized return (Ann Rtn), average turnover (AvgTO), expected tail loss (ETL95)\footnote{
	Also known as conditional value-at-risk or expected shortfall.
}
and expected tail return (ETR95), both at 95\% significance levels, and maximum drawdown (MDD).
ETR95 is the reward counterpart of ETL95, measuring the average value of those returns above the 0.95 quantile.
MDD is generally defined as the maximum drop of portfolio returns over all time intervals that can be formed as a
partition of the investment period \cite{Chekhlov_2005}.
As performance measures, we also considered the average ESG score of a portfolio over the time period,
as well as its standard deviation.
Table \ref{tab:Summary_sim} presents time-averaged values for these summary performance measures for select ($\lambda,\alpha$)
efficient frontier portfolios for both mCVaR${}_{0.99}$ and MV optimizations.

The total and annualized returns of the optimized portfolios are generally superior to that of the benchmark
for values of $\alpha \ge 0.7$ for mCVaR${}_{0.99}$ and  $\alpha \ge 0.1$ for MV optimizations.
For fixed $\lambda \ge 0.5$, as $\alpha$ increases, AvgTO tends to fall below the daily constraint value of 0.4\%.
For fixed $\alpha > 0$, AvgTO decreases with $\lambda$.
This effect is of particular importance under MV optimization, which suffers from instability of solutions
and high turnover rates which can downgrade financial performance \cite{Clarke_2002,Kirby_2021,Michaud_1989}.
For mCVaR${}_{0.99}$, except for the largest values of $\lambda$ and $\alpha$,
the magnitudes of ETL95 and ETR95 are less than that of the EWBH benchmark.
For MV, these magnitudes generally exceed that of the benchmark.
In the opimized portfolios, ETL95 always exceeds ETR95 in magnitude, with the suggestion that, for MV optimization,
at large values of $\lambda$ the magnitude difference decreases as $\alpha$ increases.
All optimal portfolios have lower MDD than the benchmark.
For optimization under mCVaR${}_{0.99}$, the MDD decreases with $\lambda$ for each fixed value of $\alpha$.
Average ESG scores for the optimized portfolios virtually always exceed that of the benchmark.
Under  mCVaR${}_{0.99}$ optimization the standard deviation of the ESG score is generally smaller than that of the benchmark;
for MV optimization the reverse is true.

\begin{figure}[h!]
	\centering
	\subcaptionbox{Price	   }{\includegraphics[width=0.48\textwidth]{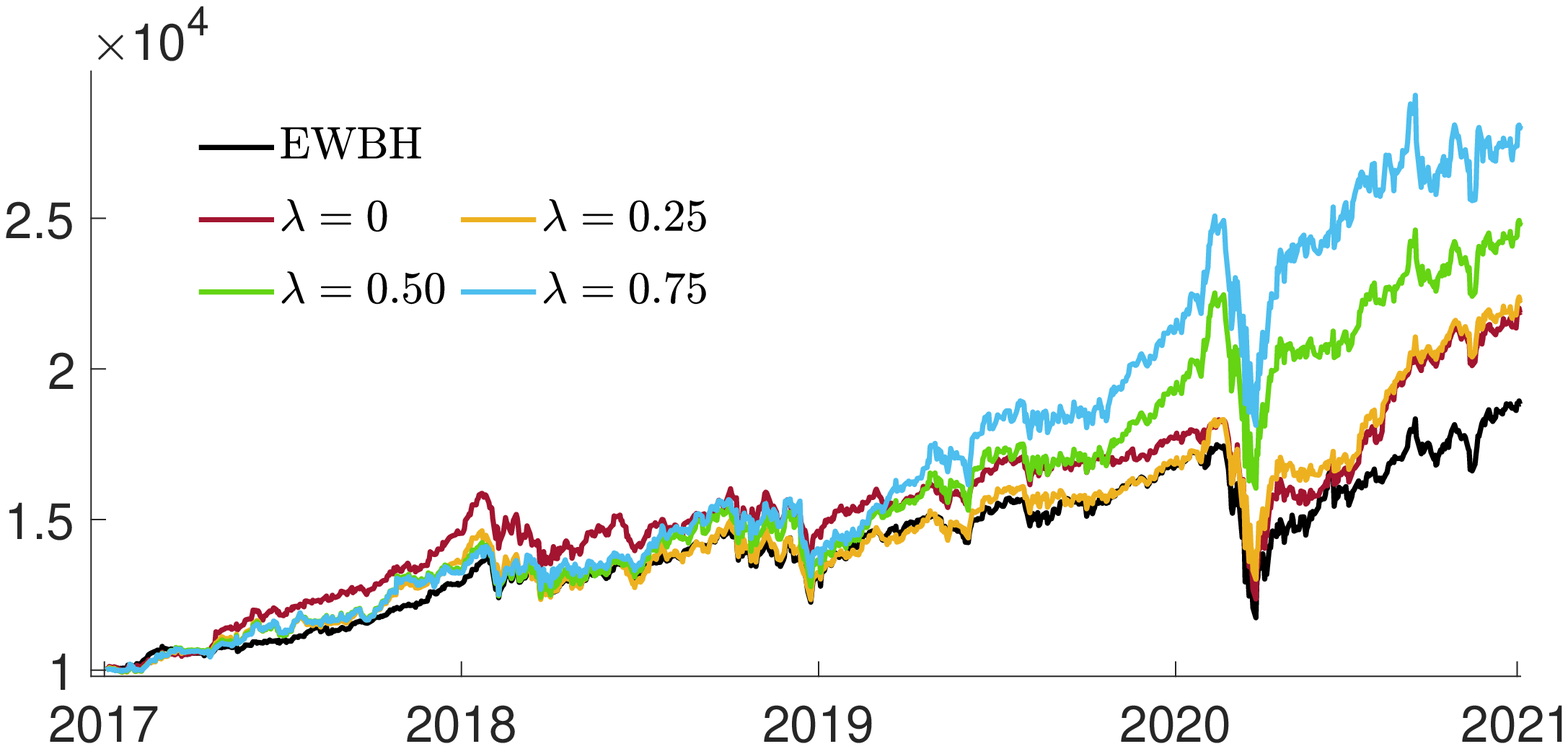}}\hspace{0em}%
	\subcaptionbox{ESG Score}{\includegraphics[width=0.48\textwidth]{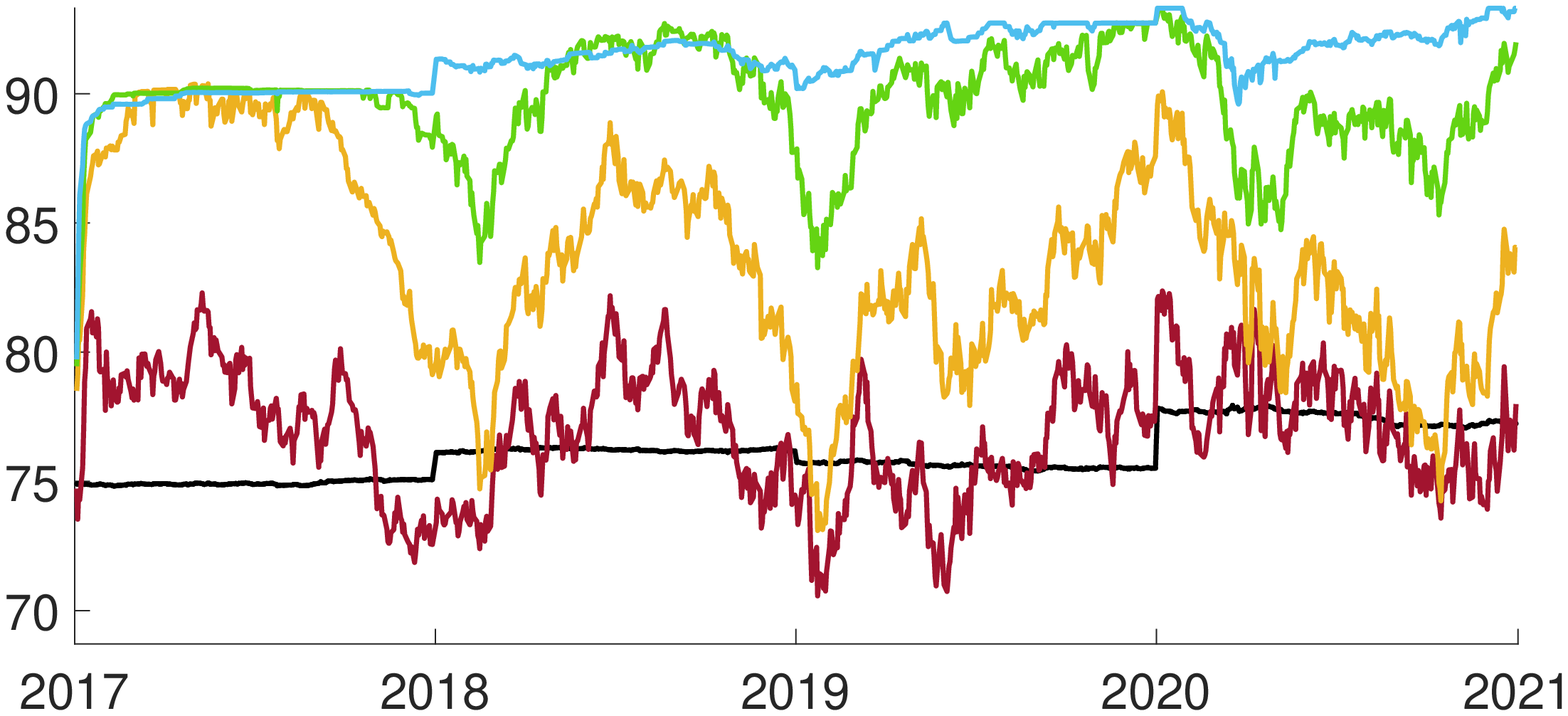}}\hspace{0em}%
	\subcaptionbox{Price	   }{\includegraphics[width=0.48\textwidth]{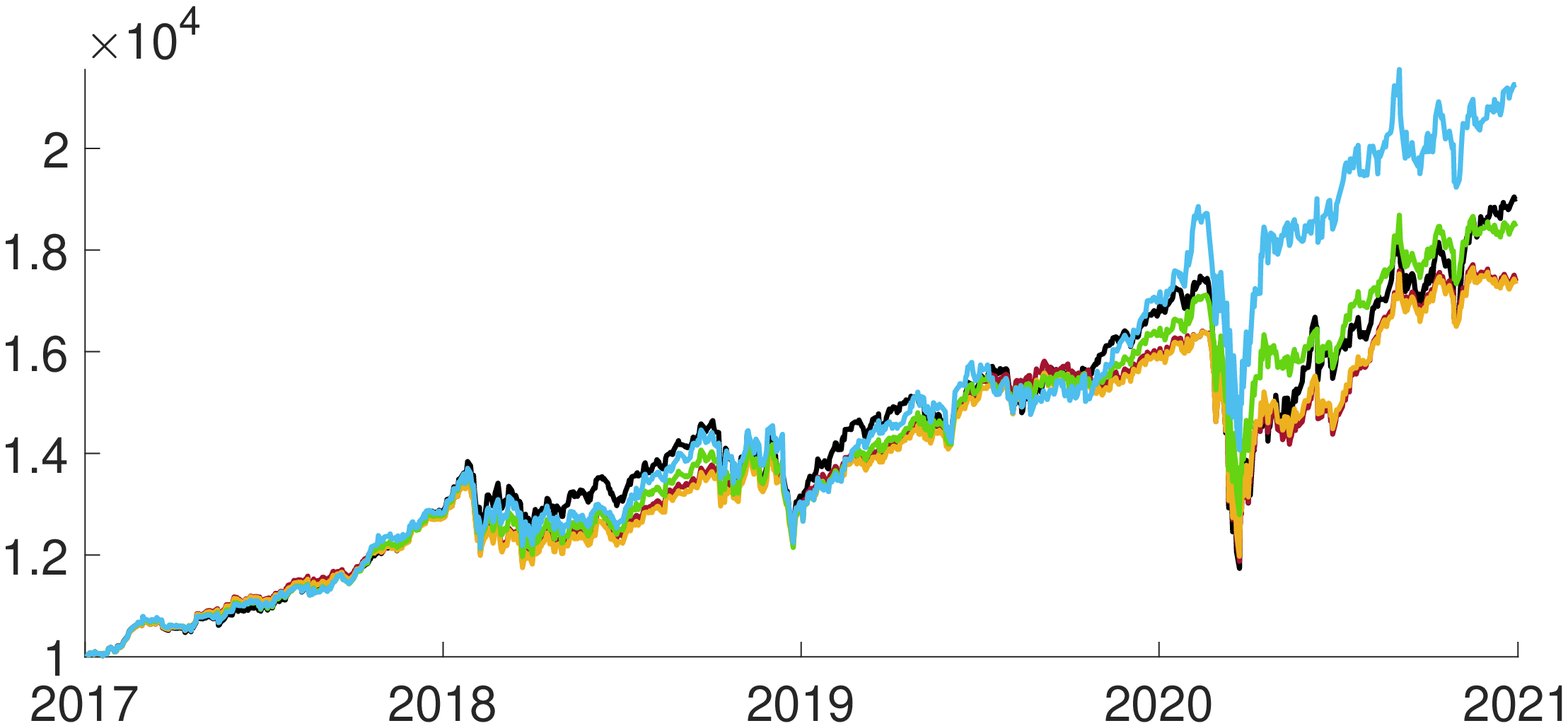}}\hspace{0em}%
	\subcaptionbox{ESG Score}{\includegraphics[width=0.48\textwidth]{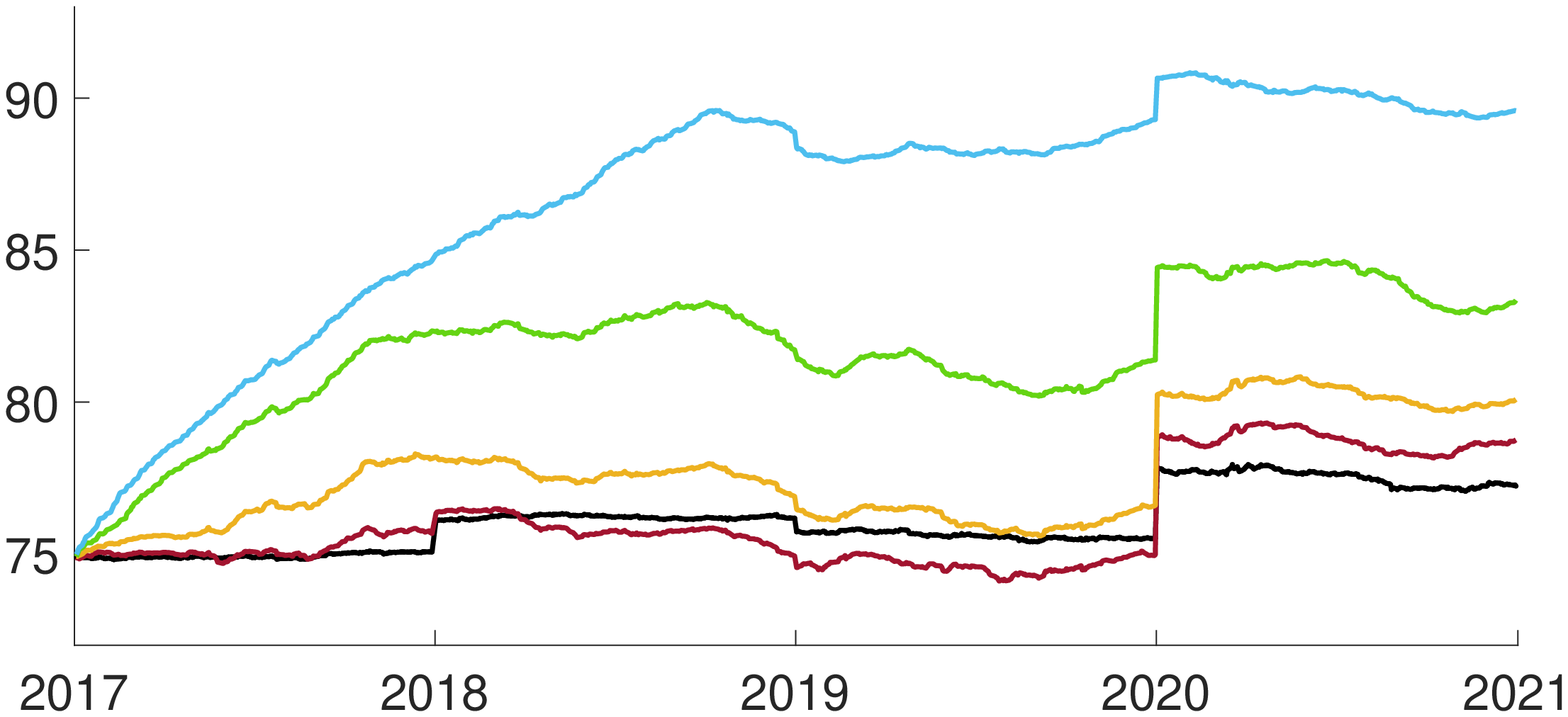}}\hspace{0em}%
	\subcaptionbox{Price	   }{\includegraphics[width=0.48\textwidth]{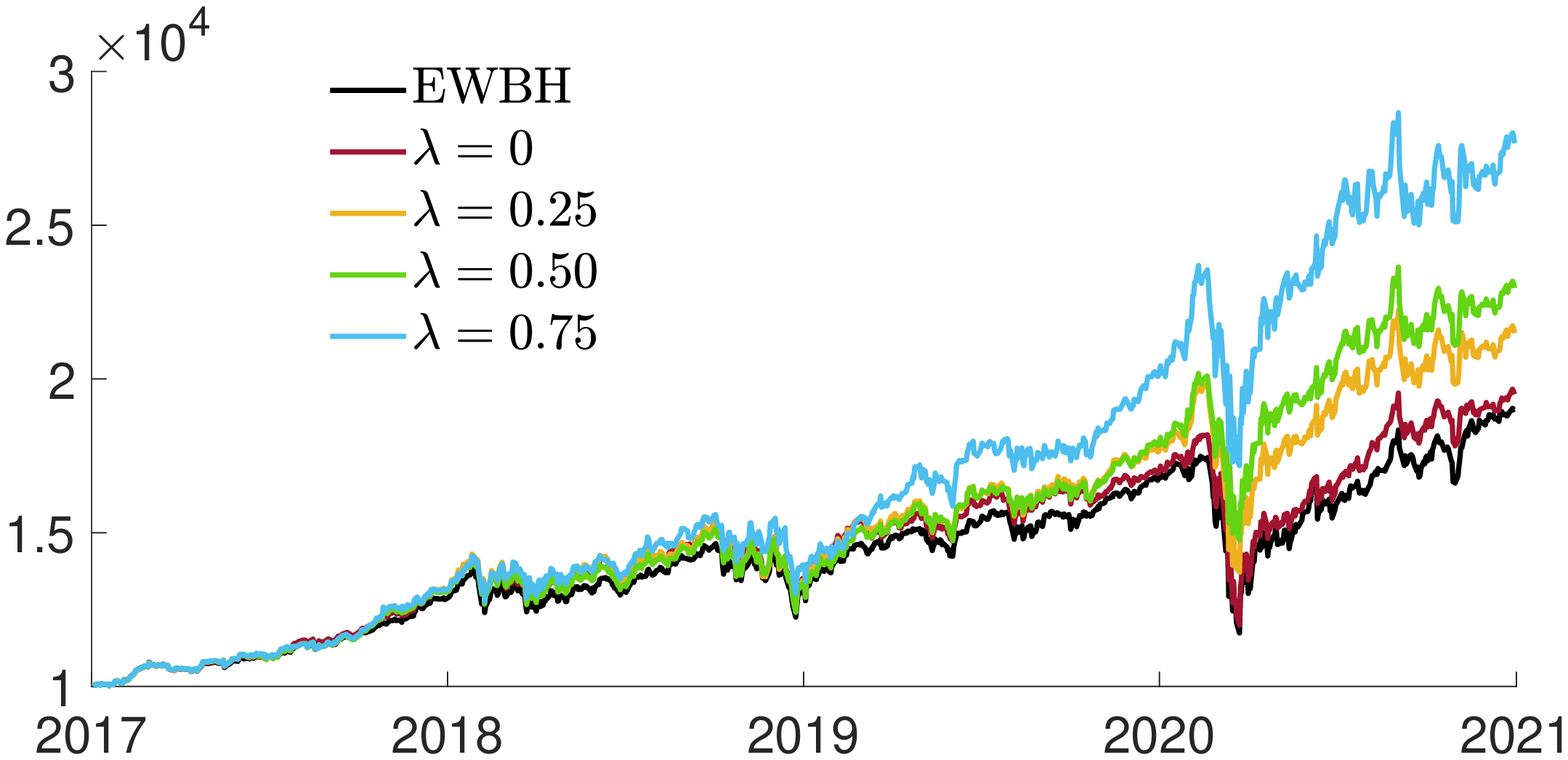}}\hspace{0em}%
	\subcaptionbox{ESG Score}{\includegraphics[width=0.48\textwidth]{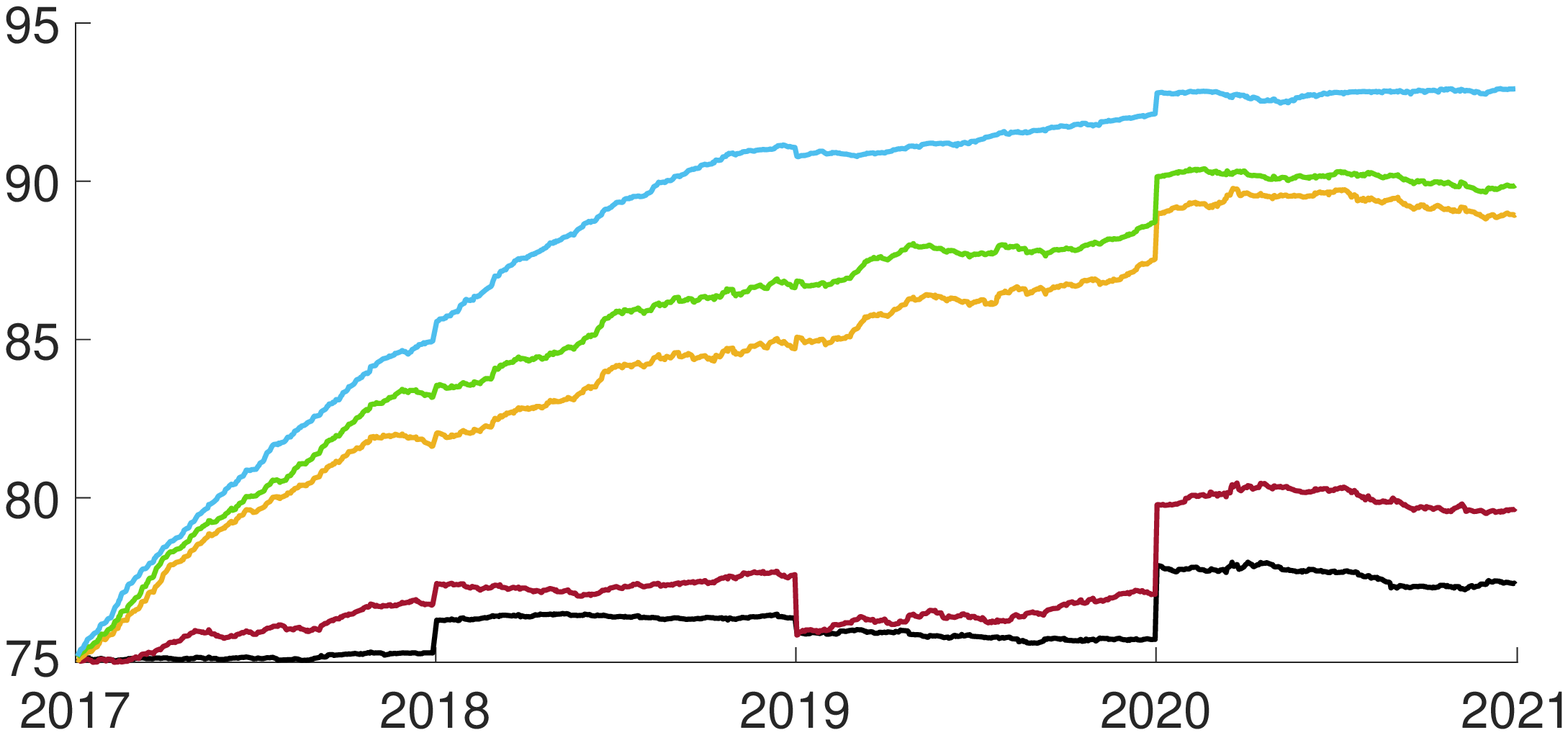}}\hspace{0em}%
	\caption{(a)-(d) Time series over the period 01/03/2017 through 12/30/2020 of price and ESG score for $\alpha = 0.7$
		efficient frontier portfolios obtained from the mCVaR${}_{0.99}$ optimization with
		(a)-(b) no and (c)-(d) 0.4\% daily turnover restriction.
		Plots (e)-(f) are for the tangent portfolios with  0.4\% daily turnover restriction.
		In each, a comparison is made against the equi-weighted porfolio EWBH.
	} 
	\label{fig:ESG_ofs}	
\end{figure}
The time series for both the price (assuming an initial investment of \$10,000) and ESG score for the
$\alpha = 0.7$ point of the $\lambda \in \{0,0.25,0.5,0.75\}$ efficient frontiers of the mCVaR${}_{0.99}$ optimized
portfolios are shown in Fig.~\ref{fig:ESG_ofs}.
The values for the benchmark portfolio, EWBH, are also shown.
As previously noted, the price and ESG time series were computed with a daily turnover constraint of 0.4\%.
For discussion purposes, these time series were also computed under optimization with no turnover constraint.
The unconstrained time series are also plotted in Fig.~\ref{fig:ESG_ofs}.
With no turnover constraints, the ESG optimized ($\lambda > 0$) portfolios outperform the benchmark,
both in cumulative price and portfolio ESG value.
With a realistic daily turnover constraint of 0.4\% to control transaction costs,
the EWBH benchmark outperforms the $\alpha = 0.7$ optimized portfolios until the Covid-19 pandemic.
Post pandemic, the $\lambda = 0.75, \alpha = 0.7$ portfolio exhibits strong recovery
while the $\lambda = 0.5, \alpha = 0.7$ portfolio displays recovery competitive with that of the benchmark.
Each turnover-unconstrained portfolio rapidly establishes a separate ESG level (around which it displays variance).
Under the turnover constraint, the portfolio ESG scores climb to their separate levels more slowly, but with much less
variance.
The noticeable discontinuity in portfolio ESG values on 01/01/2020 arises from rather large changes in the
ESG ratings of a number of these companies in the 12/31/2019 data release compared to that of 12/31/2018.
(See Table ~\ref{tab:Ref_DJ_ESG}.)

We computed moment values for the distribution of the returns for the same ($\lambda,\alpha$) selection of efficient frontier portfolios
as in Table~\ref{tab:Summary_sim}.
Specifically we consider the mean, median, standard deviation (Std), skewness (Skew), and excess kurtosis (ExKurt).
Table~\ref{tab:Moments} summarizes these values for both optimizations.
For each fixed value of $\lambda$, mean, median returns and the Std increase with $\alpha$,
much more so under MV optimization.
Skewness and ExKurt decrease as $\alpha$ increases.

We also computed the portfolios' performance relative to five popular RRRs:
the Sharpe (SR), Sortino, STAR, Rachev and Gini ratios \cite{Cheridito_2013}.\footnote{
	Here we define the Sharpe, Sortino and STAR ratios using only the portfolio returns and
	not the excess returns relative to a benchmark.}
These values are summarized in Table~\ref{tab:RRR_Ref}.
The optimized portfolios generally produce superior RRR values compared to the benchmark over all ranges of $\lambda$ and $\alpha$.
For $\lambda = 0.5$ and $0.75$, the RRRs for MV generally outperform those for mCVaR${}_{0.99}$.

When $\alpha=0$, the choice of $\lambda$ does not significantly alter the performance of the optimal portfolios as the investor is only
concerned with the risk of the investment, which is unaffected by the ESG score.   

\section{ESG-Valued Tangent Portfolios}\label{sec:ESG_TP}

The efficient frontiers in the $(\mathbb{V}[ \hat{Z}^*], \mathbb{E}[ \hat{Z}^*])$ plane
(see Fig.~\ref{fig:EF_orig_Fan_n_30122019}) can be used to identify the \textit{ESG-valued tangent portfolio}.
This approach will naturally lead to an ESG reformulation of the capital asset pricing model (CAPM), the security market line (SML),
and the two-fund separation theorem \cite{Tobin_1958}.
Determining the tangent portfolio for each ESG-valued efficient frontier in this space requires identification of an appropriate
risk-free rate process $r_{f,t}$.

The appropriate risk-free rate to be used should also be ESG-valued as in \eqref{eq:esg_transform}.
This raises the issue of assigning an ESG score to a fixed-income security, in particular to any government bond.
As a first attempt to address this, we proceeded as follows.
We used the yield on the 10-year U.S. treasury bond $r_{f,t}$ (appropriate scaled to daily rates)
and assigned the government bond a maximum ESG score of 100.\footnote{
	Another possibility is to use an ESG score for the appropriate country for a government bond.
	Some agencies are beginning to provide such scores, though not necessarily in the same scale used for companies.}
The motivation behind this choice is that the risk-free rate is not  among the set of choices available to the investor;
any other ESG value for the risk-free rate will be more arbitrary than simply assigning the maximum value.
From \eqref{eq:esg_transform}, the appropriate ESG-valued riskless rate is then
\begin{equation}
\zeta_{f,t}( \lambda )= \lambda \frac{1}{c} + \left( 1-\lambda \right) r_{f,t}. \label{eq:esg_rf}
\end{equation}
The ESG-valued tangent portfolio for date $t$ will be uniquely determine by the straight line from the point
$(0,\zeta_{f,t}(\lambda))$ which is tangent to the $\lambda$-specified ESG-valued efficient frontier.

\begin{figure}[th]
	\centering
	{\includegraphics[width=0.49\textwidth]{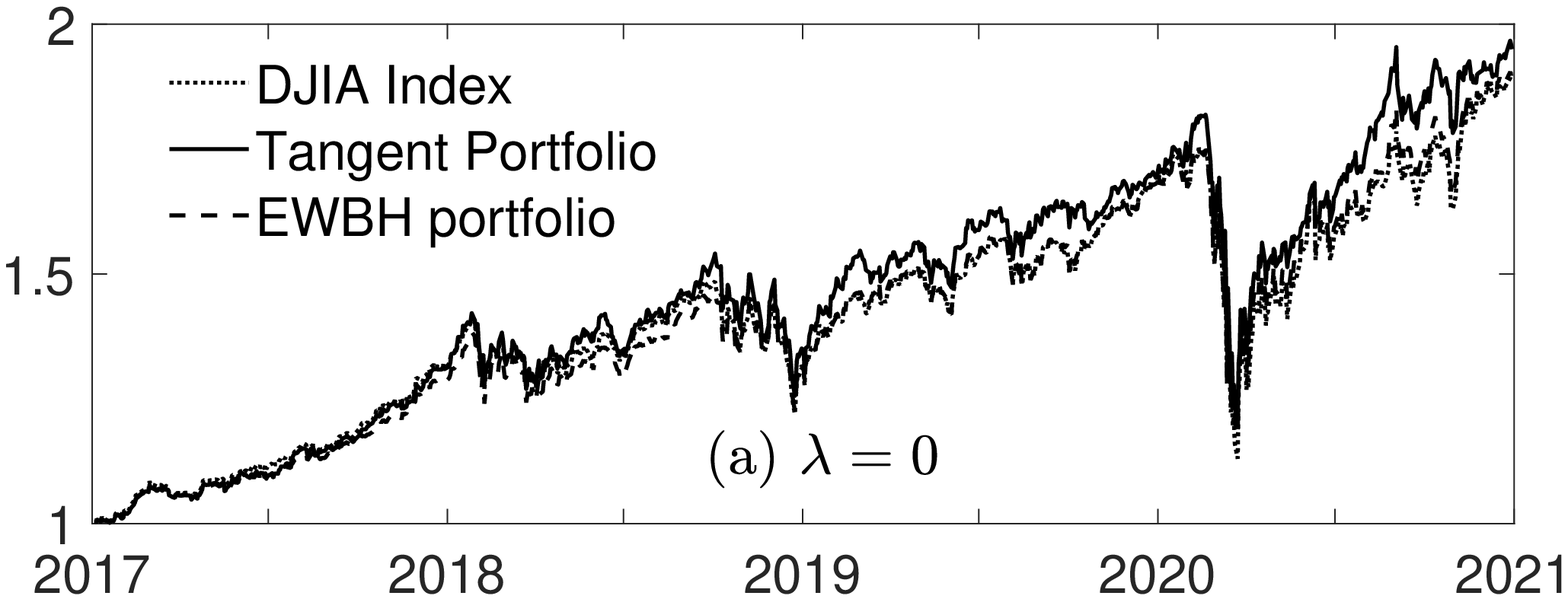}}        \hspace{0em}%
	{\includegraphics[width=0.49\textwidth]{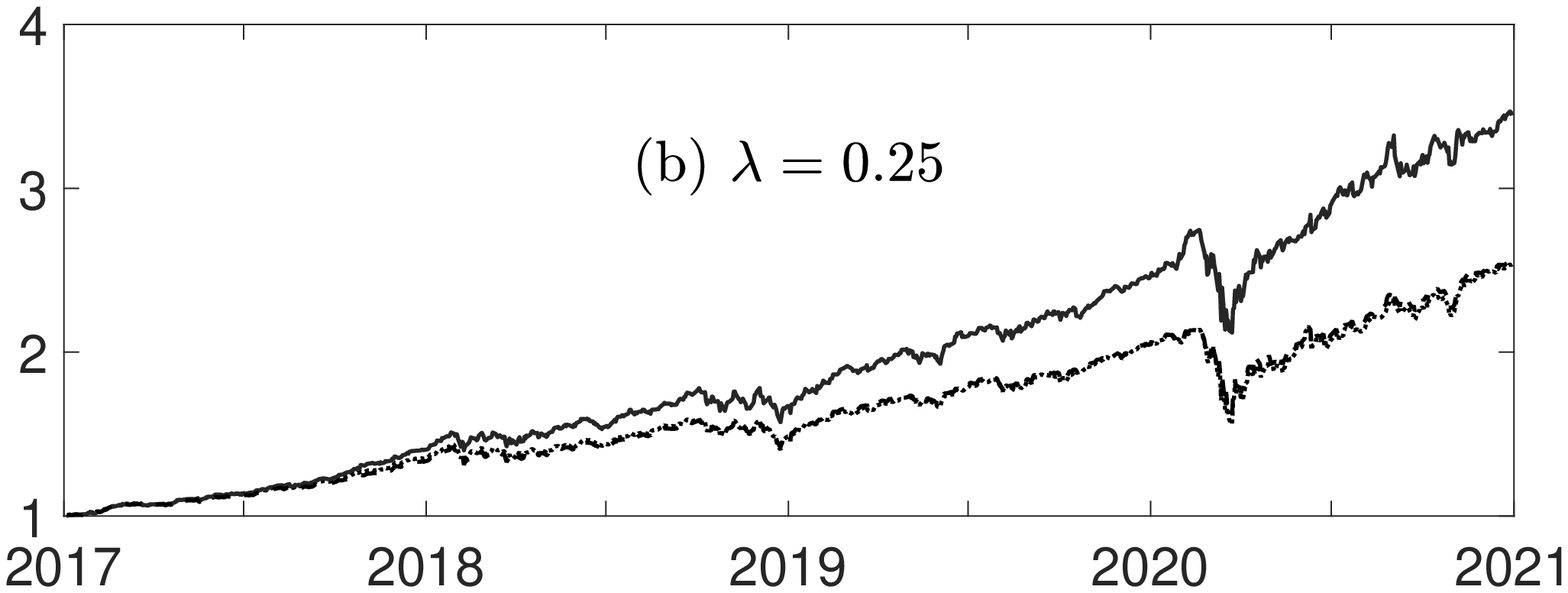}}\hspace{0em}%
	{\includegraphics[width=0.49\textwidth]{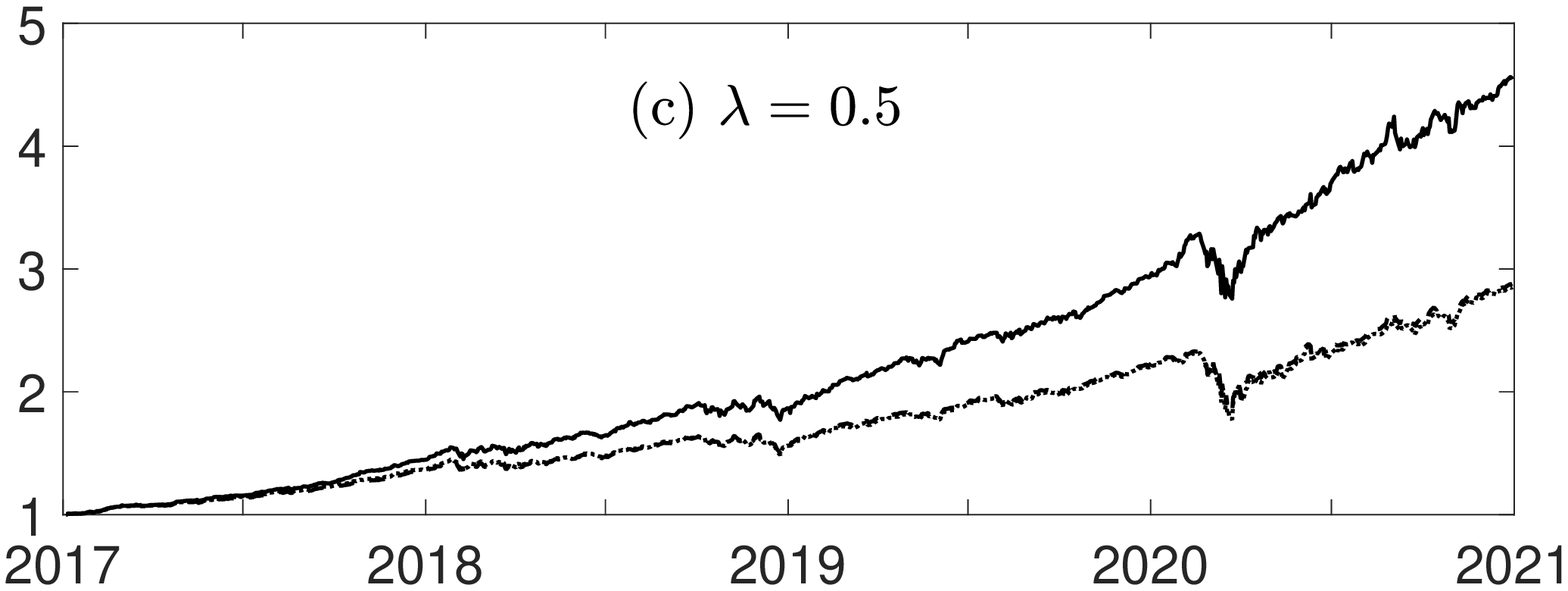}}\hspace{0em}%
	{\includegraphics[width=0.49\textwidth]{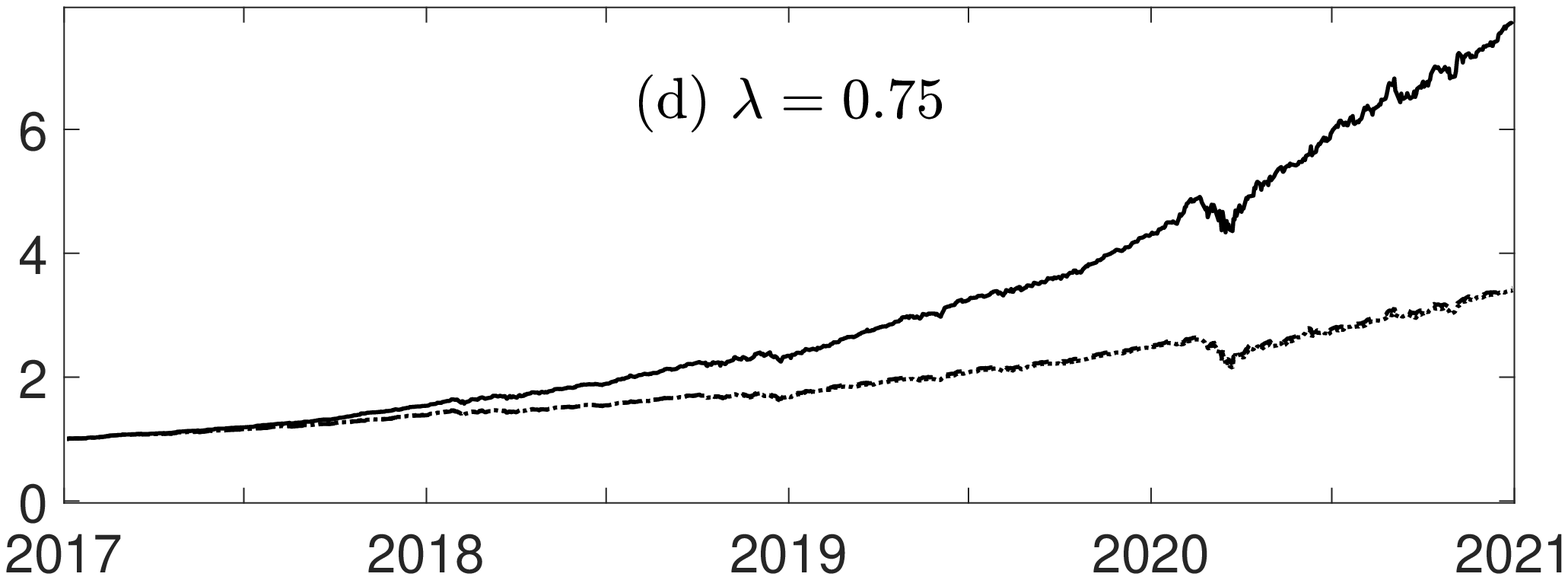}}\hspace{0em}%
	{\includegraphics[width=0.49\textwidth]{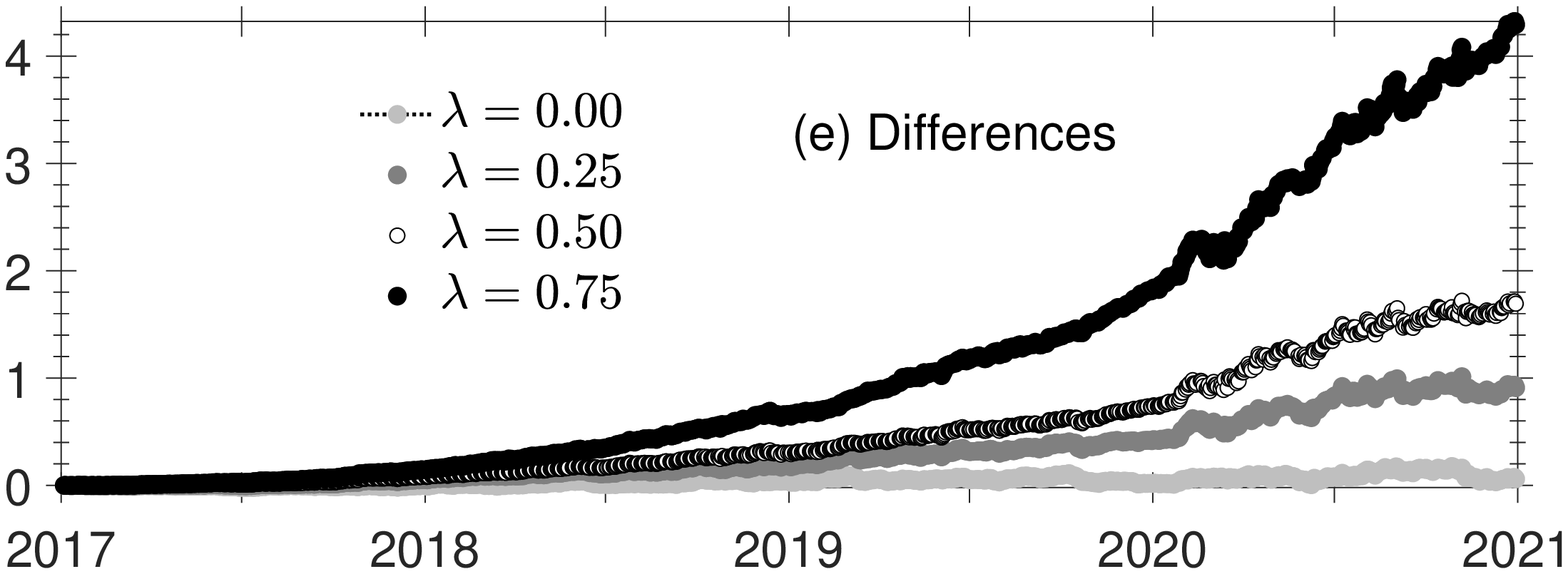}}\hspace{0em}%
	\caption{(a)-(d) ESG-valued portfolio values, $P^{(Z)}_t (\lambda)$, obtained for the tangent portfolios under
	mCVaR${}_{0.99}$ optimization,
	compared with the corresponding $P^{(Z)}_t (\lambda)$ time series for the EWBH strategy and the ESG-valued DJIA index.
	(e) Difference between the $P^{(Z)}_t (\lambda)$ values for the tangent portfolios and the ESG-valued DJIA index.}
	\label{fig:Capm-esg-adj-ofs-MCVaR}	
\end{figure}
Using \eqref{eq:esg_rf}, the ESG-valued tangent portfolio can be identified on any efficient frontier parameterized by
($\lambda, \alpha, t$).\footnote{
	It is critical to emphasize that the tangent portfolios are identified using the efficient frontiers computed in
	$(\mathbb{V}[ \hat{Z}^*], \mathbb{E}[ \hat{Z}^*], \text{ESG}^*)$ space and not in
	$(\mathbb{V}[ \hat{R}^*], \mathbb{E}[ \hat{R}^*], \text{ESG}^*)$ space.
	The optimization is performed in the former space, not the latter.
	As can be seen by comparing Figs.~\ref{fig:EF_orig_Fan_n_30122019} and \ref{fig:EF_Fan_n_30122019},
	convexity of the efficient frontiers can only be guaranteed in the former space.
}
These ESG-valued tangent portfolios were identified for the (0.4\% daily turnover constrained) portfolios computed in
Section~\ref{sec:ESG_Perf}.
Fig.~\ref{fig:ESG_ofs} shows the price performance (obtained from the returns \eqref{eq:PortRet})
as well as the ESG scores for the tangent portfolios computed with mCVaR${}_{0.99}$ optimization.
For $\lambda > 0$, the price performance of these turnover constrained tangent portfolios is
competitive with that obtained by the turnover unconstrained, $\alpha = 0.7$ efficient frontier portfolios 
(compare Figs.~\ref{fig:ESG_ofs} (a) and (e)).
Allowing for the initial transient behavior due to the turnover constraint,
for $\lambda > 0$ the ESG scores for the turnover constrained tangent portolios are superior to those
for the turnover unconstrained, $\alpha = 0.7$ efficient frontier portfolios 
(compare Figs.~\ref{fig:ESG_ofs} (b) and (f)).

The $P^{(Z)}_t (\lambda)$ time series are displayed in Fig.~\ref{fig:Capm-esg-adj-ofs-MCVaR} for these mCVaR${}_{0.99}$
optimized tangent portfolios.
$P^{(Z)}_t (\lambda)$ time-series are also shown for the EWBH benchmark as well as for an ESG-valued DJIA index.\footnote{
	With reference to equations \eqref{eq:PortRet} and \eqref{eq:esg_PortRet}, the DJIA index is defined as follows.
	$I = 29$. No scenario sets are generated;
	$\theta_i^* = w_i^{\text{DJ}}/(1-w_{\text{Dow}})$ where $w_i^{\text{DJ}}$ is the DJIA published weight for component $i$
	and $w_{\text{Dow}}$ is the published weight for Dow Inc., which is excluded from our benchmark index.
	We utilized the published DJIA weights applicable to 30/12/2020 in composing our index.
	}
Note the close agreement between the EWBH and ESG-valued index values;
both portfolios are comprised of the same stocks, with different weights applied.
The differences between  ESG-valued tangent portfolio and the ESG-valued index values are also plotted in
Fig.~\ref{fig:Capm-esg-adj-ofs-MCVaR}.
As noted earlier, the $\lambda = 0$, ESG-valued plots are just scaled versions of the corresponding price plots.
As $\lambda$ increases, the differences between the optimized portfolio ESG-valued time series and those of the benchmarks
(Fig.~\ref{fig:Capm-esg-adj-ofs-MCVaR} (e)) grow more rapidly than the corresponding differences between optimized portfolio
and benchmark price series (not shown, but can be inferred from Fig.~\ref{fig:ESG_ofs}).

Table~\ref{tab:Summary_sim_tang} provides the performance, moment, and RRR measures for these four tangent portfolios.
At the higher values of $\lambda$, the tangent portfolios outperform the benchmark.
Compared to the benchmark, total (and hence annualized) return almost doubles for the tangent portfolio as $\lambda$ increases.
With the exception of one Gini value, the value for every RRR considered is better than the benchmark as $\lambda$ increases.
Trivially the tangent portfolio average ESG score improves with $\lambda$.
(The standard deviation of ESG decreases because more weight is given to the ESG score which remains constant over each year.)
The average turnover considerably decreases  as $\lambda$ increases.
 
\section{ESG-Valued Option Valuation}\label{sec:ESG_OP}

We consider the effect of ESG-valued returns on option pricing.
Paralleling equations \eqref{eq:port_value} and \eqref{eq:esg_value}, we shall refer to
the valuations given to options using ESG-valued returns as ``ESG-valued option values'' or simply
``option values'' and reserve the phrase ``option price'' to refer to the valuation of options based
on traditional financial returns.
When $\lambda = 0$, ESG-valued option values become equal to option prices.
For a given date $t$, using the discrete methodology outlined in Appendix~\ref{sec:App_C},
we computed European contingency claim option values for a sequence of expiration dates
$t+T$, $T \in [T_{\text{min}}, T_{\text{max}}]$ and strike values $K \in [K_{\text{min}},K_{\text{max}}]$ using
an ESG-valued tangent portfolio as the underlying.
We chose the date $t = 12/30/2019$ corresponding to the efficient frontiers discussed in
Sections~\ref{sec:ESG_EF} and \ref{sec:ESG_TP}.
Option values were computed for each tangent portfolio.
The same methodology was employed to obtain option values for the same date using the ESG-valued DJIA index as the underlying.
We are interested in the difference between option values based upon the adjusted DJIA index and the tangent portfolios.   

\begin{figure}[h!]
	\centering
								{\includegraphics[width=0.24\textwidth]{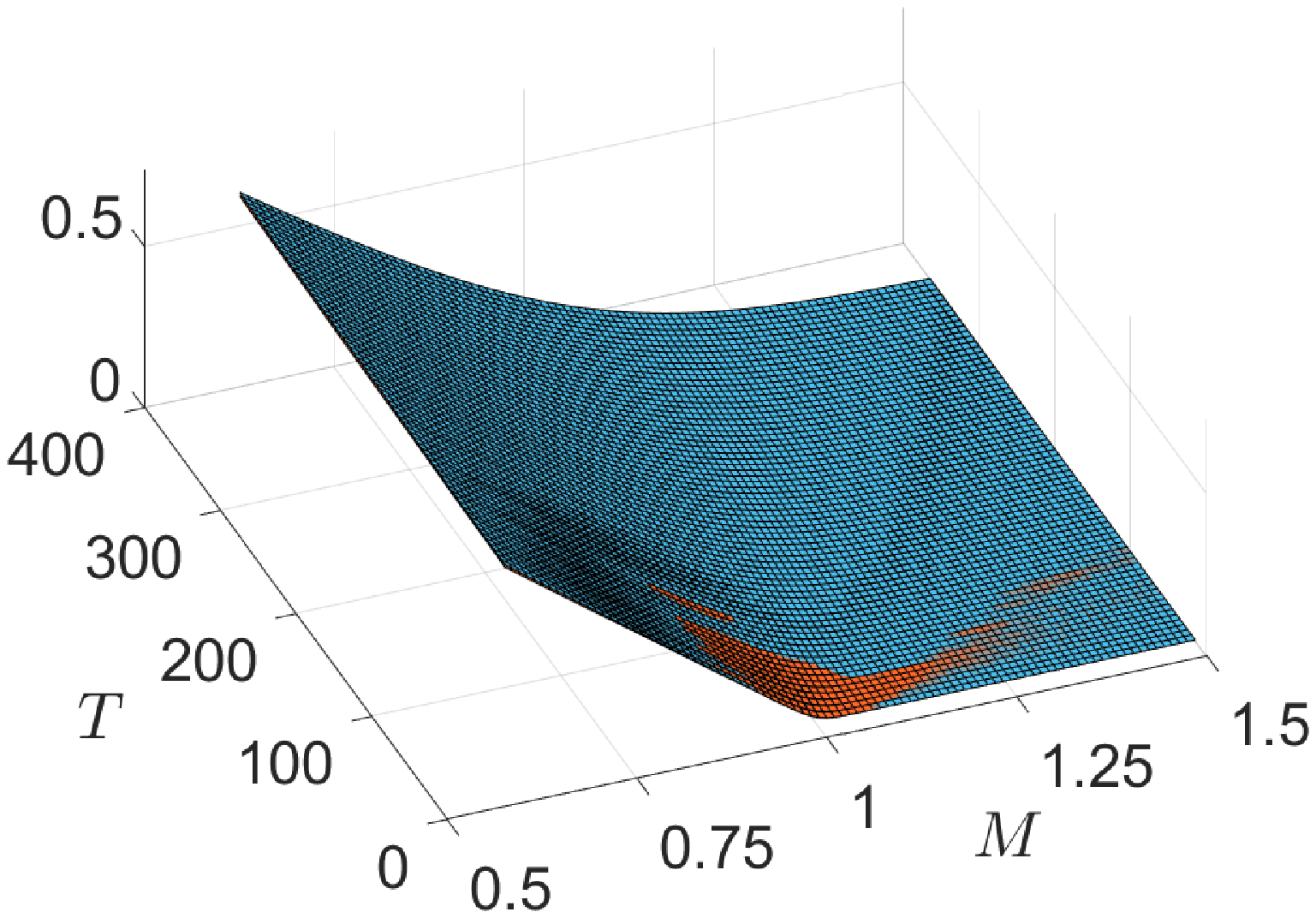}}\hspace{0em}%
								{\includegraphics[width=0.24\textwidth]{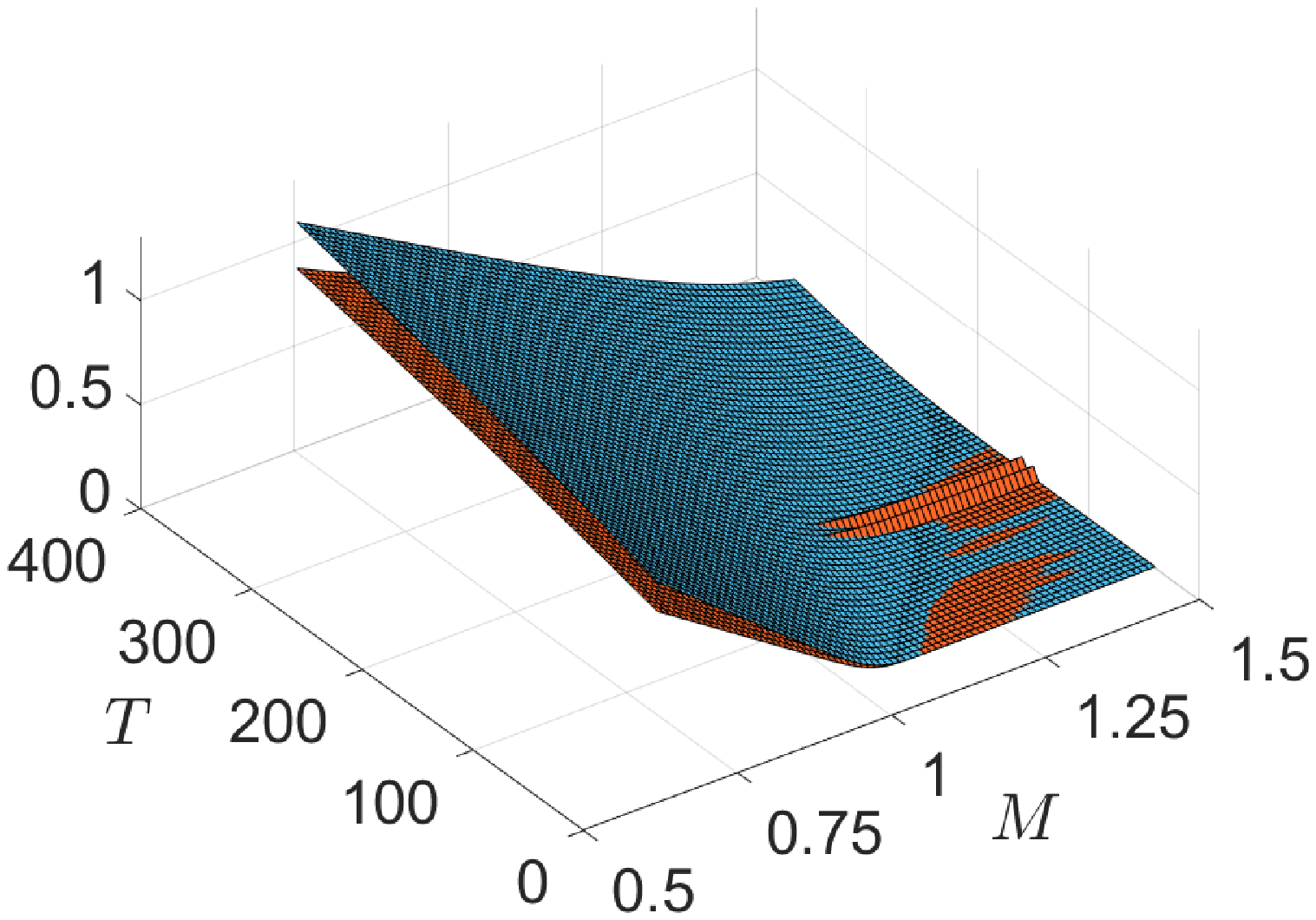}}\hspace{0em}%
								{\includegraphics[width=0.24\textwidth]{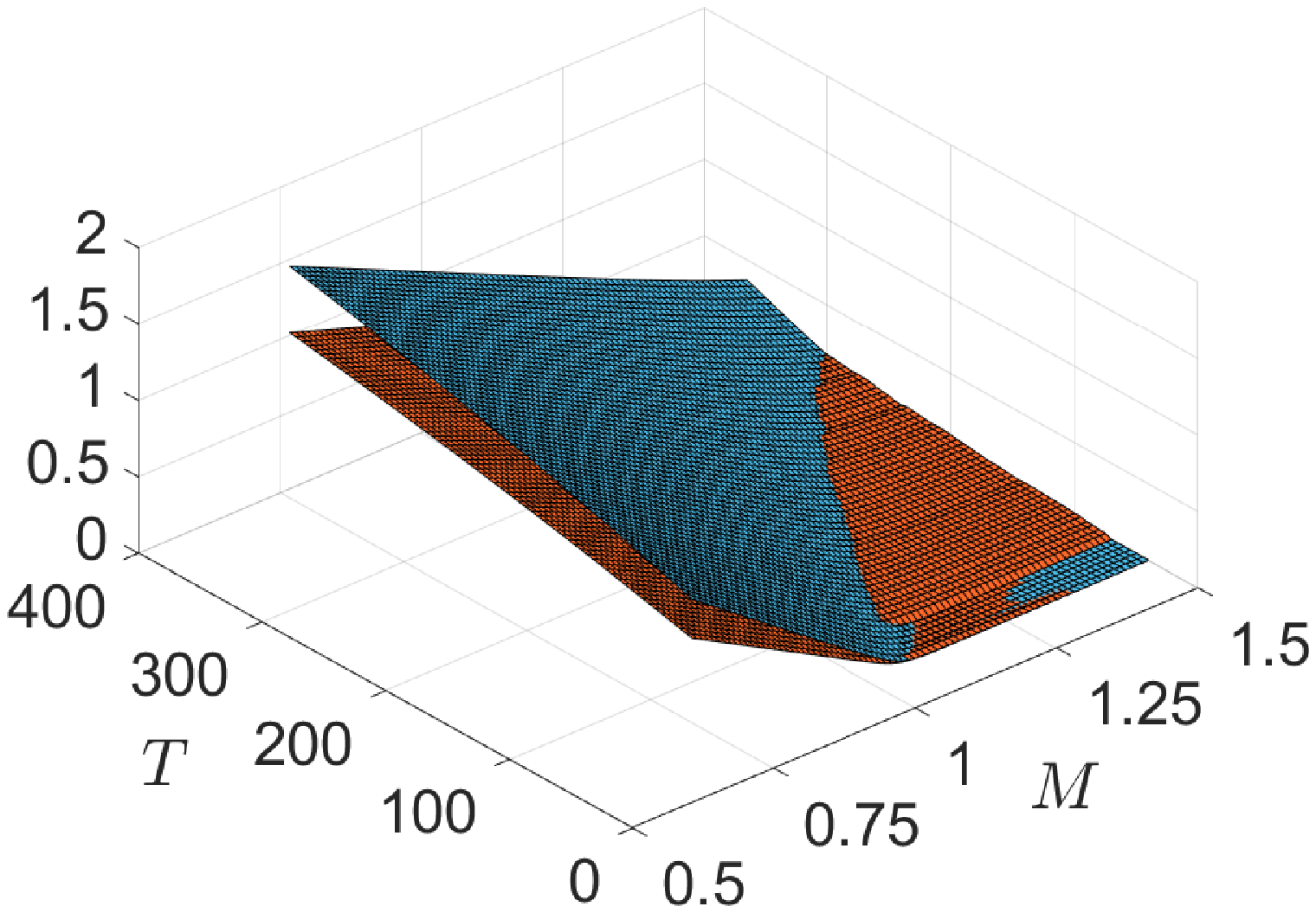}}\hspace{0em}%
								{\includegraphics[width=0.24\textwidth]{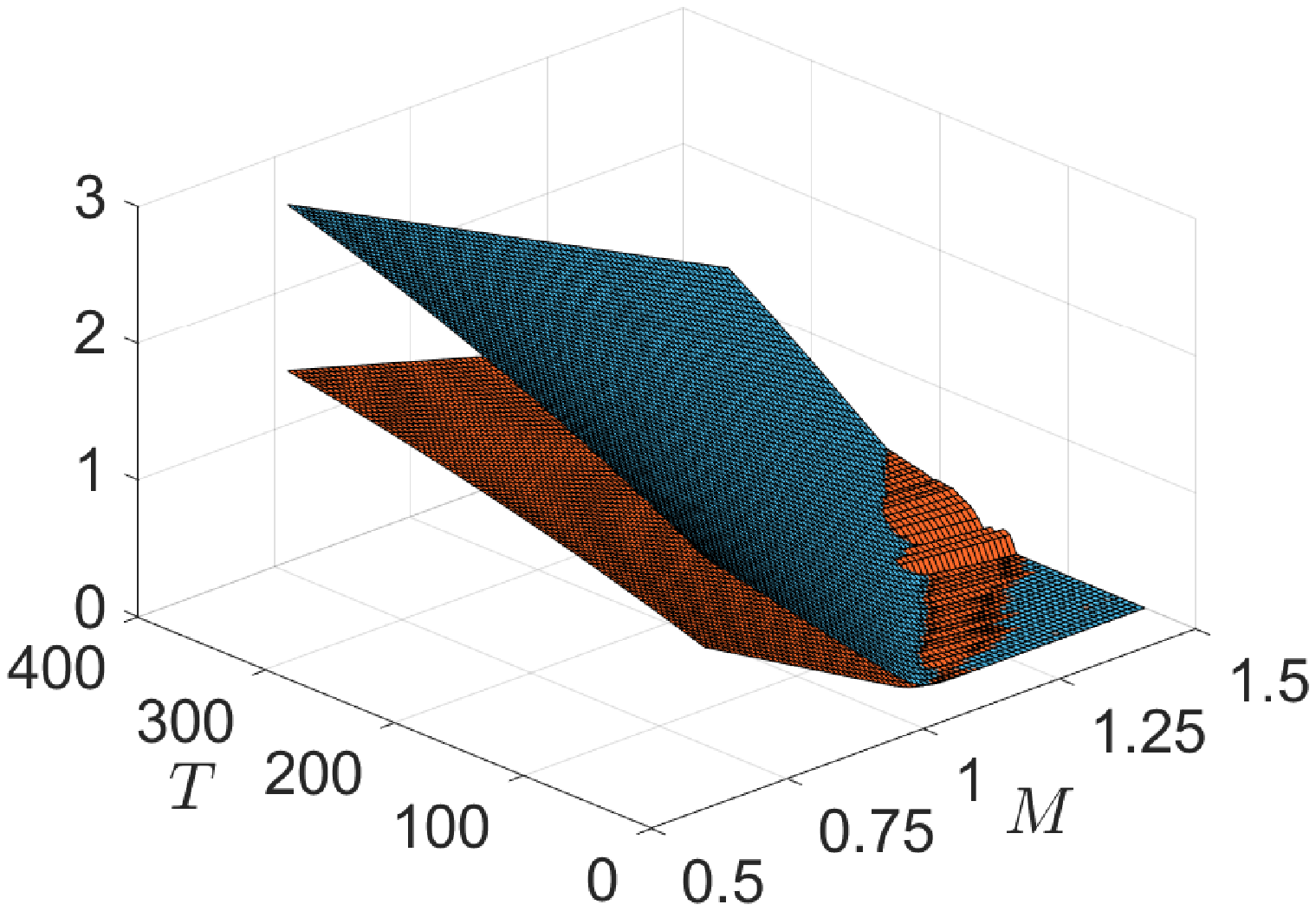}}\hspace{0em}%
	\subcaptionbox{$\lambda=0$    }{\includegraphics[width=0.24\textwidth]{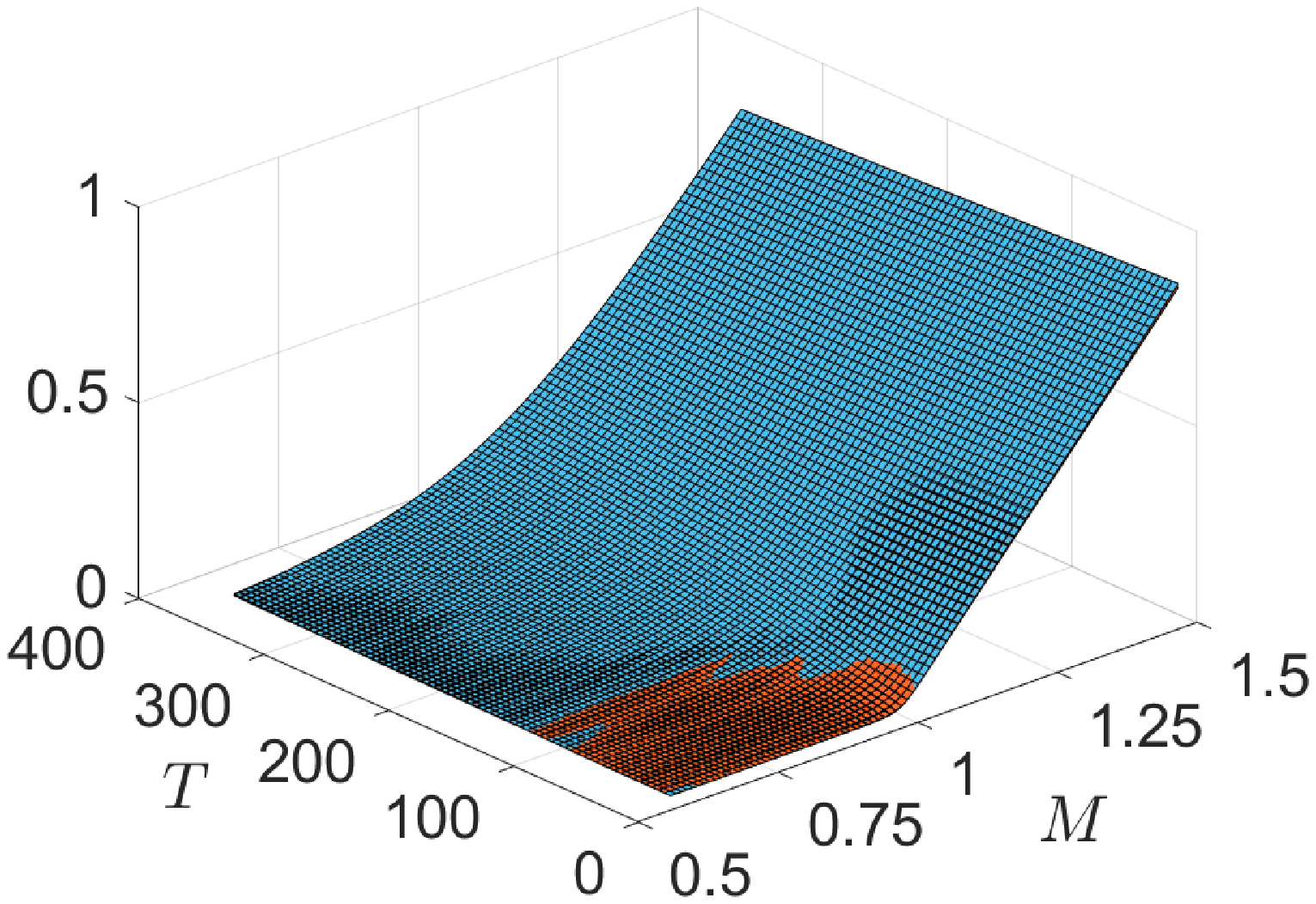}}\hspace{0em}%
	\subcaptionbox{$\lambda=0.25$}{\includegraphics[width=0.24\textwidth]{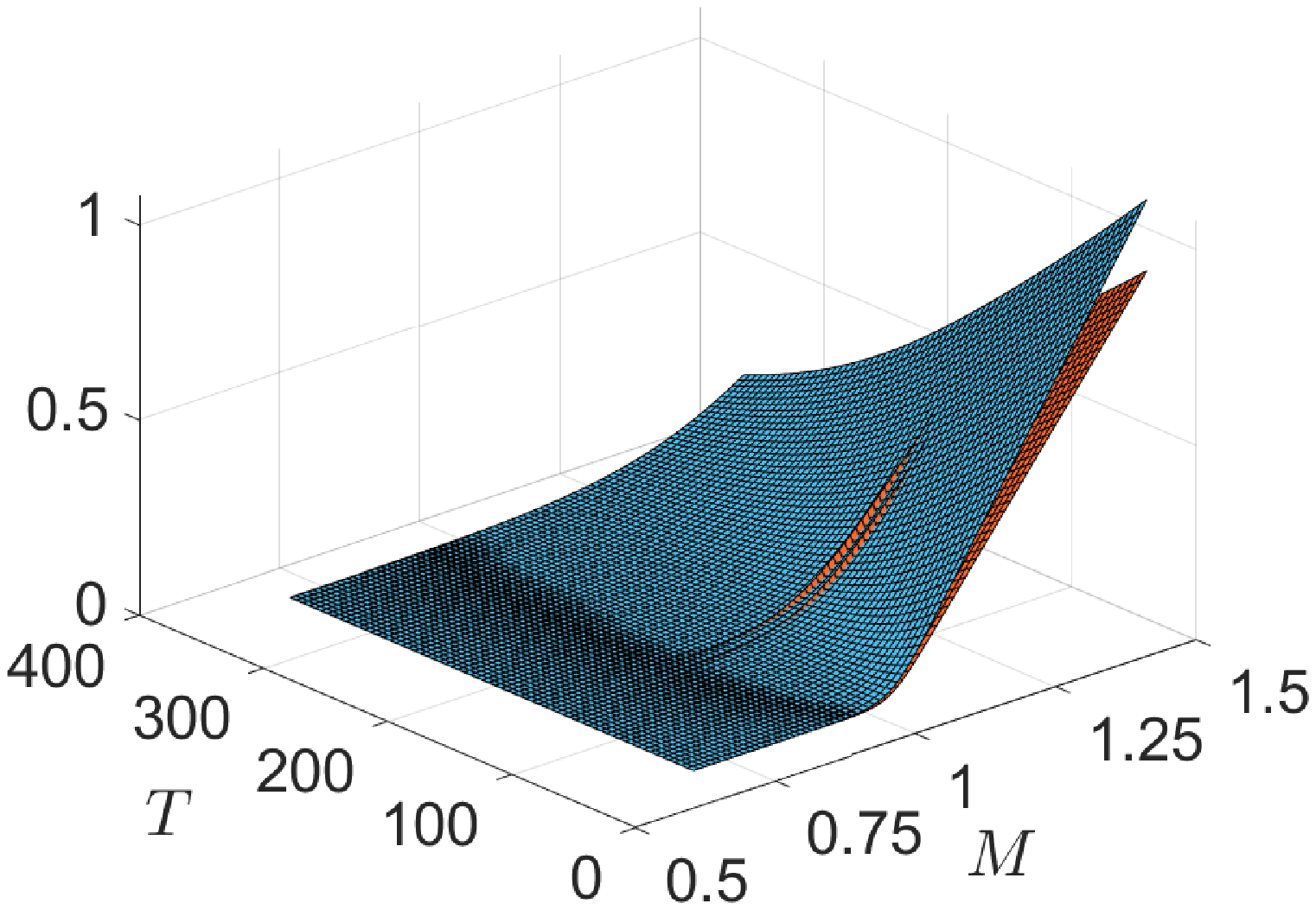}}\hspace{0em}%
	\subcaptionbox{$\lambda=0.5$ }{\includegraphics[width=0.24\textwidth]{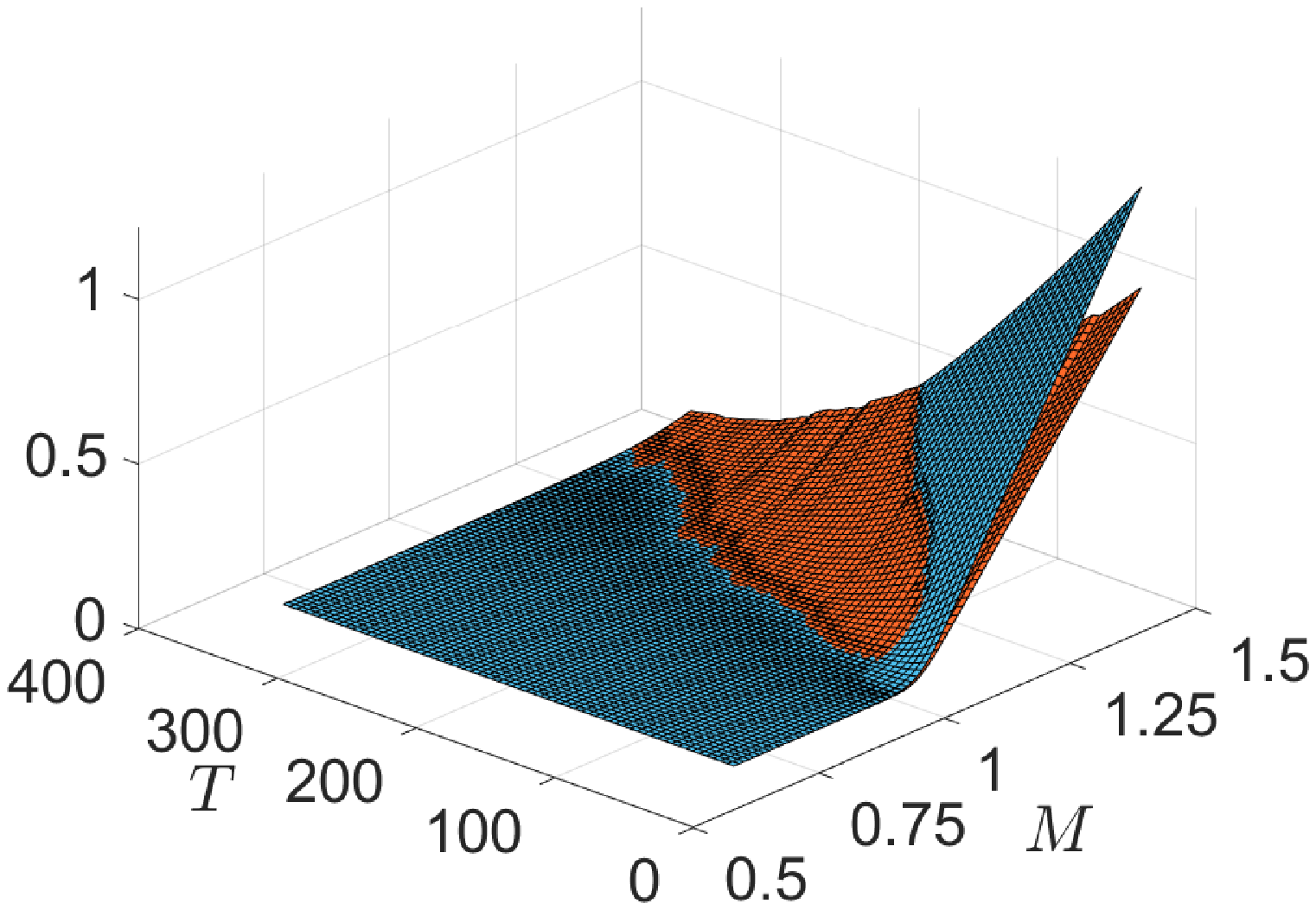}}\hspace{0em}%
	\subcaptionbox{$\lambda=0.75$}{\includegraphics[width=0.24\textwidth]{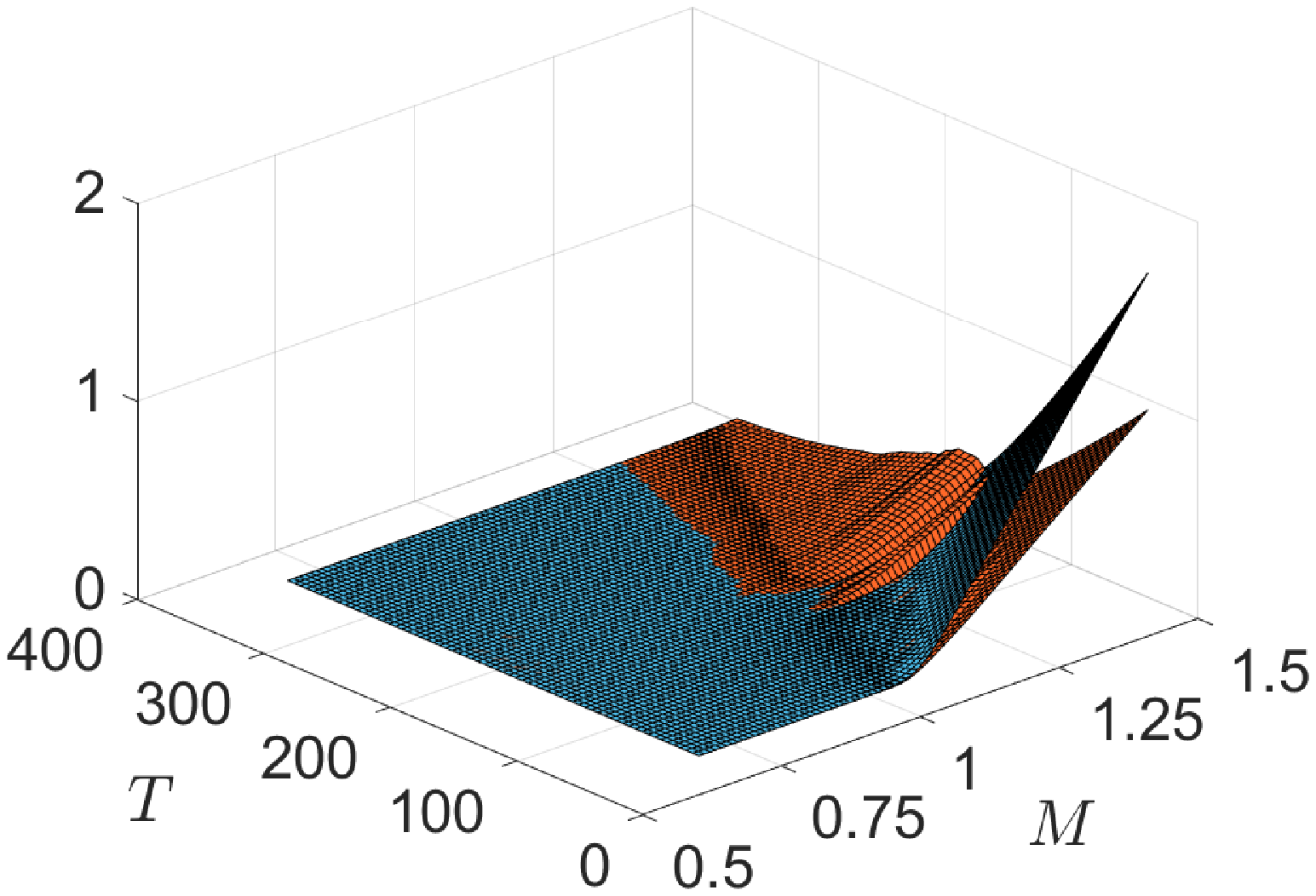}}\hspace{0em}%
	\subcaptionbox{Call}			{\includegraphics[width=0.33\textwidth]{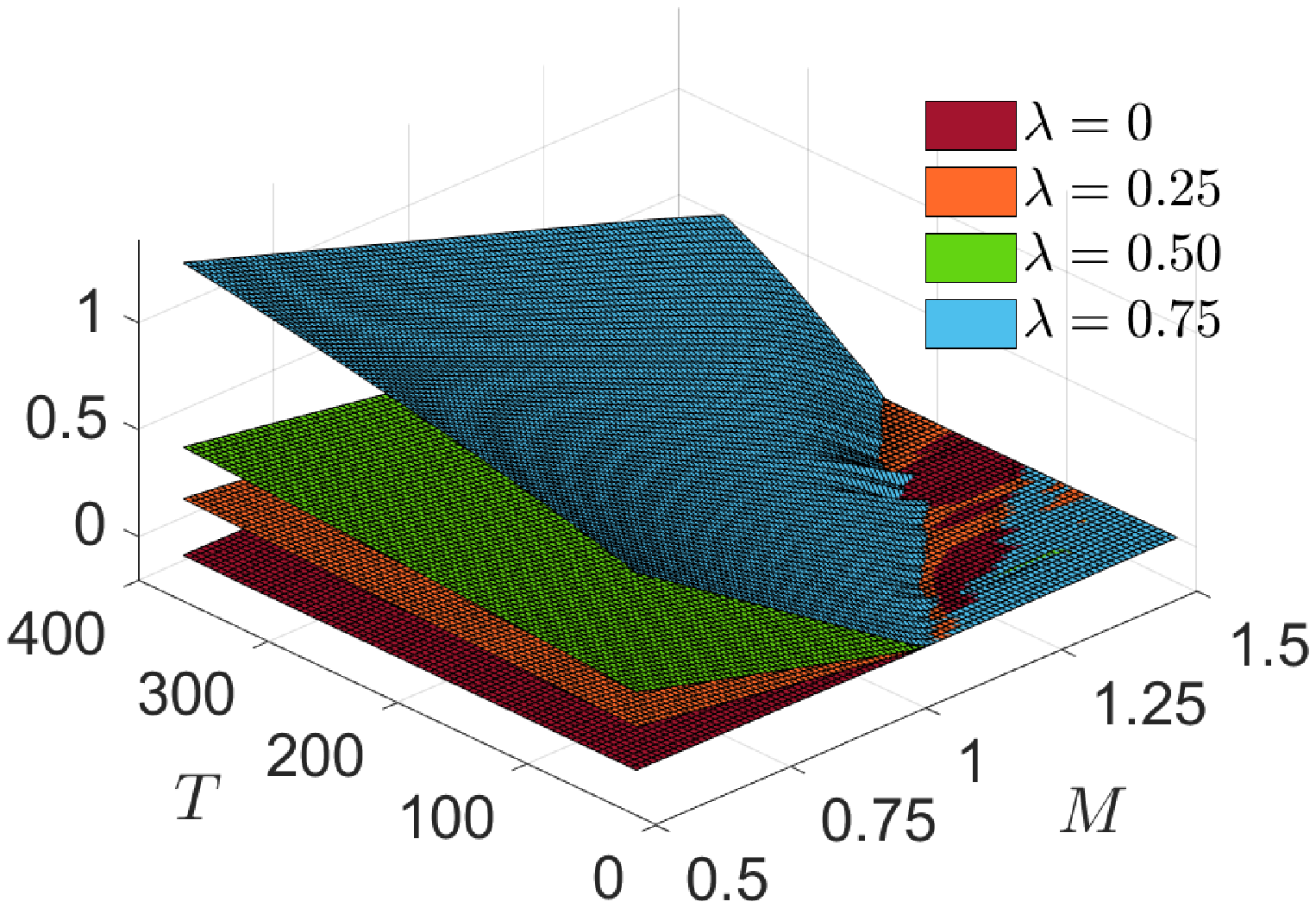}}\hspace{0em}%
	\subcaptionbox{Put}			{\includegraphics[width=0.33\textwidth]{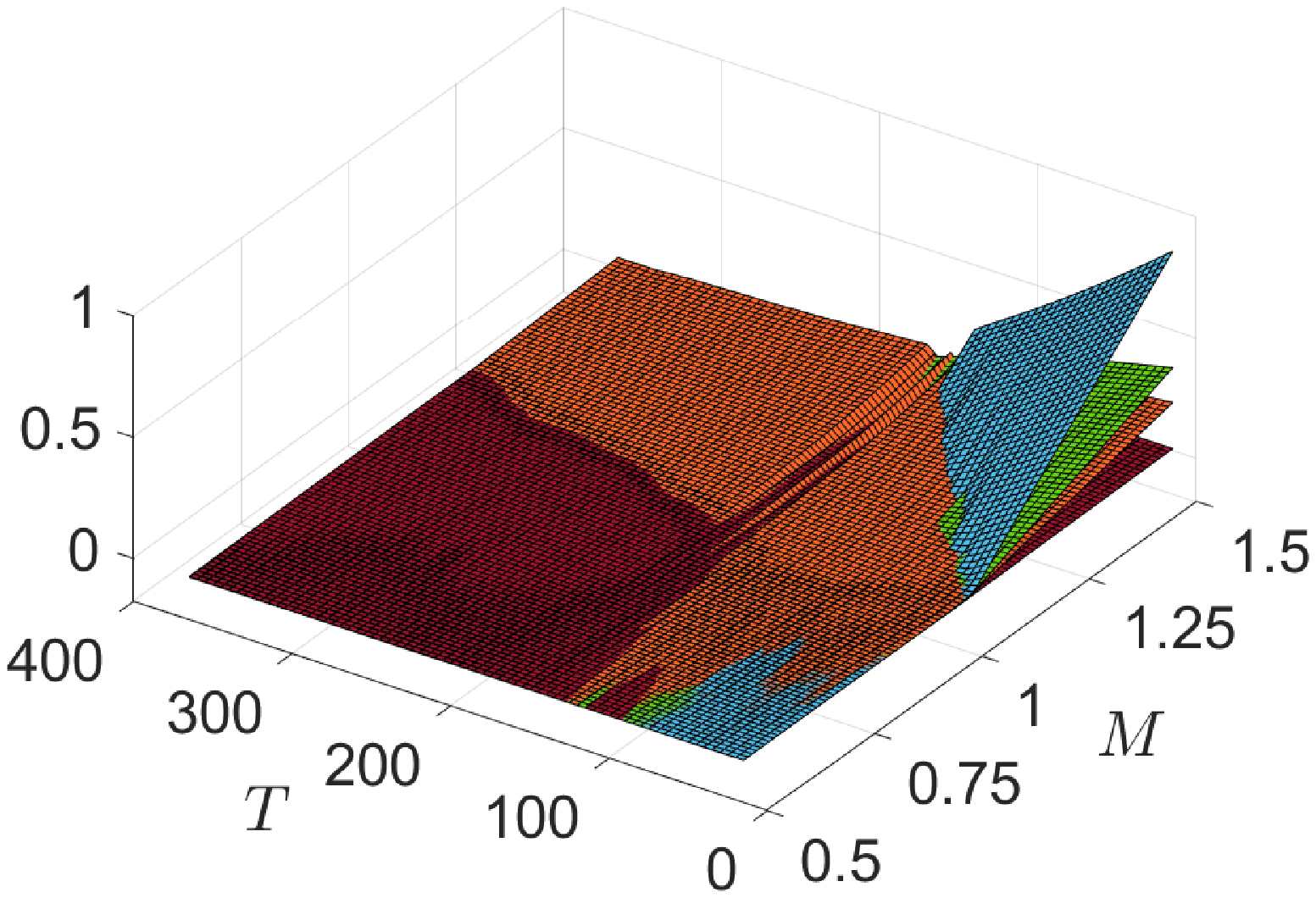}}\hspace{0em}%
	\caption{ \small (a) to (d): (top) call and (bottom) put option values for select choices of $\lambda$.
		The orange surfaces corresponds to option values for the ESG-valued DJIA index;
		blue surfaces represents option values for the mCVaR${}_{0.99}$ tangent portfolios.
		The difference in (e) call and (f) put option values between the ESG-valued DJIA index
		and the mCVaR${}_{0.99}$ tangent portfolios.
	} 
	\label{fig:Call_Put_values}	
\end{figure}

Call and put option values were expressed as functions of $T$ and ``ESG-moneyness'' $M = K/P^{Z}_t$,
where $P^{Z}_t$ is the ESG-valued price of the relevant portfolio on day $t$.
We considered ESG-moneyness values in the range $M \in [0.5,1.5]$ and maturity times $T \in [15,252]$ days.
Comparisons of the call and put value surfaces based on the ESG-valued DJIA index and on the mCVaR${}_{0.99}$
optimized tangent portfolio for selected values of $\lambda$ are given in Fig.~\ref{fig:Call_Put_values}.
The respective call and put option value differences,
$\Phi_c^{(\text{mCVaR}_{0.99})}(T,K,\lambda) - \Phi_c^{(\text{DJIA})}(T,K,\lambda)$ and
$\Phi_p^{(\text{mCVaR}_{0.99})}(T,K,\lambda) - \Phi_p^{(\text{DJIA})}(T,K,\lambda)$,
between the tangent portfolios and the DJIA index are also given in Fig.~\ref{fig:Call_Put_values}.

When $\lambda=0$, call and put values (and hence, call and put option prices)
written on the index and the mCVaR${}_{0.99}$ tangent portfolio are practically equal.
The difference increases with $\lambda$, especially as strike values move further into-the-money.
By construction, ESG-valued tangent portfolios have a higher ESG score,
which in turn implies a positive shift in their aggregate value, creating higher call and put values.
The effect of a positive $\lambda$ on call and put values is different as a function of time to maturity;
at constant in-the-money values of $K$, call values increase with $T$ while put values decrease.

\begin{figure}[h!]
	\centering		
	\subcaptionbox{$\lambda=0$}     {\includegraphics[width=0.24\textwidth]{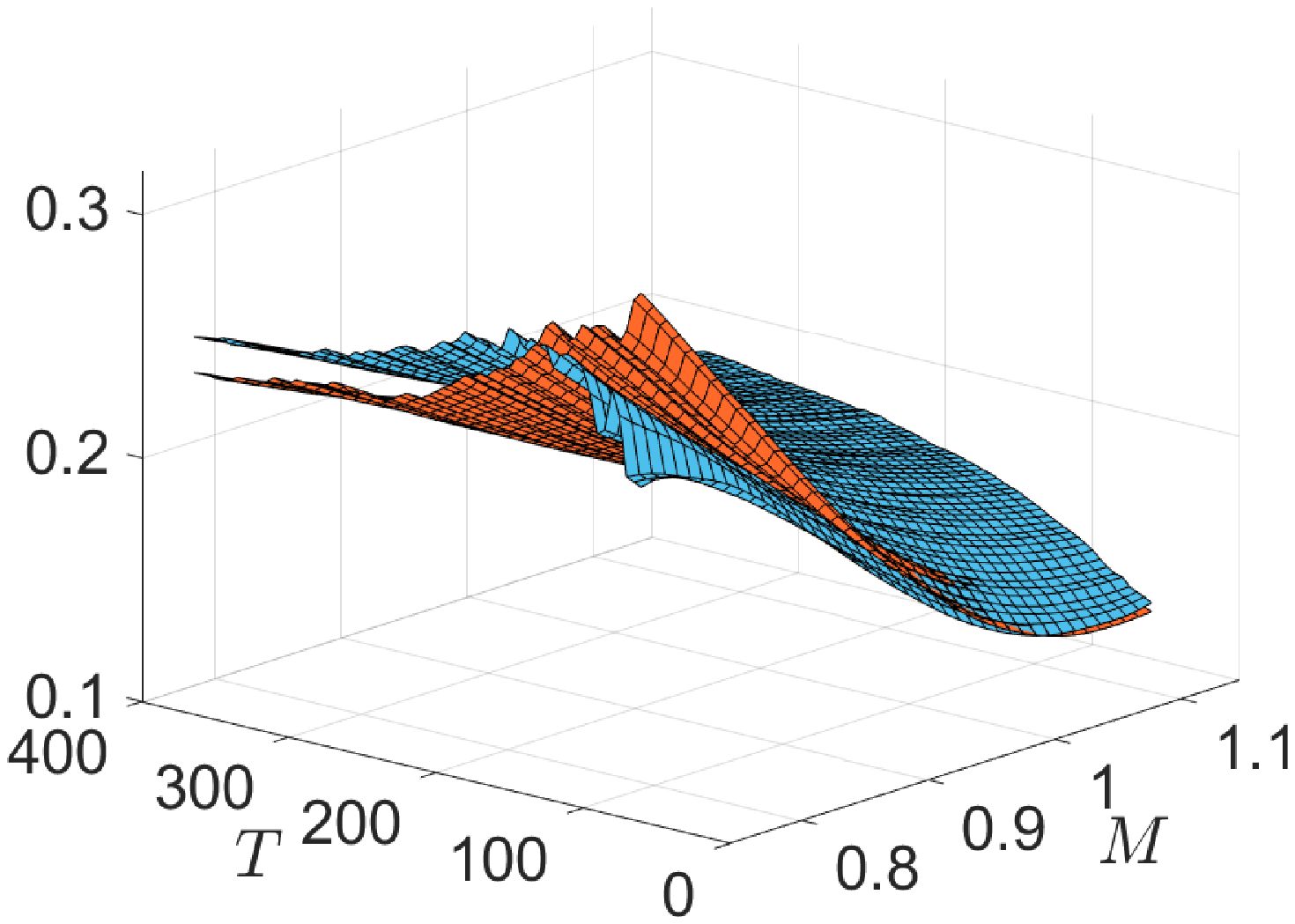}   } \hspace{0em}%
	\subcaptionbox{$\lambda=0.25$} {\includegraphics[width=0.24\textwidth]{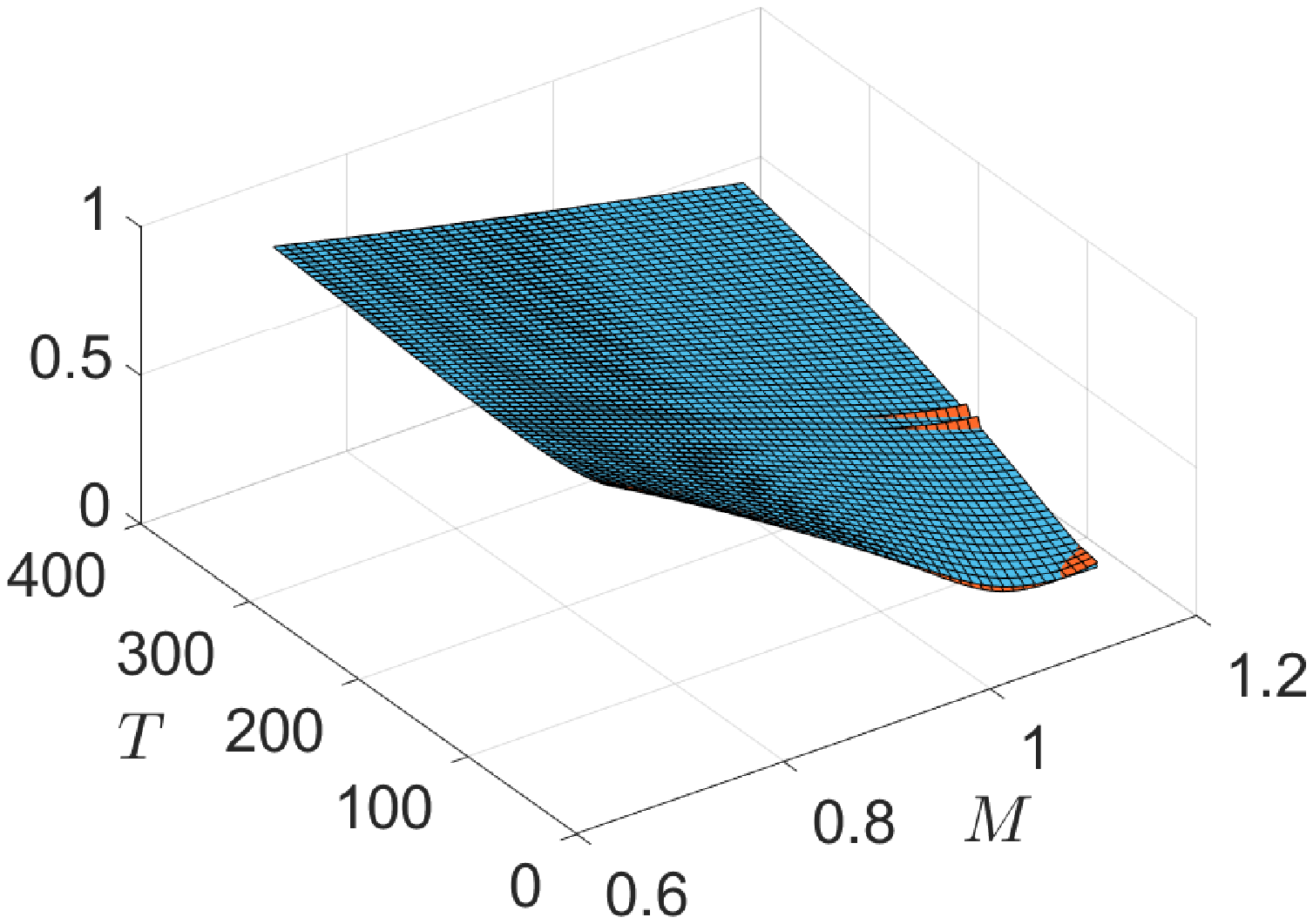}} \hspace{0em}%
	\subcaptionbox{$\lambda=0.5$}  {\includegraphics[width=0.24\textwidth]{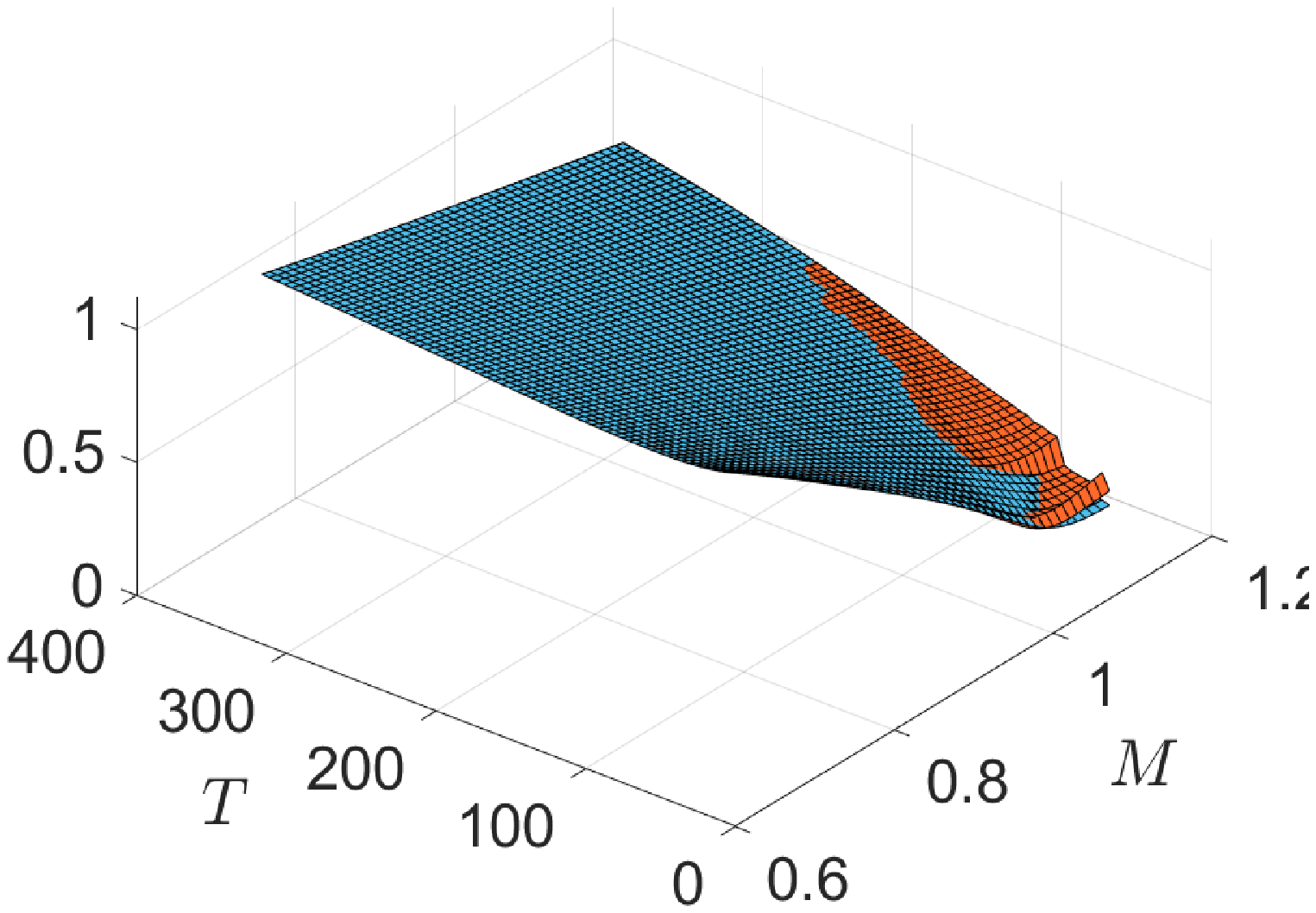}} \hspace{0em}%
	\subcaptionbox{$\lambda=0.75$}{\includegraphics[width=0.24\textwidth]{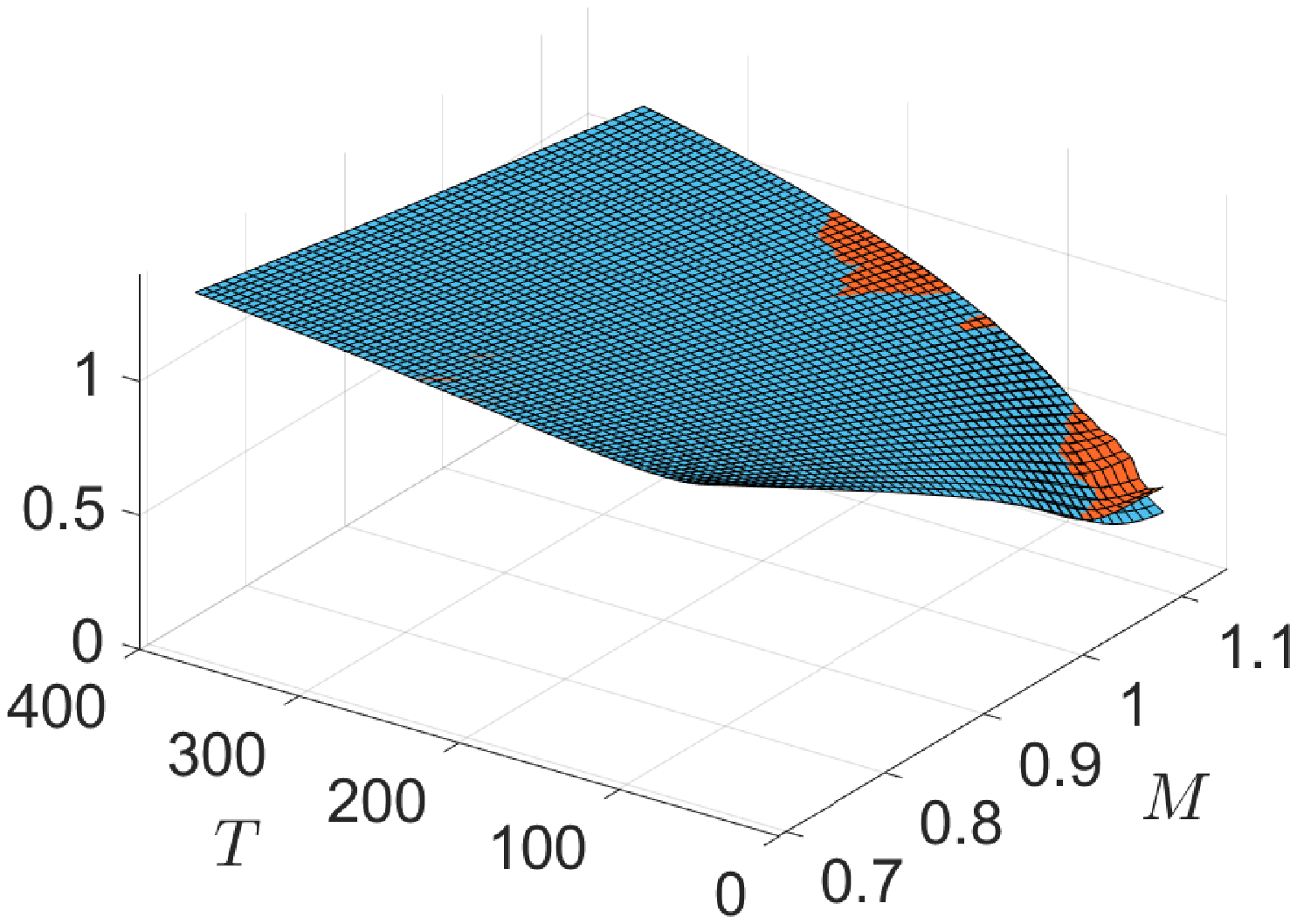}} \hspace{0em}%
	\caption{ \small Implied volatility surfaces computed from the (orange surfaces) ESG-valued DJIA option values and
		(blue surfaces) the tangent portfolios corresponding to the selected choices for $\lambda$.} 
	\label{fig:IV_Prices}	
\end{figure}
Fig.~\ref{fig:IV_Prices} compares implied volatility (IV) surfaces, computed by fitting to the Black-Scholes model,
for call options based on the DJIA index and the mCVaR${}_{0.99}$-optimized tangent portfolios.
For $\lambda = 0$ and in-the-money $K$ values, the IV surface of the DJIA index is above that of the tangent portfolio for
short maturity times.
For longer maturity dates it drops below.
As $\lambda$ increases, the two IV surfaces move upwards (overall volatility increases);
the difference between the surfaces decreases; and the volatility ``smirk'' becomes steeper.

\section{ESG-Valued Shadow Riskless Rate}\label{sec:ESG_SR}

The existence of a risk-free rate in the economy is a common and implicit assumption of modern financial theory. 
However, the situation in which the risk-free rate is not available for a market has been also considered. 
The seminal work is that of Black \cite{Black_1972}, who proposed a CAPM model without assuming the existence of a riskless rate.
Black later introduced a ``shadow real rate'' as an option on fixed-income derivatives \cite{Black_1995}.
More recently, a different approach has been  introduced by Rachev at al.  \cite{Rachev_2017},
in which a shadow riskless rate is defined as a perpetual option
on the set of risky securities selected as the investment universe.
Rachev at al. computed analytic formulas for the shadow riskless rate under different models of the stochastic process driving the
market uncertainty.
The simplest case, where diffusion is driven by Brownian motions, is summarized in Appendix~\ref{sec:App_D}.

We analyze the effect of ESG-valued returns on the shadow riskless rate (SRR) derived in \eqref{eq:LS} of Appendix~\ref{sec:App_D}
for an investment universe defined by the 29 stocks comprising our ESG-valued DJIA index.
For each trading date in the period from 01/03/2017 through 12/31/2020, using two-year moving windows
we estimated the vector $[\mu_i]$ of mean values of, and the variance-covariance matrix $\Sigma$ for, the 29 stock returns.
The asset (row) entries in $\Sigma$ were sorted in order of decreasing variance value.
The elements $\sigma_{i,j};\  i=1,\dots, N;\ j=1,\dots, N-1$ required in \eqref{eq:LS} were obtained from the
upper triangular matrix $L'$ obtained from the Cholesky decomposition $\Sigma = L L'$.
To eliminate one dimension and obtain 28 Brownian motions, the last column of $L'$ was replaced
with a linear combination of its last two columns.

\begin{figure}[h]
	\centering		
	\subcaptionbox{}{\includegraphics[width=0.48\textwidth]{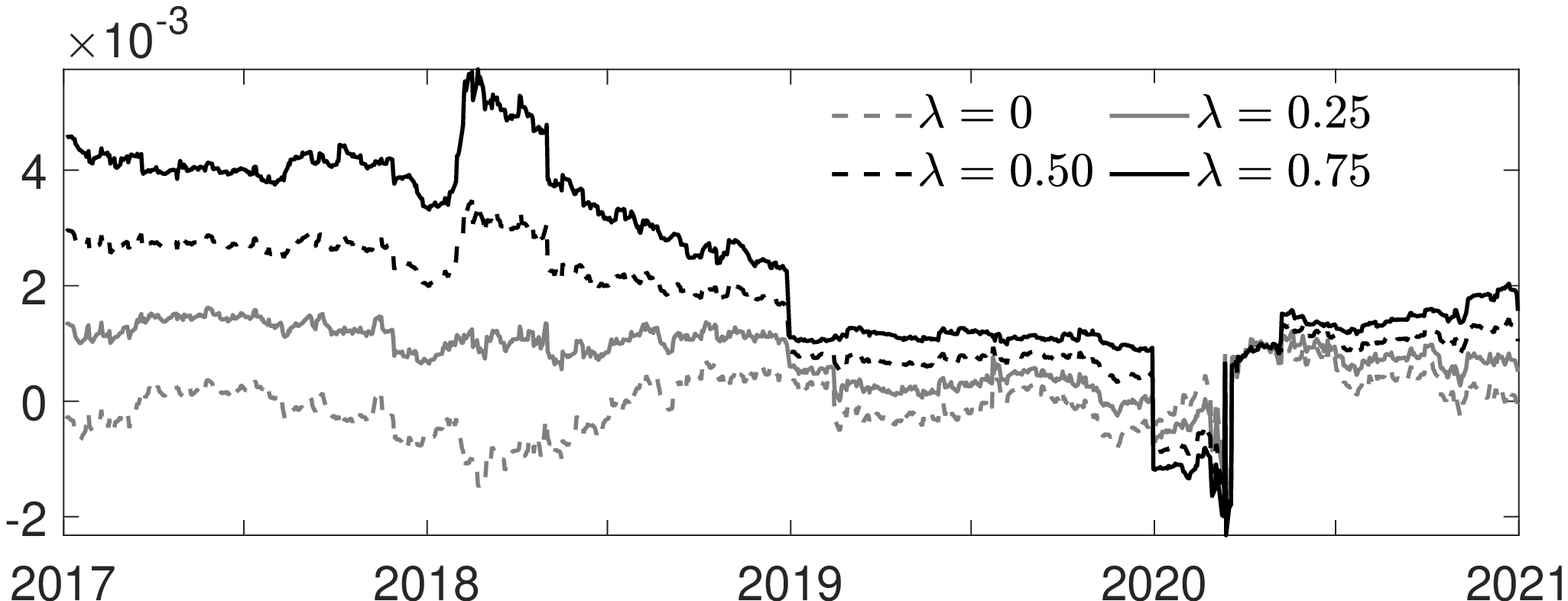} } \hspace{0em}%
	\subcaptionbox{}{\includegraphics[width=0.48\textwidth]{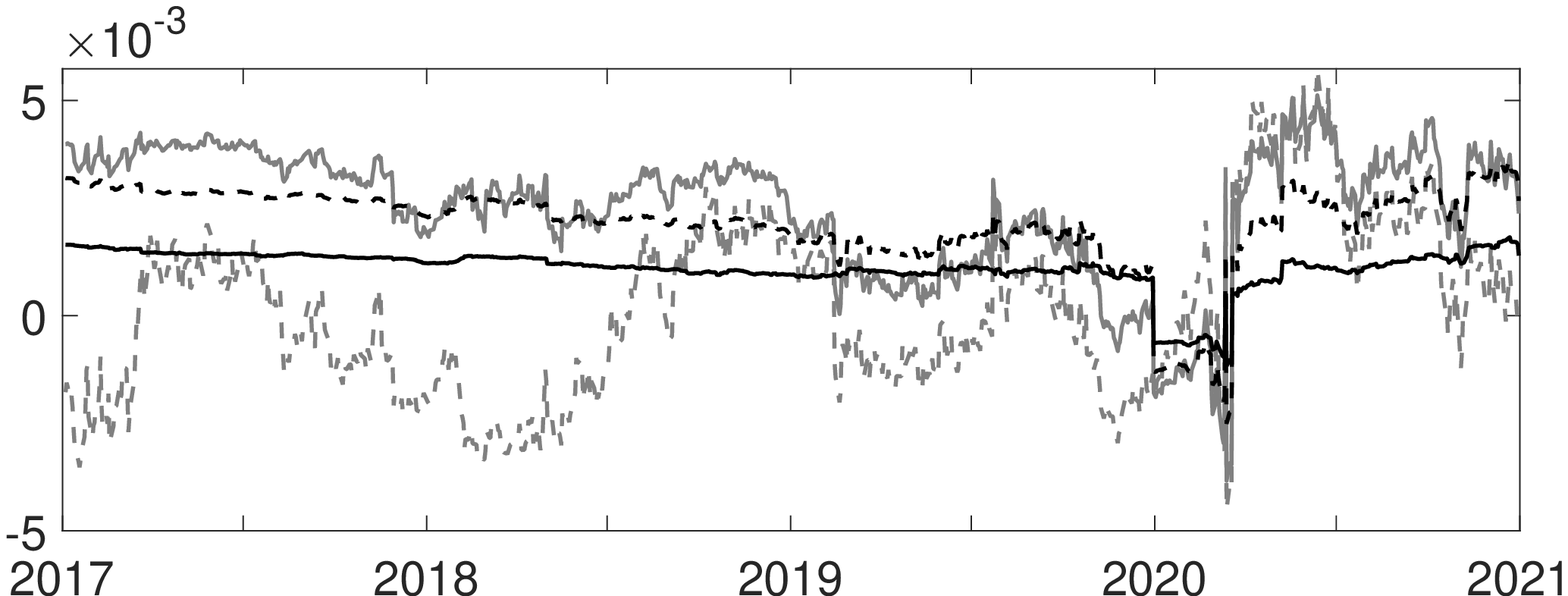} } \hspace{0em}%
	\caption{ \small  Time series from 01/03/2017 through 12/31/2020 for select values of $\lambda$ of
		(a) the SRR computed from ESG-valued returns on a universe of 29 DFJIA assets and
		(b) the information ratio of the SRR.} 
	\label{fig:SRR_Refinitiv}	
\end{figure}
The SRR computations were performed on the DJIA index for the values $\lambda \in (0, 0.25,0.5,0.75)$.\footnote{
	We perform the SRR computations on the DJIA index rather than the optimized portfolios because the
	Dow Index is conventially recognized as an indicator of the general health of the US stock market
	(though there are valid arguments against this).
}
The time series for the four SRRs are presented in Fig.~\ref{fig:SRR_Refinitiv}. 
The SRR levels show separation in rough proportion to the value of $\lambda$;
however this proportion is affected by each year-end readjustment of ESG values.
The separation between SRR curves appears to be roughly constant for the calendar years 2017 and 2019.
Over 2018 the separations between the curves narrow.
The Covid 19 crash in 2020 initially removes differences between SRR curves; after which they again begin to separate.

\begin{figure}[h!]
	\centering		
	{\includegraphics[width=0.49\textwidth]{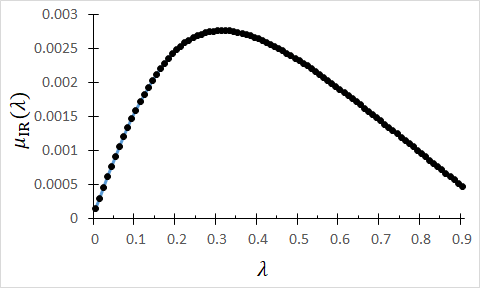} }\hspace{0em}%
	{\includegraphics[width=0.49\textwidth]{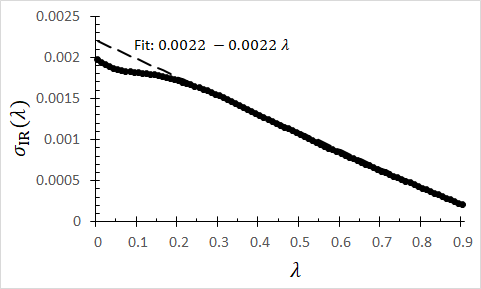} }\hspace{0em}%
	{\includegraphics[width=0.49\textwidth]{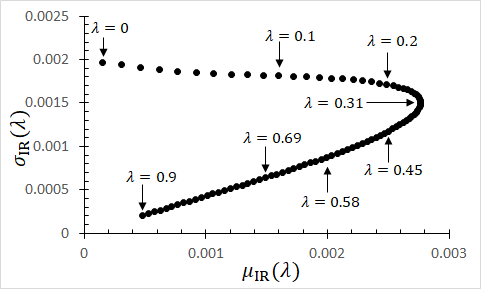} }\hspace{0em}%
	\caption{ \small The curves $\mu_{\text{IR}}(\lambda)$, $\sigma_{\text{IR}}(\lambda)$ and
		the parametric curve $(\mu_{\text{IR}}(\lambda), \sigma_{\text{IR}}(\lambda))$.
		The dashed line is a linear fit.} 
	\label{fig:IR_Lambda_Mean_Std}	
\end{figure}
We define the standard deviation of the price deflator process $\pi$ by
$\sigma_{\pi} = \sqrt{\sum_{j=1}^{N-1} \sigma_{\pi,j}^{2}}$.
This enables the definition of the SRR information ratio, IR${}_{\text{SRR}}=\mu_{\pi} / \sigma_{\pi}$.
The four time series for IR${}_{\text{SRR}}$ are also plotted in Fig.~\ref{fig:SRR_Refinitiv}.
For each time series, the average $\mu_{\text{IR}}(\lambda)$ and standard deviation  $\sigma_{\text{IR}}(\lambda)$
values of IR${}_{\text{SRR}}$ can be computed.
The behaviors of $\mu_{\text{IR}}(\lambda)$ and  $\sigma_{\text{IR}}(\lambda)$ for
$\lambda \in $\{ 0,0.01,0.02,\dots,0.89,0.9\} are presented in Fig. \ref{fig:IR_Lambda_Mean_Std}.
The value of $\mu_{\text{IR}}(\lambda)$ intially grows with $\lambda$, reaching a maximum
time-averaged IR${}_{\text{SRR}}$ value at $\lambda = 0.31$, after which the averaged SRR information ratio decreases.
The standard deviation $\sigma_{\text{IR}}(\lambda)$ decreases with $\lambda$;
for $\lambda > 0.2$, the decrease is linear.
These results quantify the nonlinear impact of $\lambda$ on the shadow riskless rate.

\section{Discussion} \label{sec:Disc}

In response to environmental changes and investor, societal, and governmental pressure,
ESG ratings are assuming an important role in financial markets.
Thus motivated,
we have introduced a consistent ESG-valued framework for the inclusion of ESG ratings into dynamic pricing theory.
There are two fundamental parameters in the framework;
in addition to the traditional financial risk aversion parameter ($\alpha \in [0,1]$),
we introduce the ESG affinity parameter $\lambda \in [0,1]$.
When $\lambda = 0$, our pricing framework reduces to traditional financial-risk versus financial-reward valuation.
As $\lambda$ increases, the framework places increased emphasis on the impact of ESG ratings on valuation.
Thus our results are presented in terms of comparisons of valuations for $\lambda > 0$ compared with $\lambda = 0$.
In this work we have applied the ESG-valued framework to portfolio optimization, risk measures, option pricing,
and the concept of shadow riskless rates.
Although the ESG-valued return \eqref{eq:esg_transform} is linearly dependent on ESG scores through $\lambda$,
other non-linearities in portfolio and option valuation ultimately result in non-linear dependence on $\lambda$.

The philosophy incorporated in our framework is that socially responsible investing places a value on an investment portfolio
that is competitive with (or ``additional to'') financial return.
Individual investor emphasis on that value is adjusted through the ESG affinity parameter.
As a consequence, our framework produces valuations (e.g. for portfolios, options, riskless rates) in terms of an
ESG-valued numeraire which reduces to a standard financial price when $\lambda = 0$.
Standardization of such a numeraire will become necessary if ESG-valued valuation is to be adopted.
This would require the development of market indices expressed in terms of a standard ESG (or SRI) numeraire.

If adopted, a framework such as that developed here will require consideration of a myriad of issues,
some of which are being pursued as described in the Introduction;
others arise from considerations encountered while implementing the framework.
In our view, one of the most important limitations  is the deterministic nature of the ESG rankings employed here;
we have ignored  the variability of agency-provided ESG scores.
As noted in the introduction, the definition of methodologies, and the consequent creation of databases related to
ESG scores has experienced huge growth in the last ten years,
resulting in a wide set of options in the choice of rating agency.
Although it is not our intention to further address that issue in this paper, as it has been extensively explored in literature cited,
we provide some insight on its impact on our ESG-valued valuation framework by considering scores from two different
providers.
In the supplement to this paper,  
we have repeated the computational results presented in this manuscript using ESG scores provided by RobecoSAM.

Despite the general disagreement of ESG scores between these two providers,
we have observed that optimal portfolios performances, and hence efficient frontiers,
on the DJIA index present relative stable results between the two scoring systems.
This in turn implies stable results for the option pricing example in Section \ref{sec:ESG_OP}.
However, optimized portfolio compositions can considerably change when computed with RobecoSAM scores.
This in turn is reflected on the shadow riskless rate implicit in the market,
even if computed under Gaussian hypotheses,
exhibiting wide variation in the SRR depending on the score system used.
This result confirms the need for a stronger effort toward a convergence of methodologies to asses ESG scores.

Given the current and foreseable variation in ESG rating systems and methodologies,
our framework accommodates (and in fact argues in favor of) ESG ratings as random variables
whose stochastic properties reflect these variations.
Evaluation of these properties will become more definitive once ESG ratings are updated on time scales shorter than
the calendar year.
Ideally, company-specific ESG ratings would change on a time scale intrinsic with the company's activity.

Our work reveals that adoption of an ESG-valued valuation framework will require resolution of critical research issues.
(i) Our model requires that ESG-ratings be rescaled to be comparable to financial returns, which requires scaling to
a bounded interval $[\varsigma^-, \varsigma^+]$ with $\varsigma^- < 0$, $\varsigma^+ > 0$.
This requires the identification of a threshold ESG value $\text{ESG}_0$ satisfying $\varsigma(\text{ESG}_0) = 0$
which effectively separates ``good'' ESG ratings from ``bad'' ESG ratings.
(ii) Traditional financial-reward : financial-risk metrics must be modified to incorporate ESG ratings, thus becoming
ESG-valued-reward : ESG-valued-risk metrics.
This raises the research question of ``coherence'', the necessary set of properties that such adjusted metrics should satisfy.
(iii) As excess return is often measured relative to a government set risk-free rate, the concept of applying ESG ratings
to governments and government-issued bonds becomes a required research focus.
(iv) The option pricing model explored here is based upon a discretization of a continuum model which captures limited
information (specifically, only the standard deviation of the financial returns of the underlying).
Thus another research avenue is the application of this ESG-valued framework to option valuation models that
are capable of capturing more of the microstructure of the underlying asset price
(and, when it becomes available, that of its ESG ratings).
(v) The methodology used here for the computation of the shadow riskless rate requires identification of the $N \times N-1$
standard deviation parameters $\sigma_{i,j}$ (equation \eqref{eq:BM} of Appendix~\ref{sec:App_D}) from the $N \times N$
elements of the variance-covariance matrix of asset returns.
There is currently no theory governing this identification; the Cholesky decomposition approach used here is ad hoc.

\begin{appendices}

\renewcommand{\thetable}{A.\arabic{table}}
\setcounter{table}{0}
\section{Tables} \label{sec:Tbl}

 \begin{table}[H] 
    \centering
 	\scalebox{0.8}{
		\begin{tabular}{ lccccccccc } \toprule
			Ticker & 12/31 & 12/31 & 12/31 & 12/30 & 12/29 & 12/31 & 12/31 & 12/31  \\
			\omit  & 2013  & 2014  & 2015  & 2016  & 2017   & 2018  & 2019  & 2020  \\
		\midrule
			MMM  & 88 & 90 & 86 & 88 & 88 & 87 & 89 & 92 \\ 
			AXP    & 50 & 51 & 51 & 68 & 73 & 73 & 78 & 78 \\ 
			AMGN & 67 & 65 & 68 & 66 & 75 & 72 & 74 & 74 \\ 
			AAPL  & 61 & 57 & 54 & 62 & 69 & 70 & 67 & 73 \\ 
			BA     & 59 & 73 & 66 & 69 & 80 & 78 & 80 & 83 \\ 
			CAT   & 75 & 69 & 75 & 73 & 65 & 69 & 69 & 72 \\ 
			CVX   & 73 & 73 & 80 & 77 & 88 & 84 & 79 & 84 \\ 
			CSCO & 83 & 85 & 85 & 87 & 87 & 88 & 88 & 89 \\ 
			KO     & 77 & 72 & 73 & 77 & 75 & 70 & 65 & 76 \\ 
			GS     & 63 & 67 & 72 & 62 & 70 & 73 & 86 & 86 \\ 
			HD     & 63 & 78 & 70 & 81 & 83 & 74 & 72 & 73 \\ 
			HON   & 68 & 53 & 63 & 67 & 68 & 74 & 75 & 83 \\ 
			IBM    & 79 & 73 & 81 & 76 & 77 & 80 & 71 & 73 \\ 
			INTC  & 86 & 91 & 91 & 90 & 87 & 88 & 88 & 88 \\ 
			JNJ     & 91 & 92 & 93 & 91 & 90 & 89 & 88 & 89 \\ 
			JPM    & 71 & 71 & 78 & 81 & 83 & 77 & 82 & 82 \\ 
			MCD   & 77 & 70 & 64 & 69 & 71 & 75 & 70 & 78 \\ 
			MRK   & 78 & 71 & 72 & 80 & 81 & 77 & 80 & 82 \\ 
			MSFT & 92 & 92 & 93 & 91 & 90 & 93 & 93 & 93 \\ 
			NKE    & 68 & 65 & 75 & 65 & 70 & 71 & 72 & 74 \\ 
			PG     & 60 & 59 & 72 & 68 & 66 & 65 & 71 & 73 \\ 
			CRM   & 46 & 41 & 46 & 56 & 63 & 72 & 62 & 67 \\ 
			TRV   & 42 & 49 & 43 & 36 & 44 & 63 & 70 & 66 \\ 
			UNH   & 62 & 64 & 67 & 82 & 78 & 80 & 72 & 71 \\ 
			VZ     & 72 & 69 & 60 & 57 & 66 & 68 & 75 & 76 \\ 
			V       & 21 & 51 & 52 & 71 & 72 & 71 & 53 & 54 \\ 
			WBA  & 39 & 53 & 60 & 71 & 72 & 74 & 81 & 94 \\ 
			WMT & 78 & 78 & 79 & 78 & 77 & 74 & 81 & 85 \\ 
			DIS   & 61 & 62 & 64 & 67 & 65 & 76 & 78 & 71 \\
		\bottomrule 
		\end{tabular} 
	}
	\caption{Refinitiv ESG scores for the DJIA Index components.}   
	\label{tab:Ref_DJ_ESG} 
\end{table}

\begin{table}[H] 
	\centering
	\scalebox{0.65}{
		\begin{tabular}{cccccccccc}
		\toprule
		\multicolumn{2}{c}{Model} & Tot. Ret & Ann. Ret & AvgTO & ETL95 & ETR95 & MDD & \multicolumn{2}{c}{ESG*} \\
		\   & \   & (\%) &  (\%) &  (\%) &  (\%) &  (\%) &  (\%) & avg  & std \\
		\midrule 
		\multicolumn{2}{c}{EWBH} & 90.14 & 17.68  & 0.00 & -3.41 & 2.86 & 33.00 & 72.72 & 2.97 \\
		\rule{0pt}{3ex}

			$\lambda$ & $\alpha$ & \multicolumn{5}{c}{mCVaR${}_{0.99}$} \\
			\cline{1-2}
			\rule{0pt}{3ex}
			0.00& 0.0  & 70.75 & 14.53 & 0.40 & -2.74 & 2.41 & 28.47 & 71.97 & 2.05  \\  
			  \     & 0.3  & 71.53 & 14.66 & 0.40 & -2.74 & 2.40 & 28.36 & 71.99 & 2.06 \\  
			  \     & 0.5  & 71.85 & 14.71 & 0.40 & -2.75 & 2.41 & 28.36 & 72.02 & 2.01 \\  
			  \     & 0.7  & 73.88 & 15.05 & 0.40 & -2.76 & 2.44 & 27.62 & 72.33 & 2.17 \\  
			  \     & 0.9  & 82.60 & 16.49 & 0.40 & -2.90 & 2.55 & 28.23 & 72.61 & 2.93 \\
			\rule{0pt}{3ex}

			 0.25  & 0.0 & 71.17 & 14.60 & 0.40 & -2.73 & 2.40 & 28.20 & 72.55 & 1.78  \\  
		       \   & 0.3 & 70.80 & 14.53 & 0.40 & -2.73 & 2.41 & 28.05 & 72.86 & 1.71  \\  
			\      & 0.5 & 71.60 & 14.67 & 0.40 & -2.73 & 2.41 & 27.66 & 73.17 & 1.68  \\  
			\      & 0.7 & 73.35 & 14.96 & 0.40 & -2.75 & 2.44 & 26.94 & 74.83 & 1.77  \\  
			\      & 0.9 & 92.13 & 18.00 & 0.39 & -2.96 & 2.65 & 25.71 & 78.59 & 2.53  \\
			\rule{0pt}{3ex}

		 	0.50  & 0.0  & 69.77 & 14.36 & 0.40 & -2.73 & 2.40 & 28.01 & 73.78 & 1.56 \\ 
			\      & 0.3 & 69.62 & 14.33 & 0.40 & -2.73 & 2.41 & 27.89 & 74.30 & 1.57 \\  
			\      & 0.5 & 84.62 & 16.81 & 0.39 & -2.86 & 2.57 & 25.23 & 80.36 & 2.72 \\  
			\      & 0.7 & 93.51 & 18.22 & 0.38 & -2.97 & 2.68 & 24.78 & 82.13 & 3.18 \\  
			\      & 0.9 & 112.02 & 20.98 & 0.36 & -3.20 & 2.94 & 24.37 & 84.51 & 4.14 \\
			\rule{0pt}{3ex}

			0.75   & 0.0 & 68.57 & 14.15 & 0.40 & -2.74 & 2.41 & 27.43 & 76.50 & 1.95  \\
			\      & 0.3 & 72.50 & 14.82 & 0.40 & -2.79 & 2.46 & 26.67 & 78.85 & 2.38 \\
			\      & 0.5 & 80.31 & 16.12 & 0.39 & -2.84 & 2.53 & 25.69 & 80.89 & 2.87 \\
			\      & 0.7 & 112.57 & 21.06 & 0.34 & -3.26 & 2.98 & 25.39 & 86.65 & 4.97 \\
			\      & 0.9 & 143.49 & 25.30 & 0.33 & -3.61 & 3.35 & 27.40 & 88.19 & 5.66 \\
			\rule{0pt}{3ex}
		     
			$\lambda$ & $\alpha$ & \multicolumn{5}{c}{MV } \\
			\cline{1-2}
			\rule{0pt}{3ex} 
            0.00   & 0.0  & 76.82 & 15.54 & 0.40 & -2.84 & 2.46 & 29.17 & 71.93 & 2.59 \\  
			\      & 0.05 & 84.53 & 16.80 & 0.40 & -3.06 & 2.61 & 30.62 & 72.38 & 2.80 \\  
			\      & 0.10 & 90.68 & 17.77 & 0.38 & -3.33 & 2.80 & 32.39 & 73.49 & 3.22 \\  
			\      & 0.15 & 90.64 & 17.77 & 0.39 & -3.55 & 2.96 & 35.00 & 74.17 & 3.46 \\  
			\      & 0.20 & 88.69 & 17.46 & 0.40 & -3.73 & 3.08 & 37.27 & 74.30 & 3.63 \\
			\rule{0pt}{3ex}

			0.25   & 0.0  & 85.13  & 16.90 & 0.40 & -2.95 & 2.52 & 30.21 & 72.01 & 2.81 \\  
			\      & 0.05 & 84.74  & 16.83 & 0.40 & -3.21 & 2.74 & 30.96 & 77.46 & 3.49 \\  
			\      & 0.10 & 102.24 & 19.54 & 0.40 & -3.51 & 3.08 & 30.64 & 81.10 & 4.29 \\  
			\      & 0.15 & 109.34 & 20.60 & 0.40 & -3.74 & 3.31 & 31.39 & 82.49 & 4.72 \\  
			\      & 0.20 & 114.60 & 21.36 & 0.40 & -3.87 & 3.43 & 32.07 & 83.14 & 5.01 \\
			\rule{0pt}{3ex} 

			0.50   & 0.0  & 92.87  & 18.12 & 0.40 & -3.14 & 2.63 & 31.68 & 72.00 & 2.78 \\
			\      & 0.05 & 93.35  & 18.19 & 0.40 & -3.42 & 3.01 & 29.52 & 81.58 & 4.96 \\  
			\      & 0.10 & 144.99 & 25.50 & 0.35 & -3.76 & 3.54 & 27.01 & 86.22 & 5.80 \\  
			\      & 0.15 & 170.92 & 28.74 & 0.32 & -3.95 & 3.75 & 27.47 & 87.59 & 5.94 \\  
			\      & 0.20 & 180.31 & 29.86 & 0.30 & -4.05 & 3.84 & 27.72 & 88.01 & 5.99 \\
			\rule{0pt}{3ex}  

			0.75   & 0.0  & 100.63 & 19.30 & 0.40 & -3.42 & 2.87 & 33.14 & 72.14 & 2.97 \\  
			\      & 0.05 & 101.12 & 19.38 & 0.39 & -3.57 & 3.20 & 28.20 & 83.86 & 5.66 \\  
			\      & 0.10 & 166.27 & 28.18 & 0.31 & -3.88 & 3.67 & 27.43 & 87.69 & 6.04 \\  
			\      & 0.15 & 191.69 & 31.17 & 0.27 & -4.05 & 3.87 & 27.53 & 88.51 & 6.04 \\  
			\      & 0.20 & 202.70 & 32.41 & 0.26 & -4.15 & 3.96 & 28.03 & 88.76 & 6.11 \\
			\bottomrule 
		\end{tabular} 
	}
	\caption{\small Performance measure values for the period 01/03/2017 through 12/30/2020
		for select efficient frontier portfolios optimized under mCVaR${}_{0.99}$ and MV.
		Note the difference between mCVaR${}_{0.99}$ and MV in the range of $\alpha$ provided.}   
	\label{tab:Summary_sim} 
\end{table}


\begin{table}[H] 
	\centering
	\scalebox{0.7}{
		\begin{tabular}{cc ccccc c ccccc}
			\toprule	
			\multicolumn{2}{c}{Model} & Mean    & Median    & Std      & Skew & ExKurt &
							 Model  & Mean    & Median    & Std      & Skew & ExKurt \\
					\  &\  & $\times 1^{-4}$ & $\times 1^{-4}$ & $\times 1^{-2}$ &\  &  &
						\ & $\times 1^{-4}$ & $\times 1^{-4}$ & $\times 1^{-2}$ &\  & \\
			\midrule
			\multicolumn{2}{c}{EWBH}        & 6.4 & 11.1 & 1.3 & -1.02 & 21.9 &
										&	&	    &		&		&		\\
			\rule{0pt}{3ex}
			
			$\lambda$ & $\alpha$ & \multicolumn{5}{c}{mCVaR${}_{0.99}$} &
					     $\alpha$ & \multicolumn{5}{c}{MV} \\
			\cline{1-2}\cline{8-8}
\rule{0pt}{3ex} 
            0.00  & 0.0 & 5.3 & 8.3 & 1.1 & -0.92 & 22.1  & 0.00 & 5.7 & 9.3 & 1.1 & -0.93 & 20.4 \\  
			\      & 0.3 & 5.4 & 8.0 & 1.1 & -0.92 & 22.2  & 0.05 & 6.1 & 10.7 & 1.2 & -1.08 & 22.9 \\  
			\      & 0.5 & 5.4 & 8.5 & 1.1 & -0.92 & 22.3  & 0.10 & 6.4 & 11.1 & 1.3 & -1.17 & 23.1 \\  
			\      & 0.7 & 5.5 & 9.6 & 1.1 & -0.90 & 22.4  & 0.15 & 6.4 & 11.2 & 1.4 & -1.32 & 23.9 \\  
			\      & 0.9 & 6.0 & 9.9 & 1.2 & -0.97 & 22.7  & 0.20 & 6.3 & 11.5 & 1.5 & -1.40 & 24.1 \\
			\rule{0pt}{3ex}

			0.25  & 0.0 & 5.3 & 8.7 & 1.1 & -0.88 & 21.8  & 0.00  & 6.1 & 10.1 & 1.1 & -1.02 & 21.0 \\   
			\      & 0.3 & 5.3 & 8.6 & 1.1 & -0.89 & 21.8  & 0.05 & 6.1 & 10.6 & 1.3 & -1.05 & 23.2 \\  
			\      & 0.5 & 5.4 & 8.5 & 1.1 & -0.88 & 21.8  & 0.10 & 7.0 & 12.3 & 1.4 & -0.91 & 20.8 \\  
			\      & 0.7 & 5.5 & 10.2 & 1.1 & -0.82 & 22.1  & 0.15 & 7.4 & 12.5 & 1.5 & -0.90 & 20.2 \\  
			\      & 0.9 & 6.5 & 10.6 & 1.2 & -0.68 & 20.4  & 0.20 & 7.6 & 12.0 & 1.5 & -0.88 & 19.4 \\
			\rule{0pt}{3ex} 

			0.50  & 0.0 & 5.3 & 9.6 & 1.1 & -0.83 & 21.2  & 0.0 & 6.5 & 11.7 & 1.2 & -1.08 & 22.3 \\  
			\      & 0.3 & 5.3 & 9.5 & 1.1 & -0.83 & 21.4  & 0.05 & 6.6 & 10.6 & 1.4 & -0.74 & 22.0 \\    
			\      & 0.5 & 6.1 & 11.1 & 1.1 & -0.51 & 19.0  & 0.10 & 8.9 & 13.4 & 1.5 & -0.43 & 17.5 \\  
			\      & 0.7 & 6.6 & 11.2 & 1.2 & -0.43 & 18.7  & 0.15 & 9.9 & 12.6 & 1.6 & -0.46 & 16.2 \\ 	
			\      & 0.9 & 7.5 & 11.0 & 1.3 & -0.34 & 17.3 & 0.20 & 10.3 & 11.7 & 1.6 & -0.49 & 16.0 \\
			\rule{0pt}{3ex} 

			0.75  & 0.0 & 5.2 & 9.4 & 1.1 & -0.70 & 19.8  & 0.0 & 6.9 & 12.3 & 1.3 & -1.11 & 23.3 \\  
			\      & 0.3 & 5.4 & 10.4 & 1.1 & -0.60 & 19.0 & 0.05 & 7.0 & 12.2 & 1.4 & -0.52 & 20.1 \\  
			\      & 0.5 & 5.9 & 10.8 & 1.1 & -0.47 & 18.3  & 0.10 & 9.7 & 13.3 & 1.6 & -0.43 & 16.4 \\  	
			\      & 0.7 & 7.5 & 11.5 & 1.3 & -0.23 & 15.8  & 0.15 & 10.7 & 11.8 & 1.6 & -0.48 & 15.8 \\  		
			\      & 0.9 & 8.9 & 12.8 & 1.4 & -0.36 & 16.0  & 0.20 & 11.0 & 11.8 & 1.7 & -0.52 & 16.0  \\ 
			\bottomrule
		\end{tabular} 	
	}
	\caption{ \small Moment values for the return distributions obtained
		for the period 01/03/2017 through 12/30/2020
		for select efficient frontier portfolios optimized under mCVaR${}_{0.99}$ and MV.
		Note the difference between mCVaR${}_{0.99}$ and MV in the range of $\alpha$ provided.}   
	\label{tab:Moments} 
\end{table}


\begin{table}[H] 
	\centering
	\scalebox{0.7}{
		\begin{tabular}{ccccccc cccccc}
		\toprule	
		\multicolumn{2}{c}{Model} & SR & Sortino & STAR & Rachev & Gini  &
		                          Model & SR & Sortino & STAR & Rachev & Gini  \\
		\ 	& \ 	 & (\%) & (\%) & (\%) & (\%) & (\%) & \  & (\%) & (\%) & (\%) & (\%) & (\%) \\
		\midrule
		\multicolumn{2}{c}{EWBH}        & 4.86      & 6.59            & 1.88            & 83.86           & 12.26 &
									&		&			&				&			&		\\
			\rule{0pt}{3ex}
		
			$\lambda$ & $\alpha$ & \multicolumn{5}{c}{mCVaR${}_{0.99}$} &
					     $\alpha$ & \multicolumn{5}{c}{MV} \\
			\cline{1-2}\cline{8-8}
\rule{0pt}{3ex} 
           	     0.00   & 0.0 & 4.87 & 6.69 & 1.94 & 87.79 & 12.42 & 0.0  & 5.09 & 6.96 & 2.00 & 86.56 & 12.97 \\ 
		       \      & 0.3 & 4.91 & 6.75 & 1.96 & 87.84 & 12.55 & 0.05 & 5.09 & 6.92 & 1.99 & 85.10 & 13.03 \\ 
			\      & 0.5 & 4.92 & 6.76 & 1.96 & 87.76 & 12.55 & 0.10 & 4.92 & 6.66 & 1.93 & 83.94 & 12.35 \\ 
			\      & 0.7 & 4.98 & 6.85 & 2.00 & 88.42 & 12.66 & 0.15 & 4.62 & 6.21 & 1.81 & 83.29 & 11.47 \\
			\      & 0.9 & 5.16 & 7.08 & 2.07 & 88.03 & 13.11 & 0.20 & 4.34 & 5.81 & 1.69 & 82.50 & 10.68 \\
			\rule{0pt}{3ex} 

			0.25   & 0.0 & 4.91 & 6.76 & 1.96 & 87.99 & 12.53 & 0.0  & 5.34 & 7.28 & 2.08 & 85.36 & 13.70  \\ 
			\      & 0.3 & 4.89 & 6.73 & 1.95 & 88.13 & 12.46 & 0.05 & 4.85 & 6.60 & 1.90 & 85.36 & 12.30 \\ 
			\      & 0.5 & 4.93 & 6.79 & 1.97 & 88.32 & 12.56 & 0.10 & 5.02 & 6.89 & 2.00 & 87.79 & 12.40 \\ 
			\      & 0.7 & 4.95 & 6.84 & 1.99 & 88.64 & 12.61 & 0.15 & 4.93 & 6.78 & 1.97 & 88.48 & 12.05 \\  
			\      & 0.9 & 5.45 & 7.59 & 2.20 & 89.56 & 13.77 & 0.20 & 4.93 & 6.78 & 1.96 & 88.78 & 11.96 \\
			\rule{0pt}{3ex}

			0.50   & 0.0 & 4.85 & 6.67 & 1.93 & 87.92 & 12.33 & 0.0  & 5.38 & 7.31 & 2.08 & 83.94 & 13.87\\  
			\      & 0.3 & 4.83 & 6.65 & 1.92 & 88.02 & 12.27 & 0.05 & 4.82 & 6.66 & 1.92 & 88.23 & 12.03 \\  
			\      & 0.5 & 5.31 & 7.43 & 2.13 & 89.90 & 13.41 & 0.10 & 5.84 & 8.26 & 2.37 & 94.17 & 14.38 \\
			\      & 0.7 & 5.49 & 7.72 & 2.21 & 90.18 & 13.81 & 0.15 & 6.18 & 8.76 & 2.51 & 94.89 & 15.14 \\ 
			\      & 0.9 & 5.80 & 8.21 & 2.34 & 91.87 & 14.46 & 0.20 & 6.24 & 8.84 & 2.53 & 94.94 & 15.26 \\
			\rule{0pt}{3ex}
    
			0.75   & 0.0 & 4.78 & 6.60 & 1.90 & 88.23 & 12.05 & 0.0  & 5.22 & 7.08 & 2.03 & 83.98 & 13.32 \\
			\      & 0.3 & 4.90 & 6.80 & 1.95 & 88.27 & 12.36 & 0.05 & 4.87 & 6.80 & 1.95 & 89.64 & 11.97 \\  
			\      & 0.5 & 5.19 & 7.24 & 2.07 & 89.17 & 13.07 & 0.10 & 6.20 & 8.80 & 2.51 & 94.65 & 15.28 \\  
			\      & 0.7 & 5.77 & 8.17 & 2.30 & 91.53 & 14.39 & 0.15 & 6.48 & 9.20 & 2.63 & 95.40 & 15.94 \\ 
			\      & 0.9 & 6.13 & 8.69 & 2.45 & 92.86 & 15.27 & 0.20 & 6.54 & 9.28 & 2.66 & 95.63 & 16.10 \\
		\bottomrule	
		\end{tabular}
	}
	\caption{ \small RRR values obtained for the period 01/03/2017 through 12/30/2020
		for select efficient frontier portfolios optimized under mCVaR${}_{0.99}$ and MV.
		Note the difference between mCVaR${}_{0.99}$ and MV in the range of $\alpha$ provided.}
	\label{tab:RRR_Ref}  
\end{table}

\begin{table}[H] 
	\centering
	\scalebox{0.8}{	
		\begin{tabular}{c ccccc ccccc}
			\toprule	
			Model & Tot. Ret & Ann. Ret & AvgTO & ETL95 & ETR95 & MDD & \multicolumn{2}{c}{ESG*} & \omit & \omit \\
			\ 	  & (\%)	    & (\%)	& (\%)	 &(\%)	& (\%)	& (\%) & avg       & std	& \omit & \omit  \\
			\midrule
			EWBH & 90.14   & 17.69     & 0.00    & -3.41   & 2.86    & 32.87 & 72.72     & 2.97  & \omit & \omit \\
			\rule{0pt}{3ex}
			$\lambda$ & \multicolumn{8}{c}{ mCVaR${}_{0.99}$ } & \omit & \omit \\
			\cline{1-1}
			\rule{0pt}{3ex} 
			0.00 & 95.01  & 18.45 & 0.40 & -3.48 & 2.92 & 34.06 & 74.04 & 3.49 & \omit & \omit \\
			0.25 & 114.91 & 21.40 & 0.35 & -3.80 & 3.38 & 31.09 & 82.98 & 4.67 & \omit & \omit \\
			0.50 & 129.63 & 23.46 & 0.33 & -3.60 & 3.36 & 26.86 & 84.58 & 4.69 & \omit & \omit \\
			0.75 & 176.66 & 29.43 & 0.23 & -3.97 & 3.78 & 27.49 & 88.08 & 5.83 & \omit & \omit \\
			\rule{0pt}{3ex}
			
			$\lambda$ & \multicolumn{8}{c}{ MV  } & \omit & \omit \\
			\cline{1-1}
			\rule{0pt}{3ex} 
			0.00 & 93.80  & 18.26 & 0.42 & -3.76 & 3.13 & 37.23 & 74.64 & 3.76 & \omit & \omit \\
			0.25 & 118.44 & 21.90 & 0.33 & -4.10 & 3.60 & 34.23 & 83.64 & 5.22 & \omit & \omit \\
			0.50 & 154.95 & 26.77 & 0.25 & -4.12 & 3.83 & 30.23 & 86.09 & 5.62 & \omit & \omit \\
			0.75 & 207.56 & 32.95 & 0.16 & -4.21 & 4.03 & 28.14 & 88.51 & 6.06 & \omit & \omit \\
			\midrule
			\\
			Model    & Mean  & Median  & Std  & Skew  & ExKurt &   Mean  & Median  & Std  & Skew  & ExKurt \\
			\ 	& $\times 10^{-4}$ &  $\times 10^{-4}$ & $\times 10^{-2}$ & \multicolumn{2}{c}{\ }
				& $\times 10^{-4}$ &  $\times 10^{-4}$ & $\times 10^{-2}$ & \multicolumn{2}{c}{\ }\\
			\midrule
			EWBH & 6.4 & 11.1 & 1.3 & -1.02 & 21.9   & \multicolumn{5}{c}{\ }  \\
			\rule{0pt}{3ex}
			$\lambda$ & \multicolumn{5}{c}{ mCVaR${}_{0.99}$  } & \multicolumn{5}{c}{ MV  }  \\
			\cline{1-1}
			\rule{0pt}{3ex} 
			0.00 &   6.7 & 11.4 & 1.4 & -1.28 & 23.9 &  6.6 & 11.6 & 1.5 & -1.36 & 23.7 \\
			0.25 &   7.6 & 11.9 & 1.5 & -0.86 & 19.1 &  7.8 & 11.7 & 1.6 & -0.92 & 18.7 \\
			0.50 &   8.3 & 12.6 & 1.4 & -0.50 & 16.8 &  9.3 & 12.9 & 1.7 & -0.64 & 16.7 \\
			0.75 & 10.1 & 11.7 & 1.6 & -0.45 & 15.8 & 11.2 & 13.0 & 1.7 & -0.51 & 15.1 \\
			\midrule
			\\
			Model       & SR   & Sortino & STAR & Rachev & Gini & SR   & Sortino & STAR & Rachev & Gini \\
			\ 		  & (\%) & (\%)  & (\%) & (\%)   & (\%)& (\%) & (\%)  & (\%)  & (\%)   & (\%)\\
			\midrule
			EWBH      & 4.86  & 6.59   & 1.88   & 83.86  & 12.26 &\omit &\omit &\omit  \\
			\rule{0pt}{3ex}
			$\lambda$ & \multicolumn{5}{c}{ mCVaR${}_{0.99}$  } & \multicolumn{5}{c}{ MV  }  \\
			\cline{1-1}
			\rule{0pt}{3ex} 
			0.00 & 4.87 & 6.57 & 1.91 & 83.82 & 12.19  & 4.48 & 6.00 & 1.75 & 83.05 & 11.02 \\
			0.25 & 5.02 & 6.92 & 2.00 & 88.87 & 12.18  & 4.79 & 6.57 & 1.90 & 87.84 & 11.51 \\
			0.50 & 5.69 & 8.01 & 2.30 & 93.20 & 13.92  & 5.61 & 7.85 & 2.26 & 92.94 & 13.57 \\
			0.75 & 6.29 & 8.93 & 2.55 & 95.31 & 15.44  & 6.52 & 9.25 & 2.66 & 95.70 & 15.88 \\
			\bottomrule
		\end{tabular} 	
	}
	\caption{\small Performance measures, return moment values and RRR results obtained for the
		tangent portfolios under mCVaR${}_{0.99}$ and MV optimization during the period 01/03/2017 to 12/30/2020.}   
	\label{tab:Summary_sim_tang} 
\end{table}


\newpage

\renewcommand{\theequation}{B.\arabic{equation}}
\setcounter{equation}{0}
\section{MV and mCVaR${}_\beta$ ESG-Valued Optimization} \label{sec:App_B}

Extension of MV and mCVaR${}_\beta$ optimization to ESG-valued returns is straightforward.
Consider a universe of $I$ assets and the corresponding series of simulated ESG-valued returns
$\{ \hat{\zeta}^s_{ i, t +1} ( \lambda )$; $s = 1, \dots, S$; $i =1,\dots,I \}$.
For MV optimization, let $\mu = [ \mu_{1}, \dots, \mu_{I} ]$ (with transpose denoted by $\mu^\prime$)
denote the vector of the sample means 
and $\Sigma$ the corresponding variance-covariance matrix computed from the adjusted returns.
Let $\theta_{i} \in [0,1]$ denote the portfolio weight\footnote{
 	Without loss of generality, we consider long-only portfolio optimization.}
for asset $i$.
The ESG-valued MV optimization  is obtained as the solution of the quadratic programming problem
\cite{Markowitz_1952,Markowitz_1956}
\begin{equation}
\min_{\theta} \left\{ -\alpha \theta'\mu+ \left( 1- \alpha \right) \theta' \Sigma w \right\},  \label{eq:MV_objf}
\end{equation}
subject to the constraints
\begin{align}
	& \theta_{i} \ge 0, \text{ for } i =1,\dots,I,     \label{eq:MV_const_1}\\ 
	& \sum_{i=1}^{I} \theta_{i} = 1,  \label{eq:MV_const_2}
\end{align}
for a particular choice of $\alpha$ and $\lambda \in [0,1]$, where $\alpha$ is the risk-aversion parameter.

The ESG-valued mCVaR${}_\beta$ solution is obtained from the minimization \cite{Rockafellar_2000}
\begin{equation}
\min_{\theta,\xi} \left\{ -\alpha \theta' \mu+  ( 1- \alpha) \left[ \xi +S^{-1} \left( 1 - \beta \right)^{-1} \sum_{s=1}^S \left[ -x_s- \xi \right]^{+} \right] \right\},
\label{eq:MCVeR_objf}
\end{equation}
where
\begin{equation}
x_s =  \sum_{i=1}^I\hat{\zeta}^s_{ i, t+1} (\lambda) \theta_i, \text{ for } s =1,\dots,S,
\end{equation}
subject to
\begin{align}
& \theta_{i} \ge 0, \text{ for } i =1,\dots,I,\\ 
& \sum_{i=1}^{I} \theta_{i} = 1,\\
& \xi \in \mathbb{R} \label{eq:MCVaR_const_4},
\end{align}
for a particular choice of $\alpha$ and $\lambda \in [0,1]$.
We use 0.95 and 0.99 as the values for the significance coefficient $\beta$.
Equations (\ref{eq:MCVeR_objf})-(\ref{eq:MCVaR_const_4}) can be reduced to a linear problem \cite{Rockafellar_2000}.

\renewcommand{\theequation}{C.\arabic{equation}}
\setcounter{equation}{0}
\section{ESG-Valued Option Valuation: Discrete Model} \label{sec:App_C}

Consider the problem of pricing a European contingent claim on the underlying securuity $\mathbb{S}$ on day $t$
having strike value $K$ and maturity date $t+T$.
We utilize the following discrete option pricing approach.
Following the procedure in Section~\ref{sec:ESG_EF},
we fit an ARMA($p,q$)-GARCH(1,1) model to a  two-year window of historical returns of $\mathbb{S}$,
assuming that the innovations are distributed according to an NIG distribution.
From the NIG fit, we randomly generate a set of of $T = T_{\text{min}}, \dots, T_{\text{max}}$ innovations
which, using the fitted ARMA-GARCH parameters, is conveted into a return trajectory
$\{ \hat{r}_{t+T};\  T = T_{\text{min}}, \dots, T_{\text{max}}\}$.
We generate an ensemble of $S = 20{,}000$ such return trajectories,
$\left\{ \hat{r}^s_{t+T};\  T = T_{\text{min}}, \dots, T_{\text{max}};\  s = 1, \dots, S\right\}$.
Using the ESG score for day $t$, from \eqref{eq:esg_transform} each trajectory can then be converted to a trajectory
of ESG-valued returns
$\left\{ \hat{\zeta}^s_{t+T};\  T = T_{\text{min}}, \dots, T_{\text{max}};\  s = 1, \dots, S\right\}$
and consequently a set of ESG-valued price trajectories
$\left\{ \hat{P}^{(Z,s)}_{t+T};\  T = T_{\text{min}}, \dots, T_{\text{max}};\  s = 1, \dots, S\right\}$.\footnote{
	As noted in Section~\ref{sec:ESG_OP}, we used $T_{\text{min}}=15$, $T_{\text{max}} = 252$.}

We compute an ESG-valued price for call and put options in the standard way;
the contract value is given by the discounted expected value of the contract pay-off, conditioned to a filtration ${\cal F}_{t}$,
where the conditional expectation is with respect to a risk-neutral measure $\mathbb{Q}$ \cite{Delbaen1994},
which discounts expected values at the ESG-valued risk-free rate $\zeta_{f,t} (\lambda)$
defined in \eqref{eq:esg_rf},
\begin{equation} 
	P^{(Z)}_t (\lambda) = \mathbb{E}^{\mathbb{Q}} \left[ P^{(Z)}_{T} (\lambda) e^{- \zeta_{f,t} (\lambda)T } |  {\cal F}_{t} \right].
	\label{eq:risk_neutral_p}
\end{equation}
Given the risk neutral probability measure $\mathbb{Q}$, European call (c) and put (p) ESG-valued option values on the
underlying can be computed as
\begin{equation}
   \begin{aligned} 
	\Phi_{c}(t,T,K,\lambda) &= \mathbb{E}^{\mathbb{Q}} \left[ \max \left( \hat{P}^{(Z)}_{T}(\lambda)-K,0 \right) e^{ - \zeta_{f,t} (\lambda)T } | {\cal F}_{t} \right],\\
	\Phi_{p}(t,T,K,\lambda) &= \mathbb{E}^{\mathbb{Q}} \left[ \max \left( K-\hat{P}^{Z}_{T}(\lambda),0 \right) e^{ -\zeta_{f,t} (\lambda)T } | {\cal F}_{t} \right].
   \end{aligned}    \label{eq:value_cp}
\end{equation}	    	
In the discrete setting of this paper,
\begin{equation*}
	\mathbb{E}^{\mathbb{Q}} \left[ P^{(Z)}_T (\lambda) e^{- \zeta_{f,t} (\lambda)T } |  {\cal F}_{t} \right]
		    = \sum_{s=1}^{S} q_s P^{(Z,s)}_T (\lambda) e^{- \zeta_{f,t} (\lambda)T },
\end{equation*}
and the option valuation problem is reduced to identifying $\mathbb{Q}$ through the coefficients $q_s$.
We solve for these coefficients by minimizing the Kullback-Leibler divergence between $\mathbb{Q}$
and the natural world probability $\mathbb{P}$ \cite{Avellaneda2001}.
This involves solving the convex problem
\begin{equation} 
	\min_{\{q_s\}} \sum_{s=1}^{S} q_{s} \ln(q_s / p_s),   \label{eq:OPIM_1}
\end{equation}
subject to the constraints
\begin{align}
	& \sum_{s=1}^{S} q_{s} = 1,\\
	& \sum_{s=1}^{S} q_s P^{(Z,s)}_T (\lambda) e^{- \zeta_{f,t} (\lambda)T } =  P^{(Z)}_t,\\
	& 0 <\   q_{s} < 1; \quad s=1,\dots,S. \label{eq:OPIM_end}
\end{align}
In \eqref{eq:OPIM_1}, $p_s$ is the probability of trajectory $s$ under $\mathbb{P}$.
As each trajectory is equally likely, $p_s = 1/S$.
Option values are then computed from the discrete version of \eqref{eq:value_cp}
\begin{equation}
   \begin{aligned}
	\Phi_{c}\left(t,T,K\right) &= \sum_{s=1}^{S} q_{s} \left[ \max \left(  \hat{P}^{(Z,s)}_{T} -K,0 \right) e^{ -  \zeta_{f,t} (\lambda)T } \right],
	\\
	\Phi_{p}\left(t,T,K\right) &=  \sum_{s=1}^{S} q_{s} \left[ \max \left(  K-\hat{P}^{(Z,s)}_{T},0 \right) e^{ -  \zeta_{f,t} (\lambda)T} \right].
   \end{aligned}
   \label{eq:disc_value_cp}
\end{equation} 

\renewcommand{\theequation}{D.\arabic{equation}}
\setcounter{equation}{0}
\section{Shadow Riskless Rate} \label{sec:App_D}

Consider a set of $N$ risky securities whose prices $P_{i,t}$; $i = 1, \dots, N$ are driven by $N-1$ Brownian motions\footnote{
	 More complex pricing models have been considered \cite{Rachev_2017}; here we employ the basic Brownian motion model.}  
$[dB_{1,t}, \dots, dB_{N-1,t}]$.
Thus
\begin{equation}
	dP_{i,t} = P_{i,t} \mu_{i} dt + P_{i,t} \sum_{j=1}^{N-1} \sigma_{i,j} dB_{j,t}, \quad i=1,\dots,N. \label{eq:BM}
\end{equation}
To eliminate arbitrage opportunities in the market defined by (\ref{eq:BM}), we search for the unique state-price deflator
\begin{equation}
	d\pi_{t} = \pi_{t} \mu_{\pi} dt +\pi_{t} \sum_{j=1}^{N-1} \sigma_{\pi,j} dB_{j,t}, \label{eq:SRR}
\end{equation}
such that the discounted prices $P_{i,t}\pi_{t}$, $i = 1,\dots,N$ are martingales.
It is easy to show\footnote{
	This follows directly from It\^{o}'s formula.}
that this requirement is equivalent to solving the linear system  
\begin{equation}
	\begin{bmatrix}
		-1	   & -\sigma_{1,1} & \dots & -\sigma_{1,N-1}\\
		-1	   & -\sigma_{2,1} & \dots & -\sigma_{2,N-1}\\
		\vdots & \vdots		& \dots & \vdots\\
		-1	   & -\sigma_{N,1} & \dots & -\sigma_{N,N-1}\\
	\end{bmatrix} 
	\begin{bmatrix}
		\mu_{\pi}\\		\sigma_{\pi,1}\\	\vdots\\		\sigma_{\pi,N-1}\\
	\end{bmatrix} 
	=
	\begin{bmatrix}
		\mu_{1}\\		\mu_{2}\\		\vdots\\		\mu_{N}\\
	\end{bmatrix},
	\label{eq:LS}
\end{equation}
for $[ \mu_{\pi},\sigma_{\pi,1},\dots,\sigma_{\pi,N-1}]$.
Following the treatment in Duffie \cite[Chapter 6D]{Duffie_2001}, the risk premium for each security is
\begin{equation}
	\mu_{i} - \hat{r} = -\sum_{j=1}^{N-1}\sigma_{i,j} \sigma_{\pi,j}, \text{ for } i = 1,\dots,N,  
\end{equation}
where $\hat{r} = -\mu_{\pi}$ is the short-term, shadow riskless rate (SRR) observed in the market.

\end{appendices}

\bibliographystyle{acm}



\begin{appendices}

\setcounter{section}{19}

\renewcommand{\thefigure}{S.\arabic{figure}}
\setcounter{figure}{0}
\renewcommand{\thetable}{S.\arabic{table}}
\setcounter{table}{0}

\section*{Supplement to ESG-Valued Portfolio Optimization and Dynamic Asset Pricing}

This appendix contains supplementary material to the article ESG Optimizing Portfolios.
Here we provide further analyses on the Refinitiv data.
Additionally we replicate all analyses in the manuscript by considering ESG scores provided by the
rating agency, RobecoSAM;
enabling preliminary results on the impact of the choice of the ESG score provider.


\subsection{RobecoSAM ESG Data: Comparison with Refinitiv}\label{sec:Data_robeco}

\begin{figure} [!h]
	\centering
	\includegraphics[width = 0.33\textwidth]{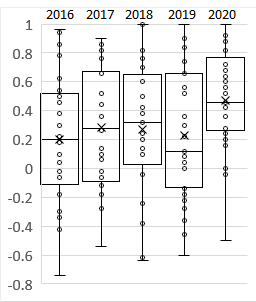}
	\caption{Box-whisker summaries of $[-1,1]$ normalized RobecoSAM data for the DJIA data listed in
			Table~\ref{tab:Rob_DJ_ESG}.}
	\label{fig:robeco_data}
\end{figure}

The ESG ratings produced by RobecoSam  are released at the end of a calendar
year and are available for application to the years 2017 through 2021.
The ratings for 29 of the companies in the the Dow Jones Industrial Average (DJIA) used in our analyses
are summarized in Table \ref{tab:Rob_DJ_ESG}.
 \begin{table}[!h] 
	\centering
	\scalebox{0.6}{
		\begin{tabular}{ l ccccc l ccccc}
		\toprule
			Ticker  & 12/30 & 12/29 & 12/31 & 12/31 & 12/31 & Ticker  & 12/30 & 12/29 & 12/31 & 12/31 & 12/31  \\
			\        & 2016  & 2017  & 2018   & 2019  & 2020  & \         & 2016  & 2017   & 2018  & 2019  & 2020 \\
		\midrule
			MMM  & 93 & 84 & 88 & 87 & 91 &  JPM  & 35 & 41 & 38 & 43 & 58 \\
			AXP    & 47 & 72 & 69 & 61 & 68 &  MCD & 69 & 64 & 60 & 53 & 76 \\ 
			AMGN & 82 & 88 & 89 & 87 & 95 &  MRK & 61 & 53 & 57 & 65 & 73 \\ 
			AAPL  & 30 & 36 & 18 & 36 & 25 &  MSFT & 97 & 93 & 91 & 92 & 96 \\ 
			BA     & 42 & 36 & 32 & 32 & 78 &  NKE & 77 & 76 & 71 & 78 & 80 \\
			CAT   & 81 & 91 & 88 & 88 & 94 &  PG   & 50 & 64 & 58 & 55 & 64 \\
			CVX   & 60 & 57 & 62 & 53 & 60 &  CRM & 65 & 83 & 85 & 83 & 91 \\ 
			CSCO & 89 & 88 & 100 & 100 & 100 &  TRV & 55 & 63 & 58 & 41 & 50 \\ 
			KO    & 41 & 41 & 39 & 39 & 64 &  UNH & 98 & 95 & 100 & 95 & 96 \\ 
			GS    & 75 & 77 & 70 & 66 & 71 &  VZ   & 33 & 23 & 19 & 20 & 49 \\
			HD   & 56 & 68 & 75 & 68 & 62 &  V    & 61 & 72 & 80 & 76 & 82 \\ 
			HON & 29 & 44 & 31 & 27 & 48 &  WBA  & 13 & 47 & 55 & 50 & 81 \\ 
			IBM  & 69 & 69 & 66 & 56 & 69 &  WMT  & 49 & 44 & 48 & 52 & 68 \\ 
			INTC & 73 & 84 & 70 & 44 & 86 &  DIS  & 52 & 55 & 56 & 54 & 73 \\
			JNJ   & 59 & 49 & 71 & 83 & 84 &  \multicolumn{5}{c}{\ } \\
		\bottomrule 
		\end{tabular} 
	}
	\caption{\small Robeco ESG scores for the DJIA Index components.}   
	\label{tab:Rob_DJ_ESG} 
\end{table}
Box-whisker summaries of the resultant normalized scores $\varsigma_{i,t}$ are shown, by year, in Fig.~\ref{fig:robeco_data}.
Compared to the Refinitiv data (manuscript, Fig.~1),
for this set of companies the yearly RobecoSAM data has smaller mean value, but larger spread.
Thus while the Refinitiv data reveals lower outliers, the Robeco data does not.
There are also instances (CSCO and UNH) where RobecoSAM assigns a ``perfect'' ESG score.

\begin{figure} [!h]
	\centering
	\includegraphics[width = 0.75\textwidth]{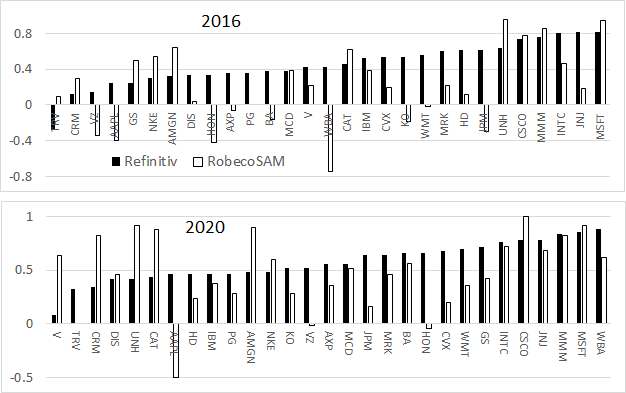}
	\caption{Normalized ESG ratings from Refinitive and RobecoSAM for 2016 and 2020.}
	\label{fig:RR_comp}
\end{figure}

 \begin{table}[!h] 
	\centering
	\scalebox{0.8}{
		\begin{tabular}{ r cccc c cccc }
		\toprule
			\   & \multicolumn{4}{c}{Kendall Tau} &\ & \multicolumn{4}{c}{Spearman Rho} \\
			\   & Ref16 & Reb16 & Ref20 & Reb20 &\ & Ref16 & Reb16 & Ref20 & Reb20  \\
		\cline{2-5}\cline{7-10}
		\rule{0pt}{3ex}
			Ref16 & 1 & 0.22 & 0.49 & \omit &\ & 1 & 0.35 & 0.64 & \omit \\
			\omit & \omit & (0.095) & ($10^{-4}$) & \omit &\ & \omit & (0.067) & ($2 \cdot 10^{-4})$ & \omit \\
			Reb16 & \omit & 1 & \omit & $-$0.08 &\ & \omit & 1 & \omit & $-$0.11 \\
			\omit & \omit & \omit & \omit &(0.55) &\ & \omit & \omit & \omit & (0.58) \\
			Ref20 & \omit & \omit & 1 & 0.15 &\ & \omit & \omit & 1 & 0.17 \\
			\omit & \omit & \omit & \omit & (0.26) &\ & \omit & \omit & \omit & (0.37) \\
		\bottomrule 
		\end{tabular}
	} 
	\caption{Rank correlation measures between Refinitiv and RobecoSAM ESG ratings released in 2016 and 2020.
		The numbers in parenthesis represent p-values.
		Ref16 = Refinitiv, 2016; etc.}   
	\label{tab:rank_test} 
\end{table}
Fig.~\ref{fig:RR_comp} presents a comparison of the normalized $\varsigma_{i,t}$ ESG ratings
from Refinitiv and RobecoSAM released at the end of 2016 and 2020.
In both plots the companies are ordered from lowest-to-highest Refinitiv ESG rating.
The plots show both the change in company rankings with time for the same rating agency
as well as the different rankings assigned by Refinitiv and RobecoSAM in the same year.
For each of these two years, the ESG rankings of these 29 companies were computed separately for the
Refinitive and RobecoSAM data.
Table~\ref{tab:rank_test} shows the results of the Kendall Tau and Spearman Rho tests on the correlation
between the rankings for the same year by the two rating agencies; 
as well as the correlation between the rankings over the two different years by the same agency.
The p-values indicate that, even at the 5\% significance level,
there is no correlation between the Refinitiv and RobecoSAM rankings for the same year.
There is a strong correlation between Refinitiv rankings for 2016 and 2020.
There is however no correlation between the RobecoSAM rankings for 2016 and 2020.

\subsection{Efficient Frontiers}\label{sec:EF_comment}

\begin{figure}[!h]
	\centering
	\subcaptionbox{}{\includegraphics[width=0.34\textwidth]{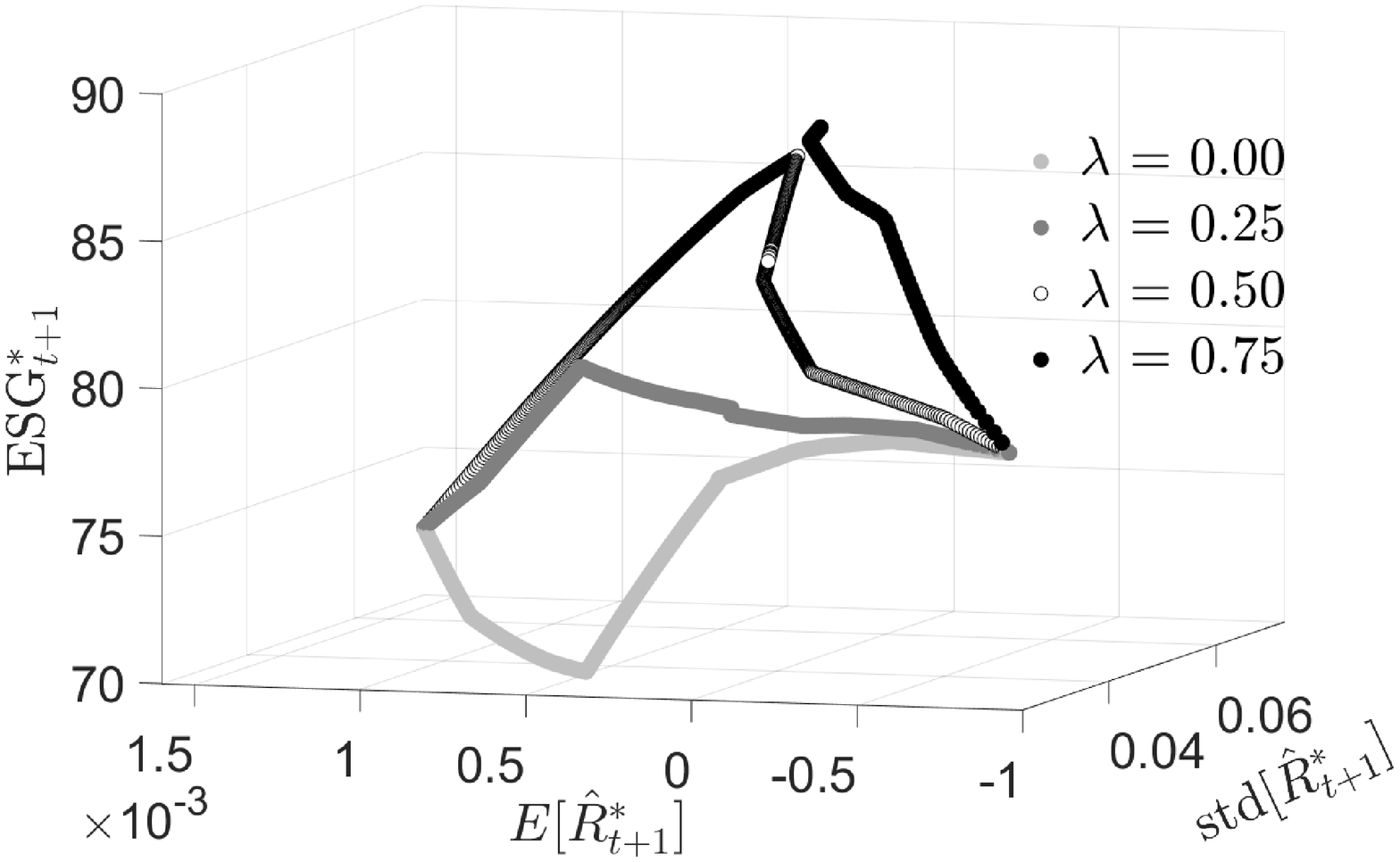}}\hspace{0em}%
	\subcaptionbox{}{\includegraphics[width=0.32\textwidth]{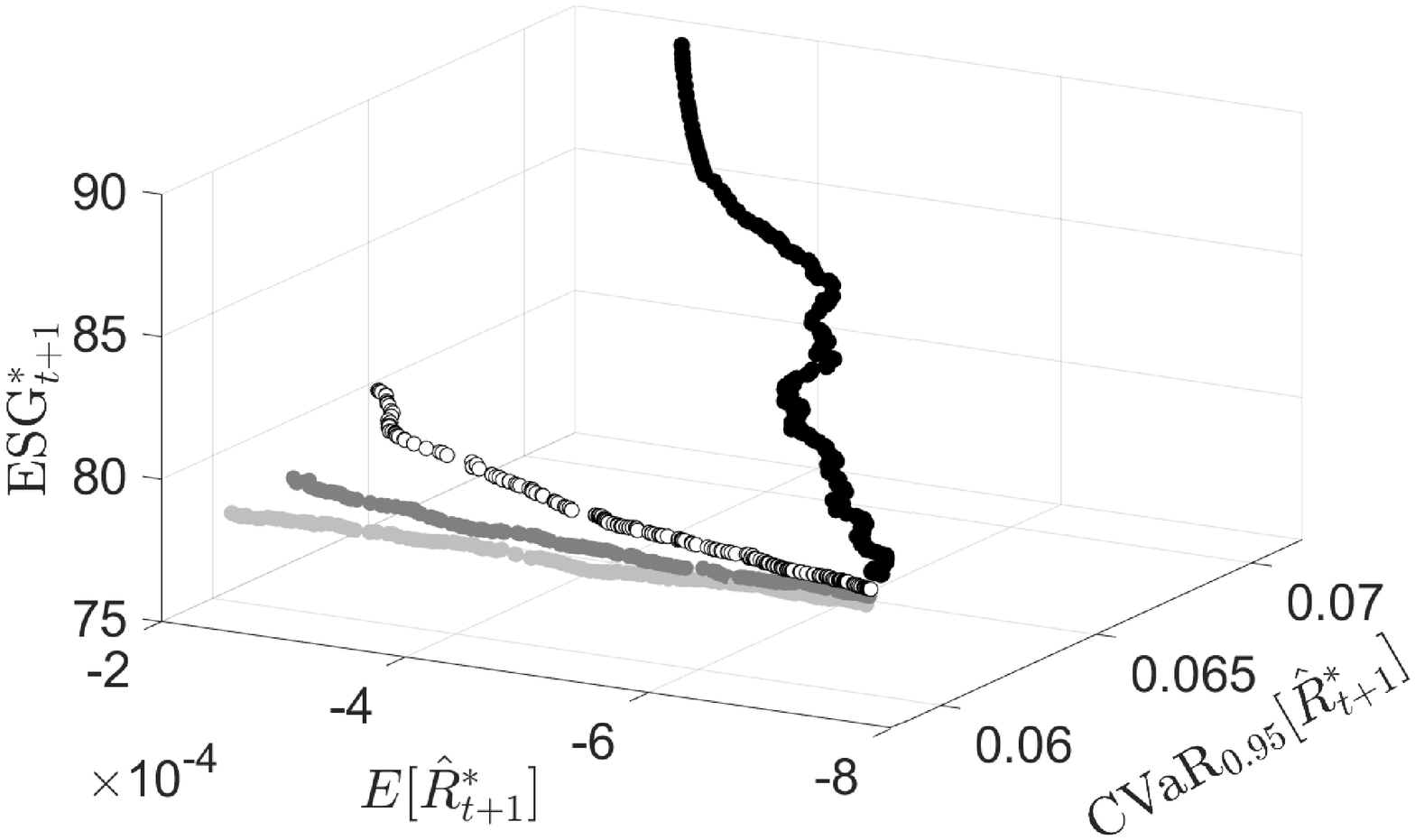}}\hspace{.2em}%
	\subcaptionbox{}{\includegraphics[width=0.32\textwidth]{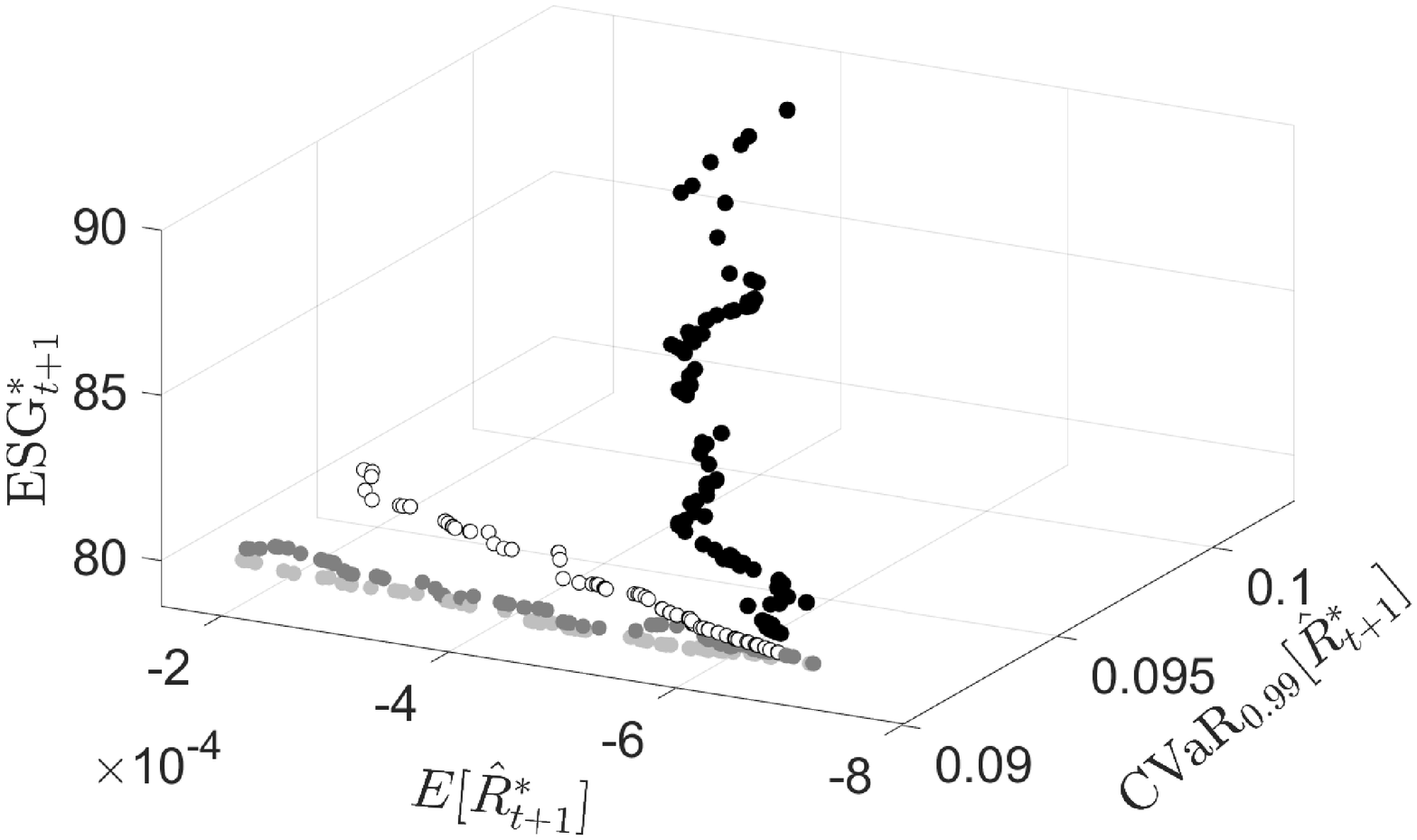}}\hspace{0em}%
	\vspace{.2em}
	\subcaptionbox{}{\includegraphics[width=0.33\textwidth]{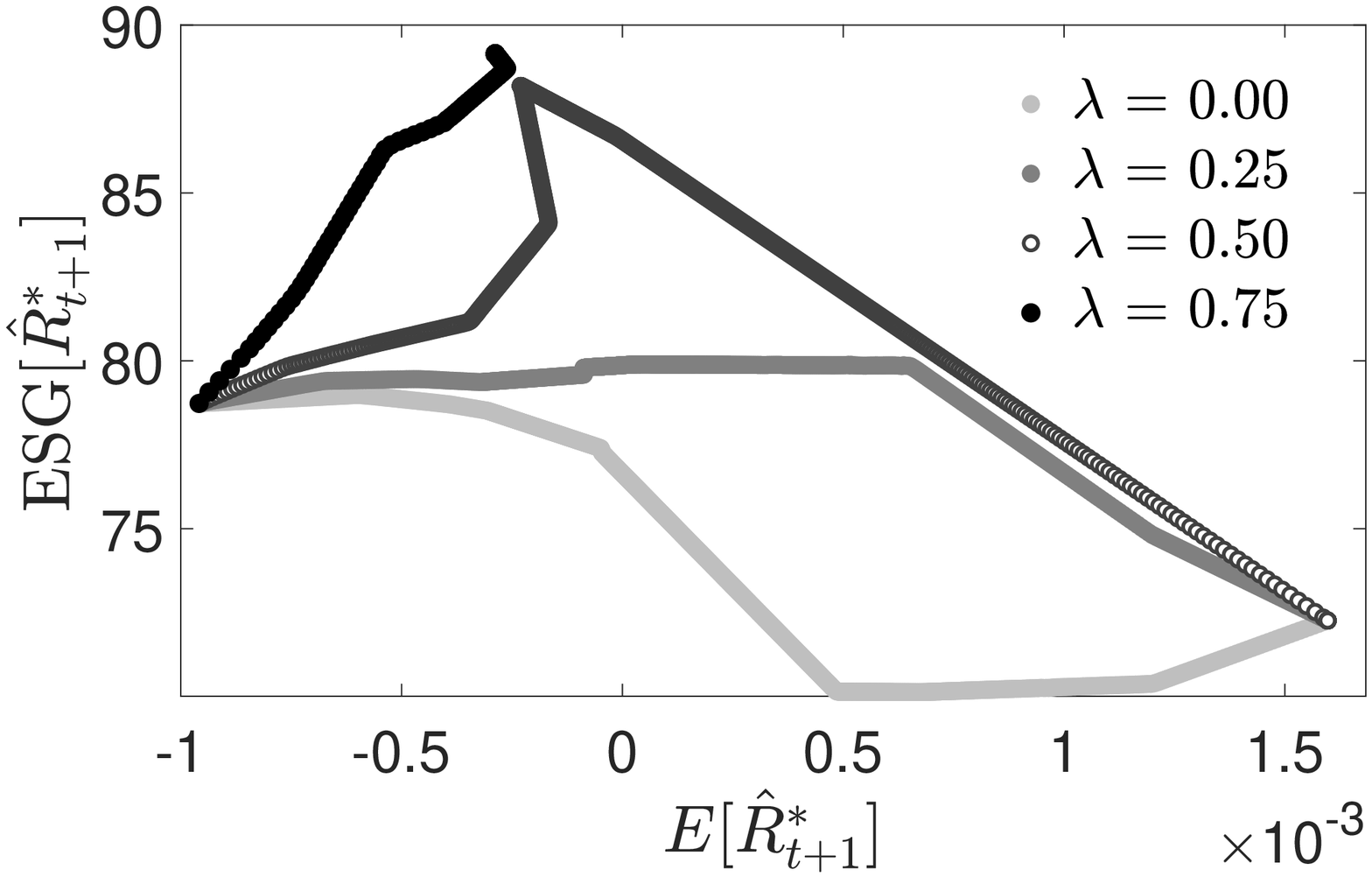}}\hspace{0em}%
	\subcaptionbox{}{\includegraphics[width=0.33\textwidth]{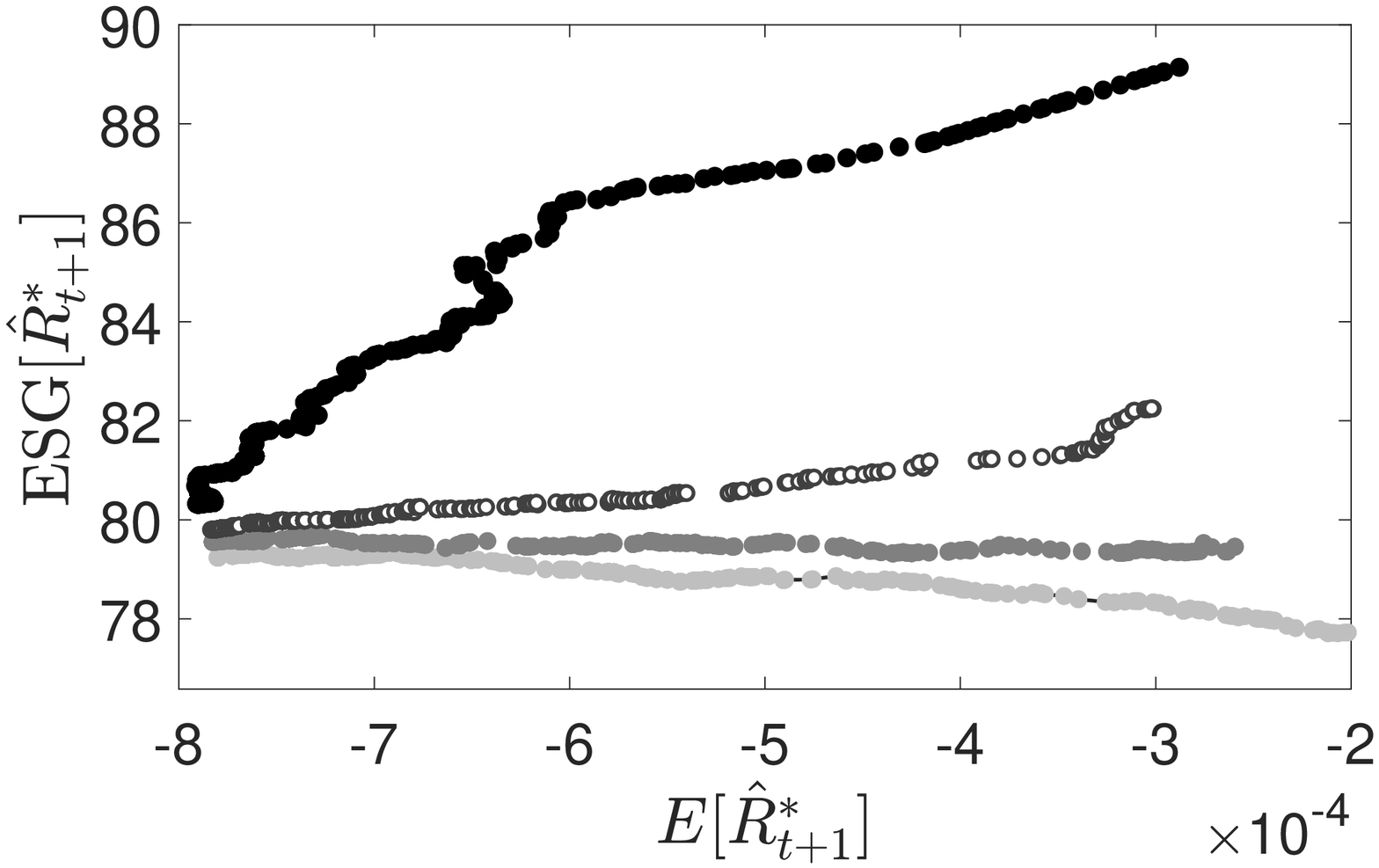}}\hspace{0em}%
	\subcaptionbox{}{\includegraphics[width=0.33\textwidth]{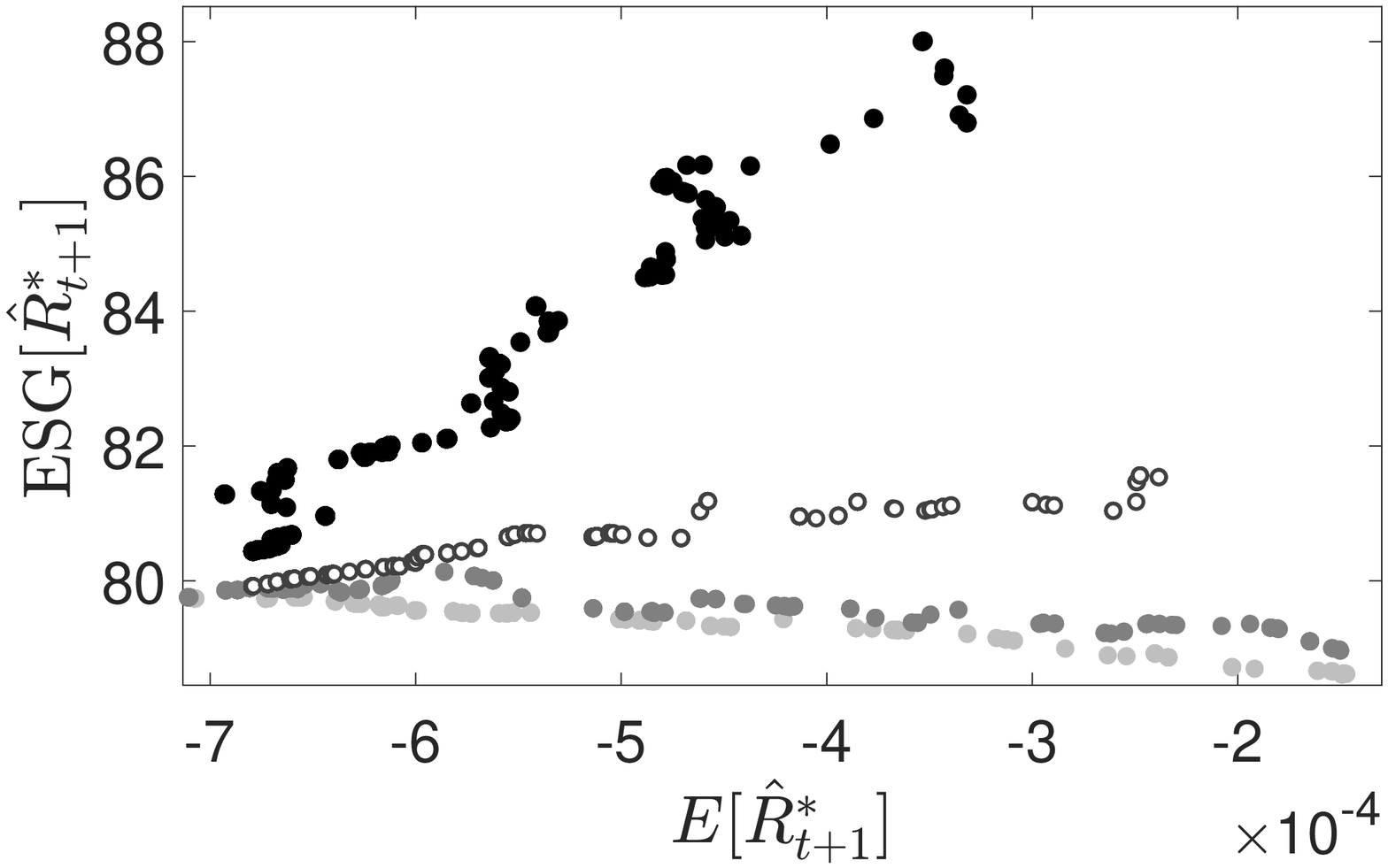}}\hspace{0em}%
	\vspace{.2em}
	\subcaptionbox{}{\includegraphics[width=0.33\textwidth]{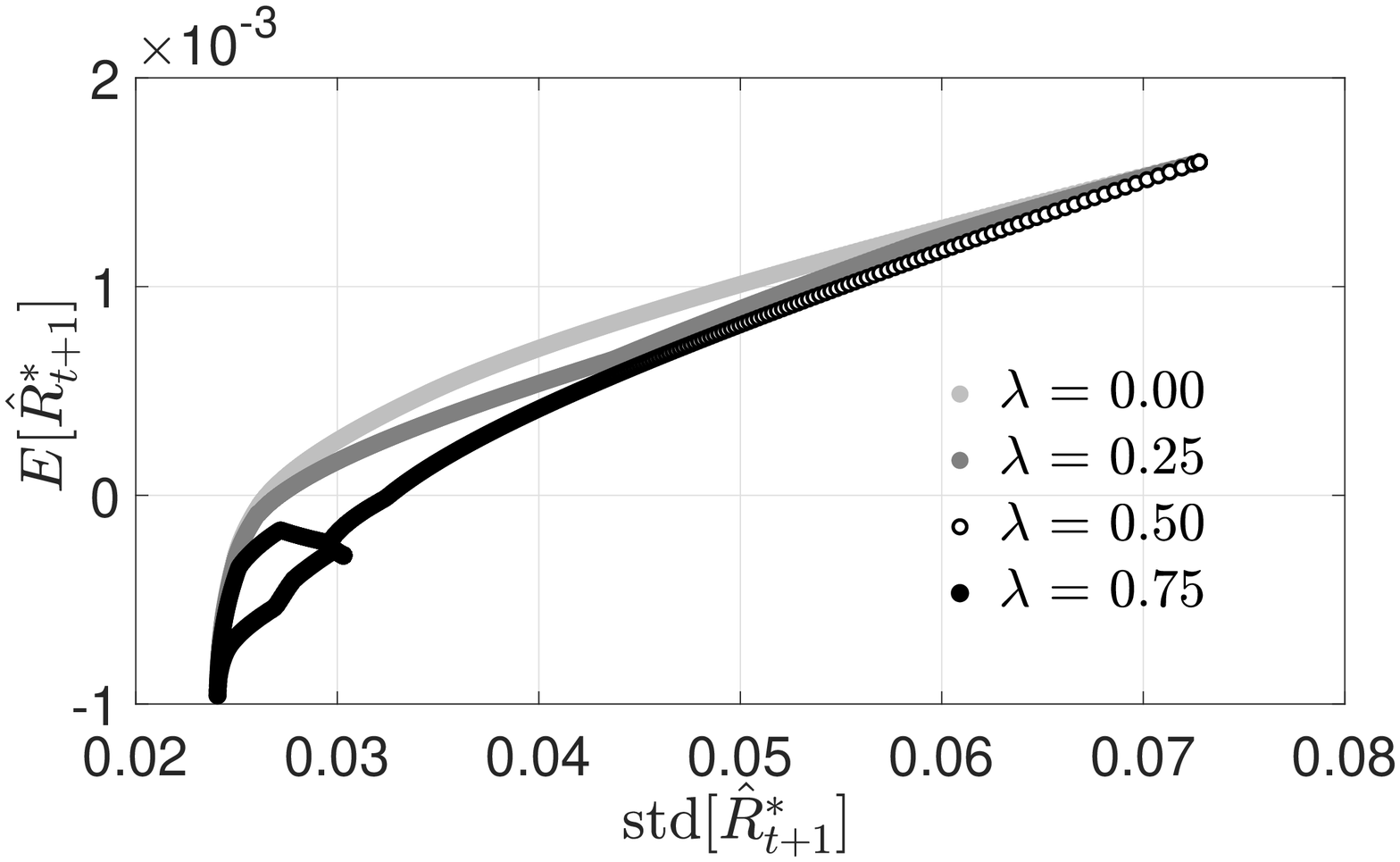}}\hspace{0em}%
	\subcaptionbox{}{\includegraphics[width=0.33\textwidth]{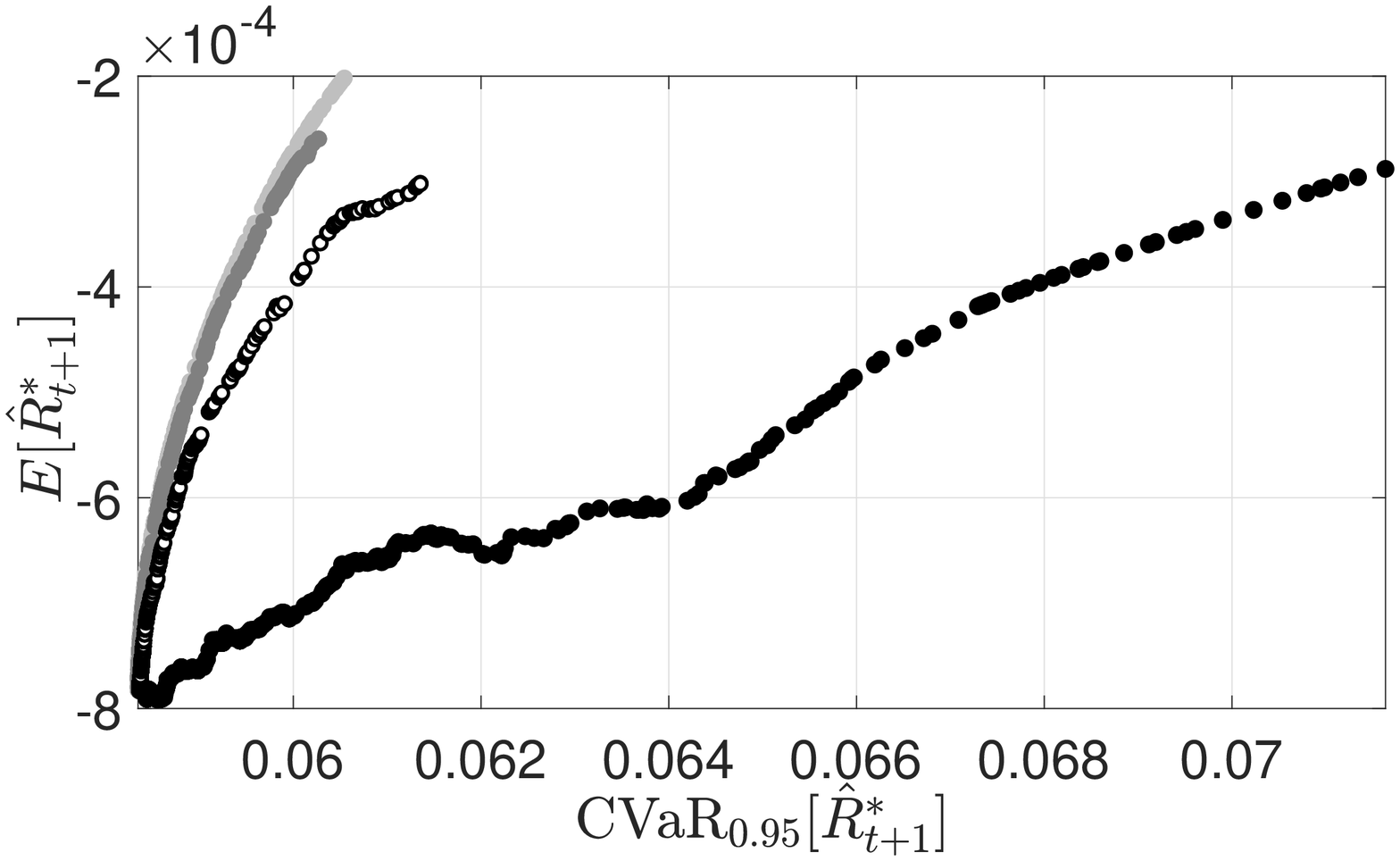}}\hspace{0em}%
	\subcaptionbox{}{\includegraphics[width=0.33\textwidth]{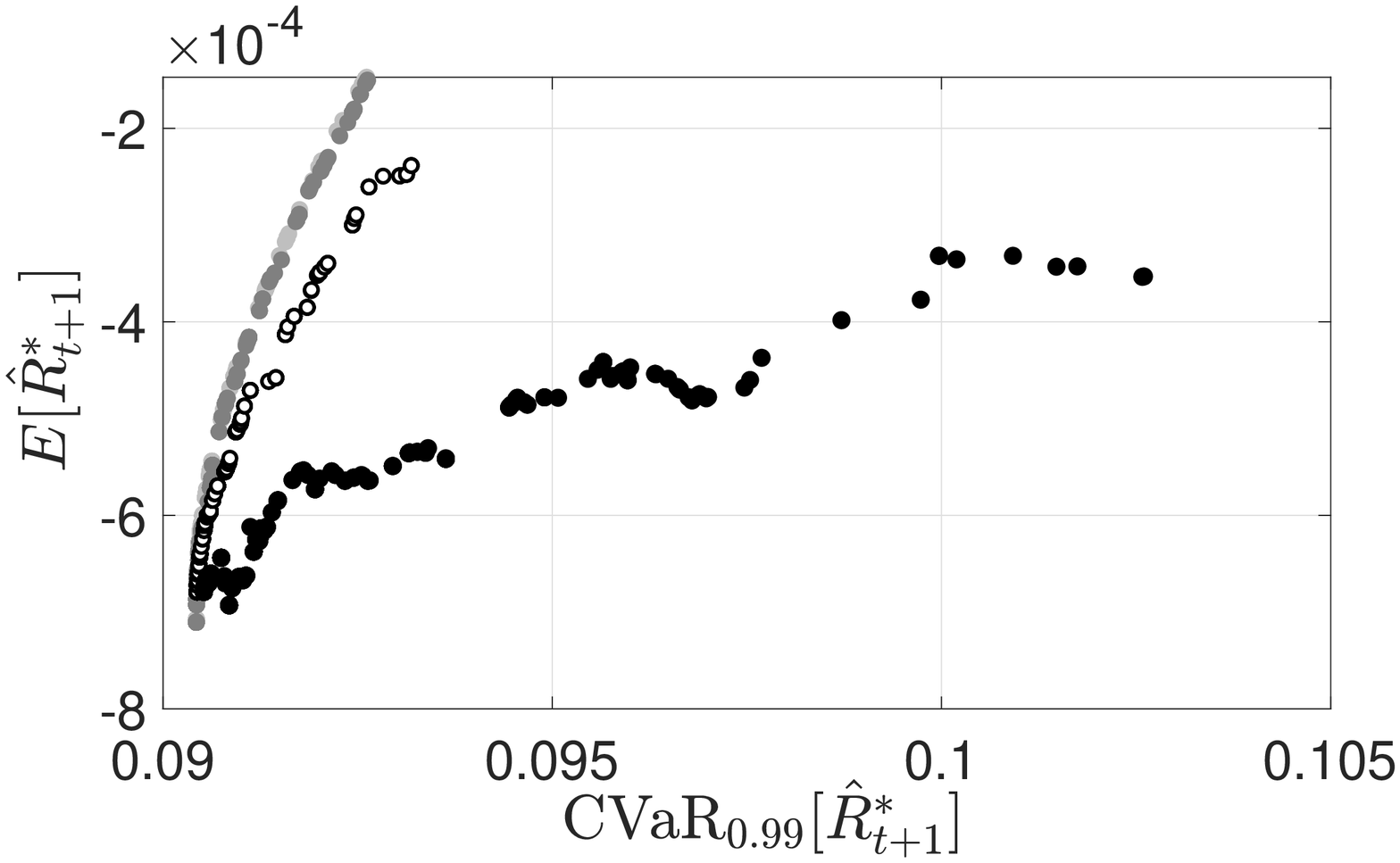}}\hspace{0em}%
	\caption{Refinitiv-based ESG-adjusted return efficient frontiers for 03/30/2020 plotted in
	 (top) $(\mathbb{V}[ \hat{R}^*_{t+1}], \mathbb{E}[ \hat{R}^*_{t+1}],\text{ESG}^*_{t+1})$ space;
	 projected onto the
	 (middle )$( \text{ESG}^*_{t+1}, \mathbb{E}[ \hat{R}^*_{t+1}] )$
	 and
	 (bottom) $( \mathbb{V}[ \hat{R}^*_{t+1}], \mathbb{E}[ \hat{R}^*_{t+1}] )$ planes.
		     From left to right the optimizations are: a,d,g) MV, b,e,h) MCVaR${}_{0.95}$ and c,f,i) MCVaR${}_{0.99}$.
	} 
	\label{fig:EF_Fan_n_30032020}	
\end{figure}

In Section 2.2, efficient frontiers were computed for the MV, mCVaR${}_{0.95}$ and mCVaR${}_{0.99}$
optimizations for the date 12/30/2019 using the ESG ratings from Refinitiv.
The efficient frontiers displayed (manscript Figs. 2 and 3) a positive, monotonic, dependence of the
optimal portfolio's ESG score with increasing value of the risk paramenter $\alpha$,
even for the ESG affinity parameter value $\lambda = 0$.
We provide here an alternate example, obtained by computing the efficient frontiers for the date
03/30/2020, again using the ESG ratings provided by Refinitiv.
These efficient frontiers are plotted in
$(\mathbb{V}[ \hat{R}^*_{t+1}], \mathbb{E}[ \hat{R}^*_{t+1}],\text{ESG}^*_{t+1})$ space
and projected onto the
$(\text{ESG}^*_{t+1}, \mathbb{E}[ \hat{R}^*_{t+1}])$
and
$(\mathbb{V}[ \hat{R}^*_{t+1}], \mathbb{E}[ \hat{R}^*_{t+1}])$ planes
in Fig.~\ref{fig:EF_Fan_n_30032020}.

The date 03/16/2020 represents the low point of the drop in the Dow Jones Industrial average
due to the Covid 19 pandemic, with the average closing at \$19,170.
By 03/30/2020, the index had regained somewhat to \$21,053.
However, it was not until 11/08/2020 that the DJI average surpassed the value of \$29,398 it
held just prior to the pandemic crash.
For 03/30/2020, only the MV optimization was capable of projecting positive expected returns;
the tail risk CVaR optimizations project negative returns.
For the optimizations on 12/30/2019, the efficient frontiers converged to the same point as $\alpha \uparrow 1$
for all four values of $\lambda$.
For the 03/30/2020 optimization, convergence only occurs for the three smaller values of $\lambda$ under MV optimization.
For the mCVaR optimizations, for each value of $\lambda$, the efficient frontiers approach different values as $\alpha \uparrow 1$.
The projections on the $( \text{ESG}^*_{t+1}, \mathbb{E}[ \hat{R}^*_{t+1}] )$ indicate radically different behaviors of
the portfolio ESG values along the efficient frontiers.
Under the MV optimization, for $\lambda = 0$, the ESG value falls and then rises.
For $\lambda = 0.25, 0.5$, the ESG value rises and then falls.
For $\lambda = 0.75$, the ESG value monotonically increases with $\alpha$.
For the mCVaR optimizations, the ESG value decreases as $\alpha$ increases for the $\lambda = 0, 0.25$ efficient frontiers;
for the other two values of $\lambda$ it increases.
For the mCVaR optimizations with $\lambda = 0.75$,
the change in $\mathbb{E}[ \hat{R}^*_{t+1}]$ with respect to $\alpha$ is non-monotonic.
This is also seen in the MV optimization with $\lambda = 0.5$ and to a small extent for $\lambda = 0.75$.

\begin{figure}[!h]
	\centering
	\subcaptionbox{}{\includegraphics[width=0.34\textwidth]{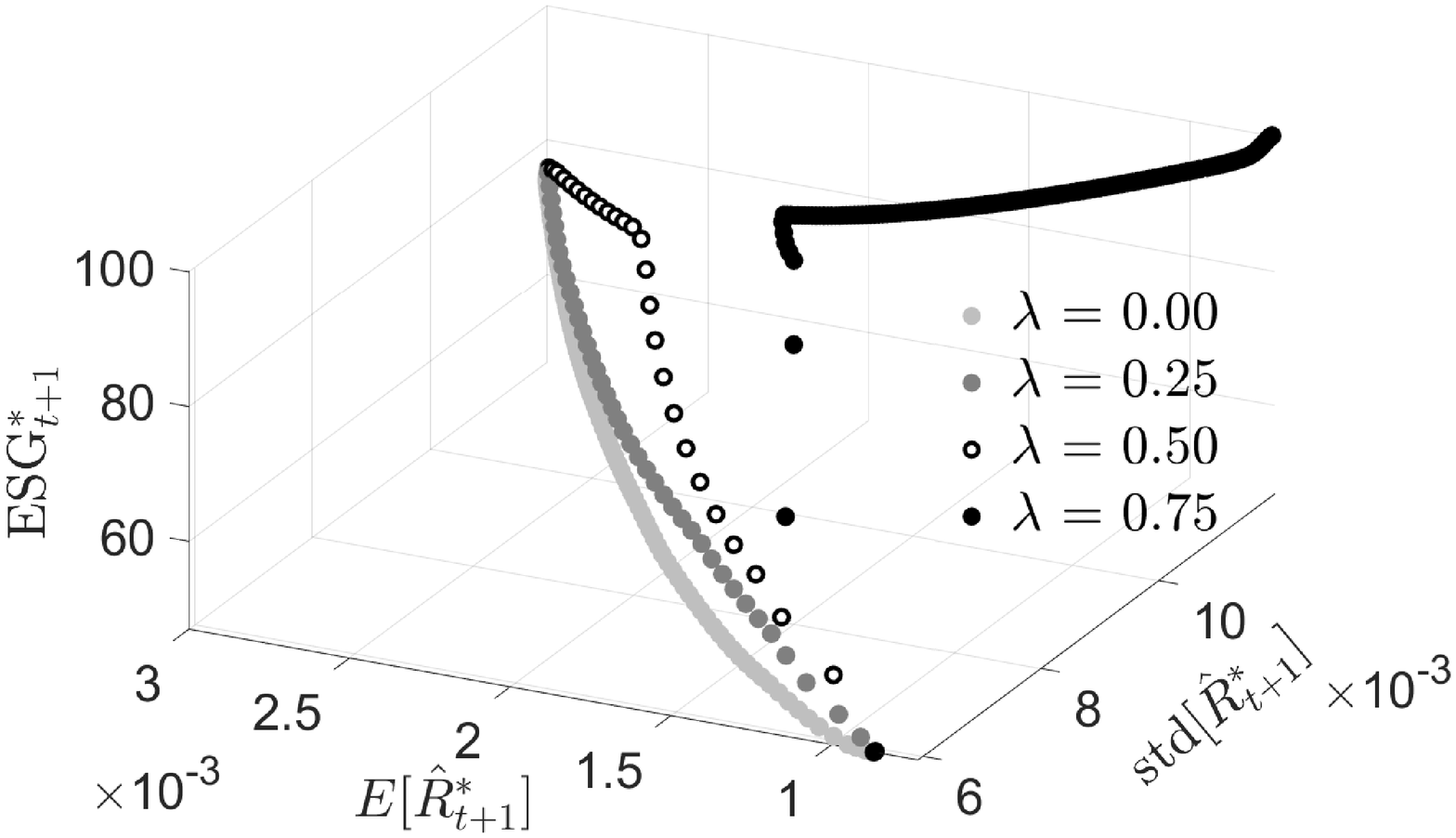}}\hspace{0em}%
	\subcaptionbox{}{\includegraphics[width=0.32\textwidth]{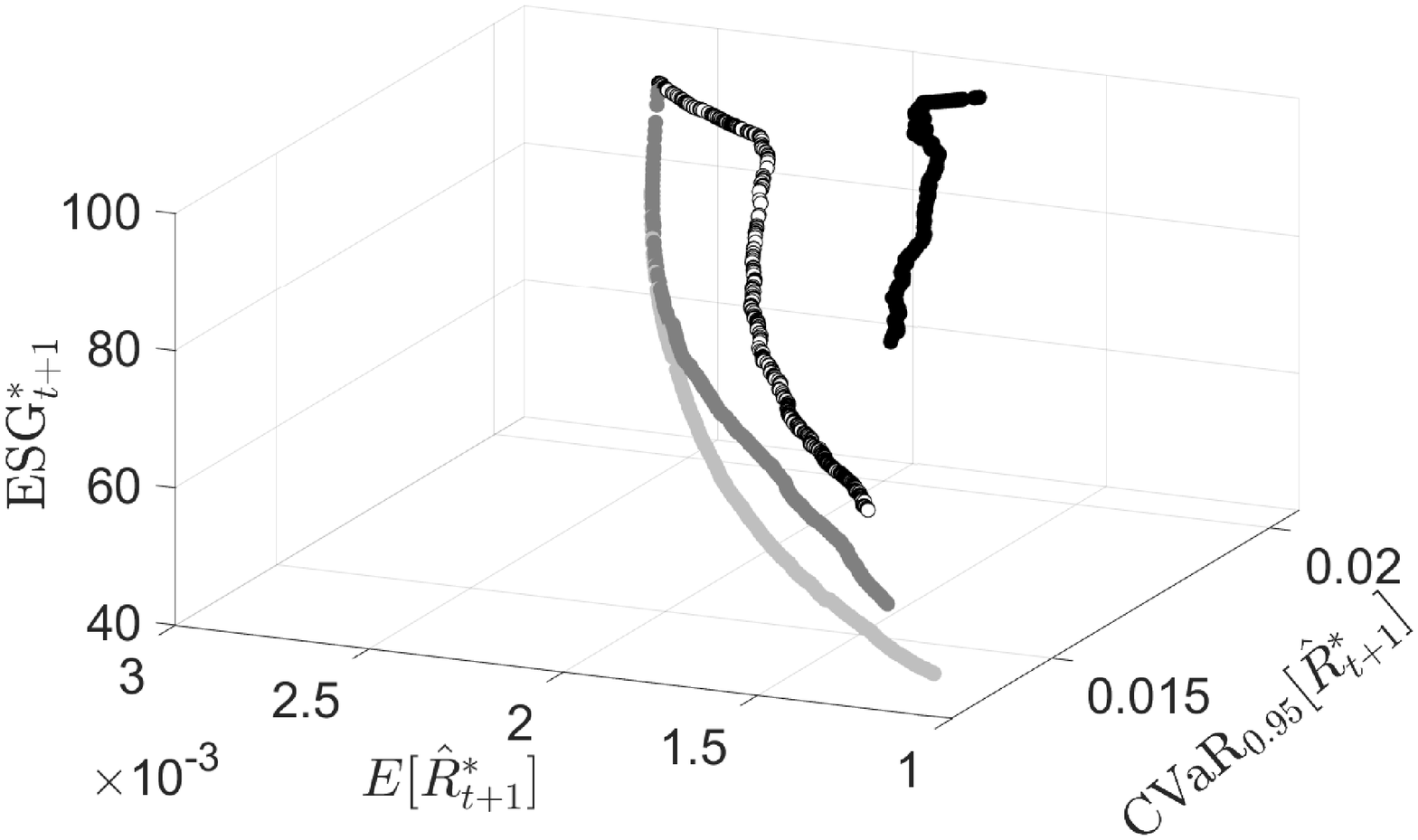}}\hspace{.2em}%
	\subcaptionbox{}{\includegraphics[width=0.32\textwidth]{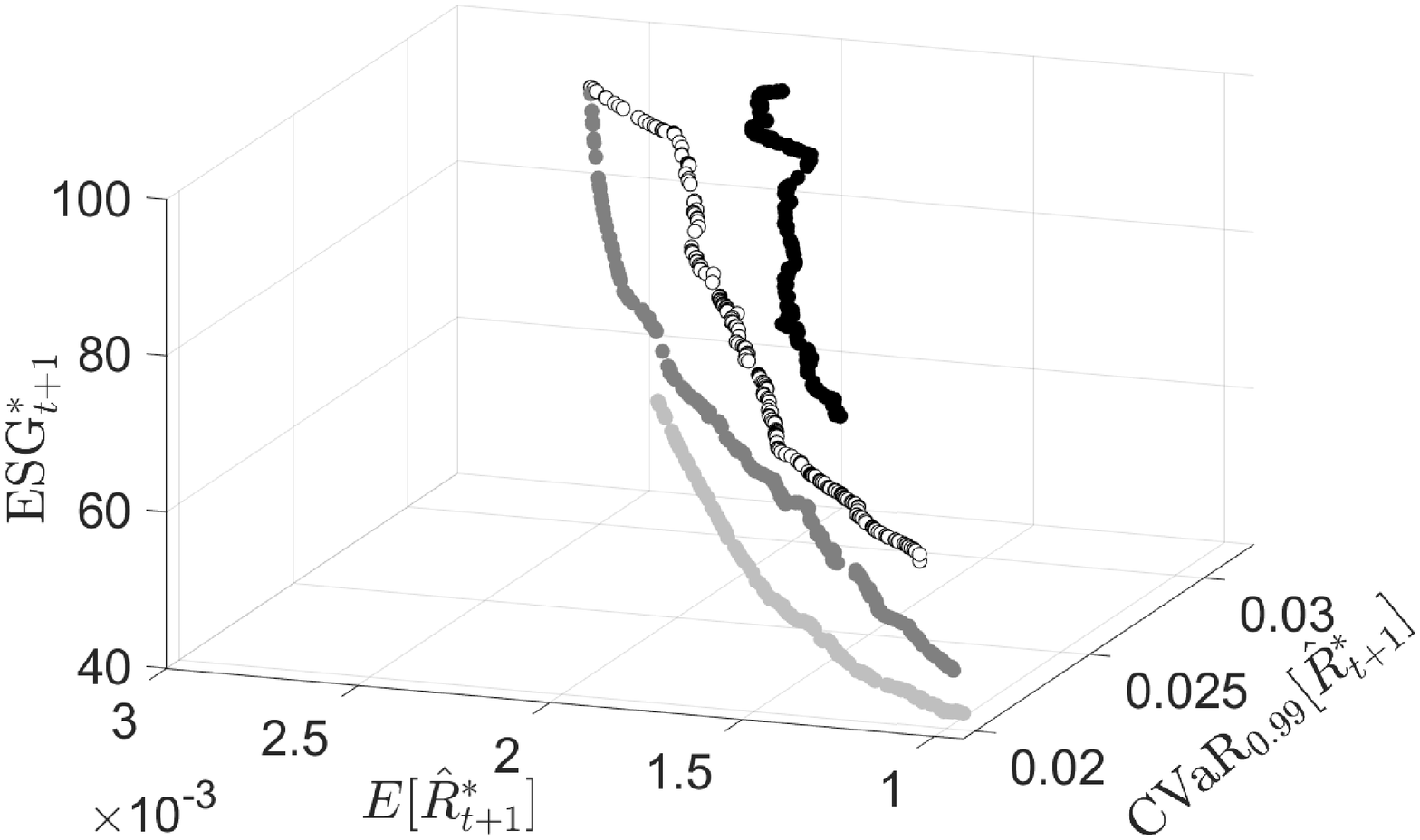}}\hspace{0em}%
	\vspace{.2em}
	\subcaptionbox{}{\includegraphics[width=0.33\textwidth]{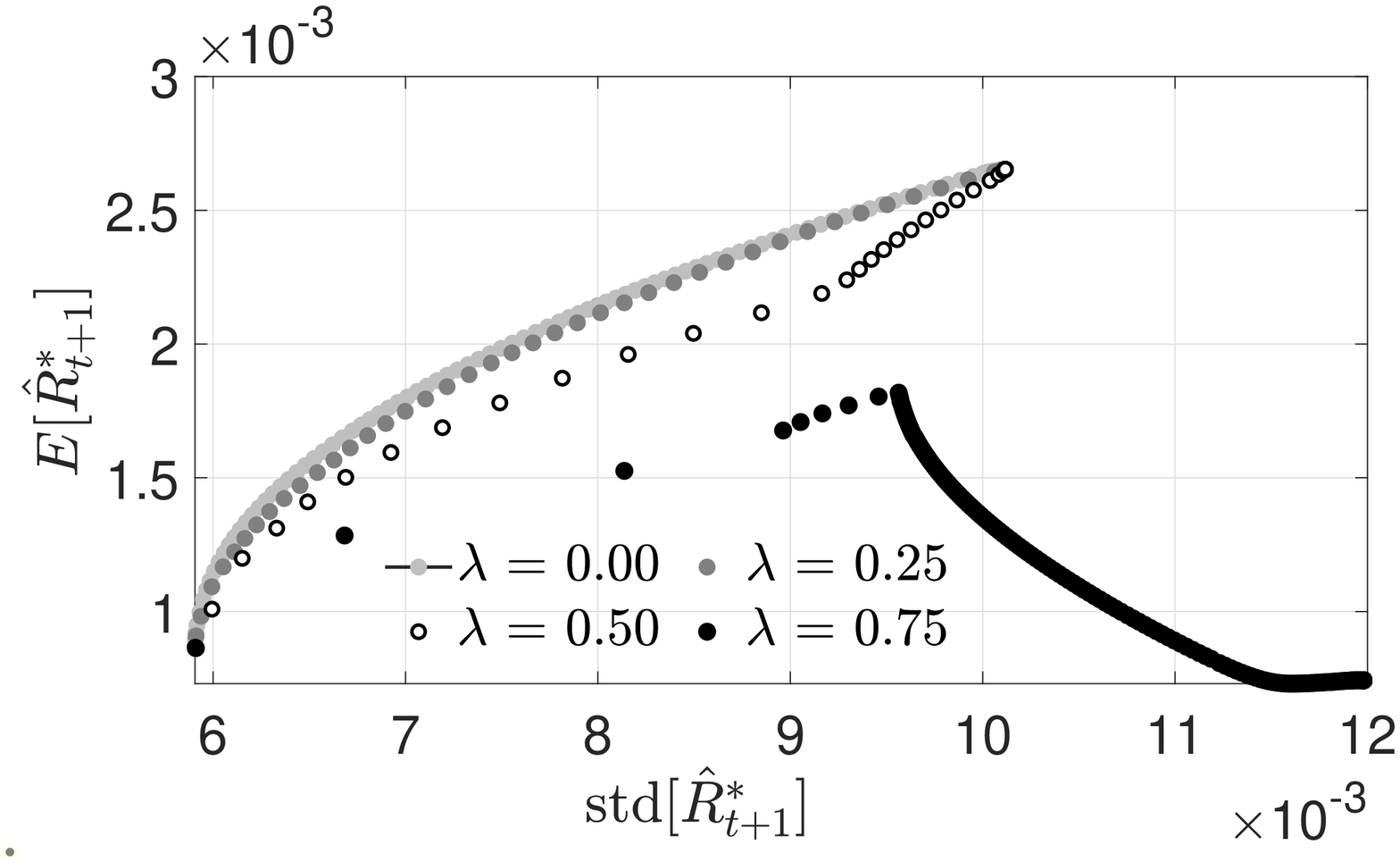}}\hspace{0em}%
	\subcaptionbox{}{\includegraphics[width=0.33\textwidth]{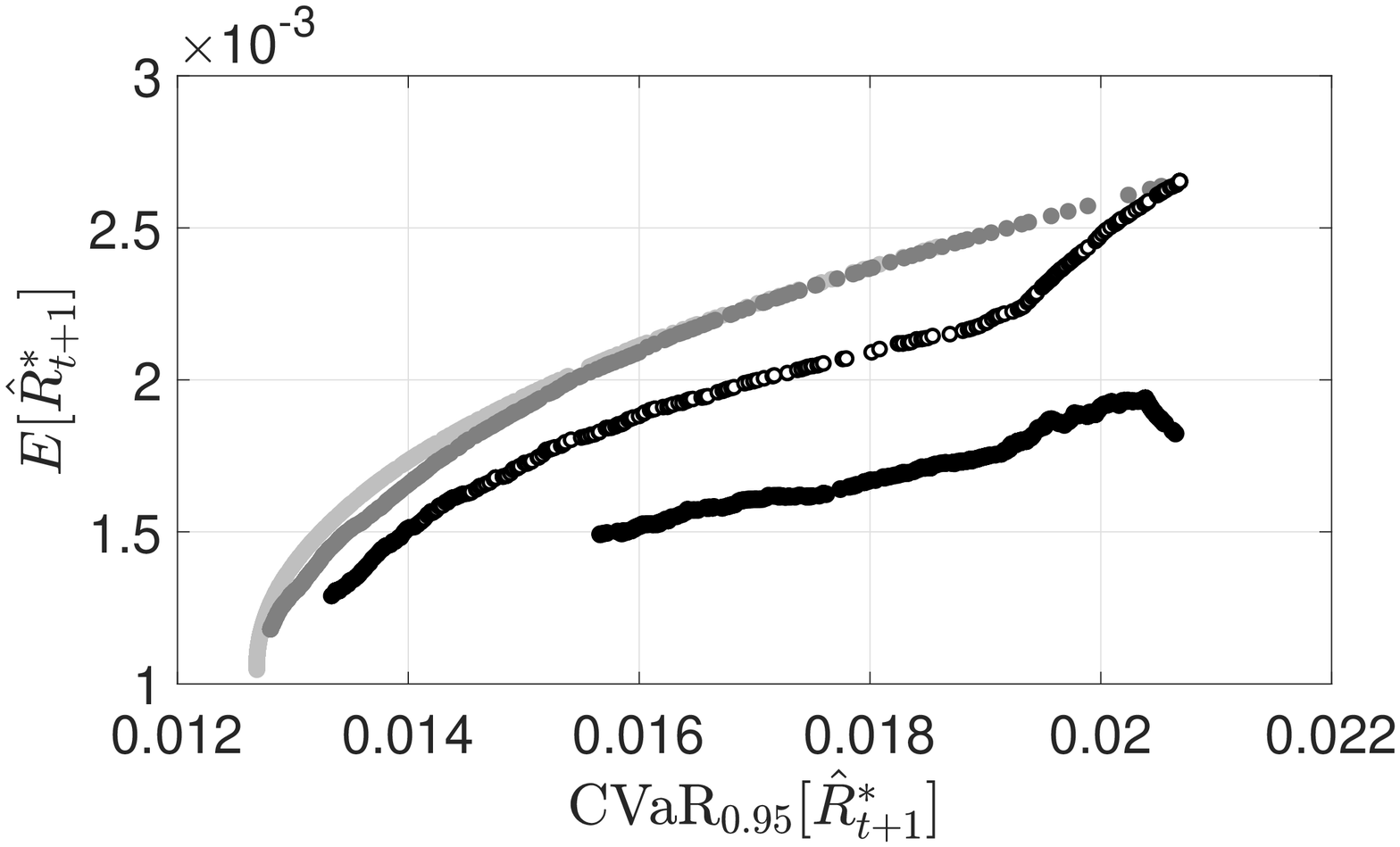}}\hspace{0em}%
	\subcaptionbox{}{\includegraphics[width=0.33\textwidth]{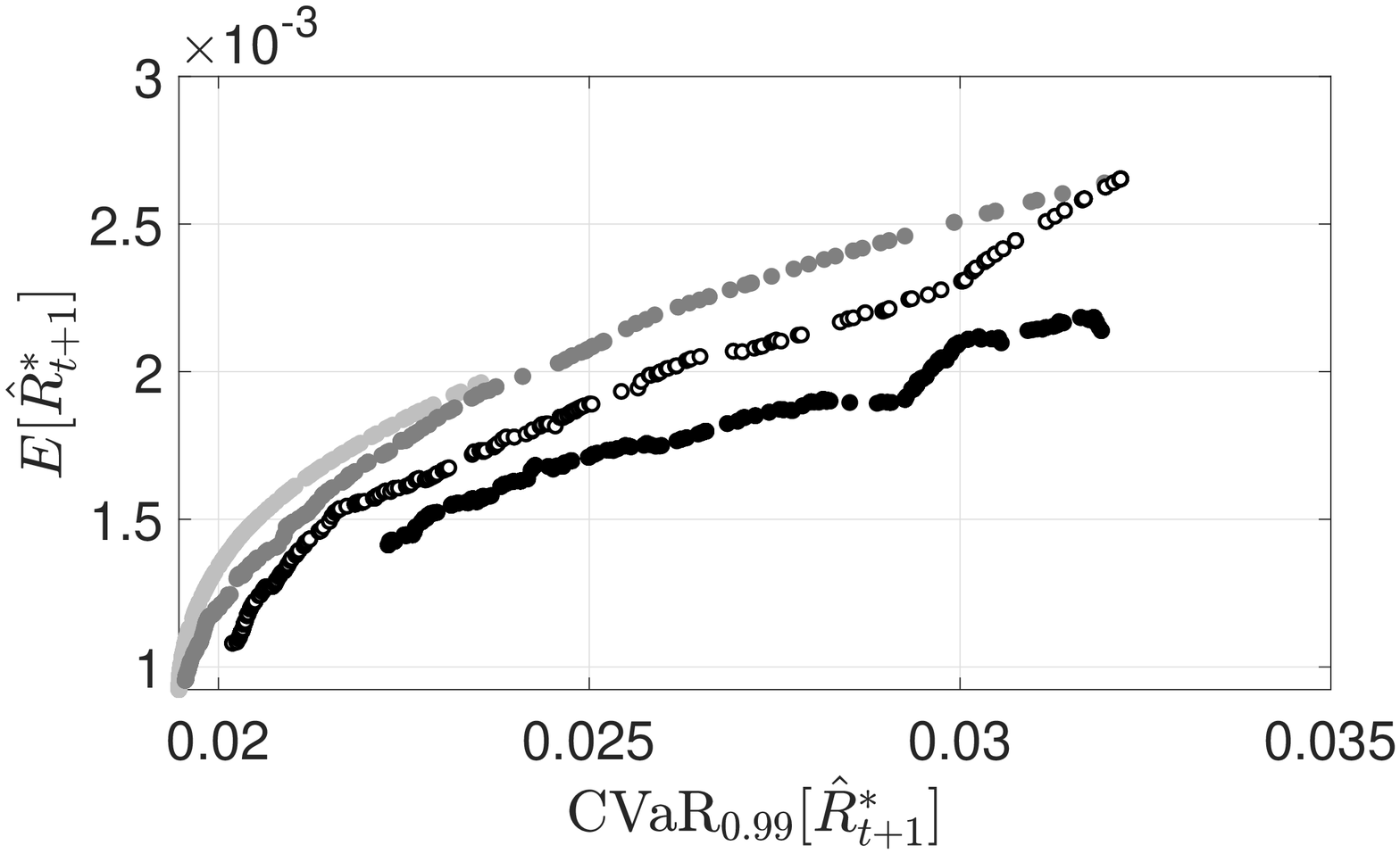}}\hspace{0em}%
	\caption{RobecoSAM-based ESG-adjusted return efficient frontiers for 12/30/2019 (top) plotted in
		$(\mathbb{V}[ \hat{R}^*_{t+1}], \mathbb{E}[ \hat{R}^*_{t+1}],\text{ESG}^*_{t+1})$ space and
		(bottom) projected onto the $(\mathbb{V}[ \hat{R}^*_{t+1}], \mathbb{E}[ \hat{R}^*_{t+1}])$ plane.
		 From left to right the optimizations are: a,d) MV, b,e) MCVaR${}_{0.95}$ and c,f) MCVaR${}_{0.99}$.
	} 
	\label{fig:EF_Fan_n_30122019_robeco}	
\end{figure}
In Fig.~\ref{fig:EF_Fan_n_30122019_robeco}, we show efficient frontiers computed for the same date 12/30/2019
as in the manuscript, but now using the RobecoSAM scores.
(Compare with manuscript Fig.~3.)
The $\lambda = 0.25$ RobecoSAM efficient frontiers display behavior comparable to that of
the $\lambda = 0.25$ and $0.5$ Refinitiv efficient frontiers.
The $\lambda = 0.5$ RobecoSAM efficient frontiers display behavior comparable to that of
the $\lambda = 0.75$ Refinitiv efficient frontiers.
However the $\lambda = 0.75$ RobecoSAM efficient frontiers display very different behaviors compared to those for
Refinitiv.
In particular, for MV optimization the $\lambda = 0.75$ RobecoSAM efficient frontier shows decreasing expected return
at hign mean standard deviation values.

\begin{figure}[h!]
	\centering
	\includegraphics[width=0.8\textwidth]{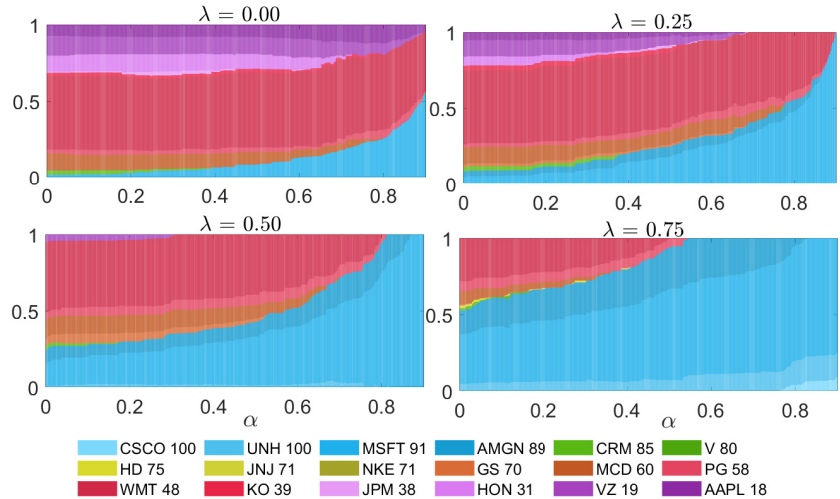}
	\caption{Variation of the weight composition along each efficient frontier (as a function $\alpha$)
		for the MCVaR${}_{0.99}$ optimal portfolios with $\lambda = 0, 0.25, 0.5, 0.75$.
		Assets with lower ESG scores are depicted with warmer colors.
	}
	\label{fig:Weights_n_MCVaR99_fan_30122019_robeco}
\end{figure}
The lack of correlation between Refinitiv and RobecoSAM ESG ratings will impact optimal portfolio composition.
For each choice of $\lambda$, the compositions, as a function of  $\alpha$, of the  mCVaR${}_{0.99}$ optimal portfolios of
Fig.~\ref{fig:EF_Fan_n_30122019_robeco}
are shown in Fig.~\ref{fig:Weights_n_MCVaR99_fan_30122019_robeco}.
Thirteen assets are represented in the mCVaR${}_{0.99}$ efficient frontier portfolios in Fig.~5 of the manuscript
while 18 are represented in Fig.~\ref{fig:Weights_n_MCVaR99_fan_30122019_robeco}.
Twelve of the assets are represented in both the Refinitiv and RobecoSAM based optimizations.
Refinitiv ESG scores for the 13 represented assets range from 65 to 83,
while the RobecoSAM scores for the 18 represented assets range from 18 to 100.

\subsection{Performance Measures}\label{sec:robeco_PM}

The performance measures Tot Rtn, Ann Rtn, AvgTO, ETL95, ETR95, and MDD,
as well as the average and standard deviation of the portfolio ESG scores,
over the time period 01/03/2017 through 12/30/2020
for select values of $\alpha$ on the RobecoSAM-based  mCVaR${}_{0.99}$ and MV optimized efficient frontiers
are presented in Table~\ref{tab:PM_sim} of Section~\ref{sec:Tbl_robeco}.
Except for the ESG scores, the performance measures for the $\lambda = 0$ efficient frontiers will be identical
whether using Refinitiv or RobecoSAM data.
For $\lambda > 0$, we note the same general trends in the performance measures as the value of $\lambda$ increases
as were observed in the Refinitiv data.
However, the performance measures obtained using the two different ESG data sets will differ,
reflecting the time-average of the types of differences displayed for the date 12/30/2019 in
Fig.~2 of the manuscript for Refinitiv data compared with Fig.~\ref{fig:EF_Fan_n_30122019_robeco} for RobecoSAM data.
For given values of $\lambda$ and $\alpha$, for mCVaR${}_{0.99}$ optimization the return performance
is generally better for RobecoSAM data, while for MV optimization the return performance is better for Refinitiv data.
The Refinitiv data based AvgTO, MDD and ESG scores are generally better under both optimizations.
Under mCVaR${}_{0.99}$ optimization, the ETL/ETR values are better for Refinitiv data;
while Robesco data produces better values under MV optimization.
There are also $\lambda$-dependant differences.
For example, while the Avg ESG score is higher by as much as 15 to 16 points for Refinitv-based optimization (compared
with RovbecoSAM) when $\lambda = 0$,
this difference decreases as $\lambda$ increases.
When $\lambda = 0.75$ the Refinitive-based optimizations produce a higher Avg ESG score by as little as 0.25 (mCVaR) to
4.5 (MV optimization).
The other pronounced $\lambda$ dependence is in the total (and annual) returns under MV optimization.
The difference in total return provided by Refinitiv-based data compared to RobecoSAM-based data
grows from 8.5 for $\lambda = 0.25$ to 42.0 for $\lambda = 0.75$.

Values for the EWBH benchmark are also given.
Relative to EWBH, the Tot. Ret, Ann. Ret, and avg ESG score of the optimized portfolios outperform that of the benchmark
as $\lambda$ increases.
Under MCVaR${}_{0.99}$ optimization, the MDD remains better than that of EWBH for all values of $\lambda$,
while the MDD for MV optimzation is better once $\lambda \ge 0.25$.
Only for the smaller values of $\lambda$ are the ETL/ETR values for the optimized portfolios better than those for EWBH.

Moments of the Robeco-SAM based return distributions for the period 01/03/2017 through 12/30/2020 are summarized in
Table~\ref{tab:Mom_sim} of Section~\ref{sec:Tbl_robeco}.
Comparing with manuscript Table~A.3 for Refinitiv based data,
mCVaR${}_{0.99}$ Mean, Std, and ExKurt values are smaller and Skew values larger for Refinitiv data;
Median values are roughly compariable.
Under MV optimization,  except for ExKurt, the moment values are generally smaller for RobecsoSAM data.
Relative to EWBH, Mean and ExKurt values of the optimized portfolios become better as $\lambda$ increases.
For mCVaR${}_{0.99}$ optimization, Skew values are always better than those for EWBH;
for MV optimization, Skew values improve beyond EWBH as $\lambda$ increases.
Std values for the optimized portfolios begin to exceed those for EWBH as $\lambda$ increases.

Risk-reward ratio values for the RobecoSAM based optimizations are summarized in
Table~\ref{tab:RRR_Rob} of Section~\ref{sec:Tbl_robeco}.
Comparing with manuscript Table~A.4 for Refinitiv based data,
we find that better RRR values are essentially split between Refinitiv and RobecoSAM data based optimizations.
The strongest exception is that Rachev ratio values are more frequently higher under MV optimization for
the Refinitiv-based data.
All RRR values for the optimized portfolios almost always exceed those for the EWBH benchmark.

\subsection{Tangent Portfolios: Capital Market Lines}\label{sec:robeco_CML}

\begin{figure}[!h]
	\centering
	{\includegraphics[width=0.49\textwidth]{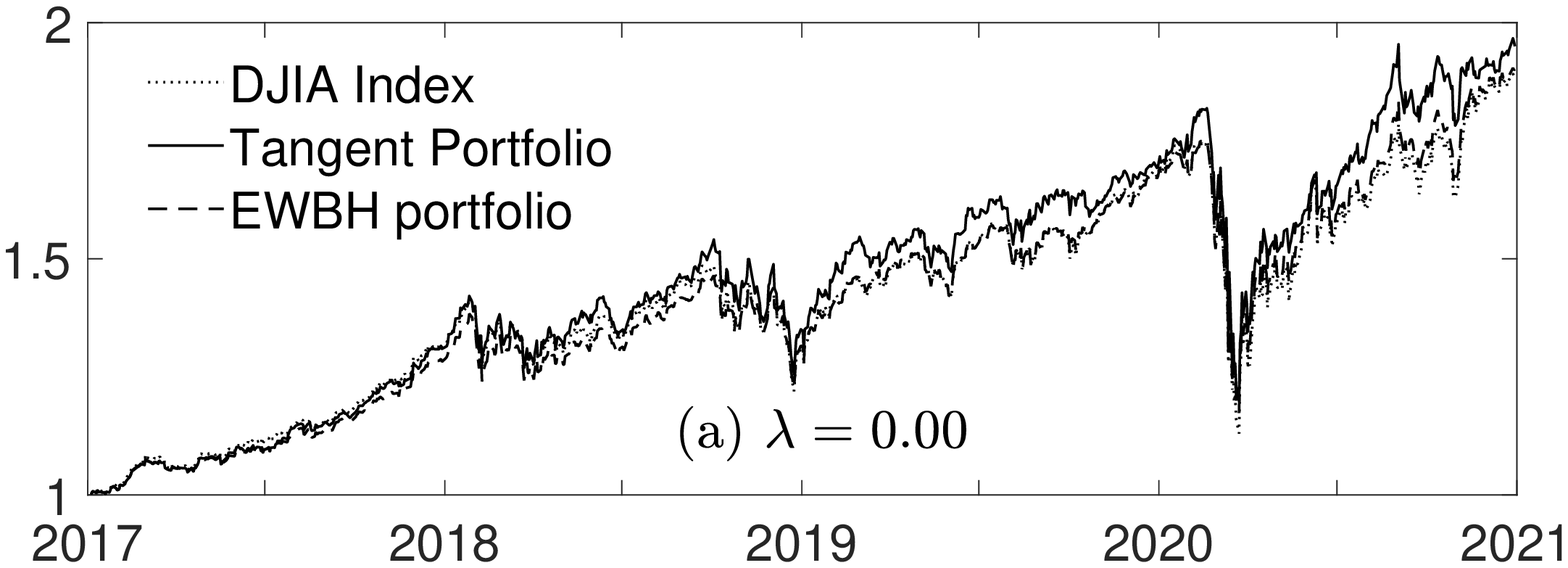}}        \hspace{0em}%
	{\includegraphics[width=0.49\textwidth]{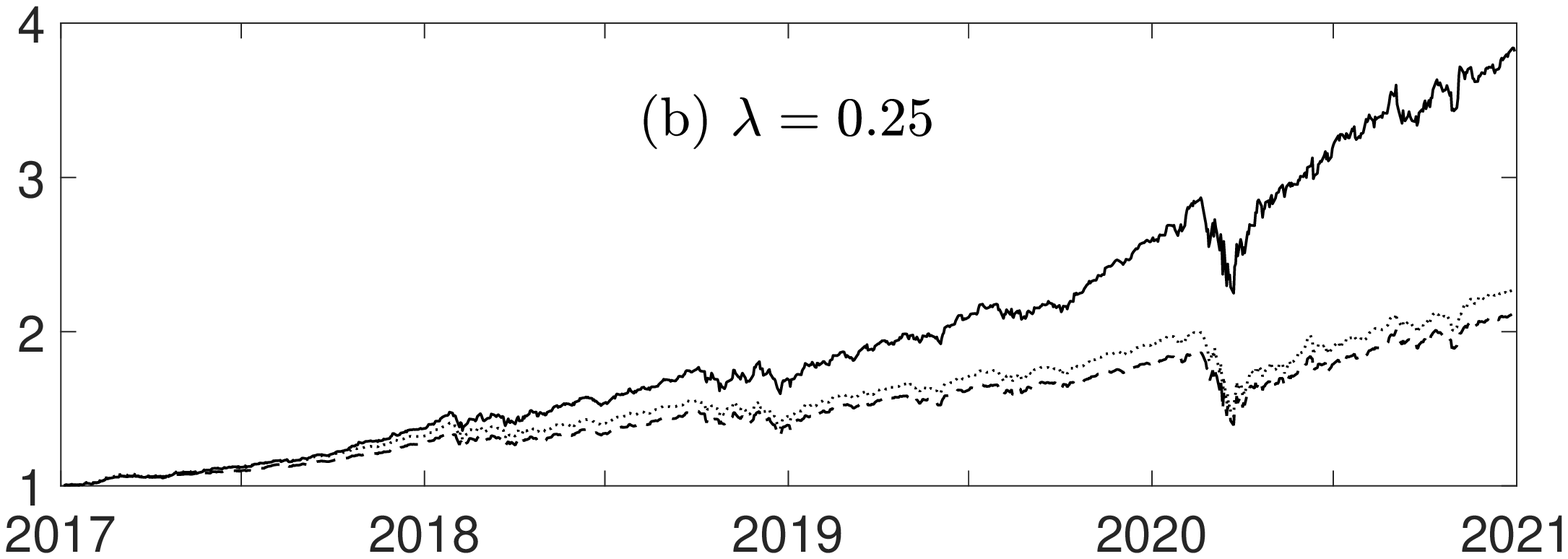}}\hspace{0em}%
	{\includegraphics[width=0.49\textwidth]{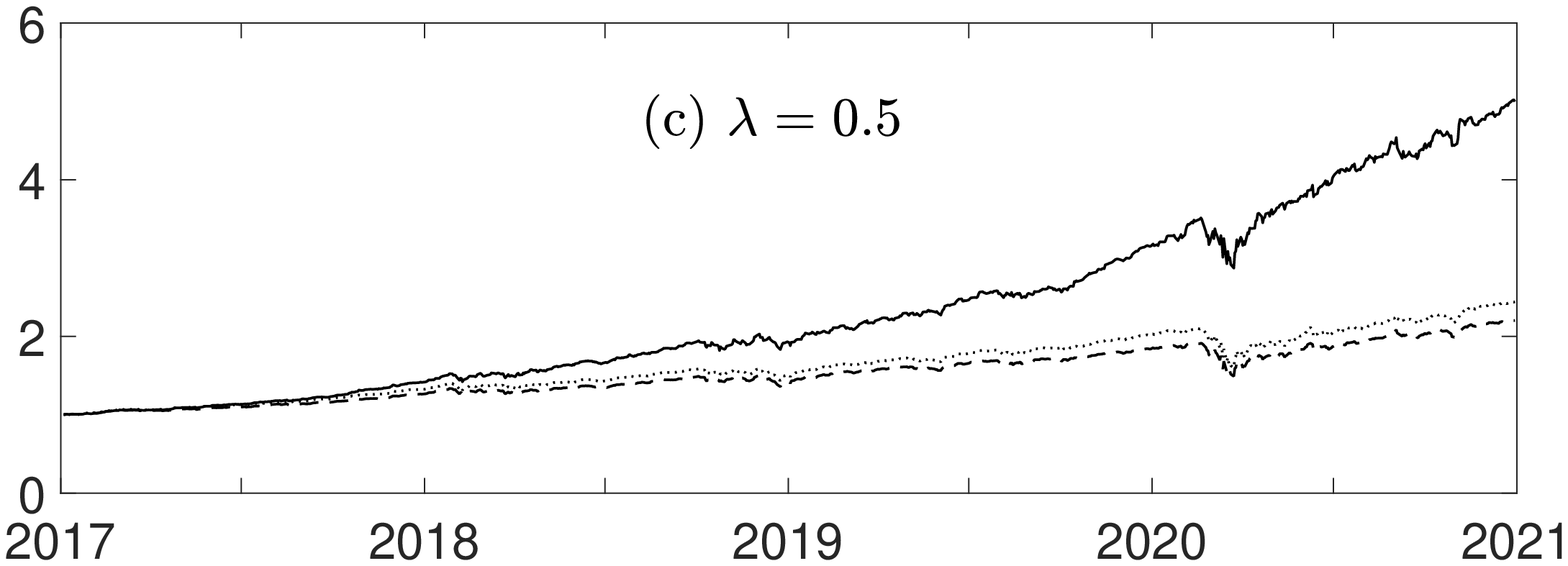}}\hspace{0em}%
	{\includegraphics[width=0.49\textwidth]{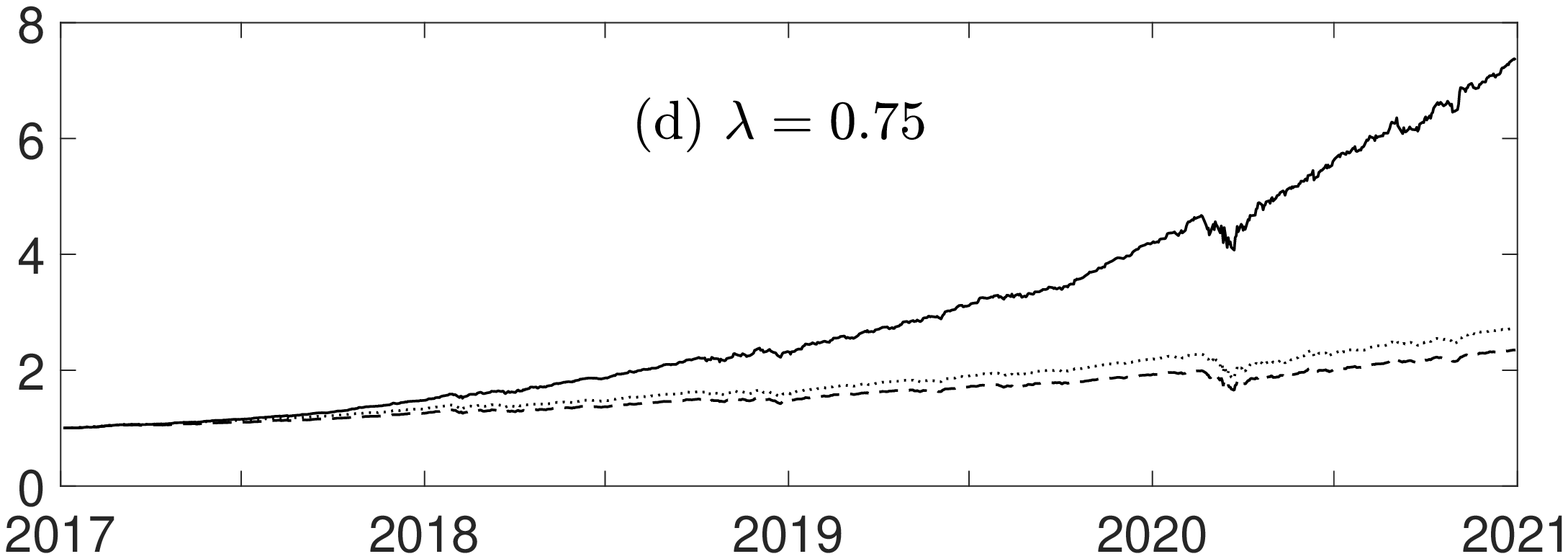}}\hspace{0em}%
	{\includegraphics[width=0.49\textwidth]{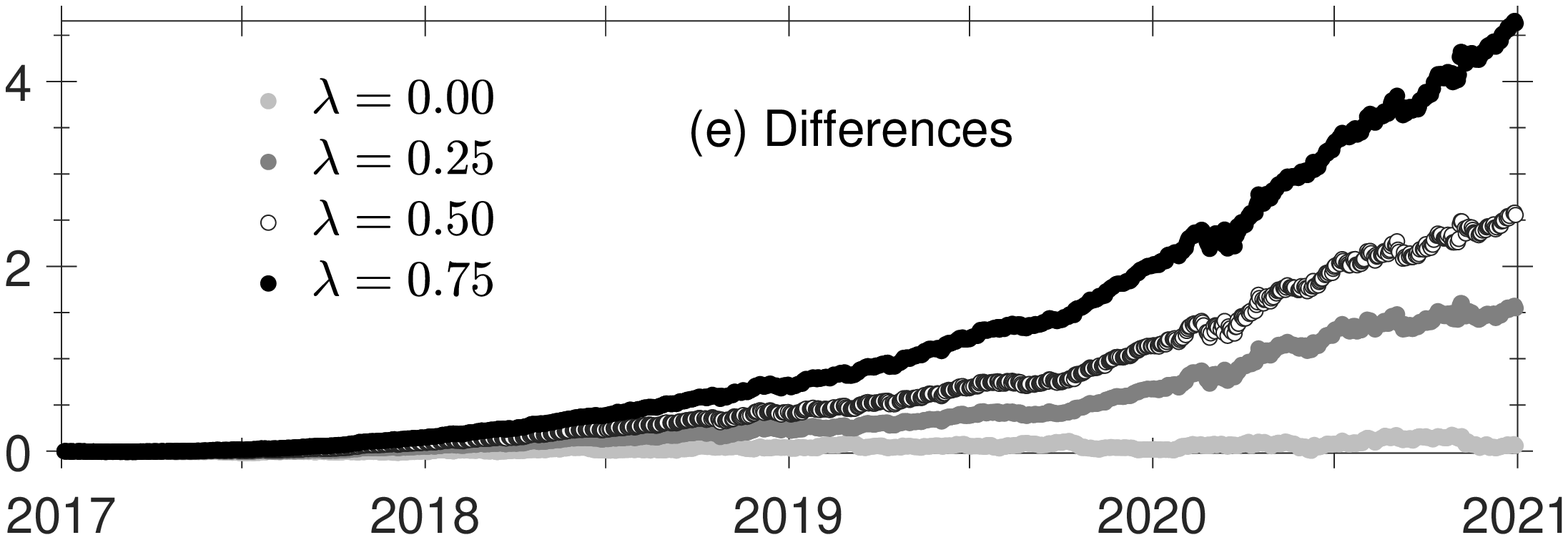}}	\hspace{0em}%
	\caption{(a)-(d) RobecoSAM-based ESG-adjusted portfolio values, $P_t^{(Z)}(\lambda)$,
	obtained for tangent portfolios under mCVaR${}_{0.99}$ optimization,
	compared with the corresponding $P_t^{(Z)}(\lambda)$ time series for the EWBH strategy and the ESG-adjusted DJIA index.
	(e) Differences between the $P_t^{(Z)}(\lambda)$ values for the tangent portfolios and the DJIA index.
	}
	\label{fig:Capm-esg-adj-ofs-MV_robeco}	
\end{figure}

The $P_t^{(Z)}(\lambda)$ time series computed from the ESG-adjusted tangent portfolios under mCVaR${}_{0.99}$ optimization
are shown in Fig.~\ref{fig:Capm-esg-adj-ofs-MV_robeco} for $ \lambda \in \left\{0,0.25,0.5,0.75 \right\}$.
Also plotted are the corresponding time series for the EWBH and DJIA index benchmarks.
All ESG-adjusted values are now computed using the RobecoSAM data.
Comparing with the results in Fig.~8 of the manscript for Refinitiv data,
we confirm that the corresponding time series must be identical when $\lambda =0$.
For $\lambda > 0$, we see a greater difference between the EWBH and DJIA index benchmarks under RobecoSAM data.
In addition, the optimized portfolio values perform slightly better under the RobecoSAM data than under the Refinitiv data.

Table~\ref{tab:PM_sim_tang_robeco} summarizes the performance measures, return distribution moments and RRR values
for the EWBH benchmark and the tangent portfolios optimized under mCVaR${}_{0.99}$ and MV.
This summary should be compared to that in manuscript Table A.5 for the Refinitiv based computations.
In Section~\ref{sec:robeco_PM} we have discussed the differences between between RobecoSAM and Refinitiv based
values for the performance measures, return distribution moments and RRRs for efficient frontier portfolios characterized
by the same values of $\lambda$ and $\alpha$.
The tangent portfolios on the corresponding efficient frontiers will have the same value of $\lambda$ but different $\alpha$ value (in general).
Thus for $\lambda > 0$, we expect some differences between the results in Tables \ref{tab:PM_sim_tang_robeco} and A.5
and those observed in Section~\ref{sec:robeco_PM}.
Under both optimizations, Tot. Ret. is better under Refinitiv data at the higher $\lambda$ values,
while RobescoSAM Tot. Ret. is better at lower $\lambda$ values.
The same is true for MDD.
AvgTO is better for Refinitiv data, whle avg ESG is better for RobecoSAM.
There are no discernible trends in the ETL/ETR results, although they are generally better for RobecoSAM data.

At the larger values of $\lambda$, Refinitiv based results are generally better for the return Mean, Median, Skew and ExKurt moments.
At $\lambda = 0.75$, RRR values for Refinitive tangent portfolios outperform thosse for RobecoSAM.
Refinitiv tangent portfolios for $\lambda =0.75$ perform well as they were heavily concentrated in Microsoft stock;
the security with the best return performance over the investment period.

\begin{figure}[!h]
	\centering
	\includegraphics[width=0.75\textwidth]{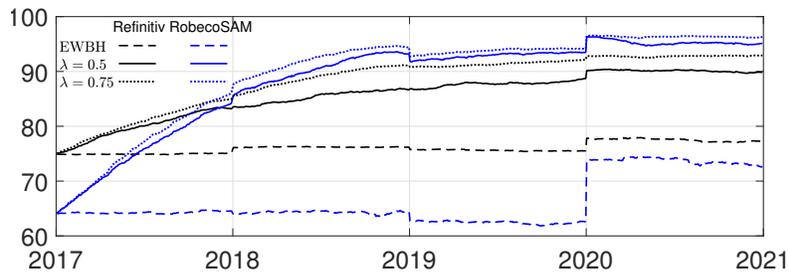}   
	\caption{ \small ESG values obtained by mCVaR${}_{0.99}$ optimized tangent portfolios and the EWBH benchmark
	under Refinitiv and RobecoSAM data.}
	\label{fig:ESG_ofs_comparison}	
\end{figure}

Fig.~\ref{fig:ESG_ofs_comparison} compares daily portfolio ESG scores (manuscript equation (14)) computed
under mCVaR${}_{0.99}$ optimization with $\lambda = \{0.5, 0.75\}$ under Refinitiv and RobecoSAM.
Also plotted is the ESG score for the EWBH benchmark.
Although the portfolio ESG scores for RobecoSAM start from a lower value in 2017, they rise more rapidly;
by 2018, RobecoSAM based portfollio ESG scores are larger than those based upon Refinitiv data.
ESG values change more noticeably under Refinitiv data than under RobecoSAM data as $\lambda$ grows from 0.5 to 0.75.
The effect produced by the yearly adjustment of RobecoSAM and Refinitiv ESG scores produces noticeable discontinuities in the ESG values.
Interestingly, for 2019 there was a noticeable downtrend in the ESG scoring by both agencies.
The effect of the large change in RobecoSAM ESG scores from 2019 to 2020 is most apparent in the EWBH benchmark results.

\subsection{Option Pricing}\label{sec:OptionPricing_robeco}

\begin{figure}[!h]
	\centering
								{\includegraphics[width=0.24\textwidth]{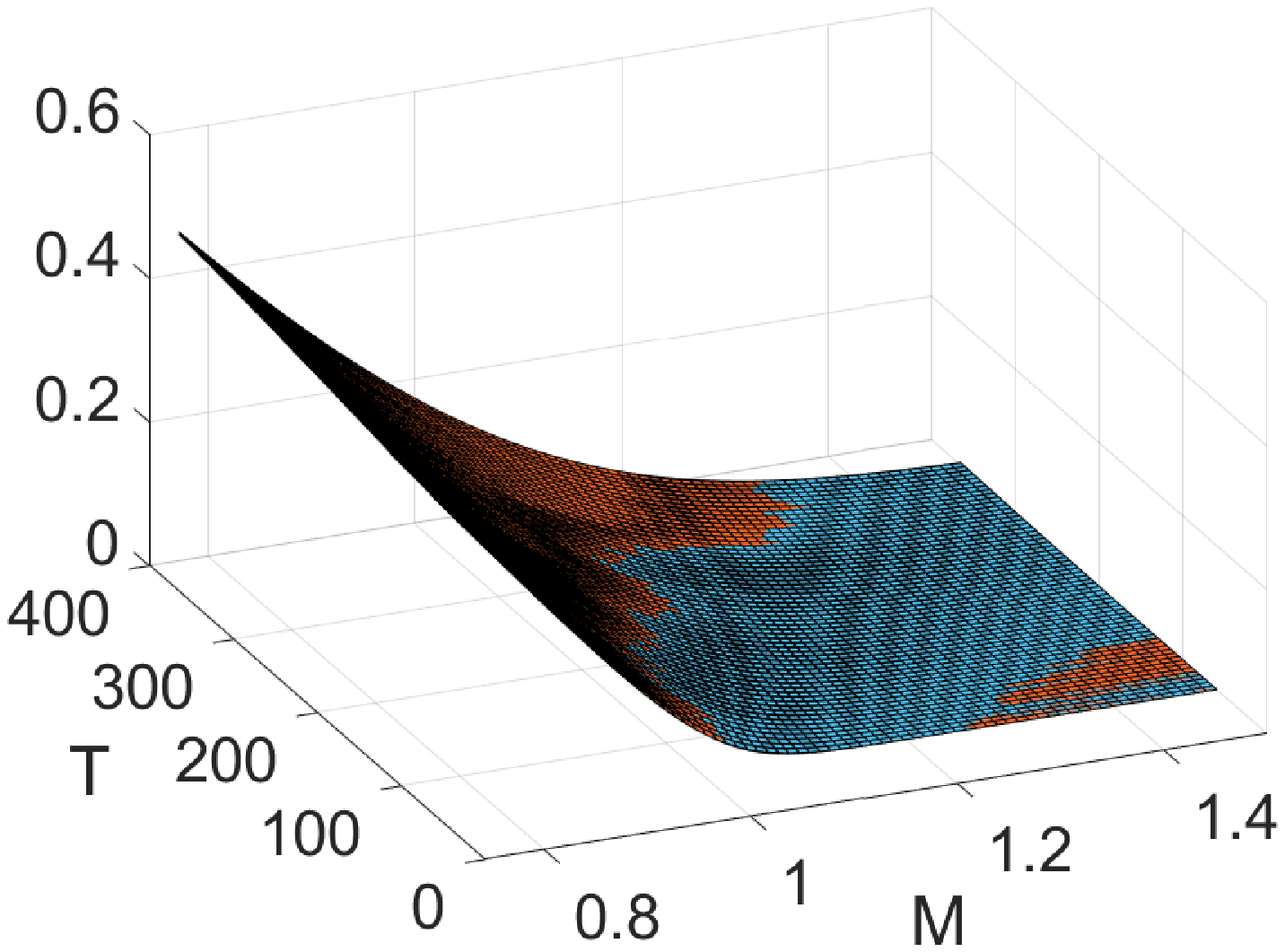}}\hspace{0em}%
								{\includegraphics[width=0.24\textwidth]{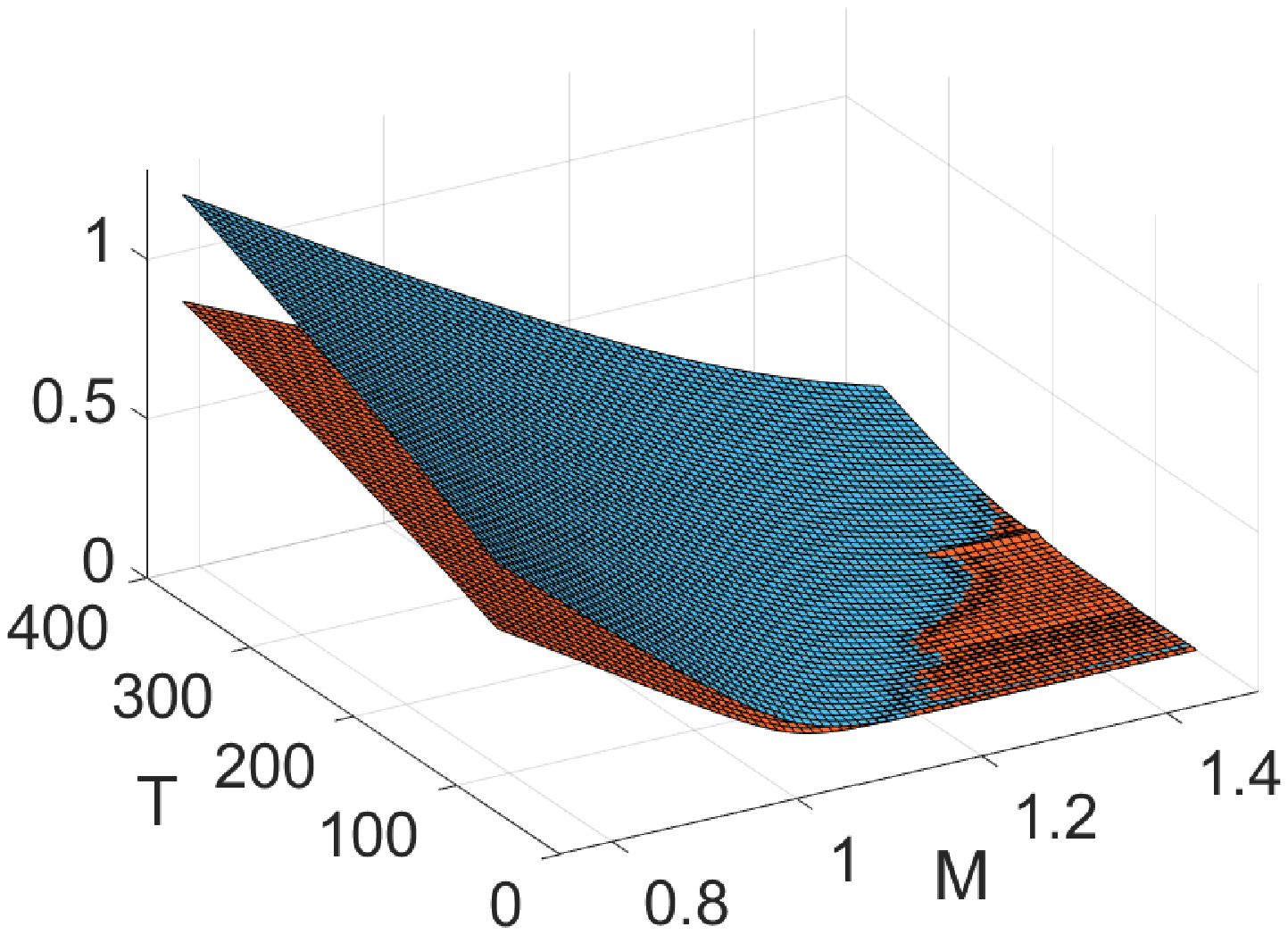}}\hspace{0em}%
								{\includegraphics[width=0.24\textwidth]{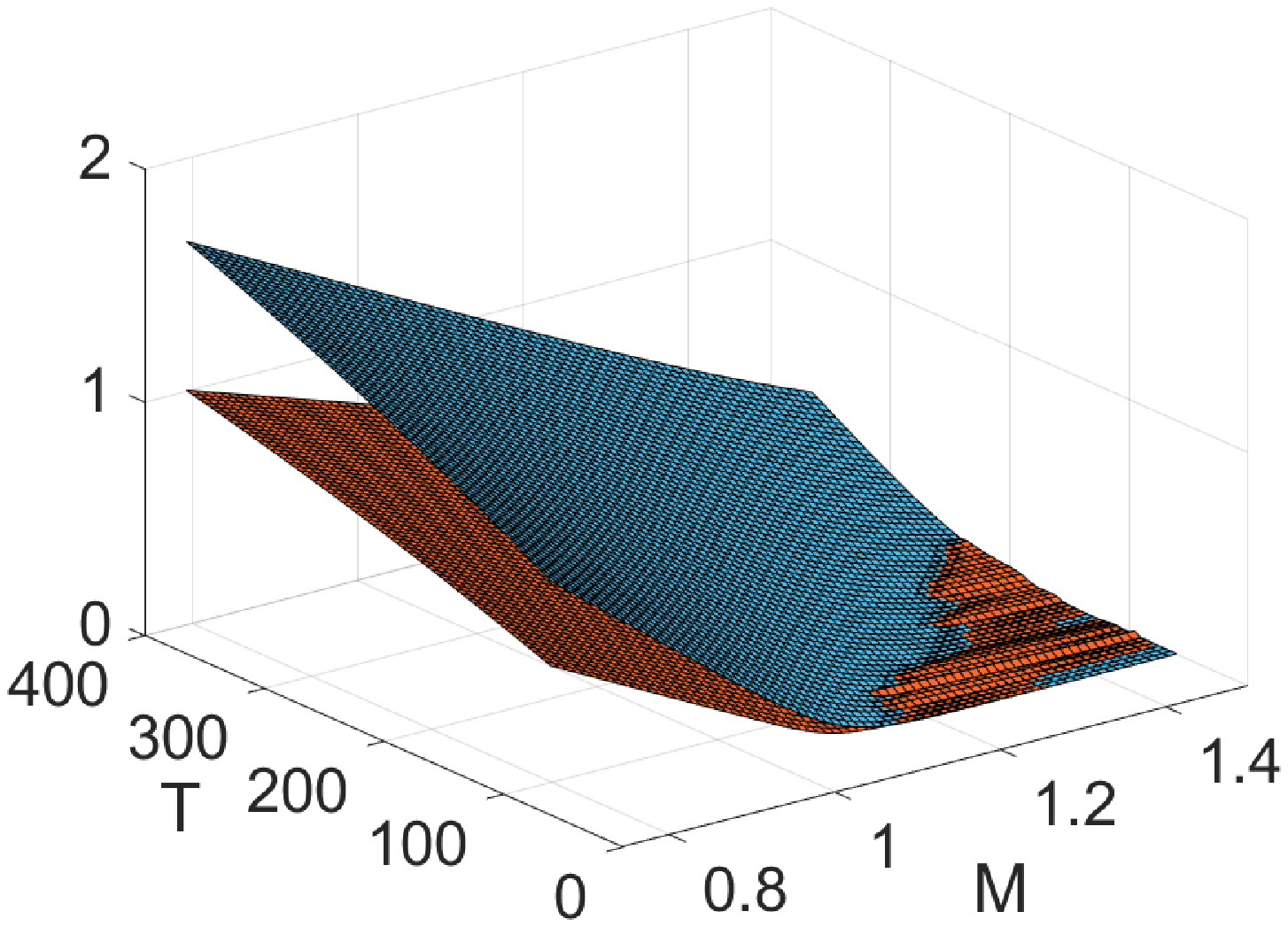}}\hspace{0em}%
								{\includegraphics[width=0.24\textwidth]{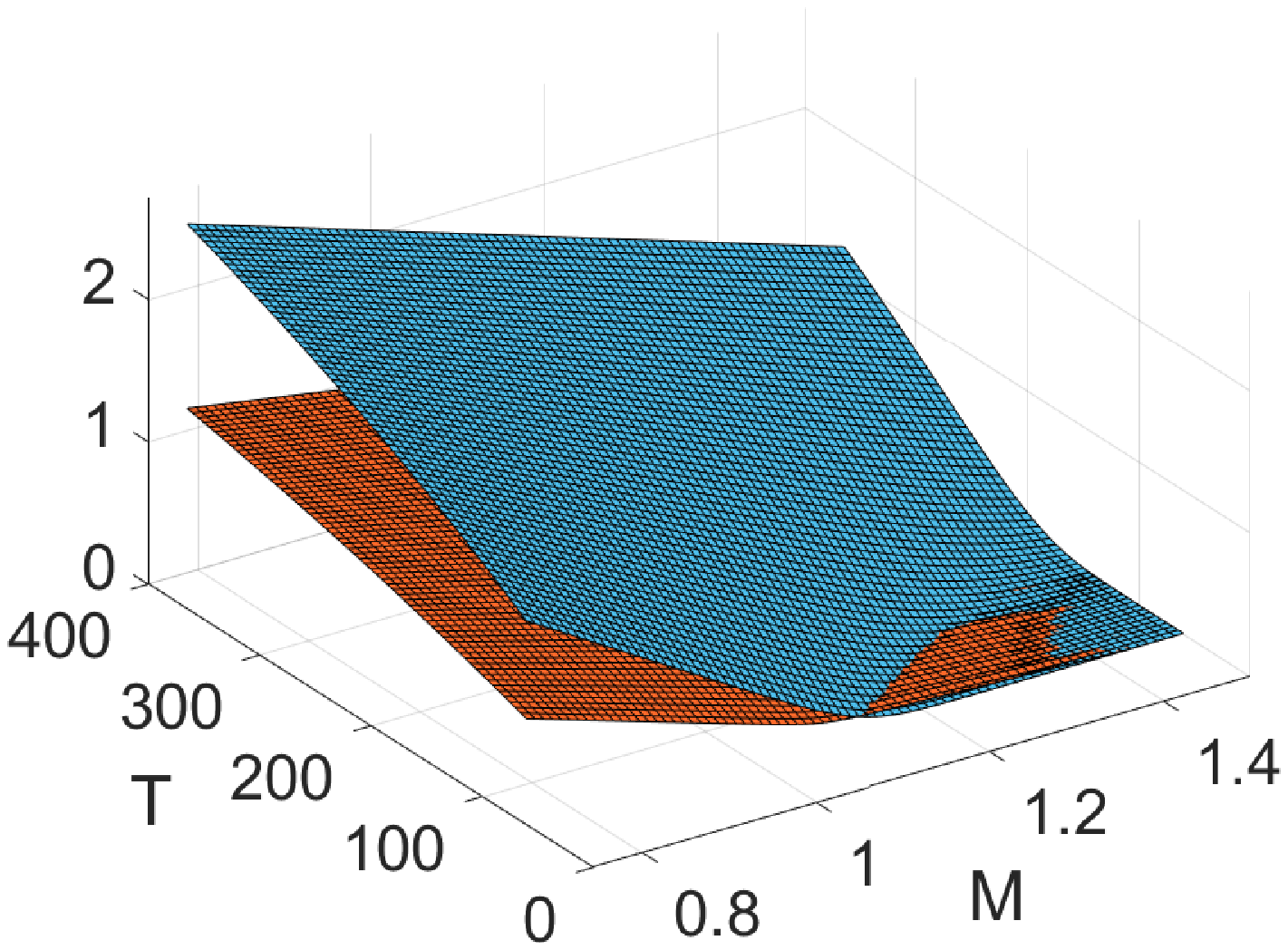}}\hspace{0em}%
	\subcaptionbox{$\lambda=0$    } {\includegraphics[width=0.24\textwidth]{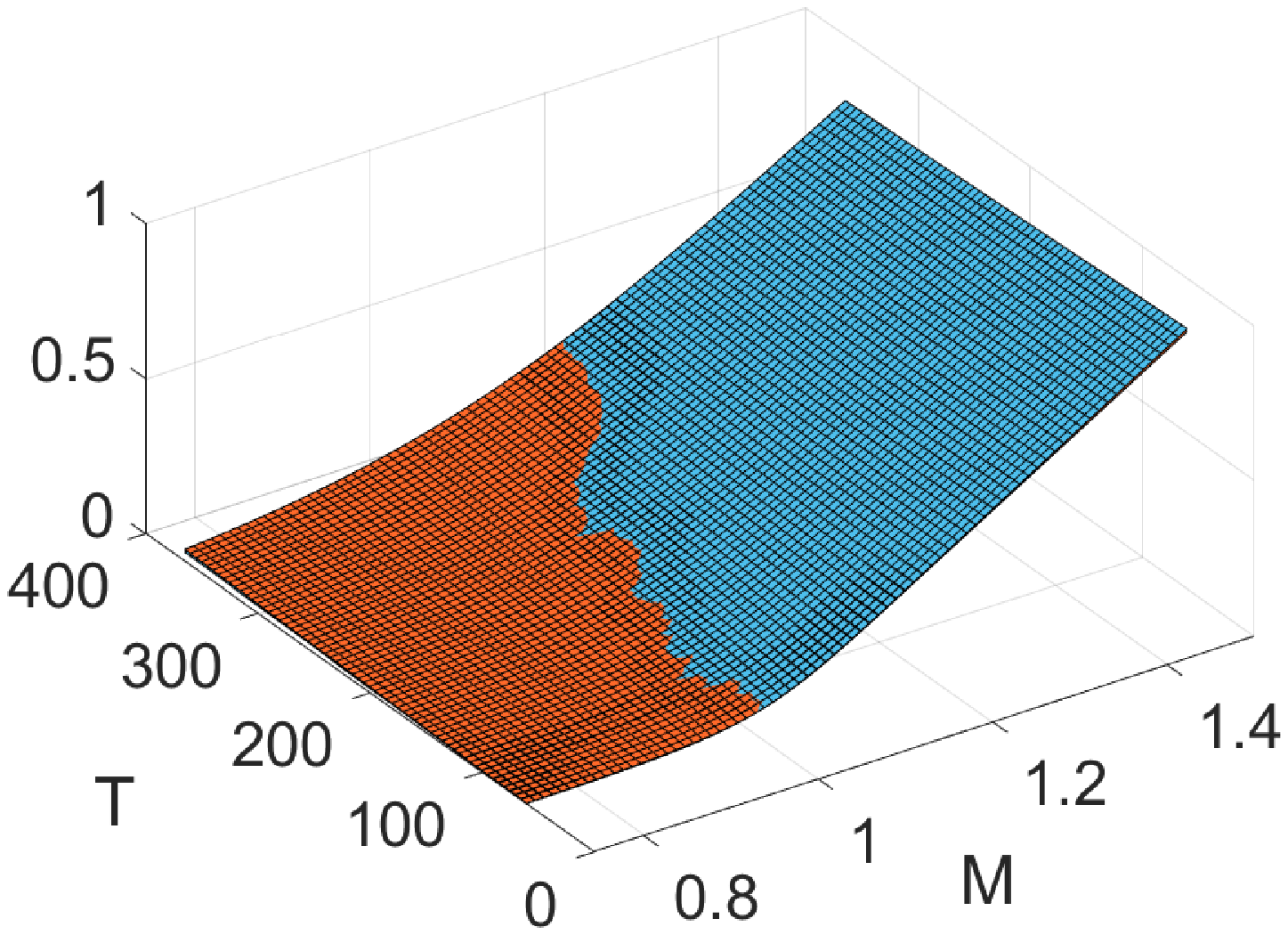}}\hspace{0em}%
	\subcaptionbox{$\lambda=0.25$} {\includegraphics[width=0.24\textwidth]{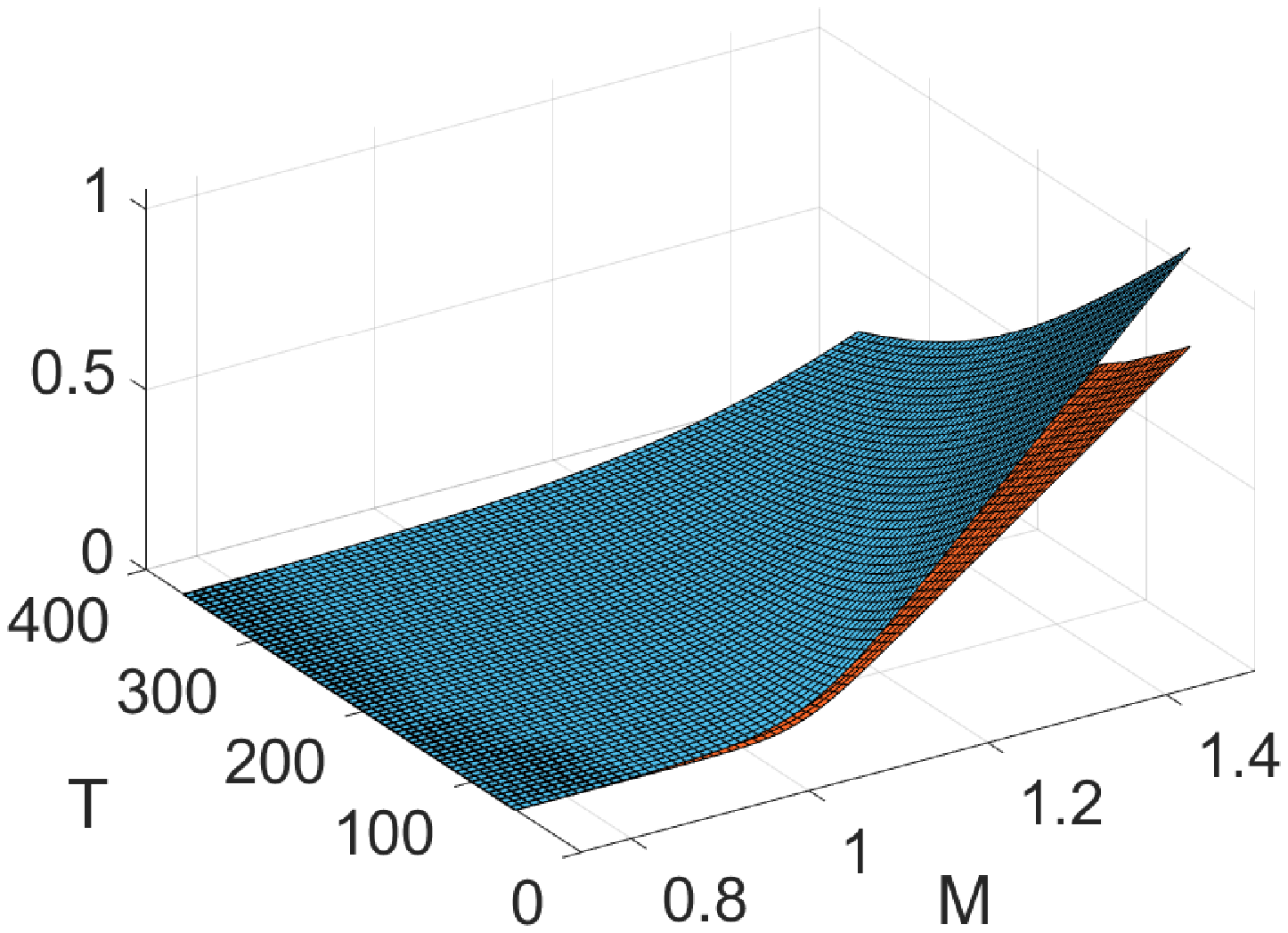}}\hspace{0em}%
	\subcaptionbox{$\lambda=0.5$ } {\includegraphics[width=0.24\textwidth]{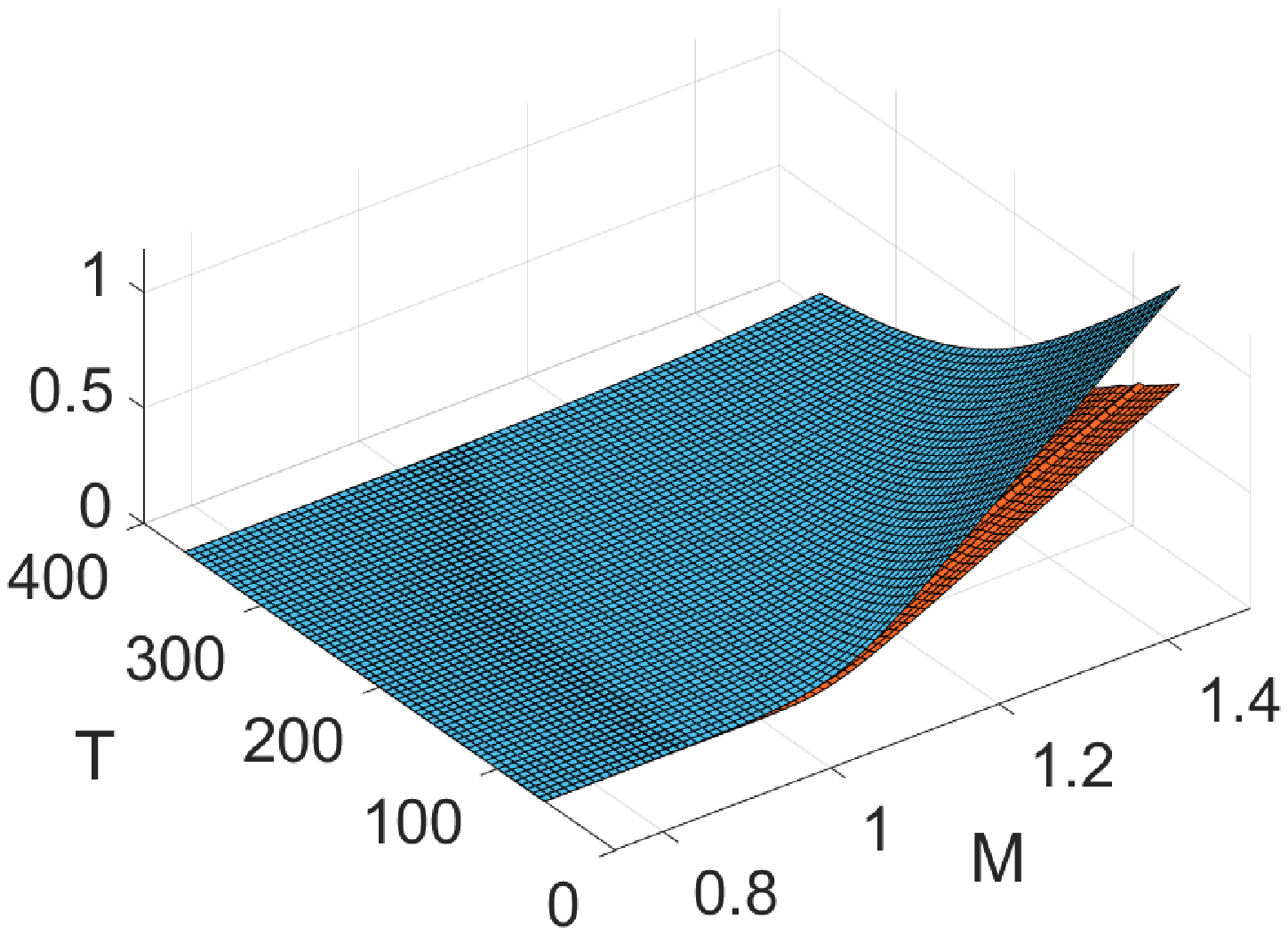}}\hspace{0em}%
	\subcaptionbox{$\lambda=0.75$} {\includegraphics[width=0.24\textwidth]{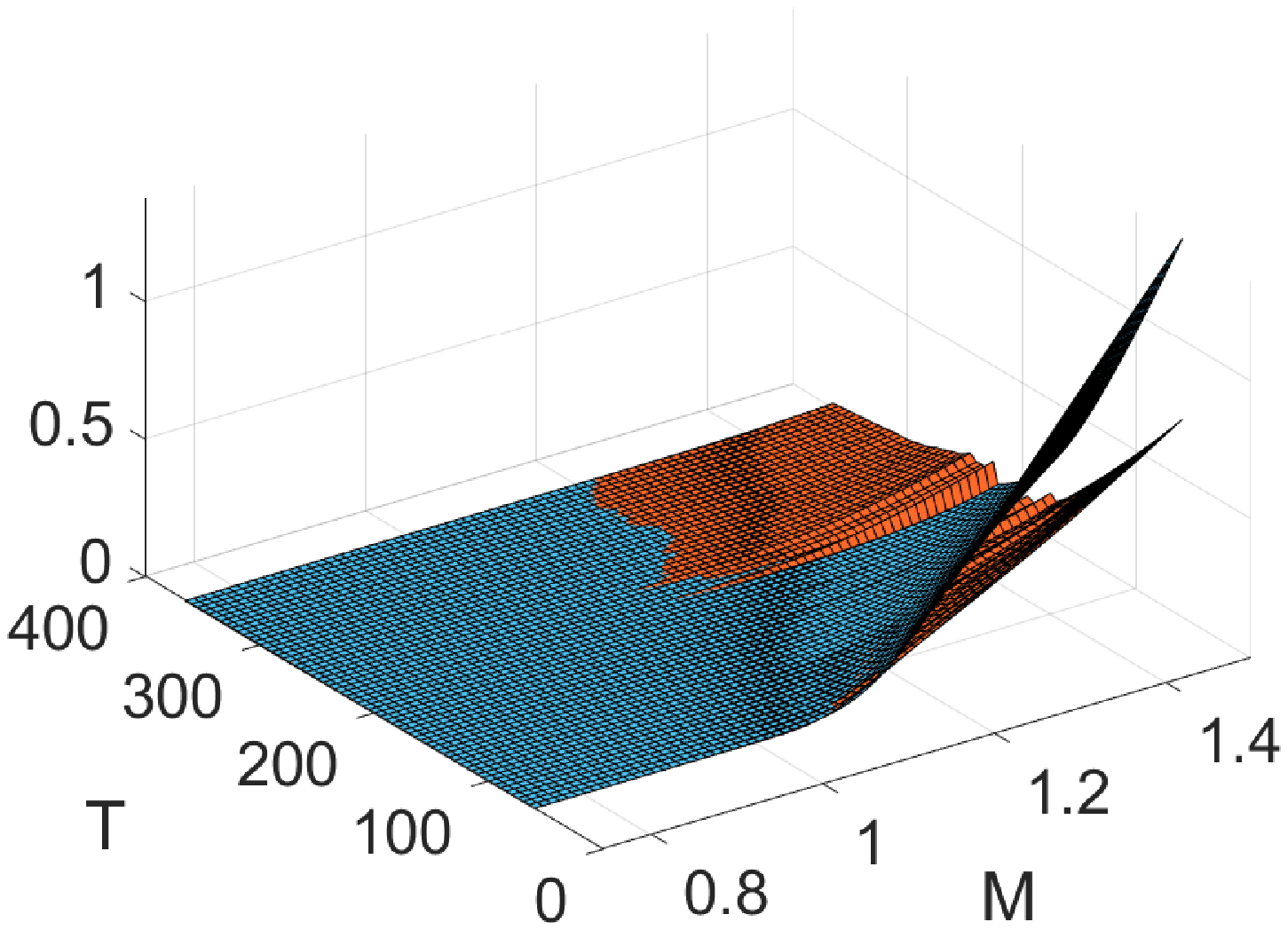}}\hspace{0em}%
	\subcaptionbox{Call} {\includegraphics[width=0.42\textwidth]{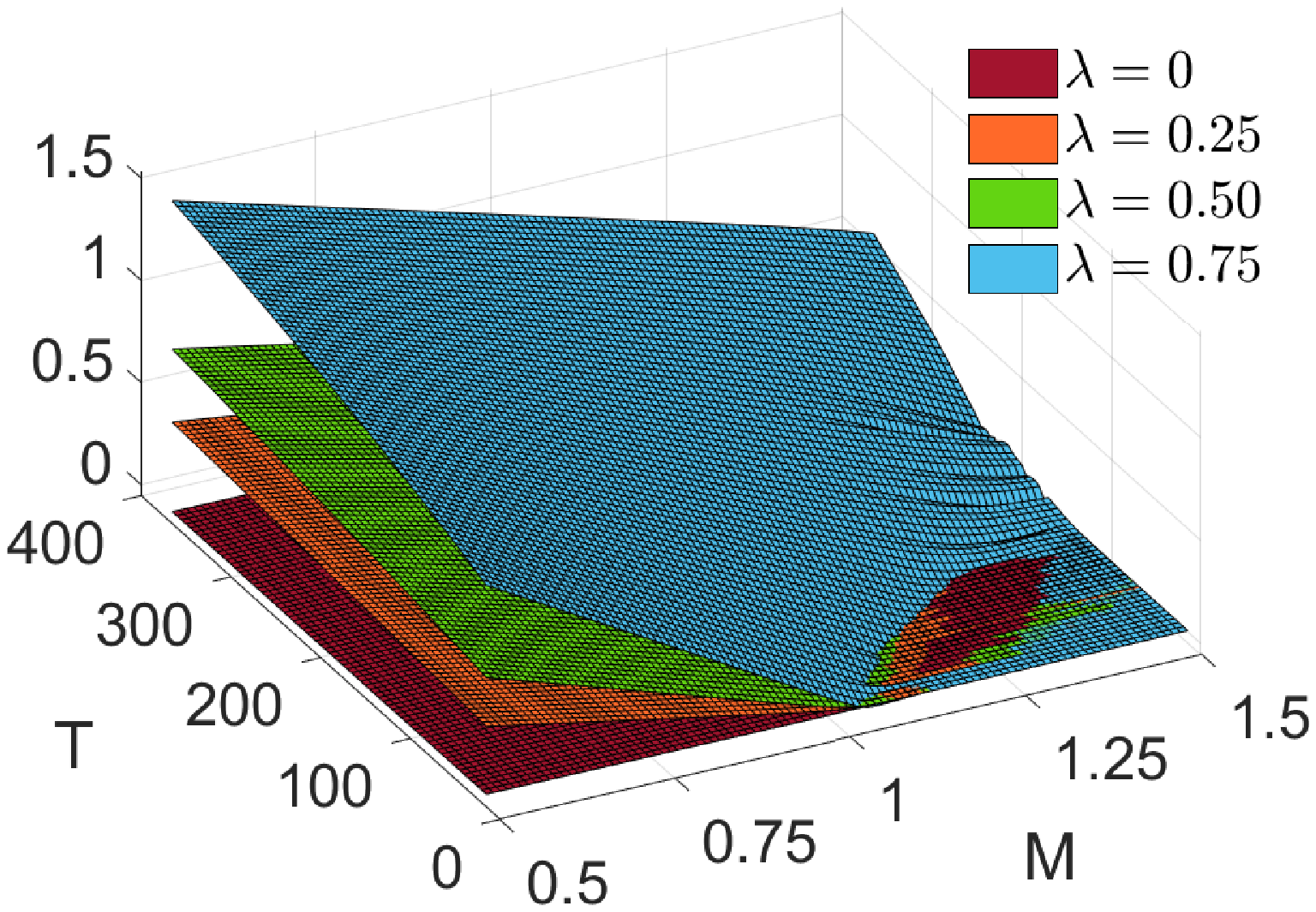}}\hspace{0em}%
	\subcaptionbox{Put} {\includegraphics[width=0.42\textwidth]{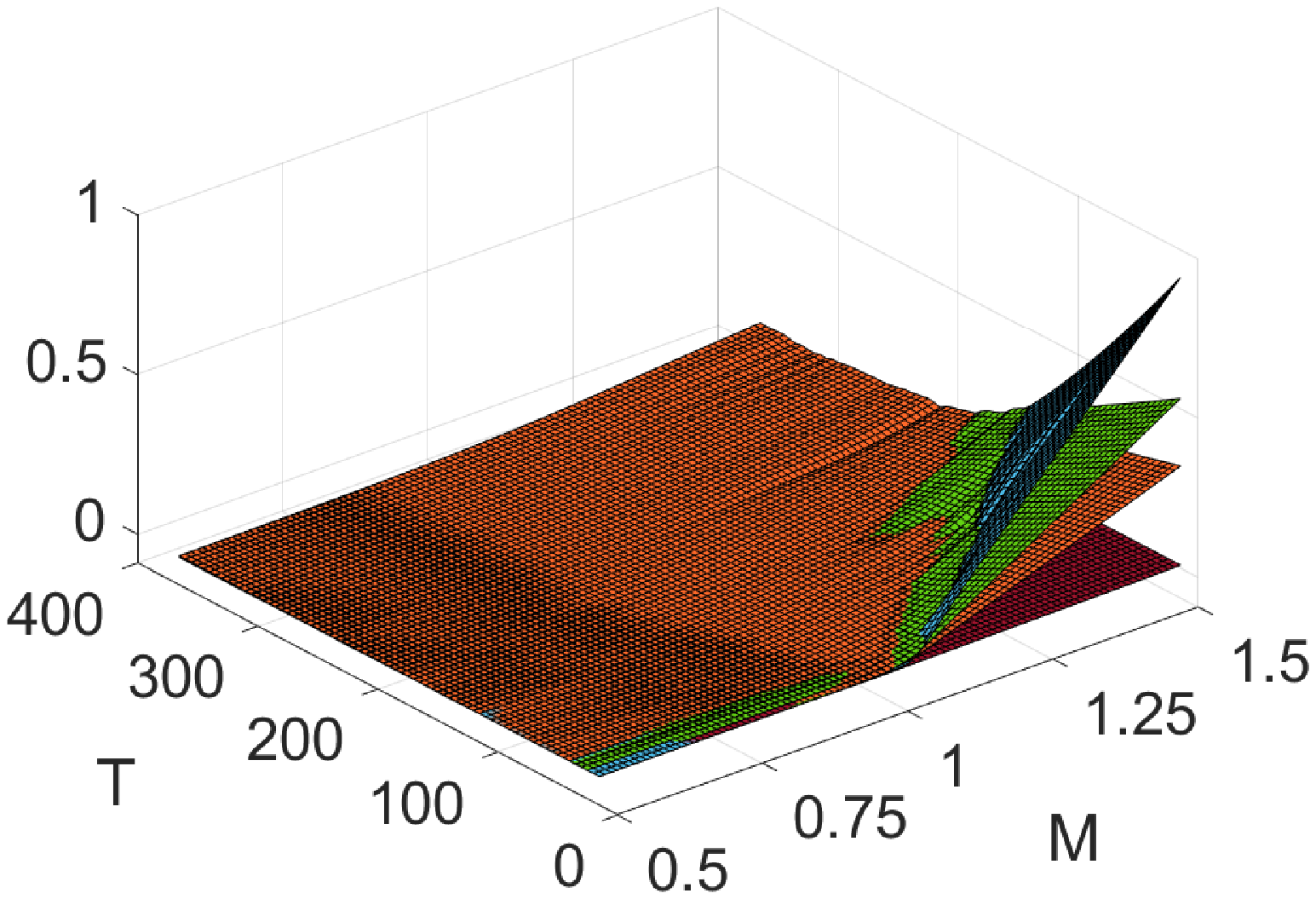}}\hspace{0em}%
	\caption{ \small Figures (a) to (d): (top) call and (bottom) put option prices for select choices of $\lambda$.
		The orange surfaces corresponds to option prices for the ESG-adjusted DJIA index;
		blue surfaces represents option prices for the tangent portfolios.
		Figures (e) and (f). The difference in call and put option prices between the ESG-adjusted DJIA index
		and the MCV tangent portfolios.} 
	\label{fig:Call_Put_Prices_robeco}	
\end{figure}

Option prices were computed for 12/30/2019 using either the RobecoSAM ESG-adjusted tangent portfolios or the DJIA index as the
underlying asset.
The  call and put price surfaces for these tangent portfolios and the DJIA index are given in Fig.~\ref{fig:Call_Put_Prices_robeco}.
Comparing the RobecoSAM results with the Refinitiv results in Fig.~9 of the manuscript,
most noticeable is the size of the difference between the tangent portfolio and DJIA index for both call and put surfaces 
for each value of $\lambda$, with the differences being larger for Refinitiv-based option pricing.

\begin{figure}[!h]
	\centering		
	\subcaptionbox{$\lambda=0$}     {\includegraphics[width=0.24\textwidth]{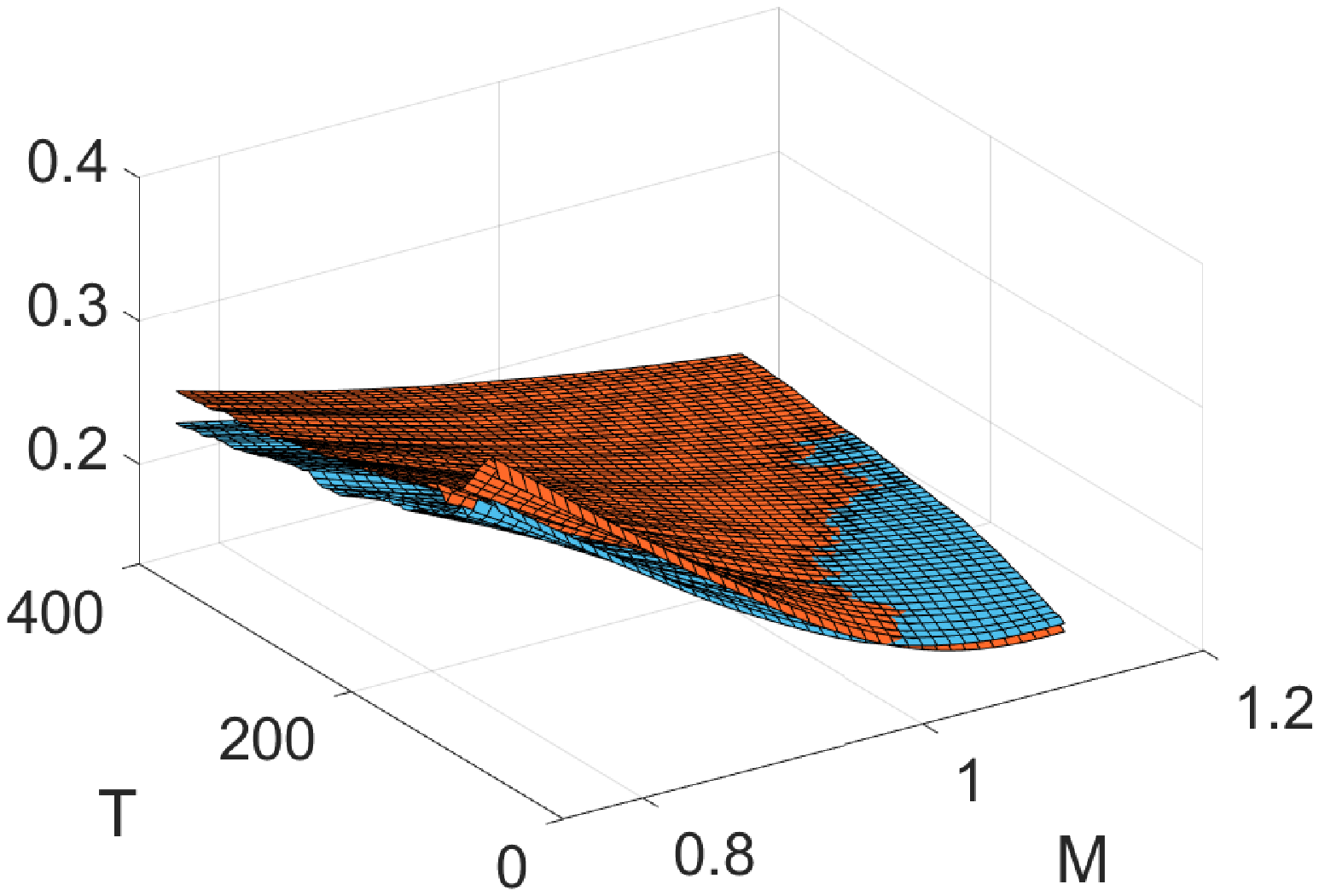}   } \hspace{0em}%
	\subcaptionbox{$\lambda=0.25$} {\includegraphics[width=0.24\textwidth]{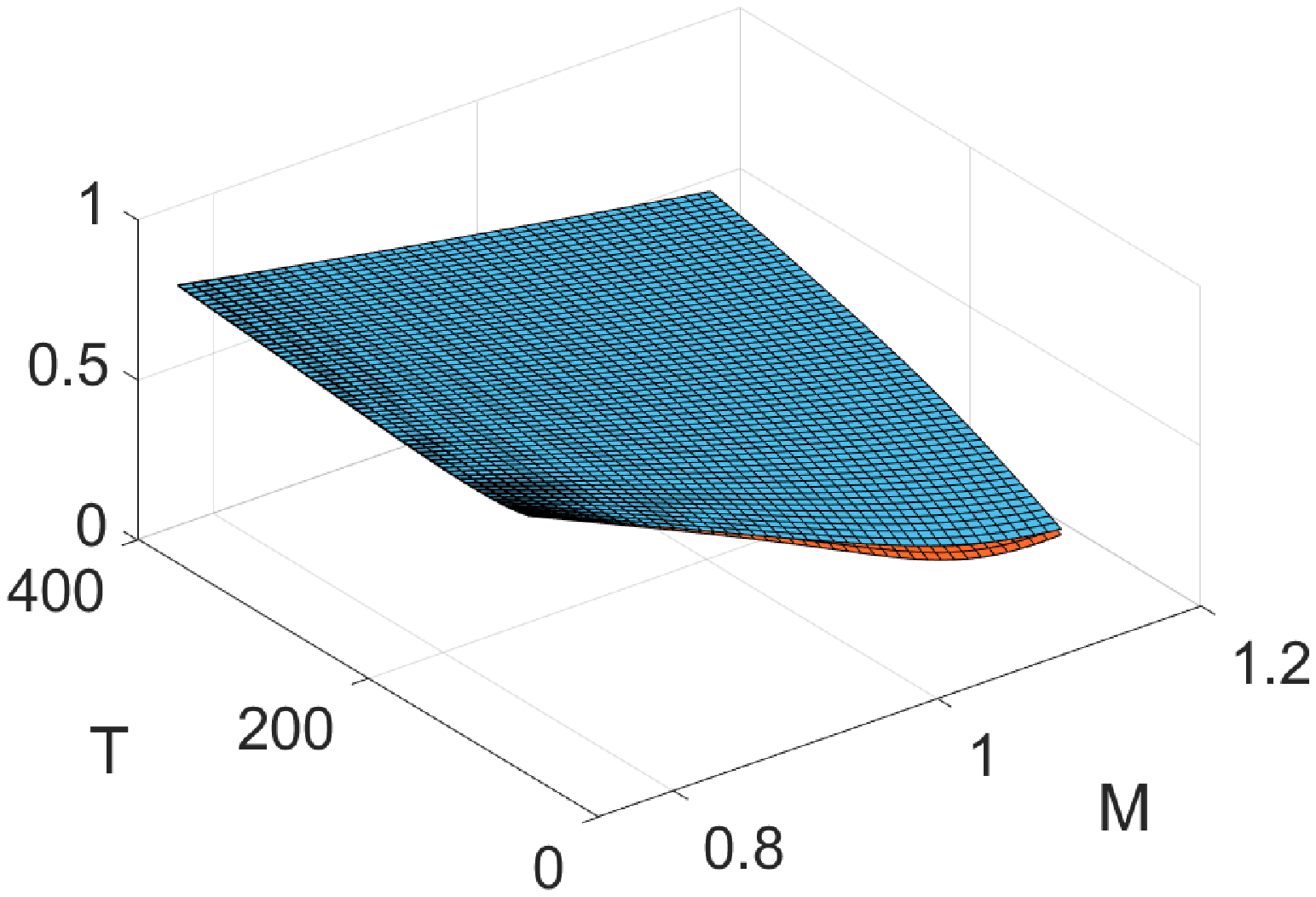}} \hspace{0em}%
	\subcaptionbox{$\lambda=0.5$}  {\includegraphics[width=0.24\textwidth]{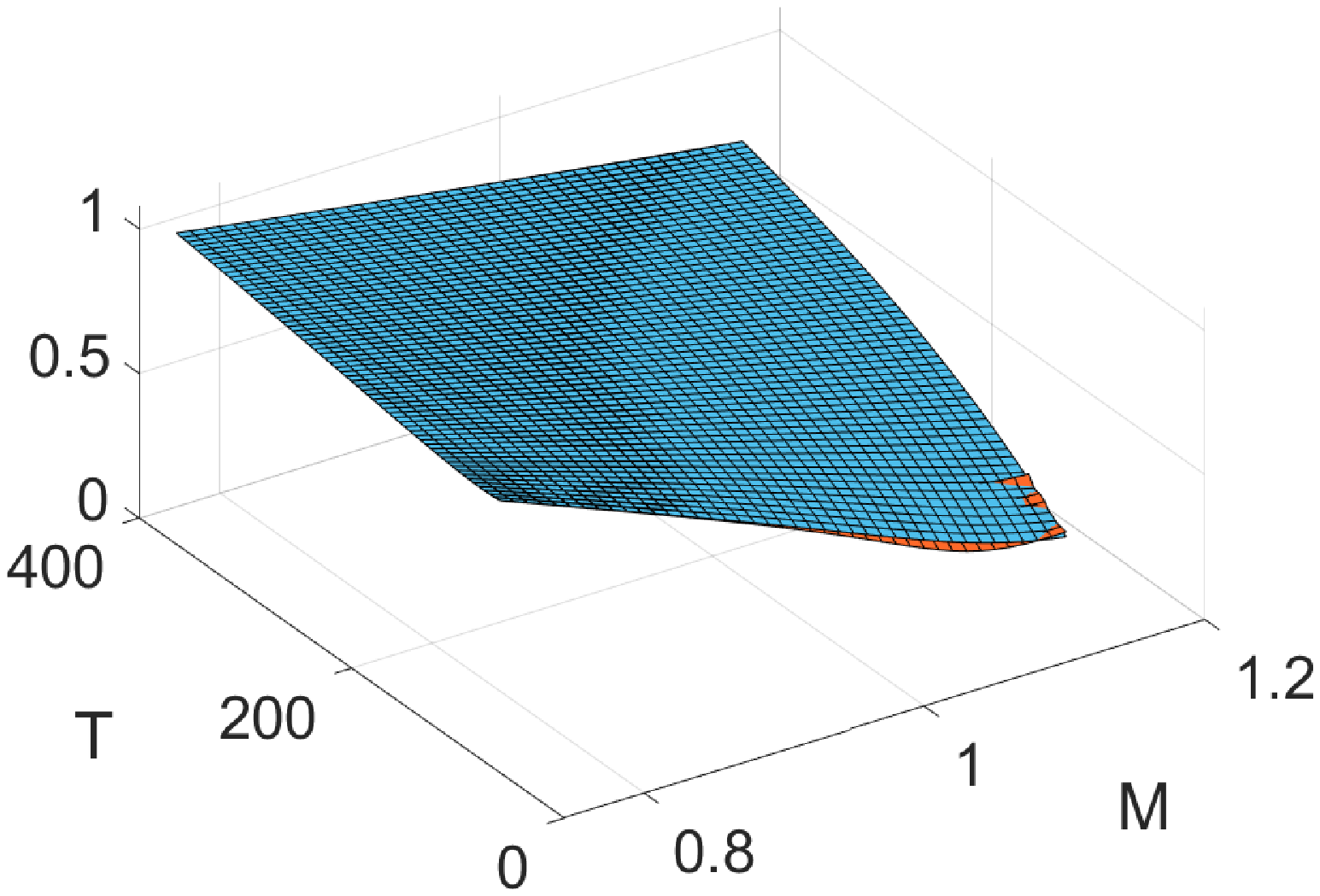}} \hspace{0em}%
	\subcaptionbox{$\lambda=0.75$}{\includegraphics[width=0.24\textwidth]{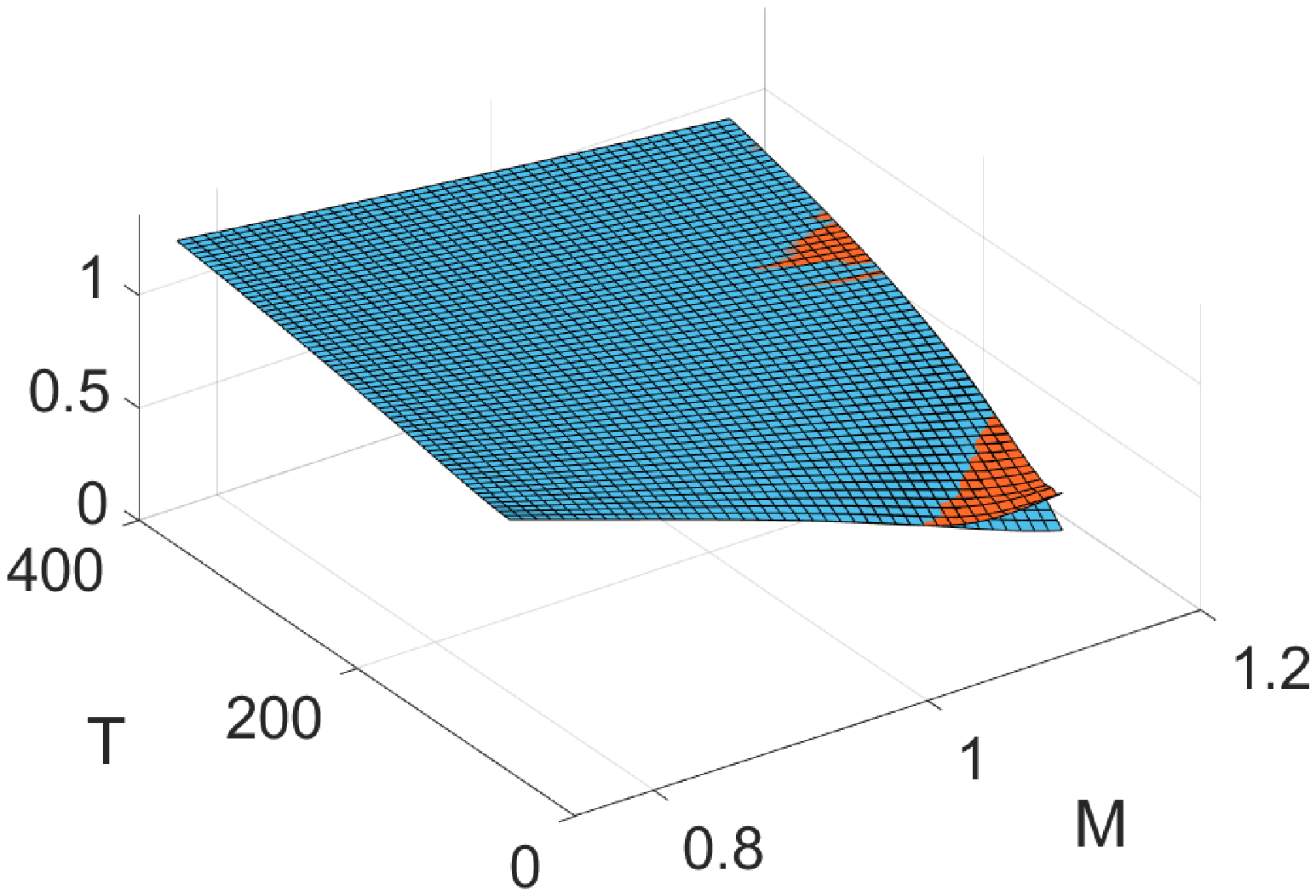}} \hspace{0em}%
	\caption{ \small Implied volatility surfaces computed from the (orange surfaces) ESG-adjusted DJIA option prices and
		(blue surfaces) the tangent portfolios corresponding to the selected choices for $\lambda$.} 
	\label{fig:IV_Prices_robeco}	
\end{figure}

Implied volatilities computed from the RobescoSAM based option prices are presented in Fig.~\ref{fig:IV_Prices_robeco}.
The results are qualitatively similar to those for Refinitiv data (manuscript Fig.~10).

\subsection{Shadow Riskless Rate}\label{sec:ShadowRate_robeco}
\begin{figure}[!h]
	\centering		
	\subcaptionbox{}{\includegraphics[width=0.48\textwidth]{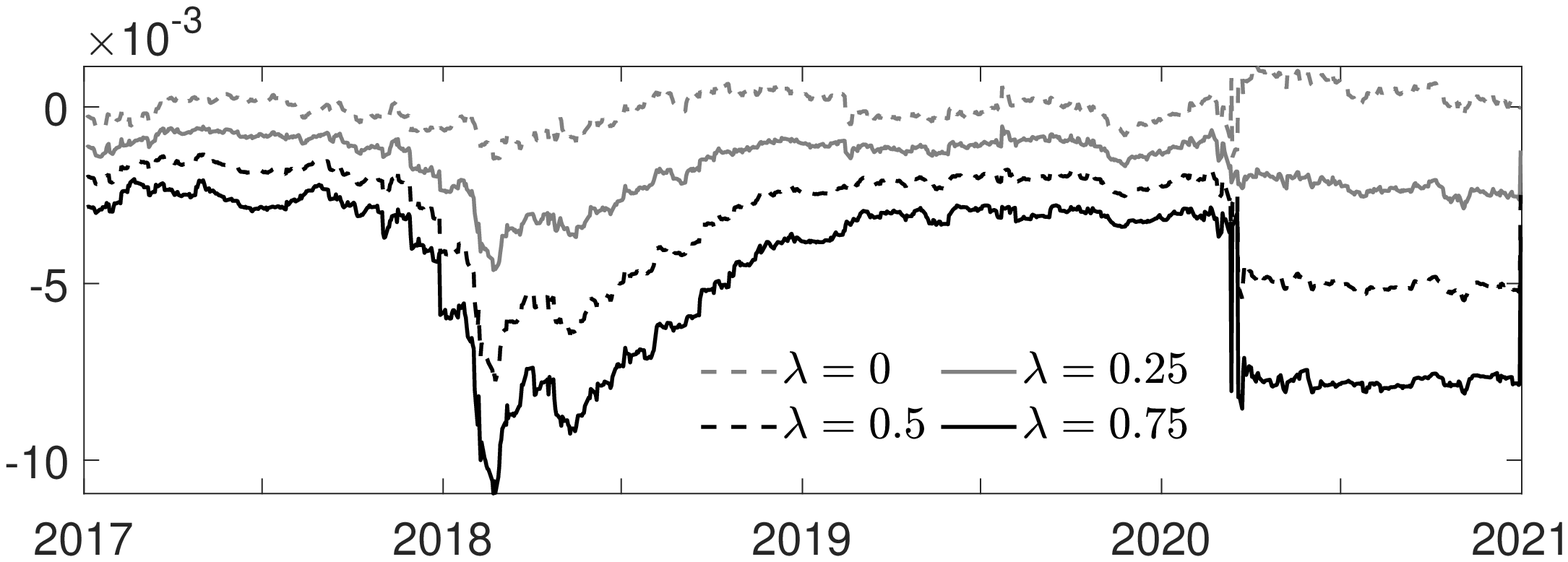} } \hspace{0em}%
	\subcaptionbox{}{\includegraphics[width=0.48\textwidth]{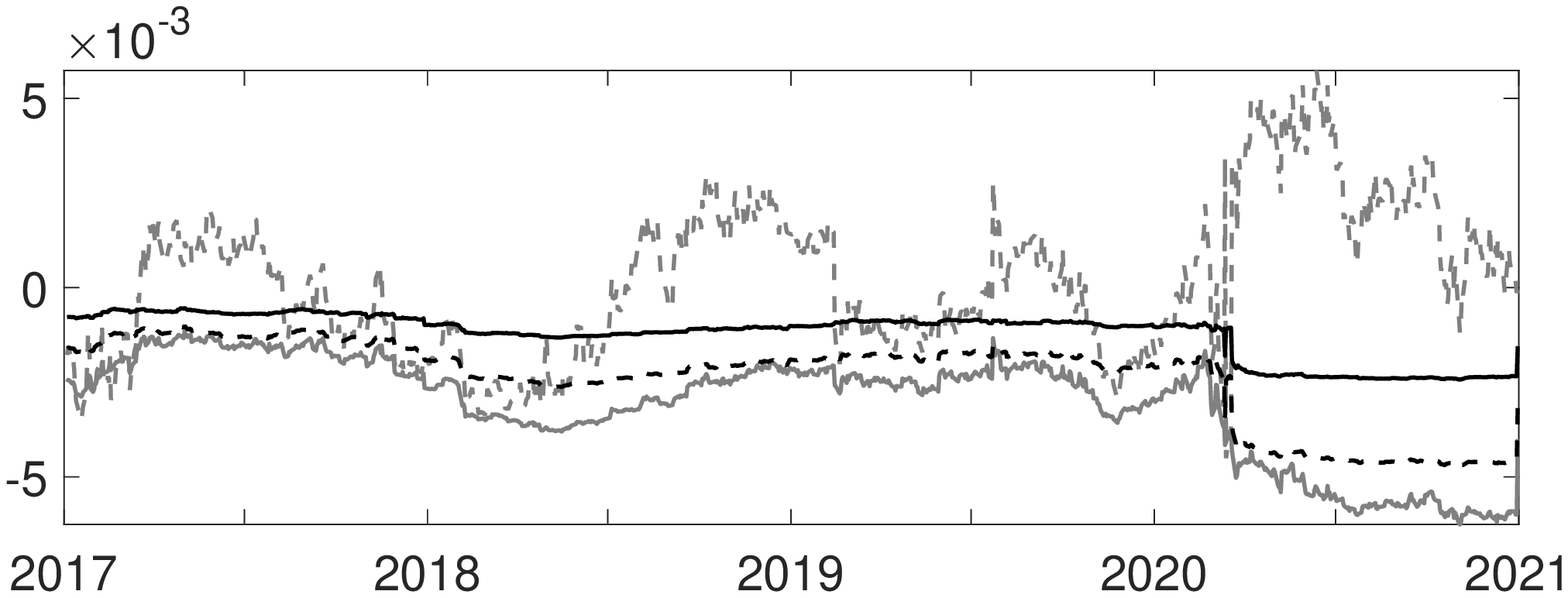} } \hspace{0em}%
	\caption{ \small  Time series from 01/03/2017 through 12/31/2020 for select values of $\lambda$ of
		(a) the SRR computed from ESG-adjusted returns on a universe of 29 DFJIA assets for select values, and
		(b) the information ratio of the SRR.} 
	\label{fig:SRR_Robeco}	
\end{figure}

The realized shadow riskless rate (SRR) time series computed using RobecoSAM data are shown in Fig. (\ref{fig:SRR_Robeco})
and are to be compared to the Refinitiv based results in Fig.~11 of the manuscript.
For $\lambda > 0$, the change in rating agencies dramatically affected the sign of the shadow rate.
Under the RobecoSAM ESG data, non-zero values of $\lambda$ drive the shadow riskless rate negative,
with a decreasing (more negative) value as $\lambda$ increases.
The reason for this behavior is not yet understood.
Similar to the results noted for SRR based on Refinitiv data (manuscript Fig.~11),
the separation between SRR values, while showing some variation with time, appears roughly linearly proportional to $\lambda$.
As a result of the negative sign for the SRR, the information ratio for RobecoSAM based data is negative.
However, similar to the behavior under Refinitiv based data, as $\lambda$ increases, the magnitude of the information
ratio tends to decrease.

\begin{figure}[!b]
	\centering		
	{\includegraphics[width=0.49\textwidth]{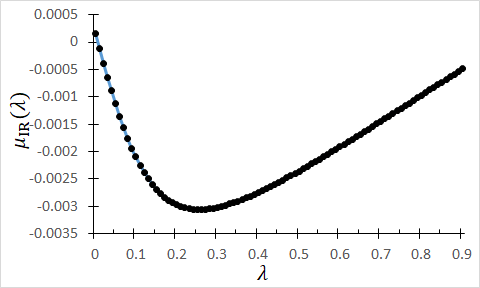} }\hspace{0em}%
	{\includegraphics[width=0.49\textwidth]{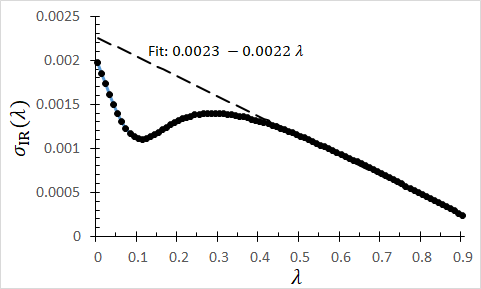} }\hspace{0em}%
	{\includegraphics[width=0.49\textwidth]{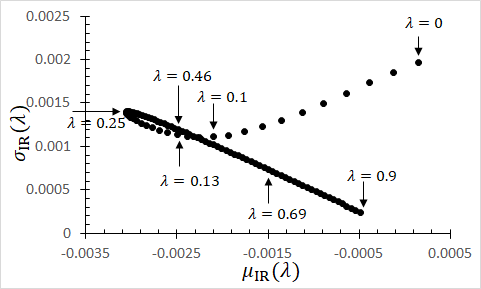} }\hspace{0em}%
	\caption{ \small The curves $\mu_{\text{IR}}(\lambda)$, $\sigma_{\text{IR}}(\lambda)$ and
		the parametric curve $(\mu_{\text{IR}}(\lambda), \sigma_{\text{IR}}(\lambda))$.
		The dashed line is a linear fit.} 
	\label{fig:IR_Lambda_Mean_Std_robeco}	
\end{figure}
The time averaged values $\mu_{\text{IR}}(\lambda)$ and $\sigma_{\text{IR}}(\lambda)$ for the RobecoSAM-based
SRR information ratio are shown in Fig.~\ref{fig:IR_Lambda_Mean_Std_robeco}.
Interesting differences and similarities emerge from the comparison with manuscript Fig.~12 for the Refinitiv data.
Again the dependence of $\mu_{\text{IR}}(\lambda)$ on $\lambda$ is ``parabolic'',
but with a local minimum (at $\lambda = 0.25$) rather than a local maximum.
Except when $\lambda = 0$, the time averaged $\mu_{\text{IR}}(\lambda)$ is negative over the range $\lambda \in [0.01,0.90]$.
The behavior of the standard deviation $\sigma_{\text{IR}}(\lambda)$ for RobecoSAM based data is qualitatively similar to that
for Refinitiv
but now with a local minimum developed at $\lambda = 0.11$, prior to following a linear trend for $\lambda > 0.4$.
This linear trend in $\sigma_{\text{IR}}(\lambda)$ for RobecoSAM based data is virtually identical to that developed for
the Refinitiv based data.

\newpage

\subsection{Tables} \label{sec:Tbl_robeco}

\begin{table}[!h] 
	\centering
	\scalebox{0.65}{
		\begin{tabular}{cc cccc cccc}
			\toprule
			\multicolumn{2}{c}{Model} & Tot. Ret & Ann. Ret & AvgTO & ETL95 & ETR95 & MDD & \multicolumn{2}{c}{ESG*}  \\
			\   & \   & (\%) &  (\%) &  (\%) &  (\%) &  (\%) &  (\%) & avg & std  \\
			\midrule 
			\multicolumn{2}{c}{EWBH} & 90.14 & 17.68 & 0.00 & -3.41 & 2.86 & 33.00 & 72.72 & 2.97 \\
			\rule{0pt}{3ex}
			$\lambda$ & $\alpha$ & \multicolumn{8}{c}{MCVaR${}_{0.99}$} \\
			\cline{1-2}
			\rule{0pt}{3ex} 0.00 & 0.0  & 70.75 & 14.53 & 0.4 & -2.74 & 2.41 & 28.47 & 54.94 & 3.69    \\  
			\                    & 0.3  & 71.53 & 14.66 & 0.4 & -2.74 & 2.40 & 28.36 & 54.94 & 3.59  \\  
			\                    & 0.5  & 71.85 & 14.71 & 0.4 & -2.75 & 2.41 & 28.36 & 55.20 & 3.74  \\  
			\                    & 0.7  & 73.88 & 15.05 & 0.4 & -2.76 & 2.44 & 27.62 & 56.32 & 3.57  \\  
			\                    & 0.9  & 82.60 & 16.49 & 0.4 & -2.90 & 2.55 & 28.23 & 58.22 & 2.88 \\
			\rule{0pt}{3ex} 
			             0.25    & 0.0 & 70.57 & 14.49 & 0.4 & -2.77 & 2.43 & 28.94 & 56.98 & 3.36  \\  
			\                    & 0.3 & 72.55 & 14.83 & 0.4 & -2.77 & 2.43 & 28.55 & 58.07 & 3.52    \\  
			\                    & 0.5 & 74.24 & 15.11 & 0.4 & -2.80 & 2.45 & 28.61 & 59.48 & 3.59  \\  
			\                    & 0.7 & 86.38 & 17.10 & 0.4 & -2.95 & 2.60 & 28.11 & 67.40 & 4.16  \\  
			\                    & 0.9 & 107.58 & 20.34 & 0.4 & -3.31 & 2.97 & 28.94 & 76.69 & 4.99  \\
			\rule{0pt}{3ex} 
			            0.50     & 0.0 & 75.82 & 15.38 & 0.4 & -2.82 & 2.47 & 28.88 & 61.48 & 3.68 \\ 
			\                    & 0.3 & 78.75 & 15.86 & 0.4 & -2.88 & 2.51 & 28.67 & 64.54 & 3.98 \\  
			\                    & 0.5 & 84.48 & 16.79 & 0.4 & -2.95 & 2.58 & 28.71 & 67.95 & 4.50 \\  
			\                    & 0.7 & 100.75 & 19.32 & 0.4 & -3.38 & 3.08 & 29.12 & 80.85 & 6.13 \\  
			\                    & 0.9 & 122.16 & 22.43 & 0.37 & -3.70 & 3.52 & 29.00 & 87.87 & 8.50 \\
			\rule{0pt}{3ex}
			            0.75     & 0.0 & 88.55 & 17.44 & 0.4 & -3.03 & 2.67 & 28.83 & 71.63 & 4.57 \\
			\                    & 0.3 & 91.38 & 17.88 & 0.4 & -3.19 & 2.85 & 29.00 & 77.06 & 5.21 \\
			\                    & 0.5 & 95.75 & 18.56 & 0.4 & -3.35 & 3.09 & 28.67 & 82.20 & 6.34 \\
			\                    & 0.7 & 110.92 & 20.82 & 0.37 & -3.65 & 3.51 & 29.73 & 88.94 & 8.87 \\
			\                    & 0.9 & 115.23 & 21.45 & 0.34 & -3.82 & 3.66 & 31.38 & 89.93 & 9.25  \\
			\rule{0pt}{3ex}	
			$\lambda$ & $\alpha$ & \multicolumn{8}{c}{MV } \\
			\cline{1-2}
			\rule{0pt}{3ex} 
			                0.00  & 0.0  & 76.82 & 15.54 & 0.4  & -2.84 & 2.46 & 29.17 & 57.21 & 3.19 \\  
			\                     & 0.05 & 84.53 & 16.80 & 0.40 & -3.06 & 2.61 & 30.62 & 57.24 & 3.28 \\  
			\                     & 0.10 & 90.68 & 17.77 & 0.38 & -3.33 & 2.80 & 32.39 & 58.01 & 2.73  \\  
			\                     & 0.15 & 90.64 & 17.77 & 0.39 & -3.55 & 2.96 & 35.00 & 58.24 & 2.68 \\  
			\                     & 0.20 & 88.69 & 17.46 & 0.40 & -3.73 & 3.08 & 37.27 & 58.78 & 2.09 \\
			\rule{0pt}{3ex} 
			                 0.25 & 0.0  & 82.99 & 16.55 & 0.4 & -2.94 & 2.52 & 30.14 & 58.30 & 2.67 \\  
			\                     & 0.05 & 85.00 & 16.87 & 0.4 & -3.08 & 2.65 & 30.05 & 65.25 & 2.14 \\  
			\                     & 0.10 & 89.48 & 17.59 & 0.4 & -3.18 & 2.74 & 30.12 & 68.38 & 3.33 \\  
			\                     & 0.15 & 95.56 & 18.53 & 0.4 & -3.27 & 2.83 & 30.23 & 72.19 & 4.58 \\  
			\                     & 0.20 & 100.63& 19.30 & 0.4 & -3.35 & 2.92 & 30.30 & 74.27 & 5.43 \\
			\rule{0pt}{3ex} 
			                0.50  & 0.0  & 92.55  & 18.07 & 0.4 & -3.13 & 2.63 & 31.79 & 60.41 & 2.02 \\
			\                     & 0.05 & 98.67  & 19.01 & 0.4 & -3.34 & 2.90 & 30.34 & 74.25 & 6.29 \\  
			\                     & 0.10 & 105.55 & 20.04 & 0.4 & -3.48 & 3.05 & 30.60 & 78.52 & 7.63 \\  
			\                     & 0.15 & 109.45 & 20.61 & 0.4 & -3.56 & 3.16 & 30.20 & 81.19 & 8.24 \\  
			\                     & 0.20 & 115.15 & 21.43 & 0.4 & -3.64 & 3.31 & 29.53 & 83.66 & 8.82 \\
			\rule{0pt}{3ex}
			                0.75  & 0.0  & 100.69 & 19.31 & 0.32 & -3.42 & 2.87 & 33.07 & 64.37 & 2.38  \\  
			\                     & 0.05 & 107.80 & 20.37 & 0.32 & -3.55 & 3.13 & 30.39 & 79.30 & 8.76 \\  
			\                     & 0.10 & 110.05 & 20.70 & 0.32 & -3.61 & 3.22 & 30.02 & 82.42 & 8.93 \\  
			\                     & 0.15 & 114.46 & 21.34 & 0.33 & -3.72 & 3.43 & 29.90 & 85.88 & 9.46 \\  
			\                     & 0.20 & 119.54 & 22.06 & 0.33 & -3.76 & 3.51 & 30.16 & 86.57 & 9.59 \\
			\bottomrule 
		\end{tabular} 
	}
	\caption{ \small Performance measure values for the period 01/03/2017 through 12/30/2020
		for select efficient frontier portfolios optimized under MCVaR${}_{0.99}$ and MV using RobecoSAM data.
		Note the difference between MCVaR${}_{0.99}$ and MV in the range of $\alpha$ considered.}   
	\label{tab:PM_sim} 
\end{table}

\begin{table}[!h] 
	\centering
	\scalebox{0.75}{
		\begin{tabular}{cc ccccc c ccccc}
			\toprule	
			\multicolumn{2}{c}{Model} & Mean    & Median    & Std      & Skew & ExKurt &
							 Model  & Mean    & Median    & Std      & Skew & ExKurt \\
					\  &\  & $\times 1^{-4}$ & $\times 1^{-4}$ & $\times 1^{-2}$ &\  &  &
						\ & $\times 1^{-4}$ & $\times 1^{-4}$ & $\times 1^{-2}$ &\  & \\
			\midrule
			\multicolumn{2}{c}{EWBH}        & 6.4 & 11.1 & 1.3 & -1.02 & 21.9  \\
			\rule{0pt}{3ex}
			$\lambda$ & $\alpha$ & \multicolumn{5}{c}{mCVaR${}_{0.99}$} &
					     $\alpha$ & \multicolumn{5}{c}{MV} \\
			\cline{1-2}\cline{8-8}
			\rule{0pt}{3ex}
				0.00  & 0.0 & 5.3 & 8.3 & 1.1 & -0.92 & 22.1  & 0.00 & 5.7 &   9.3 & 1.1 & -0.93 & 20.4 \\  
			\             & 0.3 & 5.4 & 8.0 & 1.1 & -0.92 & 22.2  & 0.05 & 6.1 & 10.7 & 1.2 & -1.08 & 22.9 \\  
			\             & 0.5 & 5.4 & 8.5 & 1.1 & -0.92 & 22.3  & 0.10 & 6.4 & 11.1 & 1.3 & -1.17 & 23.1 \\    
			\             & 0.7 & 5.5 & 9.6 & 1.1 & -0.90 & 22.4  & 0.15 & 6.4 & 11.2 & 1.4 & -1.32 & 23.9 \\  
			\             & 0.9 & 6.0 & 9.9 & 1.2 & -0.97 & 22.7  & 0.20 & 6.3 & 11.5 & 1.5 & -1.40 & 24.1 \\
			\rule{0pt}{3ex}
			     0.25   & 0.0 & 5.3 &  8.6 & 1.1 & -0.93 & 22.9  & 0.00 & 6.0 &  9.9 & 1.1 & -1.01 & 21.0 \\ 
			\            & 0.3 & 5.4 &  8.9 & 1.1 & -0.91 & 22.9  & 0.05 & 6.1 & 10.1 & 1.2 & -0.98 & 22.8  \\  
			\            & 0.5 & 5.5 &  8.8 & 1.1 & -0.91 & 23.5  & 0.10 & 6.4 & 11.0 & 1.2 & -0.95 & 23.2 \\ 
			\            & 0.7 & 6.2 &  9.8 & 1.2 & -0.89 & 24.1  & 0.15 & 6.7 & 10.6 & 1.3 & -0.90 & 23.0 \\   
			\            & 0.9 & 7.3 & 10.6 & 1.3 & -0.96 & 23.5  & 0.20 & 6.9 & 10.2 & 1.3 & -0.88 & 22.4 \\
			\rule{0pt}{3ex}
			    0.50   & 0.0 & 5.6  &  9.0 & 1.1 & -0.86 & 23.5  & 0.00  & 6.5 & 11.6 & 1.2 & -1.09 & 22.1 \\  
			\           & 0.3 & 5.8  &  9.8 & 1.1 & -0.81 & 23.3  & 0.05 & 6.8 & 11.6 & 1.3 & -0.86 & 22.3 \\   
			\           & 0.5 & 6.1  &  9.9 & 1.2 & -0.81 & 23.5  & 0.10 & 7.2 & 10.2 & 1.4 & -0.82 & 21.2 \\  
			\           & 0.7 & 6.9  & 10.4 & 1.4 & -0.78 & 22.0  & 0.15 & 7.4 &   9.7 & 1.4 & -0.83 & 20.5 \\ 	
			\           & 0.9 & 7.9  & 10.2 & 1.5 & -0.61 & 18.7  & 0.20 & 7.6 &   9.8 & 1.5 & -0.75 & 19.7 \\
			\rule{0pt}{3ex}
			   0.75    & 0.0 & 6.3  &  9.8 & 1.2 & -0.73 & 23.0  & 0.00 & 6.9 & 12.2 & 1.3 & -1.11 & 23.3 \\   
			\           & 0.3 & 6.5 & 10.4 & 1.3 & -0.66 & 22.0  & 0.05 & 7.3 & 10.8 & 1.4 & -0.82 & 20.7 \\ 
			\           & 0.5 & 6.7 & 10.8 & 1.4 & -0.58 & 19.8  & 0.10 & 7.4 &   9.7 & 1.4 & -0.76 & 19.1 \\ 	
			\           & 0.7 & 7.4 & 10.2 & 1.5 & -0.54 & 17.8  & 0.15 & 7.6 &   9.6 & 1.5 & -0.72 & 19.2 \\  		
			\           & 0.9 & 7.6 & 11.2 & 1.6 & -0.70 & 19.0  & 0.20 & 7.8 & 10.4 & 1.5 & -0.72 & 19.4 \\  
		\bottomrule 
		\end{tabular} 	
}
	\caption{ \small Moment values for the return distributions obtained for the period 01/03/2017 through 12/30/2020
		for select efficient frontier portfolios optimized under MCVaR${}_{0.99}$ and MV.}  
	\label{tab:Mom_sim} 
\end{table}

\begin{table}[!h] 
	\centering
	\scalebox{0.75}{
		\begin{tabular}{ccccccc cccccc}
		\toprule	
		\multicolumn{2}{c}{Model} & SR & Sortino & STAR & Rachev & Gini  &
		                          Model & SR & Sortino & STAR & Rachev & Gini  \\
		\ 	& \ 	 & (\%) & (\%) & (\%) & (\%) & (\%) & \  & (\%) & (\%) & (\%) & (\%) & (\%) \\
		\midrule
		\multicolumn{2}{c}{EWBH}        & 4.86      & 6.59            & 1.88            & 83.86           & 12.26 &
									&		&			&				&			&		\\
		\rule{0pt}{3ex}
		
			$\lambda$ & $\alpha$ & \multicolumn{5}{c}{mCVaR${}_{0.99}$} &
					     $\alpha$ & \multicolumn{5}{c}{MV} \\
			\cline{1-2}\cline{8-8}
			\rule{0pt}{3ex} 
			 0.00 & 0.0 & 4.87 & 6.69 & 1.94 & 87.79 & 12.40   & 0.00 & 5.09 & 6.96 & 2.00 & 86.56 & 12.97 \\   
			\      & 0.3 & 4.91 & 6.75 & 1.96 & 87.84 & 12.55  & 0.05 & 5.09 & 6.92 & 1.99 & 85.10 & 13.03 \\    
			\      & 0.5 & 4.92 & 6.76 & 1.96 & 87.76 & 12.55  & 0.10 & 4.92 & 6.66 & 1.93 & 83.94 & 12.35 \\ 
			\      & 0.7 & 4.98 & 6.85 & 2.00 & 88.42 & 12.66  & 0.15 & 4.62 & 6.21 & 1.81 & 83.29 & 11.47 \\  
			\      & 0.9 & 5.16 & 7.08 & 2.07 & 88.03 & 13.11  & 0.20 & 4.34 & 5.81 & 1.69 & 82.50 & 10.68 \\
			\rule{0pt}{3ex} 
			
		      0.25 & 0.0 & 4.80 & 6.59 & 1.92 & 87.54 & 12.29 & 0.00 & 5.25 & 7.15 & 2.05 & 85.57 & 13.45  \\   
			\     & 0.3 & 4.90 & 6.73 & 1.96 & 87.57 & 12.55 & 0.05 & 5.07 & 6.92 & 1.99 & 86.10 & 12.99 \\   
			\     & 0.5 & 4.93 & 6.78 & 1.98 & 87.55 & 12.68 & 0.10 & 5.10 & 6.99 & 2.00 & 86.19 & 13.05 \\  
			\     & 0.7 & 5.23 & 7.21 & 2.10 & 88.22 & 13.45 & 0.15 & 5.19 & 7.13 & 2.04 & 86.76 & 13.24 \\ 
			\     & 0.9 & 5.44 & 7.51 & 2.19 & 89.64 & 13.78 & 0.20 & 5.24 & 7.21 & 2.07 & 87.13 & 13.28 \\
			\rule{0pt}{3ex}
			
			0.50 & 0.0 & 4.97 & 6.84 & 1.99 & 87.52 & 12.84 & 0.0  & 5.38 & 7.30 & 2.08 & 83.89 & 13.84 \\ 
			\     & 0.3 & 5.04 & 6.94 & 2.01 & 87.29 & 13.01 & 0.05 & 5.19 & 7.15 & 2.05 & 86.78 & 13.14 \\    
			\     & 0.5 & 5.18 & 7.14 & 2.07 & 87.55 & 13.37 & 0.10 & 5.22 & 7.21 & 2.06 & 87.61 & 13.09 \\   
			\     & 0.7 & 5.09 & 7.07 & 2.05 & 91.29 & 12.70 & 0.15 & 5.21 & 7.21 & 2.07 & 88.85 & 12.95 \\   
			\     & 0.9 & 5.24 & 7.38 & 2.15 & 95.26 & 12.77 & 0.20 & 5.23 & 7.27 & 2.09 & 91.04 & 12.92 \\
			\rule{0pt}{3ex}
			
			0.75 & 0.0 & 5.22 & 7.22 & 2.09 & 88.26 & 13.43 & 0.0  & 5.23 & 7.09 & 2.03 & 83.98 & 13.34 \\  
			\     & 0.3 & 5.06 & 7.04 & 2.03 & 89.46 & 12.83 & 0.05 & 5.19 & 7.17 & 2.05 & 88.33 & 12.92 \\    
			\     & 0.5 & 4.95 & 6.92 & 1.99 & 92.28 & 12.24 & 0.10 & 5.17 & 7.16 & 2.04 & 89.25 & 12.73 \\  
			\     & 0.7 & 4.94 & 6.96 & 2.03 & 96.10 & 11.91 & 0.15 & 5.09 & 7.09 & 2.04 & 92.26 & 12.46 \\  
			\     & 0.9 & 4.86 & 6.81 & 1.99 & 95.67 & 11.68 & 0.20 & 5.16 & 7.20 & 2.08 & 93.35 & 12.65  \\
			\bottomrule	
		\end{tabular} 	
	}
	\caption{ \small RRR values obtained for the period 01/03/2017 through 12/30/2020
		for select efficient frontier portfolios optimized under MCVaR${}_{0.99}$ and MV.
		Note the difference between mCVaR${}_{0.99}$ and MV in the range of $\alpha$ provided.}
		  
	\label{tab:RRR_Rob} 
\end{table}


\begin{table}[!h] 
	\centering
	\scalebox{0.8}{	
		\begin{tabular}{c ccccc ccccc}
			\toprule	
			Model & Tot. Ret & Ann. Ret & AvgTO & ETL95 & ETR95 & MDD & \multicolumn{2}{c}{ESG*} & \omit & \omit \\
			\ 	  & (\%)	    & (\%)	& (\%)	 &(\%)	& (\%)	& (\%) & avg       & std	& \omit & \omit   \\
			\midrule
			EWBH  & 90.14  & 17.69     & 0.00     & -3.41  & 2.86    & 32.87 & 63.25 & 1.81 & \omit & \omit \\
			\rule{0pt}{3ex}
			$\lambda$ & \multicolumn{8}{c}{ mCVaR${}_{0.99}$ }  & \omit & \omit  \\
			\cline{1-1}
			\rule{0pt}{3ex} 
			0.00 & 95.01  & 18.45 & 0.40 & -3.48 & 2.92 & 34.06 & 61.50 & 1.85  & \omit & \omit \\
			0.25 & 128.28 & 23.27 & 0.35 & -3.78 & 3.54 & 29.49 & 85.82 & 8.48 & \omit & \omit \\
			0.50 & 132.40 & 23.83 & 0.35 & -3.82 & 3.62 & 29.66 & 87.68 & 8.67 & \omit & \omit \\
			0.75 & 135.19 & 24.21 & 0.34 & -3.85 & 3.69 & 29.84 & 88.97 & 8.85 & \omit & \omit \\
			\rule{0pt}{3ex}
			
			$\lambda$ & \multicolumn{8}{c}{MV  } & \omit & \omit \\
			\cline{1-1}
			\rule{0pt}{3ex} 
			0.00 & 93.80  & 18.26 & 0.40 & -3.76 & 3.13 & 37.23 & 63.10 & 2.92 & \omit & \omit \\
			0.25 & 139.50 & 24.78 & 0.33 & -3.90 & 3.64 & 29.08 & 85.98 & 8.94 & \omit & \omit \\
			0.50 & 148.41 & 25.94 & 0.32 & -3.93 & 3.72 & 29.38 & 87.85 & 8.73 & \omit & \omit \\
			0.75 & 147.24 & 25.79 & 0.33 & -3.90 & 3.73 & 29.43 & 88.70 & 8.77 & \omit & \omit \\
			\midrule
			\
			\\
			Model    & Mean  & Median  & Std  & Skew  & ExKurt &   Mean  & Median  & Std  & Skew  & ExKurt \\
			\ 	& $\times 10^{-4}$ &  $\times 10^{-4}$ & $\times 10^{-2}$ & \multicolumn{2}{c}{\ }
				& $\times 10^{-4}$ &  $\times 10^{-4}$ & $\times 10^{-2}$ & \multicolumn{2}{c}{\ }\\
			\midrule
			EWBH           & 6.4   & 11.1    & 1.3  & -1.02     & 21.9 & \multicolumn{5}{c}{\ } \\
			\rule{0pt}{3ex}
			$\lambda$ & \multicolumn{5}{c}{ mCVaR${}_{0.99}$  } & \multicolumn{5}{c}{ MV  }  \\
			\cline{1-1}
			\rule{0pt}{3ex} 
			0.00 & 6.7 & 11.4 & 1.4 & -1.28 & 23.9 & 6.6 & 11.6 & 1.471 & -1.36 & 23.7 \\
			0.25 & 8.2 & 12.0 & 1.5 & -0.76 & 19.6 & 8.7 & 12.7 & 1.586 & -0.71 & 18.5 \\
			0.50 & 8.4 & 11.5 & 1.6 & -0.75 & 19.6 & 9.0 & 12.4 & 1.607 & -0.71 & 18.5 \\
			0.75 & 8.5 & 10.9 & 1.6 & -0.72 & 19.1 & 9.0 & 11.8 & 1.607 & -0.68 & 18.2 \\
			\midrule
			\\
			Model       & SR   & Sortino & STAR & Rachev & Gini & SR   & Sortino & STAR & Rachev & Gini \\
			\ 		  & (\%) & (\%)  & (\%) & (\%)   & (\%)& (\%) & (\%)  & (\%)  & (\%)   & (\%)\\
			\midrule
			EWBH      & 4.86 & 6.59    & 1.88  & 83.86  & 12.26 &\multicolumn{5}{c}{\ }  \\
			\rule{0pt}{3ex}
			$\lambda$ & \multicolumn{5}{c}{ mCVaR${}_{0.99}$  } & \multicolumn{5}{c}{ MV  }  \\
			\cline{1-1}
			\rule{0pt}{3ex} 
			0.00 & 4.87 & 6.57 & 1.91 & 83.82 & 12.19  & 4.48 & 6.00 & 1.75 & 83.05 & 11.02 \\
			0.25 & 5.32 & 7.43 & 2.17 & 93.70 & 13.03  & 5.48 & 7.66 & 2.23 & 93.54 & 13.34 \\
			0.50 & 5.36 & 7.50 & 2.20 & 94.67 & 13.08  & 5.63 & 7.90 & 2.31 & 94.65 & 13.74 \\
			0.75 & 5.35 & 7.50 & 2.21 & 95.86 & 12.97  & 5.60 & 7.88 & 2.31 & 95.62 & 13.62 \\

			\bottomrule
		\end{tabular} 	
	}
	\caption{\small Performance measures, return moment values and RRR results obtained for the
		tangent portfolios under	mCVaR${}_{0.99}$ and MV optimization during the period 01/03/2017 to 12/30/2020.}   
	\label{tab:PM_sim_tang_robeco} 
\end{table}

\end{appendices}

\end{document}